\documentclass[12pt,twoside]{book}
\usepackage{amsmath}
\usepackage{setspace}
\usepackage{a4wide,amsthm, amsfonts, amssymb, latexsym, epsfig,natbib, fancyhdr, mathrsfs}
\usepackage{multicol}
\usepackage{graphicx}
\usepackage{textcomp}
\usepackage{float}
\usepackage{amsxtra}
\usepackage[nottoc]{tocbibind}

\usepackage{epigraph}

\usepackage{xcolor}
\usepackage{sectsty}

\usepackage{titlesec}
\titleformat{\chapter}{\normalfont\huge}{\thechapter.}{20pt}{\huge\it\bf\color{black!70}}

\usepackage{stackengine}
\usepackage{scalerel}
\usepackage{xcolor}

\usepackage{graphicx}
\usepackage{amscd}
\usepackage[titles]{tocloft}
\usepackage{calligra}
\usepackage{auncial}
\usepackage[B1]{fontenc}
\usepackage{sqrcaps}
\usepackage{LobsterTwo}
\usepackage[T1]{fontenc}
\usepackage{ifthen}
\usepackage{enumerate}

\numberwithin{equation}{chapter}
\RequirePackage{filecontents}        

\usepackage{mathrsfs}
\usepackage{fourier}
\begin{document}

\thispagestyle{empty}

\begin{center}
{ \LARGE {\bf  DYNAMICAL IMPRINT OF DARK MATTER \\
HALO AND INTERSTELLAR GAS ON \vspace{0.5cm}\\
 SPIRAL STRUCTURE IN DISK GALAXIES}}\\
\end{center}
\vspace{0.8in}
\begin{center}
{\large  A THESIS \\
 SUBMITTED FOR THE DEGREE OF\\
\vspace{0.06in}
{\LobsterTwo{\LARGE Doctor of Philosophy}\\}
\vspace{0.2in}
IN THE FACULTY OF SCIENCE}\\
\vspace{0.75in}
{ \large by}\\
\vspace{0.5cm}
{\Large \bf SOUMAVO GHOSH}\\
\vspace{0.8in}
\includegraphics[scale=2.]{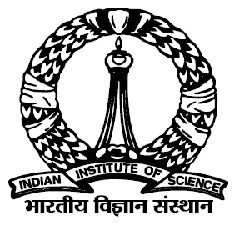} \\
\vspace{0.2in}
{\large{DEPARTMENT OF PHYSICS}}\\
\vspace{0.1cm}
{\large{Indian Institute of Science}} \\
\vspace{0.1cm}
{\large{BANGALORE \,-\, 560012}}\\
{JULY 2017}
\end{center}

\newpage

\thispagestyle{empty}
\cleardoublepage
\pagenumbering{roman}

\onehalfspacing

\chapter*{}
\begin{center}
\textcopyright {\bf Soumavo Ghosh\\
July 2017\\
All rights reserved\\}
\end{center}

\thispagestyle{empty}
\cleardoublepage

\addcontentsline{toc}{chapter}{Declaration}
\chapter*{Declaration}

\vspace*{.80in}
I hereby declare that the work reported in this doctoral thesis titled ``Dynamical Imprint of Dark Matter Halo and Interstellar Gas on Spiral Structure in Disk Galaxies'' is entirely original and is the result of investigation carried out by me in the Department of Physics, Indian Institute of Science, Bangalore -- 560012, under the supervision of Prof. Chanda J. Jog and Dr. Tarun Deep Saini.

I further declare that this work has not formed the basis for the award of any degree, diploma, fellowship, associateship or similar title of any University or Institution.

\vspace*{1.in}

\begin{flushright}
Soumavo Ghosh\\
\end{flushright}
\vspace {1cm}
July 2017\\
Department of Physics\\
Indian Institute of Science\\
Bangalore -- 560012, India\\

\newpage

\thispagestyle{empty}
\cleardoublepage

\addcontentsline{toc}{chapter}{Dedication}
\chapter*{}
\begin{center}
\vspace {-3.5cm}
{\Huge \it{To my mother...}}\\
\end{center}

\thispagestyle{empty}
\cleardoublepage

\addcontentsline{toc}{chapter}{Acknowledgement}
\chapter*{Acknowledgement}

With the thought of writing the {\it `Acknowledgement'} section of my thesis, another thought caught hold of my mind -- the very thought that {\it ``my 5-year tenure here in IISc is about get finished!''}. The day when I walked into the institute for the very first time and took the tree-shaded, dark-pitched, main stretch of the campus on a sunny afternoon to appear for the interview still feels like yesterday in my memory. Somewhere in between writing my first research paper and the draft of my thesis, I did not quite realize how time has quietly flown by!

\medskip

 In retrospect, it was quite a journey in these last 5 years here in IISc -- it enriched me in terms of experience, helped me to get matured and taught me to accept the real world the way it is. Therefore, I humbly take this opportunity to extend my heartiest gratitude to all those who have played significant role to make this 5-year long journey a memorable one.
   
\medskip

First of all, I take this very occasion to extend my heart-felt thankfulness to my PhD supervisors Prof. Chanda J. Jog and Dr. Tarun Deep Saini who introduced me to this beautiful and equally exciting world of astronomy and astrophysics. Their utmost patience and methodical guidance helped me to pass through the roller-coaster ride of the PhD life and also helped me to reach at a stage where I can think of writing a coherent story out of my 5-year long calculations. The hour-long scientific discussions with them have inculcated the {\it research instinct} in me and also enabled me to think about more problems that can be posed, investigated, and addressed. I remain highly indebted and grateful to both of them for their constant support and encouragement, without which this thesis could not have seen the light of the day.

\medskip

I would like to thank Dr. Paola Di Matteo and Dr. Misha Haywood (GEPI, Observetoire de Paris) for showing interest in my work and also for extending their generous support and cooperation towards me in submitting a research proposal for CNES ({\it Centre National d'Etudes Spatiales}) post-doctoral fellowship.

\medskip

I would like to thank the thesis examiners -- Prof. Paola Di Matteo (GEPI, Observetoire de Paris) and Prof. Mousumi Das (IIA, Bangalore) for their thorough and constructive reports. The travel supports from International Astronomical Union (IAU), GARP funding from Indian Institute of Science, and the DST-J.C. Bose Fellowship of Prof. Chanda J. Jog is acknowledged which facilitated me for attending conferences in abroad on several occasions.

\medskip
 
I would also like to thank the chairman of my department and the other faculty members of our Astronomy \& Astrophysics group for providing a wonderful, friendly and congenial working atmosphere.

\medskip

Thanks to all the office staffs of the Physics department whose sincere efforts helped the necessary documents to be at right places on several occasions.

 My special thanks to all the staffs in our Academic session ! They, over their {\it seemingly endless coffee breaks}, most generously offered one simple and yet precious lesson about research life - ``{\it Patience is the key !}''. Their {\it very loving and caring  nature made me visit there at least thrice (on an average) and wait for even longer time in order to get a 5-minutes work done!}

\medskip

My sincere thanks to all seniors -- Dr. Bidya Binay Karak, Dr. Indrani Banerjee, Dr. Sujit Kumar Nath, Dr. Upasana Das, Dr. Arpita Roy and others in astronomy \& Astrophysics division. Also thanks to our Astro family here -- Gopal, Prasun, Deovrat, Prakriti, Siddhartha, Dr. Nagendra, Dr. Naveen (Yadav), Dr. Kiran (Lakhchaura), Tushar and many others. Several healthy scientific discussions with you all have enriched me a lot.

\medskip

I would like to thank Dr. Arpita Roy and Dr. Md. Ramiz Reza for kindly providing the basic templates of the thesis on which I made some changes to make it look like what it is now. Also my sincere thanks to Dr. Arunima Banerjee whose thesis gave me an overall idea of how to draft a thesis.

\medskip

It was a nice memory to be a part of the JAP (Joint Astronomy Program) batch of 2012 (largest batch in JAP history till date!) -- Gopal, Abir, Kartick, Deovrat, Sreehari, Mohan Krishna -- you guys are being fantastic and I have learned a lot from you. The entire saga of late-night chat over coffee and solving the assignments together over the whole night was quite a sweet memory. 

\medskip

My sincere thanks goes to a {\it `team of 12'} -- Sandip Da(Mandal), Gopal (Hazra), Prasun (Dhang), Abir (Sarkar), Kartick (Sarkar), Monojit (Bhattacharya), Debabrata (Adhikari), Pradip (Bera), Somnath (Hazra), Kaushik (Goswami), Chandan (Samanta), Tathagata (Biswas) -- the people who literally sacrificed their night's sleep and comfort for several days to be present along my side in the hospital when I underwent a serious spine surgery back in 2014 and also provided the much needed mental support at that time. I remain indebted for your kind gesture which meant a lot to me.

Also my whole friend-circle (I suppressed the temptation to mention each individual name, it would probably take a full long page!) here in IISc who made me feel at home, thanks to all of you. Special mention to Tathagata (Biswas) who taught me how to cook -- a must-know skill for survival !

{\it ...And Then There Were Three...}

Gopal, Me and Prasun -- a {\it `triad'} sitting right at the core of my friend-circle here at IISc. The numerous discussion sessions with you guys on topics covering a wide variety have enriched me a lot. I got the opportunity to learn many things -- technical know-how of numerical methods, nitty-gritty about Linux, and the list goes on. I only hope that the benefits were mutual! The numerous instances of going out for trekking in the trails of the {\it Western Ghats}, restaurant hunting, many vacation trips, and the endless leg-pulling -- all these have secured a safe and sweet place in my memory.

\medskip

My sincere thanks goes to my childhood tutor Mr. Kalyan Kumar who motivated me to pursue science, my mathematics teacher Mr. Asutosh Sarkar who introduced me to the fascinating world of mathematics and inscribed a deep philosophy about life in my mind. Also, I thank my M.Sc mentor Prof. Tanuka Chattopadhyay (University of Calcutta) who encouraged me a lot to pursue a research career. I am grateful to all of them for their proper guidance and for ushering me to the point where I am right now.

\medskip

I would also like to thank another two very special {\it friend, philosopher and guide} in my life -- Mr. Somnath Guha and Mr. Partha Sarathi Nayak. Talking to both of you is always an absolute bliss for me. Your whole-hearted support in the toughest times of my life will always be remembered. Also my childhood friends Toni (Sayan), Babin (Sabyasachi) and Jishu Da for readily responding and taking care of many of my household issues on various occasions when I was away from home -- it means a lot.
\medskip

And last but not the least, my parents who bestowed unconditional love and support upon me over the years -- somebody who never questioned my capabilities and accepted me the way I am. They stood beside me rock-solid in many tough phases of my life when it was much required. I will remain highly indebted to them forever. Without their sacrifices, I could have not come this far. Regards...

\thispagestyle{empty}
\cleardoublepage

\addcontentsline{toc}{chapter}{Preface}

\chapter*{Preface}
\vspace {2.5cm}

The topic of this thesis deals with the spiral structure in disk galaxies with a specific aim of probing the influence of the dark matter halo and the interstellar gas on the origin and longevity of the spiral arms in late-type galaxies through theoretical modeling and numerical calculations. The basic theoretical model of the galactic disk used involves gravitationally-coupled two-component system (stars and gas) embedded in a rigid and non-responsive dark matter halo, i.e., the static potential of the dark matter is used in the calculations. However, at places, depending on the nature of the problem addressed, the disk is treated as consisting of only stellar component or only gas component followed by proper justifications for the assumptions. The disk is rotationally-supported in the plane and pressure-supported perpendicular to the plane of the disk. The first part of the thesis involves searching for the dynamical effect of dark matter halo on small-scale spiral structure in dwarf low surface brightness (LSB) galaxies and also some dwarf irregular galaxies which host an extended $HI$ disk. In both cases, the rotation curves are found to be dominated by the contribution of the dark matter halo over a large radial distance, starting from the inner regions of the galaxies. The next part of the thesis deals with the investigation of the possible effect of the interstellar gas on the persistence issue and the pattern speeds of the spiral structure in the disk galaxies. The last part of the thesis involves in studying the dynamical effect of dark matter halo on large-scale spiral structure. Following is the layout of the thesis.

Chapter 1 gives a general introduction to the topic of spiral structure of late-type disk galaxies, followed by a broad overview of the theoretical development of the topic and the present status of the topic. Then the thesis starts with studying the {\it small-scale} spiral features and evolves to studying the {\it large-scale} spiral features seen in disk galaxies in the following way:  Chapters 2 \& 3 deal with the effect of dark matter halo on small-scale spiral structure. Chapters 4 \& 5 focus on the dynamical effect of the interstellar gas on the spiral structure using the {\it local} dispersion relation. Chapters 6 \& 7 discuss the possible effect of dark matter halo on large-scale spiral structure in disk galaxies. Chapter 8 contains the summary of results and future plans.

\section*{Effect of dark matter halo on small-scale spiral structure}

The spiral arms in the disks of galaxies are often broken into several smaller parts or patches that create a messy visual impression when viewed from a `face-on' configuration. They are generally termed as `small-scale' or flocculent spiral arms. Several studies showed that the small-scale spiral arms are basically material arm, i.e., they can be thought of as {\it `tubes'} filled with stars and gas. Spiral arms are known to participate in the secular evolution of the disk galaxies. Since disk galaxies are believed to reside within a halo of dark matter, therefore a detailed understanding of possible effects of dark matter halo on the spiral arms is necessary. 

In {\bf Chapter 2}, we investigate the effect of dark matter halo on small-scale spiral features in the disks of LSB galaxies. Modeling the mass distribution within a galaxy from the rotation curve of a typical small LSB galaxy reveals the generic fact that for most of the radii, dark matter halo dominates over the stellar disk. This trend is found to be true from the very inner regions of an LSB disk which in turn makes the LSBs a suitable laboratory for probing the effect of dark matter halo on the dynamics of disk galaxies. Following a semi-analytic approach, and using the observationally measured input parameters for a typical superthin LSB galaxy, UGC~7321, we showed that the dominant dark matter halo suppresses the small-scale spiral structure in the disk of UGC~7321. Since UGC~7321 possesses features typical of a LSB galaxy, we argued that this finding will also hold true for other typical LSBs. The result is at par with the observational evidences for the lack of prominent, strong small-scale spiral structure in LSB galaxies.

In {\bf Chapter 3}, we employed the similar techniques for probing the effect of dark matter halo on small-scale spiral structure, except this time we took five dwarf irregular galaxies with an extended $HI$ disk as the {\it sample} for our investigation. The main important difference between these dwarf irregular galaxies with the earlier LSB galaxies is that for these dwarf irregular galaxies with extended $HI$ disk, the largest baryonic contribution comes from the interstellar gas (mainly $HI$), and not from the stars (as seen in LSBs). The extended $HI$ disks of these galaxies allow one measure the rotation curve, and hence modeling the dark matter halo parameters for a large radial range from the galactic center. Here also the rotation curves are found to be dominated by dark matter halo over most of the disk, thus providing yet another `laboratory' for testing the dynamical effect of dark matter halo on the dynamics of the disks. Using the observed input parameters for five such dwarf irregular galaxies, we showed that the dense and compact dark matter halo is responsible for preventing strong small-scale spiral structure in these galaxies, which is in fair agreement with the observations.

\section*{Dynamical effect of interstellar gas on longevity of spiral arms}

Any late-type disk galaxy contains a finite amount of interstellar gas along with the stellar component. The atomic hydrogen ($HI$) constitutes the bulk of the interstellar gas along with the molecular hydrogen ($H_2$), ionized hydrogen ($HII$), and a trace amount of heavy elements like helium. The mass fraction present in the interstellar gas in disk galaxies is found to vary with the Hubble sequence, with the amount of interstellar gas increasing from Sa type to Scd type of galaxies. Due to the lower value of velocity dispersion as compared to that of stars, gas is known to have a larger destabilizing effect in the disk. Therefore, the natural question arises about what possible role the interstellar gas could play in the origin and the persistence issue of spiral arms.

In {\bf Chapter 4}, we explored how the interstellar gas could influence the longevity of the spiral arms in late-type disk galaxies by treating the spiral structure as {\it density waves} in the disk. The disk is modeled as a gravitationally coupled stars plus gas (two-component) system, where the stars are modeled as a {\it collisionless} system and the gas treated as a {\it fluid} system. Using the appropriate local dispersion relation for the above mentioned model for the disk of galaxy, we calculated the group velocity of a wavepacket of density wave and then studied the variation of the group velocity with increasing amount of interstellar gas in the system. We showed that the group velocity of a wavepacket in a Milky Way-like disk galaxy decreases steadily with the inclusion of gas, implying that  the spiral pattern will survive for a longer time-scale in a more gas-rich galaxy by a factor of few.

In {\bf Chapter 5}, we investigated the role of interstellar gas in obtaining a stable density wave corresponding to the observed pattern speed for the spiral arms. The underlying local dispersion relation remains same as that is in Chapter~4. Using the observationally measured pattern speed and the rotation curves for three late-type disk galaxies we showed that the presence of interstellar gas in necessary in order to maintain a stable density wave corresponding to the observed values for pattern speeds. Also we proposed a method to determine a range of pattern speed values at any particular radius, corresponding to which the density wave can be stable. We applied this method to the same three late-type galaxies which we used in the earlier part of this chapter. We found that, for these three galaxies, the observed pattern speed values indeed fall in the predicted range.

\section*{Imprint of dark matter halo on large-scale spiral structure}

Along with the small-scale spiral arms, there also exists another type of spiral arms -- the large-scale spiral structure, like what we see M~51 or in NGC~2997, which occupy almost the entire outer optical disk in the galaxy. These spiral arms are termed as `grand-design' spiral structure. One of the competing theories, namely, {\it Density wave theory} proposes that the large-scale structure is basically a density wave in the disk and the pattern exhibits a rigid-body rotation with a definite constant pattern speed. In the earlier part this thesis (Chapters 2 \& 3), it was shown that the small-scale spiral structure gets damped by the dominant dark matter halo. Therefore, a natural question arises whether dominant dark matter plays any role on these large-scale spiral structure; and if yes, to what extent it affects the large-scale spiral structure.

In {\bf Chapters 6 \& 7}, we investigated how the large-scale structure in disk galaxies gets affected when the disk galaxy hosts a dark matter halo that dominates over most of the disk regions. We again chose the LSB galaxies as laboratory for this study. In {\bf Chapter 6}, we modeled the stellar component as a {\it fluid} system and in {\bf Chapter 7}, we treated the stellar system as more realistic {\it collisionless} system. In both cases, global spiral modes are identified from the appropriate dispersion relations via a novel quantization rule, and they are used as a {\it `proxy'} for the large-scale spiral structure. Using the input parameters for UGC~7321, in Chapter~6 we showed that the fluid representation of stellar system failed to make an impression in suppression of the global spiral modes. However, when stellar component is treated as a more realistic collisionless system, we found that the dark matter halo suppresses the large-scale spiral features as well  in the disks of LSB galaxies, in fair agreement with the observations.

Finally, in {\bf Chapter~8}, the thesis concludes with a summary of main results and a brief discussion of the scope for future work.

\thispagestyle{empty}
\cleardoublepage

\addcontentsline{toc}{chapter}{List of Publications}
\chapter*{List of Publications}

\vspace {1.5cm}

\begin{itemize}
\item{{\it ``Suppression of gravitational instabilities by dominant dark matter 
halo in low surface brightness galaxies''} {\bf{Ghosh S.}} \& Jog C. J., 2014, {\bf MNRAS, 439,} 929}
\vspace{0.3cm}
\item{{\it ``Role of gas in supporting grand spiral  structure''} {\bf{Ghosh S.}} \& Jog C. J., 2015, {\bf MNRAS, 451,} 5868}
\vspace{0.3cm}
\item{{\it ``Effect of dark matter halo on global spiral modes in galaxies''} {\bf{Ghosh S.}}, Saini T. D., \& Jog C. J., 2016, {\bf MNRAS, 456,} 943}
\vspace{0.3cm}
\item{{\it ``Dynamical effect of gas on spiral pattern speed in galaxies''} {\bf{Ghosh S.}} \& Jog C. J., 2016, {\bf MNRAS, 459,} 4057}
\vspace{0.3cm}
\item{{\it ``Effect of dark matter halo on global spiral modes in a collisionless galactic disk''} {\bf{Ghosh S.}}, Saini T. D., \& Jog C. J., 2017, {\bf New Astronomy, 54,} 72}
\vspace{0.3cm}
\item {{\it ``Dwarf Irregular galaxies with extended $HI$ disks: suppression of small-scale spiral arms by dark matter halo''} {\bf{Ghosh S.}} \& Jog C. J., 2018,  {\bf New Astronomy, 63}, 38}

\end{itemize}

\thispagestyle{empty}
\cleardoublepage

\tableofcontents \cleardoublepage

\pagestyle{fancy}
\renewcommand{\chaptermark}[1]{\markboth{\textsl{\thechapter.\ #1}}{}}
\renewcommand{\sectionmark}[1]{\markright{\textsl{\thesection.\ #1}}}
\fancyhf{}
\renewcommand{\headrulewidth}{.05pt}
\fancyhead[LE]{\thepage} \fancyhead[RO]{\thepage}
\fancyhead[RE]{\leftmark} \fancyhead[LO]{\rightmark}

\fancypagestyle{plain}

\pagenumbering{arabic} \setcounter{chapter}{0}

\thispagestyle{empty}

\chapter[Introduction]{Introduction}
\chaptermark{\it Introduction}
\vspace {2.5cm}

\section {Spiral Galaxies}
This thesis deals with disk galaxies with a particular emphasis on the study of spiral structure in disk galaxies. As the very name suggests, these galaxies contain a disk morphology where it exhibits various non-axisymmetric features like bars and spiral arms. In the Hubble tuning fork diagram, they lie towards the right end following the `ellipticals', and the gas-rich ones are termed often as `late-type' galaxies.

A typical disk galaxy consists of a supermassive Black Hole (BH) sitting right at the center of the galaxy and controlling some of the exotic events like launching relativistic jets, a central mass concentration, termed as `bulge', a disk morphology, and the whole system is believed to be embedded in a large and extended dark matter (DM) halo. Stars form out of the interstellar medium (ISM) mainly due to gravitational collapse of massive gas cloud, they evolve in that environment, and finally they end up being a `compact object' (namely white dwarfs, neutron stars, BH etc.) while returning most of their material to the ISM.

As far as the dynamics of the disk galaxy is concerned, the disk is supported by the rotation in the plane of the disk, and in the vertical direction the self-gravity is balanced by the pressure-support of the disk. In general, the stellar component dominates the baryonic mass, and the mass fraction due to the stars is an order of magnitude higher than that due to the gas. However, the total `dynamical mass' of a disk galaxy is primarily due to the dark matter envelope around the disk, which is typically required to the ten times more massive than the whole baryonic disk (stars and gas taken together) of the galaxy. Besides that, there also exists magnetic fields and the cosmic rays particles spiraling along the magnetic field lines, but their contribution in the secular evolution and dynamics of the galaxy is now been considered as only `secondary'. The main force that prevails over the dynamics of the disk and decides the `fate' of the disk galaxy is the gravitational force.

Study of spiral structure in galaxies has drawn attention for almost over five decades now, and a great deal of analytical, observational and $N$-body simulation studies have been devoted to unveil the mystery of the origin and the survival of the spiral structure in the disk galaxies. Although much of insight was gained from these studies, still there are subtle issues regarding the generation and effect of the spiral structure which are not well-understood till date, and that requires more study in those areas. Over the past decade, there is an increasing amount of surveys of galaxies that revealed many exciting aspects of disk galaxies. To give a brief account of that: SDSS (Sloan Digital Sky Survey) has provided the most detailed three-dimensional map of our universe, APOGEE (APO Galactic Evolution Experiments) can record very high-resolution, and high signal-to-noise (S/N) infrared spectroscopy to probe the structure of the bulge and the disk of our Milky Way, SEGUE (Sloan Extension for Galactic Understanding and Exploration) can measure the spectra of huge number of stars with varying spectral types to investigate the Milky Way structure, MaNGA (Mapping Nearby Galaxies at APO) records the spectral measurements across the face of $\sim$ 10,000 nearby galaxies (previous SDSS surveys could measure the spectra only at the center of the targeted galaxies) with a goal to trace the evolutionary track of the present-day galaxies.
 Along with that, there are some more surveys like GAIA (data release 2-- DR2) lined up in the coming years, with the promise of dissecting our Milky Way (and some other nearby spiral galaxies) with unprecedented detail and providing a better understanding of the dynamical phenomena of the spiral galaxies and the spiral structure. These rapidly growing observational studies call for a parallel theoretical development in modeling realistic disk galaxies for interpreting these observational trends in their full glory.

\subsection{Historical Note}

On a dark moon-less night, far away from the light pollution of the civilization, the majestic bright sweep that covers a part of the night sky mesmerized the human being from the very ancient time. 
Several literatures from that ancient time bear the testimony of it. To mention a few, in Greek mythology, it was described as the milk split when {\it Hera, the wife of Zeus,} was feeding {\it Heracles}. The term `galaxy' came from the Greek word for milk. The present-day name `Milky Way' was derived from the Latin word {\it `Via Lactea'}. According to the Hindu mythology, in `{\it Bhagavata Purana'} this was coined as {\it `Akasaganga'} which means the {\it `The Ganges River of the Sky'}.

Several meticulous studies carried out by various astronomers, starting from Galileo (1620) to Edwin P. Hubble (1925), brought out the realization that the {\it`bright band of light'} is basically a huge conglomeration of many faint stars which are the part of our Galaxy, with the solar system situated in it. For a detailed account of this long journey of discovery, see \citet{BT87,BM98}. Hubble measured the distance of the Andromeda galaxy (M~31) by using the Cepheid variable stars and he measured the distance to be around 300 kpc, a distance too high to be a part of Milky Way. This established the fact that the Milky Way is one of many galaxies in the local universe. Hubble also classified galaxies in the local universe into `ellipticals' and `spirals', and further subdivided into classes, and it is known as the Hubble Tuning Fork diagram \citep[for details see e.g.][]{BM98}. Although this classification helped to gain a better understanding of the spiral galaxies, back then and even now, still a {\it theoretical foundation} to study the origin and nature of these spiral structure was largely missing.

The theoretical development in studies of origin of the spiral structure mainly started with works of Bertin Linblad, who first addressed the spiral arms in terms of Maclaurin ellipsoids (flattened spheroids rotating in an equilibrium state) \citep{Lin27}. Later, \citet{Lin35} derived a condition for gravitational instability, and thereby the spiral arms, in a series of rotating spheroids. He addressed the spiral arms in terms of individual stellar orbits rather than a collective process. However, identifying the spiral arms as {\it wave} was largely missing until in 1960s \citep[for detailed discussion see e.g.][]{DB14}. 

Around 1960s, the theory of spiral structure received a major boost. In 1964, C. C. Lin and F. Shu proposed their theory about spiral arms in which they treated the spiral arms as {\it stationary density waves}. According to this theory, the spiral arms are treated as the over-dense regions in the disk that exhibits a rigid body rotation in the stellar disk and the pattern rotates with a single constant pattern speed, and the pattern remains largely unchanged over a long time \citep{LS64,LS66}. For the first time, the Lin--Shu hypothesis provided the much needed theoretical foundation to study the spiral structure in the disk galaxies in great detail. It also allowed one to furnish a number of predictions about the spiral arms that can be tested observationally.

Around the same time, P. Goldreich and D. Lynden-Bell carefully examined the response of a differentially rotating fluid disk when any small-scale disturbance is introduced. Surprisingly, it turned out that in spite of the disk being stable to local axisymmetric perturbations, the disk responds remarkably to these small-scale non-axisymmetric perturbations -- the initial disturbances get amplified, grew for a limited time, and then ultimated get sheared by the differential rotation of the disk \citep{GLB65}. At the same time, W. Julian and A. Toomre studied the response of a differentially rotating stellar disk to a point mass orbiting on a circular path in the disk. They showed that the gravitational field produced a strong spiral-shaped wave in the stellar disk \citep{JT66}.

 Thus, these studies taken together, marked the beginning of formal theoretical studies of spiral structure in disk galaxies. Since then, it has become one of the prime topic of interest in galactic dynamics.


\subsection {Variety of spiral structure}

The very presence of spiral structure made the late-type disk galaxies distinguishable from elliptical galaxies. However, there exists a huge scatter in the morphology, nature, and the pitch angle of the spiral arms within the the class of disk galaxies.

 As for the morphology of spiral arms, sometimes spiral arms appear to be smooth, regular, and extends up to almost the edge of the optical disk. The spiral arms seen in galaxies like M~51, NGC~2997 are typical example for these large-scale spiral arms, which is often termed as a {\it `grand-design'} spiral pattern. Also there are many late-type disk galaxies where the spiral structure appear to be patchy and broken into many smaller parts. These small-scale spiral features are often called {\it `flocculent'} spiral arms. Recent surveys on galaxy morphology \citep[e.g. see][]{Elm11} also revealed that the spiral arms are generally found to be of three main types, namely, grand-design spiral arms, flocculent spiral arms and multiple spiral arms.

 The pitch angle which serves as a measure of how tightly the spiral arms are wound in the disk, is found to increase along the Hubble sequence, with Sa-type of galaxies have the smallest pitch angle while Scd-type of galaxies have the highest pitch angle. However, \citet{Ken91} showed that trend holds only in the average sense, and there is a considerable amount of scatter in the values of pitch angle within any specific sub-type in the Hubble sequence.

Also sometimes the appearance of the spiral structure changes when the galaxy is viewed in a different wavelength. This happens because the optical light is mainly dominated by the newly-formed young stars and the $HII$ regions ({\bf blue arm}), whereas the near-infrared (NIR) view is mainly dominated by the older stellar population ({\bf red arm}) which constitutes almost 70 per cent of the total stellar population. Thus, images taken at different wavelengths essentially trace very different stellar populations, and thus a multi-wavelength study of the spiral structure of a disk galaxy will provide more insight about the spiral arms. The existence of such spiral structure in older stellar population was first discovered by \citet{Zwi55}, and later \citet{Sch76} followed it up and extended this work. This coexistence of red arms and blue arms essentially tell that the whole stellar population takes part in the spiral structure.

\subsection {Small-scale spiral structure -- Swing Amplification}

In a disk of spiral galaxy, stars have their individual epicyclic motion. Also the differential rotation of the disk introduces a shear in the system whose sense coincides with that of the epicyclic motion of stars. And there is self-gravity of the disk along with the pressure of the stars (treated as fluid). All these can lead to a finite growth of the non-axisymmetric perturbations in the disk under certain circumstances. \citet{Too81} coined this as {\it `Swing amplification'} as the growth of the non-axisymmetric perturbation occurs when the wave {\it swings} from the leading position to the trailing position. 

 Here, a brief account of the swing amplification in a fluid disk is presented, for a detailed description see \citet{GLB65}. 

The galactic disk is taken to be infinitesimally thin, and the stellar component is modeled as {\it isothermal fluid}, characterized by the surface density $\Sigma$ and the one-dimensional velocity dispersion or the sound speed $c$. The fluid assumption makes the mathematical formulation simpler. The non-axisymmetric perturbations are treated to be as {\it local} i.e. the wavelength of the perturbation is small as compared to the galactocentric radial distance, and the perturbation is taken to be planner which requires the minimum wavelength of the perturbations is greater than the vertical scaleheight of the fluid.

The galactocentric cylindrical coordinates ($R$, $\phi$, $z$) are used throughout in this section.

The Euler equation of motion in a uniformly rotating frame is 
\begin{equation}
\frac{\partial {\bf v}}{\partial t}+ ({\bf v \cdot \nabla v}){\bf v}= - \frac{c^2}{\Sigma} {\bf \nabla \Sigma}-{\bf \nabla \Phi} - 2{\bf \Omega} \times{\bf v}+{\bf \Omega^2 R}
\label{intro-Euler}
\end{equation}
\noindent where {\bf $ \Omega$ }= $\Omega {\bf z}$ is the angular velocity of the rotating frame. $-2\bf {\Omega \times v} $ is the coriolis force and $-\bf {\Omega^2 R}$ is the centrifugal force associated with the rotating frame. 

Consider a point $(R_0, \phi_0)$ which is corotating in the above frame of reference. Then a local patch around that point is chosen to carry out the local, non-axisymmetric perturbation analysis.

Choose a Cartesian coordinate system $(x, y, z)$ with center at $(R_0, \phi_0)$ and with the unit vectors {\bf i, j, k} such that the {\bf i} coincides with the initial outward radial direction. Then, the non-axisymmetric perturbation analysis is carried out in that local patch.

The coriolis force term in equation~(\ref{intro-Euler}) will be balanced by the unperturbed gravitational force \citep[for details see][]{GLB65}.

The perturbed Euler equations of motion in $(x, y, z)$ coordinates will be\\

\begin{equation}
\frac{\partial v_{x}}{\partial t} +2Ax \frac{\partial v_{x}}{\partial y}-2\Omega v_{y} = -\frac{\partial \Phi}{\partial x}-\frac{c^2}{\Sigma_0}\frac{\partial}{\partial x}(\delta \Sigma)
\label{intro-Euler_x}
\end{equation}

\begin{equation}
\frac{\partial v_{y}}{\partial t} +2Ax \frac{\partial v_{y}}{\partial y}+2Bv_{x} = -\frac{\partial \Phi}{\partial y}-\frac{c^2}{\Sigma_0}\frac{\partial}{\partial y}(\delta \Sigma)\,,
\label{intro-Euler_y}
\end{equation}
\noindent where $v_{x}$ and $v_{y}$ are the perturbed velocity components. $\Sigma_{0}$ and $\delta \Sigma$ denote the unperturbed and perturbed surface density, respectively, and $\Phi$ is the gravitational potential of the fluid disk. A, B are the Oort constants and their functional forms are given as \citep[for details see e.g.][]{BT87,BM98}
\begin{equation}
A(R) = \frac{1}{2}\left[\frac{v_{\rm c}(R)}{R}-\frac{d v_{\rm c}(R)}{dR}\right] \mbox{;} \hspace{0.2 cm}
B(R) =-\frac{1}{2}\left[\frac{v_{\rm c}(R)}{R}+\frac{d v_{\rm c}(R)}{dR}\right]\\\,,
\end{equation}
\noindent where $v_{\rm c}(R)$ is the circular velocity at radius $R$.

Similarly, the perturbed equation of continuity is
\begin{equation}
\frac{\partial}{\partial t}(\delta \Sigma)+2Ax\frac{\partial}{\partial y}(\delta \Sigma)+\Sigma_0\left(\frac{\partial v_x}{\partial x}+\frac{\partial v_y}{\partial y}\right)=0\,,
\label{intro_conti_pertur}
\end{equation}
\noindent and the perturbed Poisson equation is 
\begin{equation}
\left(\frac{\partial^2}{\partial x^2}+\frac{\partial^2}{\partial y^2}+\frac{\partial^2}{\partial z^2}\right)(\delta \Phi)=4 \pi G \delta \Sigma \delta(z)\,,
\label{intro-pertur_Poisson}
\end{equation}
\noindent where $\delta(z)$ is the Dirac delta function.

The differential rotation present in the disk of the galaxy induces a shear in the system, therefore, to take account of that, we introduce sheared coordinates system $(x', y', z', t')$ defined as the following
\begin{eqnarray}
x'=x,\: y'=y-2Axt,\: z'=z, \: t'=t\,.
\end{eqnarray}


A trial solution of the form exp[$i(k_x x' + k_yy')$] is introduced for the independent perturbed quantities, e.g. perturbed surface density $\delta \Sigma$.

We define $\tau$ as:\\
\begin{equation}
\tau \equiv 2At'-k_x/k_y \hspace{0.3 cm} \mbox{, for a wavenumber} \hspace{0.1 cm} k_y \ne 0
\end{equation}
In the sheared coordinates, $\tau$ is a measure of time, and it becomes zero when the modes becomes radial, i.e., where $x$ is along the initial radial direction. 


The local, linearized perturbation equations~(\ref{intro-Euler_x})--(\ref{intro-pertur_Poisson}), when written in the sheared coordinates will take the forms as given below \citep[for details see][]{GLB65} 

\begin{equation}
\frac{\partial v_x}{\partial \tau}-\frac{\Omega}{A}v_y=-i\frac{k_y}{2A}\tau\left[-\delta\Phi -\frac{c^2}{\Sigma_0}(\delta \Sigma)\right]
\label{intro-shear_vx}
\end{equation}

\begin{equation}
\frac{\partial v_y}{\partial \tau}+\frac{B}{A}v_x=-i\frac{k_y}{2A}\tau\left[-\delta\Phi -\frac{c^2}{\Sigma_0}(\delta \Sigma)\right]
\end{equation}

\begin{equation}
\frac{\partial}{\partial \tau}(\delta \Sigma)-i\frac{k_y}{2A}\tau \Sigma_{0}v_x+i \frac{k_y}{2A}\Sigma_{0}v_y=0
\end{equation}
\noindent and,

\begin{equation}
\left[-k^2_y(1+\tau^2)+\frac{\partial'^2}{\partial z^2}\right](\delta \Phi)=4 \pi G \delta\Sigma \delta(z')
\label{intro-shear_Poisson}
\end{equation}

Now we define $\theta$, the dimensionless measure of the density perturbation as\\
\begin{equation}
\theta = \delta \Sigma/ \Sigma_0
\label{intro-theta}
\end{equation}
\noindent where $\Sigma_0$ denotes the unperturbed surface density and  $\delta \Sigma$ denotes the variation in surface density.

After some algebraic simplifications along with the usage of equation~({\ref{intro-theta}}), equations~({\ref{intro-shear_vx}})--(\ref{intro-shear_Poisson}) reduce to

\begin{equation}
\begin{split}
\left(\frac{d^2\theta}{d\tau^2}\right)-\left(\frac{d\theta}{d\tau}\right)\left(\frac{2\tau}{1+\tau^2}\right)+\theta\Bigg[\frac{\kappa^2}{4A^2}+\frac{2B/A}{1+\tau^2}
+\frac{k_y^2c^2}{4A^2}(1+\tau^2)\\
-\Sigma\left(\frac{\pi G k_y}{2A^2}\right)(1+\tau^2)^{1/2}\Bigg]=0\,,
\end{split}
\label{intro-swing_final}
\end{equation}
\noindent where $\kappa$ is the local epicyclic frequency.
The four terms within the square bracket of equation (\ref{intro-swing_final}) are due to the epicyclic motion, the unperturbed shear flow, the gas pressure, and the self-gravity, respectively.

 The systematic behavior of $\theta$ is as follows:

 When $\tau$ is large, the pressure term dominates over other terms, and hence the solution will be oscillatory in nature. But when $\tau$ is small, the epicyclic motion term and the unperturbed shear flow dominate over the pressure term and they cancel each other completely for a flat rotation curve. This results in setting up a kinematic resonance. In addition, if the self-gravity term dominates over the pressure term then the duration of kinematic resonance increases and the mode undergoes a swing amplification while evolving from radial position ($\tau =0$) to trailing position ($\tau > 0$) \citep[for details see][]{GLB65,Too81}.


\subsection {Large-scale spiral structure -- Density Wave theory}

According to the {\it Quasi stationary spiral structure (QSSS)} hypothesis \citep{LS64,LS66}, at least the large-scale (global) spiral arms are the global density wave -- the regions of over-densities in the galactic disk, and the pattern will rotate rigidly in the disk with a single, constant pattern speed. The pattern will be maintained by the self-gravity of the disk component (stars, gas), and it will largely remain constant, and will evolve slowly over time, thus justifying the nomenclature of {\it quasi-stationary}. For a detailed description of density wave theory and its development see \citet{Roh77,Pas04a,Pas04b}.

The advantage of this wave-like picture over the material arm description of the spiral arms is that, it saves the spiral arms from the so-called {\it `winding problem'}, and the spiral arms can be long-lived, rather from getting wound up on a few dynamical time-scale which is much shorter than the age of the galaxy.

The dispersion relation for a fluid disk was derived by \citet{LS64} and the dispersion relation for a collisionless stellar disk was derived by \citet{LS66} and \citet{Kal65}, and they are commonly known as {\bf Lin--Shu} (LS) and {\bf Lin--Shu--Kalnajs} (LSK) dispersion relations, respectively.


\subsubsection{Dispersion relations for fluid and stellar disk in WKB limit}


The basic assumptions are as follows:

{\bf Linear perturbations :} The spiral arms are treated only as the small perturbations in the underlying axisymmetric disk (fluid or collisionless, whichever applicable), and consequently a linear perturbation analysis is carried out for the equations of motion, the continuity equation and the Poisson equation. It is inherently assumed that the perturbed quantities are very small as compared to their unperturbed counterparts.\\

{\bf Tight-winding approximation of the WKB limit :} The major problem of carrying out a global modal analysis in a galactic disk is that one has to deal with the self-gravity of the disk. Since gravitational force is long-range force in nature, and hence all the perturbations in the different parts of the galactic disk gets coupled, making it very complicated to deal with it \citep[for details see e.g. ][]{BT87}. The tight-winding approximation or the WKB  (Wentzel - Kramers - Brillouin) approximation turned out to be an effective tool to investigate the properties of the density waves in a differentially rotating disk. It makes the gravitational field {\it local}, so that the dispersion relation can be written in terms of the local quantities \citep{BT87}. For example, if the radial variation of any perturbed quantity is written in terms of its amplitude and phase as given below
\begin{equation}
A(R)=\Phi(R)e^{i f(R)}
\end{equation}
\noindent then the tight-winding approximation assumes that the phase ($f(R)$) varies rapidly as compared to the amplitude ($\Phi(R)$) \citep[for detail see][]{DB14}.

Taking these assumptions, the resulting dispersion relation for an infinitesimally thin fluid disk becomes

\begin{equation}
(\omega-m\Omega)^2=\kappa^2-2 \pi G \Sigma|k|+c_{\rm s}^2k^2\,,
\label{intro-disp_flu}
\end{equation}
\noindent where $\omega$ and $k$ are the frequency and radial wavenumber of the perturbations, respectively. $\kappa$ denotes the local epicyclic frequency, and $\Sigma$ and $c_{\rm s}$ denote the surface density and the sound speed of the fluid, respectively.

Division of both sides of this equation~(\ref{intro-disp_flu}) by $\kappa^2$ will lead to

\begin{equation}
s^2=1-|x|+Q_{\rm s}^2 x^2\,,
\label{intro-disp_flufinal}
\end{equation}
\noindent where $s$ (= $(\omega-m\Omega)/\kappa$) and $x$ (=$k/k_{\rm crit}$) are the dimensionless frequencies, and $k_{\rm crit}$ (=$\kappa^2/(2\pi G \Sigma)$) is the critical wavenumber for which it is hardest to stabilize the system \citep{BT87}. $Q_{\rm s}$  (= $\kappa c_{\rm s}/\pi G \Sigma_{\rm s}$) is the standard Toomre $Q$ parameter which denotes the stability of the disk against the local, axisymmetric perturbation. $Q_{\rm s} > 1$ denotes the stability of the disk, $Q_{\rm s} < 1$ denotes the instability, and $Q_{\rm s} = 1$ denotes the neutral stability of the disk, respectively \citep{Too64}. $m \ge 0$ is any integer, and it denotes the $m$-fold rotational symmetry ($m=2$ for two-armed spirals) of the perturbation.

Similarly, the dispersion relation for a collisionless stellar disk is given as \citep[for details see][]{BT87}
\begin{equation}
(\omega-m\Omega)^2=\kappa^2-2 \pi G \Sigma|k|{\mathcal F}\left(\frac{\omega-m\Omega}{\kappa}, \frac{\sigma_R^2 k^2}{\kappa^2}\right)
\label{intro-disp_stellar}
\end{equation}
\noindent where ${\mathcal F} \le 1$ is the reduction factor, and it takes account of the reduction in the self-gravity of the disk due to the velocity dispersion ($\sigma_R$) of the stellar component.

Expressing this equation~(\ref{intro-disp_stellar}) in terms of the dimensionless frequency ($s$) and dimensionless wavenumber ($x$) we get

\begin{equation}
s^2=1-|x|+Q_{\rm s}^2 x^2 {\mathcal F}(s, \chi)\,,
\label{intro-disp_colfinal}
\end{equation}
\noindent where, $\chi$= ${k^2\sigma^2_{\rm R}}/{\kappa^2}$ = $0.286 Q_{\rm s}^2 x^2$.

The form for ${\mathcal F}(s, \chi)$ for a razor-thin disk whose stellar equilibrium state is described by the Schwarzchild distribution function is given by \citep{BT87}: 
\begin{equation}
{\mathcal F}(s, \chi)=\frac{2}{\chi}\exp(-\chi)(1-s^2)\sum_{n=1}^\infty\frac{I_n(\chi)}{1- s^2/n^2}\,,
\end{equation}
\noindent where $I_n$ is the modified Bessel function of first kind.

A careful inspection of the forms of equations~(\ref{intro-disp_flufinal}) and (\ref{intro-disp_colfinal}) will lead to important insights about the density wave in a galactic disk.
\begin{itemize}

{\item When $s=0$, then the frequency of the perturbation (or the pattern speed $\Omega_{\rm p}$ = $\omega/m$) and the rotational frequency ($\Omega$) become the same, and the corresponding radius is termed as Corotation radius (CR). Also when $|s|=\pm 1$, the forcing frequency seen by a particle, (orbiting with a angular frequency $\Omega$) $m(\Omega_{\rm p}-\Omega)$, and the natural radial frequency $\kappa$ become equal and thus will be able to set up a steady wave, these radii are called Outer Linblad resonance (OLR) and Inner Linblad resonance (ILR),respectively \citep[for details see][]{BT87}.}


{\item For a stellar disk, $|s|$ can never be greater than 1, but for the fluid disk the value of $|s|$ can exceed 1 (see expressions for $|s|$ in equations~(\ref{intro-disp_flufinal}) and (\ref{intro-disp_colfinal})). The physical reason behind this is the following: in a fluid disk, the pressure term can provide further support so that the waves exist even when $|s| > 1$, but the collisionless stellar disk does not have any pressure support, and hence waves cannot exist for $|s| >1$.}

\end{itemize}

\subsubsection{Propagation of tightly-wound density wave packet}
The density wave, although taken as quasi-stationary in the density wave theory, will actually propagate radially with its group velocity towards the increasingly short-wavelength in the disk.

For an inhomogeneous medium, the group velocity at a fixed radius is defined as \citep{Wit60,Lit65}
\begin{equation}
c_{\rm g}(R)= \frac{\partial \omega(k,R)}{\partial k}\\.
\end{equation}

Using this definition, the resulting group velocity for the fluid disk and the stellar disk become \citep{BT87}
\begin{equation}
c_{\rm g}(R)=sgn(k) \frac{|k|c_{\rm s}^2-\pi G \Sigma}{\omega-m\Omega}\,,
\end{equation}
\noindent and,
\begin{equation}
c_{\rm g}(R)=-\frac{\kappa}{k}\frac{1+2\frac{\partial ln {\mathcal F}(s, \chi)}{\partial ln \chi}}{\frac{\partial}{\partial s}ln\frac{{\mathcal F} (s, \chi)}{1-s^2}}\,.
\end{equation}

\citet{Too69} studied the propagation of such wavepacket and calculated the group velocity of a density wavepacket using the typical values for the solar neighborhood. Taking the pattern speed ($\Omega_{\rm p}$) $\sim$ 12.5 km s$^{-1}$ kpc $^{-1}$, $Q_{\rm s} = 1$, and $\sigma_{R}$ $\sim$ 35 km s$^{-1}$, he found the group velocity ($c_{\rm g}$) $\sim$ -10 km s$^{-1}$ (the negative sign implies the inward motion of the wavepacket), and even with this value of the group velocity, the wavepacket can travel a distance of 10 kpc in about $10^9$ years \citep{Too69}. Thus, the wavepacket will be destroyed at the center of the galaxy, and the stellar density wave will have a short lifetime. This clearly posed a serious problem for the prediction of {\it ever-lasting} density wave picture.

However, this can be avoided if there exists some reflecting/refracting boundary in the central region, so that before the wave reaches the ILR, it can reflect/refract the wave back to its wave cycle, and thus a standing wave in the disk can be established. A plausible mechanism is that if the Toomre $Q$ parameter gets very high (known as {\it $Q$ barrier}), so that the wave can be reflected back before it gets absorbed at ILR \citep[for details see][]{DB14,SaEl16}. Also at corotation (CR), short trailing stellar density waves can be excited by {\it `Wave amplification by stimulated emission radiation'} \citep[(WASER)][]{Mark74,Mark76} or by swing amplification \citep{GLB65,JT66,Too81}. Thus, a standing wave can be set up in the galactic disk between a reflecting/refracting boundary in the inner region and the CR.

\subsection {Other mechanisms to trigger spiral features}

In the literature, some other mechanisms are also shown to trigger the spiral arms in the disks of late-type galaxies. For example, bars have been proposed to generate spiral arms in the disks of barred-spiral galaxies. The motivation came from the observational fact that in many barred-spiral galaxies, the two-arm spirals start just from the end position of the bar. In the past, several studies have been carried out to study the gaseous spiral arms driven by bars \citep[e.g. see][]{SH76,CG85,Ath92}. Also, $N$-body simulation study by \citet{STT10} showed that the bar can heat up the stellar disk and also can induce spiral structure in the stellar disk.

Tidal interaction is shown to be another mechanism which can induce spiral structure in the galactic disk. M~51 is one such plausible case where the companion (M~52) is thought of as the driver for the spiral arms seen in M~51. A comprehensive study by \citet{ToTo72} showed how the interaction between two such galaxies can produce spiral arms, galactic bridges and tails for a wide range of possible alignments of the two such galaxies. For a review on dynamics of the interacting galaxies, see e.g. \citet{BH92}

\subsection {Impact of spiral arms on disk evolution}

Spiral arms play an important role in the secular evolution of disk galaxies. Due to the non-axisymmetric perturbations such as spiral arms, stars can gain or lose their angular momentum \citep{LBK72}, and this leads the stars to {\it migrate} in the outward or inward directions from their birth location - this phenomenon is called radial migration. Thus the chemical pattern in disk of a galaxy gets blurred and consequently the process of studying the star formation history in the disk galaxy becomes difficult. Several $N$-body simulations have studied the effect of transient spiral arms in the radial migration process \citep[see e.g.][]{SelBi02,SchoBi09,Min11}.  The pattern speed of the spiral arms set the location of resonance points where the angular transport phenomenon is shown to happen, and thus spiral structure has a direct implication in shaping up the evolution of disk galaxies.

\section {Presence of Interstellar gas}

It is well known that any late-type disk galaxy contains a finite amount of interstellar gas along with the stellar component in the disk of the galaxy. The interstellar medium (ISM) is a multiphase system consisting of components which have different temperature, density and ionization states. Based on these characteristic properties, the ISM can be broadly divided into three components whose details are given below \citep[for details see e.g.][]{DyWil95,Dra11}.

\subsection*{ Neutral atomic hydrogen gas}

The neutral atomic hydrogen gas consists of two distinct phases, namely, Cold neutral medium (CNM) and Warm neutral medium (WNM) where the hydrogen number density and the temperature differ significantly. For the CNM, the cold diffuse $HI$ clouds are found with temperature $\sim$ 100 $K$, and density ($n_H$) $\sim$ 50 cm$^{-3}$. The {\it 21 cm} $HI$ emission line is used to trace this population. For the WNM, the intercloud gas is found with temperature $\sim$ 5000--6000 $K$ and density ($n_H$) $\sim$ 0.2-0.5 cm$^{-3}$. Here also, {\it 21 cm} $HI$ emission line and optical UV absorption lines are used to trace this population \citep{KulHei87}

\subsection*{ Ionized gas}

The ionized gas, also termed as Warm Ionized medium (WIM), consists of gas where the hydrogen has been photoionized by ultraviolet photons from hot stars. This phase has a low density ($n_H$ $\sim$ 0.1 cm$^{-3}$) and a high temperature (T $\sim$ 8000 $K$). This WIM can be traced by the optical and UV ionic absorption line and thermal radio continuum.

\subsection*{ Molecular hydrogen gas}

In disk galaxies like our Milky Way, most of the molecular hydrogen is found in giant molecular clouds with typical sizes $\sim$ 400 pc and mass a few times $10^{5}$, column density $\ge$ $10^{3}$ cm$^{-3}$, and temperature $10-20 K$. Unlike the other phases of ISM, they are gravitationally bound system. Since the bulk of molecular hydrogen is cold, and therefore it is hard to detect them. The first rotational level, accessible only through a quadrupolar transition, is more than 500 $K$ above the fundamental level. Thus, the presence of molecular hydrogen is inferred essentially from the other tracer molecules, like $CO$ which is the most abundant molecule after molecular hydrogen. Its dipole moment is small and hence $CO$ can be easily excited, the emission of $CO(1-0)$ at 2.6mm is ubiquitous in disk galaxies. These molecular clouds are the sites where the star formation takes place \citep[see e.g.][]{DyWil95,Dra11}.


\subsection{Extent \& mass of interstellar gas}

The extent of the gas disk is mainly determined by the extent of the $HI$ disk as the distribution of molecular hydrogen ($H_2$) is more centrally concentrated than $HI$. \citet{Swa99} showed that the ratio of the radial $HI$ extent to the optical size in dwarf irregular galaxies is $\sim 1.8$ similar to the ratio of $\sim 1.5$ for larger galaxies. In extreme cases, some late-type dwarf irregular galaxies like DDO~154, NGC~3741 host an $HI$ disk which is 4 and 8 times larger than their Holmberg radius, respectively \citep{CF88,BCK05,Gen07}. However, the extent of the $HI$ disk also depends on the limiting $HI$ column density, as the latter sets the limit up to which one can trace the $HI$ extent. For example, for the recent $HI$ survey for the nearby galaxies, THINGS ({\it The $HI$ Nearby Galaxy Survey}), the limiting value is typically 4 $\times$ 10$^{19}$ cm$^{-2}$ \citep{Wal08}. For a detailed discussion of the size of the $HI$ disk see section 12.2 in \citet{GioHay88}. The extended $HI$ disk is crucial in studying the distribution of dark matter in disk galaxies as an extended $HI$ disk allows one to measure the rotation curve, and hence the deduction of the dark matter halo parameters far out from the stellar disk. In many cases, the outer regions of the $HI$ disk display the signatures of lopsidedness and warps. 

 Also the ratio of the gas mass ($M_{\rm gas}$) to the total dynamical mass ($M_{\rm dyn}$, as inferred from the observed rotation curve) varies with the Hubble sequence. To explain, the ratio of the gas mass to the total dynamical mass increases monotonically as one moves from Sa-type disk galaxies to the Scd-type disk galaxies, i.e. the early-type disk galaxies are less gas-rich as compared to the late-type disk galaxies. The median value of M$_{\rm gas}$/M$_{\rm dyn}$ varies from 4 per cent for Sa-type galaxies to 25 per cent for Scd-type galaxies, although there is scatter within a specific Hubble type \citep[e.g. see in][]{YoSc91,BM98}. The ratio of total neutral gas (M$_{HI}$ and M$_{H_2}$) to the stellar mass (M$_{*}$) also varies with the Hubble type. For example, the gas fraction ($\epsilon$ = $M_{\rm gas}$/$M_{*}$) value for Sa-Sab type type is typically 5 per cent whereas for Scd-type of galaxies the value of $\epsilon$ becomes $\sim$ 30 per cent \citep[e.g. see ][]{YoSc91}.

\subsection {Radial surface density profiles for interstellar $HI$ and $H_2$ gas}

The radial surface density profiles for $HI$ and $H_2$ do not follow any specific universal functional form, unlike the stellar component, and they show a wide variation in their radial distribution in the disk galaxies. The distribution of molecular hydrogen ($H_2$) shows a more centrally concentrated behavior than that for $HI$, and only for a minority of galaxies it shows a depression in the central regions, like what we see in case of the Milky Way \citep{BM98}. On the other hand, the radial surface density for the atomic hydrogen ($HI$) increases monotonically from the inner radii, and in the outer parts of the disk galaxies it displays more or less a constant values. Sometimes, the distribution of $HI$ also shows a depression in the central regions, like what we see for the Sb-type galaxies NGC~2841 and NGC~7331 \citep[e.g. see][]{BM98}.

A recent study by \citet{BB12} using the $CO$ observations from HERACLES ({\it  HERA CO--Line Extragalactic Survey}) and $HI$ observations from THINGS showed for 33 nearby, late-type spiral galaxies that the total neutral gas (atomic hydrogen and the molecular hydrogen) surface density profile obeys a well-constrained universal exponential distribution outside $0.2\times R_{25}$, and $R_{25}$ denotes the radius where the $B$-band surface brightness drops to 25 mag arcsec$^{-2}$.

\subsection {The gas content of the Milky Way}

The observed surface density profiles for $HI$ and $H_2$ in the Milky Way are irregular in nature. The $H_2$ distribution shows (after a gap in the very central region) a ring-like structure with mean radius of $\sim$ 4.5 kpc and full width at half maximum $\sim$ 2 kpc where the maximum of $\Sigma_{\rm H_2}$ reaches $\sim$ 20 M$\odot$ pc$^{-2}$ around 5 kpc and then falls off exponentially \citep[see e.g. ][]{ScoSan87,BM98}. The $HI$ distribution is relatively regular; after 4 kpc it gradually increases and then saturates to a value 5-6 M$_\odot$ pc$^{-2}$ in the outer parts \citep{BG78,CSS88,DL90,YoSc91,BM98}, but it shows a depression in the very inner part where the surface density of $HI$ suddenly goes to almost zero at R $\sim$ 1.5 kpc \citep{BM98}. Recent study of $HI$ distribution revealed that the radial surface density
distribution of $HI$ follows an exponential fall-off for 12.5~$\le$~R~$\le$ 30 kpc. For R~$\le$~12.5 kpc the surface density saturates at approximately $\Sigma_{\rm inner}$ $\sim$ 10 M$_\odot$ pc$^{-2}$ \citep{KD08}.

\subsection {Past results of effect of interstellar gas on dynamics of disk galaxies}

Several past studies have addressed the effect of the interstellar gas, having a lower velocity dispersion as compared to the stellar component, in different contexts of the dynamics of the galaxy. For example,  it was shown that the addition of the low velocity dispersion component, namely, gas in a gravitationally coupled two-fluid (stars plus gas) system pushes the system towards being more unstable against the local axisymmetric perturbations. In other words, even when the constituent systems (namely gas and stars) are stable when treated individually, the inclusion of gas makes the corresponding two-component system more prone to become unstable against the local axisymmetric perturbations \citep{JS84a,JS84b,BR88,Jog96}. This trend also holds for the gravitationally coupled two-component (star plus gas) system, even if the stellar component is treated as a collisionless system \citep{Raf01}. Also, \citet{Jog92} showed that owing to the low velocity dispersion, non-axisymmetric perturbations in gas get more swing amplified and are found to be more tightly wound than the non-axisymmetric perturbations present in the stellar component.

\section {Dark Matter in disk galaxies}

Studies of dark matter in the contexts ranging from the structure formation to galactic scale is a very active topic of research since last couple of decade or before. Since this component does not emit light, it is impossible to detect it directly via observations. However, this unseen component makes its presence felt by influencing the dynamics of the system via gravitational interaction. According to the latest survey carried out by Planck satellite, the current estimates for the dark matter component is $\sim$ 23 per cent of the total matter in the universe while the baryonic component constitutes only $\sim$ 4 per cent of the total matter of the universe \citep{Ade16}

 After the recent observational detection of gravitational wave in 2015, the next challenging thing in astronomy is to either detect directly/indirectly the presence of such unseen matter or to nullify the proposition of such  component. Although several studies have been carried out (including this thesis!) to quantify and understand the possible role of the dark matter on the dynamics of systems in the galactic scale, simultaneously it is also required to set up experiments to understand the exact nature of this unseen component and also to propose missions for either detecting directly/indirectly the existence of this dark matter or dismiss conclusively the very existence of the dark matter. 
Several missions are already proposed/functional such as PAMELA (Payload for Antimatter Matter Exploration and Light-nuclei Astrophysics), Fermi-LAT (Fermi-Large Angle Telescope), for direct/indirect detection of dark matter.
 Till date (at the time of writing this thesis!) no conclusive detection of dark matter has been reported so far, but it should be also borne in mind that more improvement in both techniques and sensitivity is required for detecting the very existence of dark matter.

\subsection {Historical development}

\citet{Oort32} measured the average matter density near the solar neighbourhood using the statistics for $K$ giant stars. He found in the solar neighbourhood, the average matter density ($\rho$) $\sim 0.15$ $M_{\odot}$ pc$^{-3}$. On the other hand, when only the visible stars are taken into account, the corresponding average density is only about half of what was deduced from observation. That clearly indicated the possibility of having some unseen matter in the solar neighbourhood also. The first detection of {\it significant mass discrepancies} between the luminous mass and the dynamical mass was shown by \citet{Zwi33} for the Coma clusters. \Citet{Zwi33} applied the virial theorem to 7 galaxies in the Coma cluster to determine the total mass of the cluster. The measured mass--to--light ($M/L$) ratio in that cluster was found to be $~ 400$.

The evidence for the presence of dark matter in the case of individual galaxies was brought into the light by the work of \citet{RF70}. Theoretically, the rotation curve  of a stellar disk is expected to rise in the inner region and then falling off in a Keplerian fashion. However, the observed rotation curves were not seen to fall-off in such a Keplerian fashion, and remained nearly flat in the outer parts, in sharp contrast to the theoretical predictions. However, \citet{RF70} used the optical data which was limited to a few kpc in the inner region, therefore the mass-discrepancies predicted by \citet{RF70} was only by a factor a 2--3. The $HI$ rotation curve allowed one to measure the rotation curve to farther distances and beyond the optical disk which in turn allowed one to measure the mass discrepancies to a large distances from the galactic center. The final convincing proof of this mass discrepancy came from the work by \citet{Bos78}. Using the 21 $cm$ emission line of atomic hydrogen ($HI$), \citet{Bos78} measured the rotation curve for 20 disk galaxies out to several disk scalelengths. These observed flat rotation curves extended beyond the optical disk and these showed the presence of mass discrepancies, thus putting the proposition for existence of dark matter in disk galaxies on a firm basis.



\subsection {The {\it `Core--Cusp'} problem}
The discrepancy between the observed flat rotation curve and the theoretically calculated rotation curve with Keplerian fall-off at larger radius for luminous matter had led to the proposition for existence of the dark matter component in the galactic scale. Although in many galaxies, the inner region is still dominated by the luminous (baryonic) matter, however in the outer regions of disk galaxies, the dark matter component contributes the most to the rotation curve. 

As the rotation velocity remains almost constant in the outer parts and dark matter dominates in the outer parts and the total mass increases linearly with radius, therefore, it is suggestive that the mass density profile of dark matter in the outer part is likely to follow a distribution similar to isothermal sphere. In other words, observations suggest that $\rho$ $\sim$ $r^{-2}$. However, the exact behavior of the mass density in the inner region remains somewhat elusive.

The rotation velocity due to the dark matter component is found to rise with radius in a linear fashion in the inner parts of a disk galaxy. This solid-body rotation is indicative of a presence of a density distribution which remains constant in the inner region, in other words it has a {\it `core'} in the inner parts of the galaxy. In the past, some `cored' density profile for dark matter halo like pseudo-isothermal halo, non-singular isothermal halo were tried to model the observed rotation curves, and they produced good fit to the observed rotation curves \citep{Ath87,Beg91,Bro92}. Therefore, from the observational point view, a `cored' density profile for the dark matter halo is more favored.

On the other hand, early $N$-body simulations of Cold dark matter (CDM) along with the more recent $N$-body simulations of $\Lambda$CDM predicted that the density profiles for the galactic dark matter halos should have a {\it `cusp'} at the central region, i.e. the density of the dark matter will display a {\it `spike'} in the central region \citep{DC91,Nav96,Nav97}. In other words, cosmological $N$-body simulations favored a `cuspy' density profile for the galactic dark matter halo. 
This discrepancy is popularly known as {\it `Core--Cusp'} problem or {\it ``the small-scale crisis in cosmology''}. 

In the past, several studies aimed to reconcile between these core/cusp problem and tried to put forward some mechanism that might be held responsible for making a cuspy dark matter halo into a cored dark matter halo. Also, some other studies aimed to find some observational artifacts that might create an illusion of a cored dark matter halo while it reality they might be actual cuspy. To mention a few, using numerical simulations, \citet{Na96} studied the effect of star formation on baryons and the dark matter and found that if a large fraction of baryonic component is expelled in the dark matter halo, it can produce a core distribution of the dark matter halo. Similarly \citet{GS99} also tried a detailed study of the effect of this blow-out process on the shape of the dark matter halo. Another study by \citet{Mash06} showed that the random bulk motions of gas in small primordial galaxies can make a cuspy density distribution flat in a short time-scale ($\sim$ $10^8$ years), thus can give rise to a cored density distribution. In observations also, there are some uncertainties present that can affect the process of identifying the core/cusp distribution of dark matter. For example the process `beam-smearing' effect can decrease the observed velocities, and for observations affected severely by this beam-smearing process can make a steeply rising rotation curve into a slowly rising rotation curve, thus an inherently cuspy density may appear as a cored density distribution for the dark matter halo \citep[e.g. see discussions in][]{deB96}. However, \citet{deB97} modeled the effect of this beam-smearing and concluded that the effect is not so strong to affect the appearance of the dark matter density distribution completely. For a comprehensive discussion of the core/cusp problem, see \citet{deB10}.

\subsection {Different profiles for DM halo}

In the past literature, several profiles for the dark matter halo have been used to derive mass models from the observed rotation curves for the disk galaxies. While some are motivated from the high-resolution studies of  cosmological simulations, some other profiles are obtained which produce a  good fit to the observed rotation curve. Below, I list some of these dark matter halo profiles that are used extensively for fitting the observed rotation curves.

\subsubsection{Pseudo--Isothermal profile}

The density profile ($\rho_{\rm DM}$) for the pseudo-isothermal halo is given as
\begin{equation}
\rho_{\rm DM}(R, z)= \frac{\rho_0}{1+\frac{R^2}{R_{\rm c}^2}}\,,
\end{equation}
\noindent where $\rho_0$ and $R_{\rm c}$ are the core density and the core radius, respectively.
As seen from the expression, it has a {\it `core'} as the radius ($R$) approaches the center of the galaxy. In the past, it has been extensively used to model the observed rotation curves \citep[e.g. see][]{MCS98,deB08,Oh08} especially for late-type and dwarf galaxies.

\subsubsection{NFW profile}

Motivated by the high-resolution simulation study of cold dark matter structure by \citet{Nav96}, NFW (Navarro--Frenk--White) profile is also used to model the observed rotation curves of the disk galaxies \citep[e.g. see][]{Sof16} and also in $N$-body simulations of isolated disk galaxies \citep[e.g. see][]{Don13}.

The density profile ($\rho_{\rm DM}$) for the NFW profile is given as
\begin{equation}
\rho_{\rm DM}(r)= \rho_{\rm crit}\frac{\delta_{\rm char}}{(r/r_{\rm s})(1+r/r_{\rm s})^2}\,,
\end{equation}
\noindent where $r_{\rm s}$ is the scale radius and $\delta_{\rm char}$ is the characteristic overdensity. As is evident from the expression, it has a {\it `cusp'}, the density tends to become infinite as the radius ($r$) approaches the center of the galaxy.



\subsubsection{de Zeeuw \& Pfenniger profile}

The density profile ($\rho_{\rm DM}$) for the  de Zeeuw \& Pfenniger profile is given as \citep{dePf88}
\begin{equation}
\rho_{\rm DM}(R, z)= \frac{\rho_0}{\left(1+\frac{R^2+\frac{z^2}{q^2}}{R^2_{\rm c}}\right)^{p}}\,,
\end{equation}
\noindent where $\rho_0$ is the core density, $R_{\rm c}$ is the core radius, $p$ is the density index, and $q$ represents the vertical--to--planar axis ratio. In contrast to the earlier profiles listed above, this profile allows one to include `prolate' or `oblate' shapes for modeling the dark matter halo depending on whether $q$ is $> 1$ or $< 1$, respectively.

\section {Our way of treating Dark matter halo}

In the work presented in this thesis, we have treated the dark matter halo to be {\it gravitationally inert}, i.e. it is assumed to be non-responsive to the gravitational force of the perturbations in the disk for simplicity. In other words, the static potential for dark matter halo obtained from the mass-modeling of the observed rotation curve is used. Also we assume that the dark matter halo is {\it concentric} to the stellar disk, so that the net rotation curve can be obtained by adding the respective contributions of stellar and the dark matter halo in quadrature.

 In all the cases studied, we have used a pseudo--isothermal profile for the dark matter halo. The best-fit models for the dark matter halo of the galaxies that we have used in this work of thesis are taken from the observations which makes our analysis more realistic.

At this point, we mention that in this thesis throughout, we have assumed a non-rotating dark matter halo. In the past, people sometimes used a rotating/spinning dark matter halo in the $N$-body simulations. For example, \citet{SaNa13} showed that a spinning dark matter halo can facilitate the growth of bars through resonant gravitational interactions. Another study by \citet{KiLe13} presented the first nymerical evidence for the hypothesis that the late-type LSBs form in the haloes with large angular momenta. However, we caution that it is extremely difficult to use a spinning/rotating dark matter halo in the semi-numerical approach that we have followed throughout the thesis.

The question remains how to extract the dynamical imprints of the dark matter halo on the spiral structure of different scales seen in the disk galaxies? To achieve that goal, we first considered the  disk component and theoretically calculated the corresponding rotation curve, and some related quantities like angular frequency ($\Omega$), local epicyclic frequency ($\kappa$) for it. Then using these values, we calculate the dispersion relations or carried out the swing amplification analysis, whichever is applicable. Next we include the dark matter halo in the calculation and the contributions of the stellar and dark matter component are added in quadrature in order to get the net rotation curve, i.e.,

\begin{equation}
v^2_{\rm c, net}=v^2_{\rm c, disk}+v^2_{\rm c, DM}
\end{equation}

Although, we point out that this scheme of treating the stellar disk alone (disk alone case) and then adding the dark matter (stars plus dark matter halo case) is a little stylized, however, this novel scheme allows us to identify the exact role of dark matter halo on the spiral structure at different scales.

\thispagestyle{empty}

 \setcounter{chapter}{1}
\chapter[Suppression of gravitational instabilities by dominant dark matter 
halo in low surface brightness galaxies]{Suppression of gravitational instabilities by dominant dark matter 
halo in low surface brightness galaxies {\footnote {Ghosh \& Jog, 2014, MNRAS, 439, 929}}}
\chaptermark{\it Suppression of instabilities by halo }
\vspace {2.5cm}

\section{Abstract}
The low-surface-brightness galaxies are gas-rich and yet have a low star formation rate; this is a well-known puzzle.
The spiral features in these galaxies are weak and difficult to trace, although this aspect has not been studied much.
These galaxies are known to be dominated by the dark matter halo from the innermost regions.
 Here, we do a stability analysis for
the galactic disk of UGC~7321, a  low-surface-brightness, superthin galaxy, for
which the various observational input parameters are available. We show that the disk is stable against local, linear axisymmetric
and non-axisymmetric perturbations. The Toomre $Q$ parameter values are found to be large ($\gg 1$) mainly due to the low disk surface density, and
the high rotation velocity resulting due to the dominant dark matter halo, which could explain the observed low star formation rate.
For the stars-alone case, the disk shows finite swing amplification but the addition of dark matter halo suppresses that amplification almost completely. Even the inclusion of the low-dispersion gas which constitutes a high disk mass fraction 
does not help in causing swing amplification.
This can explain why these galaxies do not show strong spiral features. 
Thus, the dynamical effect of a halo that is dominant from inner regions can naturally explain why star formation and spiral features are largely suppressed
in low-surface-brightness galaxies, making these different from the high-surface-brightness galaxies.

\section{Introduction} 
\label{intro1}

It is well known that the class of galaxies known as the low-surface-brightness (LSB) galaxies are characterized by a low stellar disk surface density \citep{deBMc96,deBMcRu01} and a low star formation rate \citep{IB97}. 
Despite having a high 
mass fraction of the disk in gas, these do not appear to have had much star formation  and are thus
unevolved \citep{BIMc97}.  This has been a long-standing basic puzzle about these galaxies. 
Many LSB galaxies do show spiral structure but it is fragmentary or incipient, and is extremely faint and difficult to trace \citep{Schom90,McSB95}, also \citet[][see Figs. 5 and 6 there]{SMMc11}.

The lack of star formation in the LSBs is not well-understood. One common reason suggested for this is the low surface density that lies below the onset for threshold \citep{van93}, or the lack of molecular gas which normally forms the site of star formation \citep{OSH03,Jog12}. Further, these galaxies are seen in isolated regions \citep{MMcB94} or at the edges of voids \citep{Ros09}. Thus, there is a lack of galaxy interaction that could have triggered
star formation \citep[e.g.,][]{Mih97} or spiral features \citep[e.g.,][]{No87}.

It is well established in the literature that the LSB galaxies are dark matter dominated down to their central regions \citep{BIMc97,deBMc97,deBMcRu01,Com02,Ban10}. We note that this is different from the dominance of dark matter at large radii as deduced from the flat rotation curves that are seen in all galaxies. In the LSBs, within the optical disk, the dark matter constitutes about 90 per cent of the total mass and the rest (10 per cent) is in baryons, while within the same radius, the two are comparable in the `normal' or high-surface-density (HSB) galaxies \citep[e.g.,][]{deBMcRu01,Jog12}. It has sometimes been
claimed that the dark matter halo prevents star formation though the details  are not well understood. 

Early work using numerical simulations has indicated \citep{Mih97} that the LSBs are stable against the growth of global non-axisymmetric modes such as bars due to low surface density and large dark matter content. However, observations show that bars are common in LSBs and can be off-centered as in irregular galaxies \citep{MG97}.  This needs to be followed up in future work. The discrepancy could be partly due to different definitions of bars used in these papers.

Some of the dynamical properties of LSB galaxies such as stability against a bar mode 
\citep{Mih97}, and the superthin nature of a particular subset of LSB galaxies as arising due to a dense and compact halo \citep{BJ13}, have been studied so far. However, the absence of strong spiral features in LSB galaxies has not drawn the much-deserved attention so far.

In this paper, we present a dynamical study of local axisymmetric and non-axisymmetric 
perturbations in an LSB, superthin galaxy, UGC 7321,
for which the observational parameters are known. A previous study of this galaxy has revealed it is dark matter dominated at all radii starting from the innermost regions \citep{Ban10}. Also the de-projected radial $HI$ surface density lies well below the critical $HI$ surface density for star formation for all radii \citep{UM03}.
 Thus, UGC 7321 is a suitable candidate to probe the effect of dark matter halo on both axisymmetric and non-axisymmetric local perturbations. We find that this galaxy is highly stable against both these mainly due to the low disk surface density and the high fraction of dark matter from the innermost regions which dominates its rotation curve.

We obtain the values for the $Q$ parameter for local stability first for gas-alone \citep{Too64} and then for gravitationally coupled stars and gas as in Jog (1996). 
The value of $Q < 1$ is taken to indicate onset of star formation \citep{Kenn89}. We also study the local non-axisymmetric linear perturbations using the idea of swing amplification as a mechanism for generating local spiral features \citep{GLB65,Too81}. This amplification is temporary and arises as a mode swings past the radial direction, and occurs due a kinematic resonance between the epicyclic motion and the trailing sense of differential rotation in a disk, when supported by the self-gravity in the disk.
While using swing amplification as a tool, people often use some educative guess values for the  $Q$ parameter, but in this paper we calculate the actual values of $Q$ parameters obtained from the observed quantities and use these as input to study the local axisymmetric stability and also the swing amplification. 
 We show that in a galaxy like UGC 7321, the dark matter halo dominates in mass from the innermost regions, which increases the $Q$ values and hence suppresses star formation and swing amplification nearly completely.

~\S~\ref{formu1}  presents the details of the formulation of the problem,~\S~\ref{nu-sol1} presents the details of input parameters, and the numerical solution and the results.~\S~\ref{dis1} and~\S~\ref{conclu1} contain the discussions and the conclusions, respectively.

\section{Formulation of the Problem}
\label{formu1}

\subsection{Axisymmetric case}
\label{axi-case}

A disk supported by rotation and random motion is stable to local, axisymmetric perturbations if the following $Q$ criterion is satisfied (Toomre 1964)
\begin{equation}
Q \: = \: \kappa c / \pi G \mu > 1 
\end{equation}
\noindent where $\kappa$ is local epicyclic frequency, $c$ is the one-dimensional random velocity dispersion and $\mu$ is the surface density of the disk. 

This was extended for a two-component disk consisting of gravitationally coupled stars and gas by \citet{Jog96}, 
where the local stability parameter  $ Q_{\rm s-g}$ for axisymmetric  two-fluid case is defined as 
\begin{equation}
\left[\frac{2{\pi}G k{\mu}_{\rm s}}{{\kappa}^2+k^2c_{\rm s}^2}+\frac{2{\pi}Gk{\mu}_{\rm g}}{{\kappa}^2+k^2c_{\rm g}^2}\right]_{\rm {at} \: k_{\rm min}} \equiv \frac{2}{1+(Q_{\rm s-g})^2}\,,
\end{equation}
\noindent where  $k$ is the wavenumber of the perturbation and $k_{\rm min}$ is the wavenumber at which it is hardest to stabilize the two-fluid system, and $c_{\rm s}$ and $c_{\rm g}$ denote the random velocity dispersion in stars and gas respectively. 
While this is difficult to solve analytically, a semi-analytical approach to solve this was given by \citet{Jog96}
 which is summarized next.

Three dimensionless parameters are used, $Q_{\rm s}$ and $Q_{\rm g}$ the standard $Q$ parameters for local stability of  stars-alone and gas-alone cases, respectively, and $\epsilon = \mu_{\rm g} / (\mu_{\rm g} + \mu_{\rm s})$ the gas mass fraction in the disk.
In terms of the above three dimensionless parameters, the above condition reduces to 

\begin{equation}
\begin{split}
\frac{(1-\epsilon)}{l_{\rm s-g}\{1+[Q_{\rm s}^2(1-\epsilon)^2]/(4l_{\rm s-g}^2)\}}+\frac{\epsilon}{l_{\rm s-g}[1+Q_{\rm g}^2{\epsilon}^2/(4l_{\rm s-g}^2)]}\\
\equiv \frac{2}{1+(Q_{\rm s-g})^2}\\\,,
\label{find-kmin}
\end{split}
\end{equation}
\noindent where 
$l_{\rm s-g}= [{\kappa}^2/{2{\pi}Gk_{\rm min}({\mu}_{\rm s}+{\mu}_{\rm g})}]$ is the dimensionless wavelength at which it is hardest to stabilize the two-fluid system. In analogy with the one-component case, the
 disk is shown to be stable, marginally stable or unstable  against axisymmetric perturbations depending on whether the $Q_{\rm s-g}$ value is $> 1$,  = 1, or $ < 1$ respectively \citep{Jog96}.

A normal-mode linear perturbation analysis 
of the two-fluid (stars coupled with gas) system supported by random motion and rotation \citep[see][]{JS84} showed that a radial mode $(k,{\omega})$ obeys the dispersion relation 
\begin{equation}
{\omega}^2(k)=\frac{1}{2}\left \{({\alpha}_{\rm s}+{\alpha}_{\rm g})-\left[({\alpha}_{\rm s}+{\alpha}_{\rm g})^2-4({\alpha}_{\rm s}{\alpha}_{\rm g}-{\beta}_{\rm s}{\beta}_{\rm g})\right]^{1/2}\right \}, 
\end{equation}
\noindent where \\
          
\begin{equation}
\begin{split}
{\alpha}_{\rm s} = {\kappa}^2+k^2 c_{\rm s}^2-2{\pi}G k{\mu}_{\rm s}\\
{\alpha}_{\rm g} = {\kappa}^2+k^2 c_{\rm g}^2-2{\pi} G k{\mu}_{\rm g}\\
{\beta}_{\rm s}=2{\pi} G k{\mu}_{\rm s}\hspace{0.3cm} {\mbox and } \hspace{0.3cm}{\beta}_{\rm g}=2{\pi} G k{\mu}_{\rm g}\,,
\end{split}
\end{equation}
\noindent  where $k$ is the wavenumber (=$ 2 \pi / \lambda$), where $\lambda$ is the wavelength
 and $\omega$ is the angular frequency of the perturbation.

For a given set of three dimensionless parameters introduced above,
we numerically find the minimum of the dimensionless dispersion relation ${\omega}^2 (k) /\kappa^2$ function, 
and thus obtain the corresponding $k_{\rm min}$.
To do this we vary the function between the wavelengths for the minima for the gas-alone and
stars-alone cases ($l_{\rm g} = Q_{\rm g}^2\epsilon^2/2$ and $l_{\rm s} = Q_{\rm s}^2(1-\epsilon)^2/2$) since the minimum of the dispersion relation will occur at a wavelength that lies between these two values \citep{JS84}.  Then substituting this value of $k_{\rm min}$ in equation (\ref{find-kmin}) yields the value of $Q_{\rm s-g}$ for a particular choice of values for the set $Q_{\rm s}$, $Q_{\rm g}$ and $\epsilon$.

\subsection{Non-axisymmetric case}
\label{nonaxi-case}
We consider a local, non-axisymmetric linear perturbation analysis of the galactic disk, where the perturbation is taken to be planar. The treatment follows as in GLB for a one-component disk, and for a two-fluid case in Jog (1992).
 The disk is taken to be thin for the sake of simplification of calculation as in \citet{Jog92}.
We next introduce the sheared coordinates $(x',y',z',t')$ defined by
\begin{eqnarray}
x'=x,\: y'=y-2Axt,\: z'=z, \: t'=t
\end{eqnarray}
\noindent These were introduced by GLB to study the dynamics of a local patch in a differentially rotating disk. A trial solution of the form exp$[i(k_xx'+k_yy')]$ for the perturbed quantities including the density
$\delta \mu$ is introduced.
Next, we define ${\tau}$ (for a wavenumber $k_y\ne 0$):
\begin{equation}
\tau \equiv 2At'-k_x/k_y
\end{equation}
\noindent where $x$ is along the initial radial direction. In the sheared frame, $\tau$ is a measure of time and it has different zeros for
each mode characterized by a different $k_x/k_y$ \citep{Jog92}. In each case, $\tau = 0$ when the 
wave-vector is along the radial direction. 
Further, $\delta \mu$ represents the density variation or amplification with time in the sheared frame,
while in the non-sheared frame centered on the galactic center, it gives the density for a mode of wavenumber $[k_y (1 + \tau^2)^{1/2}]$
that is sheared by an angle $\gamma = \tan^{-1} \tau$ with respect to the radial direction, and $\mu$ is the unperturbed disk surface density.

Define ${\theta}_{ i}$, the dimensionless perturbation surface density to be 
\begin{equation}
\theta_i={\delta \mu_{i}}/{\mu_{i}}
\end{equation}
Using the above trial solution and making use of the definition of ${\tau}$,  the linearized perturbation equations for the Euler equation, continuity equation and the joint Poisson equation combine to give the following two coupled differential equations which describe the evolution with $\tau$ in $\theta_i$ 
\begin{equation}
\begin{split}
\left(\frac{d^2\theta_i}{d \tau^2}\right)-\left(\frac{d\theta_i}{d\tau}\right)\left(\frac{2\tau}{1+\tau^2}\right)+\theta_i\left[\frac{\kappa^2}{4A^2}+\frac{2B/A}{1+\tau^2}+\frac{k_y^2}{4A^2}(1+\tau^2){c_i^2}\right]\\
\!\!\!\!\!\!\!\!\!\!=(\mu_{s}\theta_s+\mu_{g}\theta_g)\left(\frac{\pi G k_y}{2A^2}\right)\!\!(1+\tau^2)^{1/2},
\end{split}
\label{swing-gen}
\end{equation}
 where $i = s, g$ correspond to stars and gas, respectively.
 The three terms
in the parentheses on the L.H.S describe the effect of the epicyclic motion, the sheared motion and the pressure term.

\noindent The analogous equation for the one-fluid (say stellar case) case is obtained by setting $\mu_{\rm g}$ = 0 in equation (\ref{swing-gen}):
\begin{equation}
\begin{split}
\left(\frac{d^2\theta_s}{d \tau^2}\right)-\left(\frac{d\theta_s}{d\tau}\right)\left(\frac{2\tau}{1+\tau^2}\right)+\theta_s\!\!\left[\frac{\kappa^2}{4A^2}+\frac{2B/A}{1+\tau^2}+\frac{k_y^2}{4A^2}(1+\tau^2){c_s^2}\right]\\
-\mu_{s}\theta_{\rm s}\left(\frac{\pi G k_y}{2A^2}\right)(1+\tau^2)^{1/2}=0\,.
\end{split}
\end{equation}
Following the two-fluid approach by \citet{Jog92}, we introduce the dimensionless parameters as $Q_{\rm s}, Q_{\rm g}$ the $Q$ factors for stars-alone and gas-alone, respectively, $\epsilon$, the gas mass fraction in the disk, $\eta={2A}/{\Omega}$
a measure of differential rotation in the disk, and $X = \lambda_{y} / \lambda_{\rm crit}, $ where $\lambda_{\rm crit} = 4 \pi^2 G (\mu_{\rm s} + \mu_{\rm g})/ \kappa^2$.

Using these dimensionless parameters, the above coupled equations reduce to
\begin{equation}
\begin{split}
\left(\frac{d^2\theta_{\rm s}}{d \tau^2}\right)-\left(\frac{d\theta_{\rm s}{\rm }}{d\tau}\right)\left(\frac{2\tau}{1+\tau^2}\right)+{\theta_{\rm s}}\left[{\xi^2}+\frac{2(\eta-2)}{\eta(1+\tau^2)}+\frac{(1+\tau^2)Q_{\rm s}^2(1-\epsilon)^2\xi^2}{4X^2}\right]\\
=\frac{\xi^2}{X}(1+\tau^2)^{1/2}[\theta_{\rm s}(1-\epsilon)+\theta_{\rm g}\epsilon]\\
\end{split}
\label{swing-star}
\end{equation}

\begin{equation}
\begin{split}
\left(\frac{d^2\theta_{\rm g}}{d \tau^2}\right)-\left(\frac{d\theta_{\rm g}{\rm }}{d\tau}\right)\left(\frac{2\tau}{1+\tau^2}\right)+\theta_{\rm s}\left[{\xi^2}+\frac{2(\eta-2)}{\eta(1+\tau^2)}+\frac{(1+\tau^2)Q_{\rm g}^2\epsilon^2\xi^2}{4X^2}\right]\\
=\frac{\xi^2}{X}(1+\tau^2)^{1/2}[\theta_{\rm s}(1-\epsilon)+\theta_{\rm g}\epsilon]\\\,,
\end{split}
\label{swing-gas}
\end{equation}

\noindent where ${\xi^2}={\kappa^2/4A^2}= 2(2-\eta)/\eta^2$.\\
\noindent Similarly, the one-fluid analog for the stars-alone case is given by 
\begin{equation}
\begin{split}
\left(\frac{d^2\theta_{\rm s}}{d\tau^2}\right)-\left(\frac{d\theta_{\rm s}}{d\tau}\right)\left(\frac{2\tau}{1+\tau^2}\right)+{\theta_{\rm s}}\left[{\xi^2}+\frac{2(\eta-2)}{\eta(1+\tau^2)}+\frac{(1+\tau^2)Q_{\rm s}^2\xi^2}{4X^2}\right]\\
=\frac{\xi^2}{X}\theta_{\rm s}(1+\tau^2)^{1/2}\\\,.
\end{split}
\label{swing-onecomp}
\end{equation}

To bring out the sole effect of non-axisymmetric perturbations, we solve the above equations in the special case when the system is stable against the axisymmetric perturbation.
The necessary condition for axisymmetric stability is \citep[see][]{JS84}
\begin{equation}
\frac{(1-\epsilon)}{X'\{1+[Q_{\rm s}^2(1-\epsilon)^2/4X'^2]\}}+\frac{\epsilon}{X'[1+(Q_{\rm g}^2\epsilon^2/4X'^2)]} < 1
\label{axi-cond}
\end{equation}
\noindent $X'=\lambda_{\rm a}/\lambda_{\rm crit}$ and $\lambda_{\rm a}$ is the wavelength of the axisymmetric perturbation.
Thus, for each set of parameters, equations~(\ref{swing-star}) and~(\ref{swing-gas}) are to be solved while ensuring that the inequality given by equation (\ref{axi-cond}) is satisfied for all wavelengths ranging from $Q_{\rm g}^2\epsilon^2/2$ to $Q_{\rm s}^2(1-\epsilon)^2/2$.

\section{Numerical Solution }
\label{nu-sol1}
\subsection{Input parameters}
For the stellar disk and halo parameters, we assume the values obtained observationally and by modeling, respectively \citep{Ban10}.
The central surface density of the stellar disk is obtained as 50.2 M$_{\odot}$ pc$^{-2}$ and the value of the exponential disk scalelength 
$R_{\rm d}$ is  $2.1$ kpc. This gives a radial variation of the stellar disk surface density required to calculate the $Q_{\rm s}$ parameters for the stellar disk. Note that this is very low, nearly a factor of 12 smaller than the central value for the Galaxy \citep[e.g.,][]{NJ02}.
 For the dark matter halo, a pseudo-isothermal density profile is obtained (see equation (\ref{den-dm})), where the core radius, $R_{\rm c}$, is obtained to be $2.5$ kpc and the core density $\rho_o$ is  0.057 M$_{\odot}$ pc$^{-3}$.
The gas surface density as a function of radius is taken from \citet[][see Fig. 13 in that paper]{UM03}.

For the stars, the one dimensional velocity dispersion in the $z$-direction is taken to vary with radius as $c_{\rm s}= (c_{\rm s})_0 \:$ exp$(-R/2R_{\rm d})$ \citep{Ban10}, where$(c_{\rm s})_0$ is  equal to 14.3  km s$^{-1}$ . For the solar neighborhood, it is observationally found that, the ratio of $(c_{\rm s})_z$ to the radial dispersion is $\sim$ 0.5 \citep[e.g.,][]{BT87}. Here we assume the same conversion factor for all radii in this galaxy.

\citet{Ban10} found that to get the best fit to the $HI$ scale height data, a higher value of gas velocity dispersion is needed in the inner parts whereas a slightly lower value is required for the outer regions of UGC~7321. Here, we assume the same values they used, thus imposing a small gradient in the gas velocity dispersion by letting it vary linearly between 9.5 km s$^{-1}$ at $R=7$ kpc and 8 km s$^{-1}$ at $R=12.6$ kpc.

\subsection{Solution of Equations and Results}
\subsubsection{Axisymmetric case}

 Based on the above input parameters,   we next calculate the  values of $Q_{\rm s}$ and $Q_{\rm g}$ at different radii for UGC~7321. The value of $\kappa$ is obtained from observed rotation curve which already includes the effect of the dark matter halo and the gas.
For a given set of values of input parameters $Q_{\rm s}$, $Q_{\rm g}$ and $\epsilon$, we obtain the stability parameter $Q_{\rm s-g}$ for a two-component disk (see the procedure outlined in~\S~\ref{axi-case}).
The corresponding values are given in Table~\ref{res-axi}.

It is clear from Table~\ref{res-axi} that at all radii, 
the joint $Q_{\rm s-g}$ values are well above $1$, implying that the galaxy is highly stable against the two-fluid axisymmetric perturbations.
For gas alone case, the $Q$ parameter is $\gg 1$ due to the low gas density, and a high $\kappa$ which reflects the effect of the dominant halo
in setting the undisturbed rotational field. While the star-gas gravitational interaction does lower the two-component values, these are still $> 3$ at all radii.
 Thus, as a whole the galaxy is stable all the way starting from the inner part up to the very outer region. This is opposite to the picture for a high surface brightness galaxy such as the Milky Way. The joint $Q_{\rm s-g}$ values in the inner part are close to $1$ in that case, indicating that the Galactic disk is close to onset of
 two-fluid axisymmetric instability \citep[see Table 1 of][]{Jog96}.
Thus, it is the dark matter halo that is dominant from the inner regions in the LSB galaxy UGC~7321 studied here that helps to stabilize the disk against one-fluid as well as two-fluid local, axisymmetric perturbations.

Note that $Q < 1$ is routinely used as a criterion for the onset of star formation \citep{Kenn89}. Hence, we can argue that the
high values of the $Q$ parameter even for a two-component case, as obtained from observational input parameters for UGC~7321 
indicates  
why the LSB galaxies have a low star formation rate.
 Thus, our result confirms the previous result on the lack of star formation  by \citet{van93}, and \citet{UM03}, and puts it on a more firm quantitative ground.
\nopagebreak
\begin{table*}
\centering
   \caption{$Q$ values at different radii of UGC 7321}
\begin{tabular}{ccccc}
\hline
$R/R_{\rm d}$ & $\epsilon$  & $Q_{\rm s}$ & $Q_{\rm g}$ & $Q_{\rm s-g}$ \\
& (gas fraction)\\
\hline
1.0 & 0.33 & 4.6 & 9.7 & 3.8 \\
1.5 & 0.39 & 4.4 & 6.4 & 3.4 \\
2.0 & 0.42 & 4.0 & 5.3 & 3.0\\
2.5 & 0.50 & 4.3& 5.3 & 3.0 \\
3.0 & 0.54 & 5.2 & 6.5 & 3.5 \\
3.5 & 0.50 & 6.1 & 11.6 & 4.4 \\
4.0 & 0.50 & 7.1 & 15.4 & 5.2 \\
\hline
\end{tabular}
\label{res-axi}
\end{table*}

\subsubsection{Non-axisymmetric case}

We first consider a stars-alone galactic  disk for which the epicyclic frequency $\kappa$ is obtained theoretically. The definition of $\kappa$
at the mid-plane for a general potential $\Phi$ is defined by \citep{BT87}:
\begin{equation}
{\kappa}^2=\left[\frac {{\partial}^2 \Phi}{\partial R^2}+\frac{3}{R}\frac{{\partial}\Phi}{\partial R}\right]_{z=0}\,.
\label{pot-midplane}
\end{equation}
The potential $\Phi(R,0)$ for an exponential disk in the equatorial plane is given by \citep[see equation~(2.168) in][]{BT87}
\begin{equation}
{\Phi}(R,0)=-{\pi}G{\Sigma}_0R[I_0(y)K_1(y)-I_1(y)K_0(y)]\,,
\label{pot-disk}
\end{equation}
\noindent where ${\Sigma}_0$ is the disk central surface density, y is the dimensionless quantity given by $y=R/2R_{\rm d}$, $R$ being galactocentric radius and $R_{\rm d}$ being disk scalelength and $I_n$ and $K_n$$(n=0,1)$ are the modified Bessel function of the first and second kind, respectively.
For this choice of  potential, equation (\ref{pot-midplane}) gives
\begin{equation}
{\kappa}^2_{\rm disk}=\frac{{\pi}G{\Sigma}_0}{R_d}[4I_0K_0-2I_1K_1+2y(I_1K_0-I_0K_1)]\,.
\label{kappa-disk}
\end{equation}
\noindent Using this expression, we calculate the $Q_{\rm s}$ values for the disk treated as a stars-alone case.
At $R=3R_{\rm d}$ which is used for illustrative purposes here, the resulting value of $Q_{\rm s}$ = 1.2.

In presence of a dark matter halo, the unperturbed rotational velocity and hence $\kappa$ and hence $Q$ are higher. This is particularly 
expected to be true for LSB galaxies like UGC 7321  when the dark matter dominates from the inner regions.
This fact is normally not realized since $\kappa$ is obtained using the observed rotation curve which already includes the effect of the halo.

\begin{figure}
\centering
\includegraphics[height=3.0in,width=4.0in]{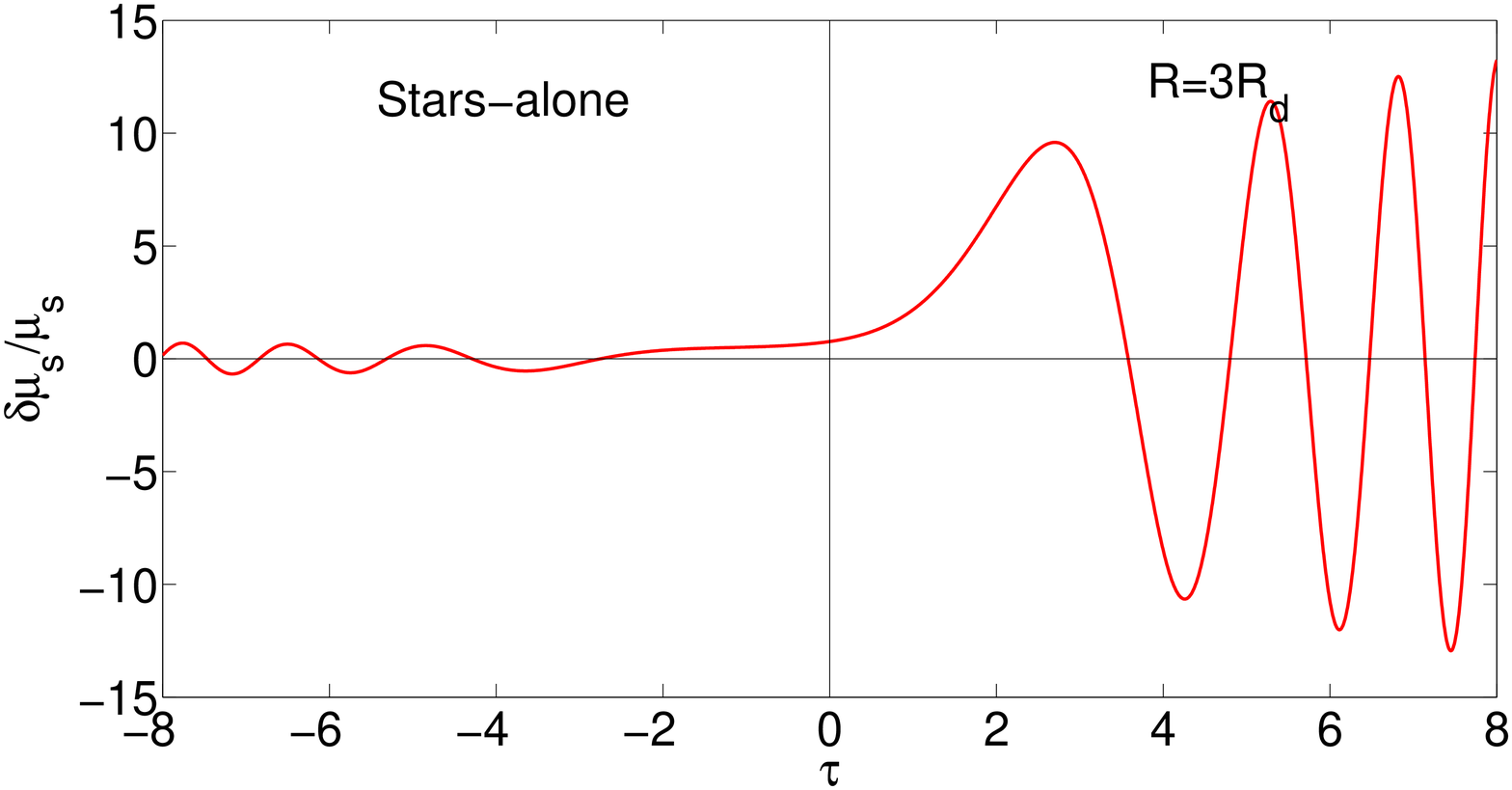}
\medskip
\includegraphics[height=3.0in,width=4.0in]{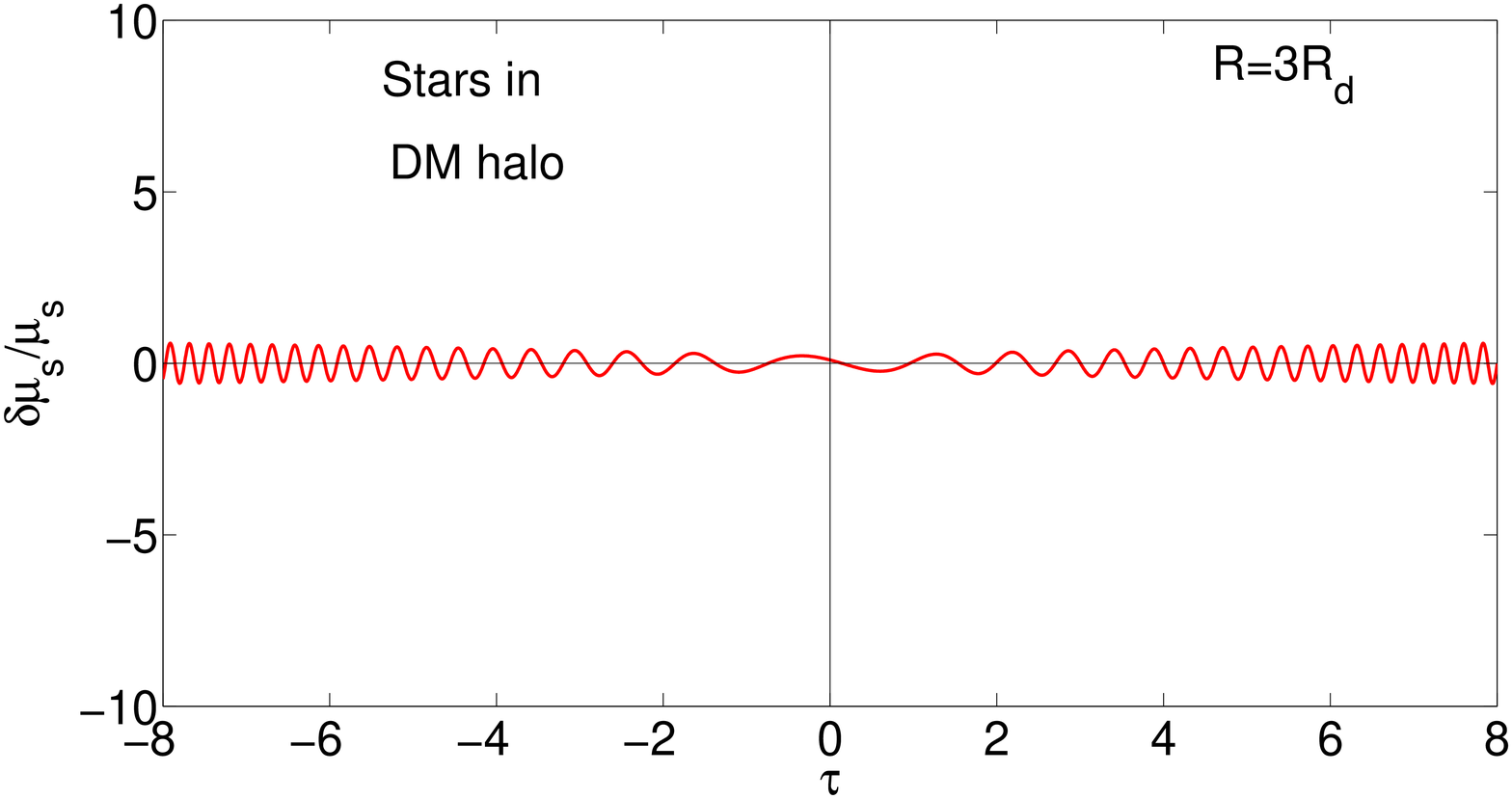}
\caption{Variation in $\theta_{\rm s}=\delta\mu_{\rm s}/\mu_{\rm s}$, the ratio of the perturbation surface density to the unperturbed surface density,with $\tau$, dimensionless time in the sheared frame. The top panel shows 
the stars-alone case ($Q_{\rm s}=1.2$, $\eta=1$, and $X=1$) at the radius $R= 3R_{\rm d}$ and the lower panel shows
the case of stars in the dark matter halo  ($Q_{\rm s}=5.0$, $\eta=1$, and $X=1$) at the same radius. The stars-alone disk shows a finite amplification, but when the stars are taken in the field of the dark matter halo, the amplification is almost completely damped. This is due to the higher rotational velocity and $Q$ values in presence of the dominant dark matter halo.}
\label{ugc-fig}
\end{figure}

To illustrate the effect of considering a disk in halo, we theoretically calculate the net epicyclic frequency.
In this case, the net ${\kappa}^2$ is given by adding the values for the disk alone and the halo in quadrature
\begin{equation}
{\kappa}^2_{\rm net} = {\kappa}^2_{\rm disk} + {\kappa}^2_{\rm halo}\,.
\label{kappa-net}
\end{equation}
A spherical, pseudo-isothermal halo is found to give the best fit while modeling the halo parameters of UGC 7321 using the $HI$ scale height data and observed rotation curve as simultaneous constraints \citep{Ban10}. To keep parity with that work here also we take an pseudo-isothermal halo as characterized by
\begin{equation}
{\rho}(r)=\frac{{\rho}_0}{(1+r^2/R_{\rm c}^2)}\,,
\label{den-dm}
\end{equation}
\noindent\ where ${\rho}_0$ 
 is the core density and $R_{\rm c}$ is the core radius of the dark matter halo.
Using this profile we solve the Poisson equation  to obtain the potential and then converting it into standard galactic cylindrical co-ordinates $(R,\phi,z)$, the final expression for potential due to the dark matter halo is
\begin{equation}
\begin{split}
{\phi}_{\rm halo}=(4{\pi}{\rho}_0 R_{\rm c}^2) \Bigg[\frac{1}{2}\log(R_{\rm c}^2+R^2+z^2)+\Bigg(\frac{R_{\rm c}}{(R^2+z^2)^{1/2}}\Bigg) \\
\times \tan^{-1}\Bigg(\frac{(R^2+z^2)^{1/2}}{R_{\rm c}}\Bigg)-1\Bigg]\,.
\end{split}
\end{equation}
\noindent  Using this in equation (\ref{pot-midplane}) yields
\begin{equation}
\kappa_{\rm halo}^2= 4{\pi}G {\rho}_0 R_{\rm c}^2\bigg[\frac{2}{R_{\rm c}^2+R^2}+\left(\frac{R_{\rm c}^2}{R^2}\right)\frac{1}{R_{\rm c}^2+R^2}-\frac{R_{\rm c}}{R^3}{\tan}^{-1}\left(\frac{R}{R_{\rm c}}\right)\bigg]\,.
\label{kappa-dm}
\end{equation}
The value of $\kappa_{\rm net}$ is obtained by combining the disk and the halo contributions (equations~\ref{kappa-disk},~\ref{kappa-dm}) as in equation (\ref{kappa-net}).
At $R=3R_{\rm d}$ which is used for illustrative purposes here, using this value of $\kappa_{\rm net}$ the  resulting value of $Q_{\rm s}$ = 5.0. Note that this is much higher than for
the stars-alone disk ($Q_{\rm s} = 1.2$) and this leads to suppression of swing amplification in presence of a halo as shown in Figure~\ref{ugc-fig}.

Though the rotation curve for UGC~7321 is not perfectly flat, instead slightly rising right after $2R_{\rm d}$, but for sake of simplicity of calculation we assume a flat rotation curve for from $2R_{\rm d}$ for this galaxy. Hence we set $\eta=1$ and $\xi^2=2$ in equations (\ref{swing-star}~--~\ref{swing-onecomp}) for the cases discussed here.

Here we consider the analysis  at $R=3R_{\rm d}$.
We first consider the stars-alone case
and then solve by adding the effect of dark matter halo as described above. This scheme is used to check how strong is the effect of dark matter halo on the swing amplification (see Figure~\ref{ugc-fig}).

For a given set of  parameters, we have to solve the second-order linear differential equation~(\ref{swing-onecomp}) for the one-fluid case. We treat it as a set of two coupled first order linear differential equation in $\theta_{\rm s}$ and ${d\theta_{\rm s}}/{d\tau}$ and then solve them numerically in an iterative fashion using Fourth order Runge-Kutta with the initial values at $\tau_{ini}$, the initial value of $\tau$. Following \citet{GLB65}, the possible set of initial values for these two variables at 
$\tau_{\rm ini}$ are (1,0) and (0,1).
As the values of $(\theta_{\rm s})_{\rm max}$ and MAF$(=(\theta_{\rm s})_{\rm max}/(\theta_{\rm s})_{\rm ini})$ depend crucially on the choice of $\tau_{\rm ini}$ \citep{Jog92}, we varied both the $\tau_{\rm ini}$ and initial values of the variables simultaneously to get the maximum swing amplification possible for that given set of parameters.

A typical solution starting at large negative $\tau$ values shows an oscillatory behavior due to the dominance of the pressure term.
As the mode goes past the radial direction the shearing term and the epicyclic term are in a temporary kinematical resonance, the duration of which is enhanced due to the self-gravity of the perturbation, this gives rise to the swing amplification as the mode swings past the radial direction (or $\tau =0$ in the sheared frame). At large $\tau$ values the pressure term dominates again. This explanation \citep{GLB65,Too81} describes the schematic behavior of the solution seen in Figure~\ref{ugc-fig} (top panel). 
\noindent When the stars-alone disk is considered then the disk shows finite swing amplification, but the addition of the dark matter halo suppresses the swing amplification completely. Thus this galaxy will not exhibit strong spiral features.

For the sake of completeness, we next also
 include the gas and treat the galactic disk as a gravitationally coupled two-fluid system to make our approach more realistic.
Here while calculating the ${\kappa}$ we use the best fit of observed rotation curve of that galaxy obtained by \citet{Ban10}.
This includes the effect of dark matter halo as well as stars and gas in determining the undisturbed rotational velocity in the galactic disk. At $R=3R_{\rm d}$, this gives $Q_{\rm s} = 5.2$ and $ Q_{\rm g} = 6.5$ (see Table~\ref{res-axi}). Note that this value of $Q_{\rm s}$ obtained from the observed rotation curve is close to that obtained by treating a disk in a dark matter halo potential ($Q$=5.0), this shows the small change due to the inclusion of gas in the real case.

Given a set of parameter values, we are to solve  equations~(\ref{swing-star}) and~(\ref{swing-gas}) simultaneously for the two-fluid case. These are treated as 
a set of four first order linear coupled differential equations in $\theta_{\rm s}$, ${d\theta_{\rm s}}/{d\tau}$, $\theta_{\rm g}$, and ${d\theta_{\rm g}}/{d\tau}$ and then solve them numerically in a  way similar to what was done in \citet{Jog92}.

\begin{figure}
\centering
\includegraphics[height=3.0in,width=4.0in]{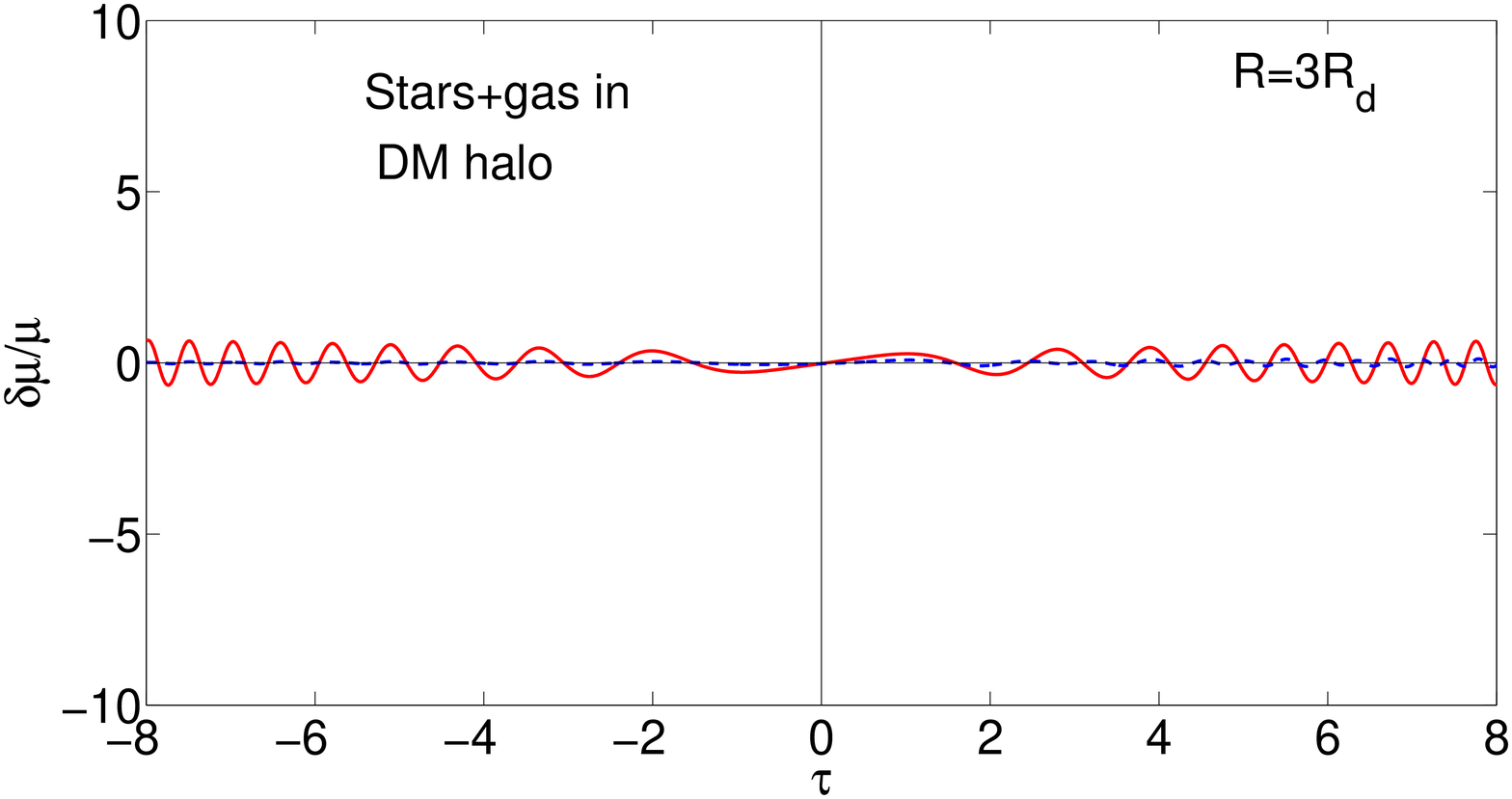}
\caption{Variation in $\theta=\delta\mu_ i/\mu_ i$, the ratio of the perturbation surface density to the unperturbed surface density,with $\tau$, dimensionless time in the sheared frame treating galactic disk as a two-fluid gravitationally coupled system consisting of stars and gas in the field of the dark matter halo at $R= 3 R_{\rm d}$. $Q_{\rm s}=5.2$, $Q_{\rm g}=6.5$, $X=1$, $\eta=1$ and the gas fraction $\epsilon=0.54$.
Despite the high gas fraction, the addition of gas does not make any drastic change, hardly any swing amplification is seen for both stars and gas.}
\label{ugc-2comp}
\end{figure}

 From Figure~(\ref{ugc-2comp})  it is clear that the inclusion of gas makes hardly any impression  as there is no sign of swing amplification in both stars and gas. This is in contrast to the strong effect of gas shown in a typical HSB galaxy \citep{Jog92}. This is because of the low disk surface density and the large dark matter content in UGC~7321, that prevents the supply of necessary disk self-gravity to amplify the non-axisymmetric dynamical seeds. This supports a similar conclusion by \citet{Mih97} for the bar instabilities.

As the rotation curve for UGC 7321 is not exactly flat, instead a slightly rising beyond $2R_{\rm d}$ \citep[see Fig.1 of][]{Ban10} 
then the value of $\eta$ will be different from $1$. We 
calculated the value of $\eta$ from observed rotation curve and found this to give $\eta =0.36$
 and re-did the above calculation for the two-fluid case. Even the consideration of actual $\eta$ value does not make any drastic change, hardly any swing amplification is seen for both stars and gas.
 
\section {Discussion}
\label{dis1}
 \subsection {Weak spiral structure in LSBs and its origin}

Some LSB galaxies do show spiral structure although in general it is  
 weak and difficult to trace (~\S~\ref{intro1}). However, in view 
of the analysis in the present paper which shows a suppression of 
instabilities, the surprise is that the LSB disks do show some 
structure at all albeit feeble. In this section, we try to address 
this interesting question in terms of the amplitude of spiral features 
and the mechanisms that could give rise to the weak features.

A quantitative measurement by a modal analysis on lines of 
that done routinely for the HSB galaxies for $m=2$ (spiral arms or bars) 
or $m=1$ (lopsidedness) \citep{RZ95,Bou05} has not been done for the LSB galaxies. But, a visual
inspection shows the spiral features in LSBs to be much fainter and 
fragmentary than in the HSB galaxies \citep{McSB95,SMMc11}. A modal analysis for the LSBs needs to be carried out to 
measure the Fourier amplitudes of the various modes quantitatively.

In this paper, we have focused on swing amplification as a mechanism for 
generation of local spiral features. It has been shown \citep{Too81,SC84} that this mechanism stops being effective 
for $Q > 2-2.5$ or $X > 3$ (where $X$ is the dimensionless wavelength (see ~\S~\ref{nonaxi-case}). 
Beyond this range, strong swing amplification does not occur \citep{BT87}. However, this does not rule out
the possibility of spiral features occurring outside this range of 
parameters. The disk could still support occasional, weak spiral features.
This is because a differentially rotating disk is always susceptible to the growth of 
non-axisymmetric features due to the kinematic resonance between the 
epicyclic motion and the shear, both of which have the same sense of 
motion. Even if the disk gravity is feeble, this resonance could last 
for about dynamical time-scale ($\sim 10^ 8$ yr), and the resulting 
non-axisymmetric feature would have a low amplitude. This could explain 
some of the putative spiral features seen in  LSB galaxies. 
If the disk
self-gravity is important, then the resonance lasts much longer and this results 
in a regular swing amplification process as discussed in \citet{Too81} 
where the amplitude (the fractional increase in the surface density, or $MAF$) 
will be much higher. Such high-amplitude swing amplified features are 
suppressed in the LSBs due to the low disk surface density and the 
dominant dark matter halo as shown in this paper.

We caution that even the LSB galaxy, UGC 7321, that we have studied, 
may possibly show feeble spiral structure. It is seen edge-on with an 
inclination of 88$^0$ \citep{Matt00}, so we do not know if it would 
exhibit spiral structure if it were to be seen face-on. It appears to 
have only nominal dust content and patchy star formation \citep{Matt00}, 
but both are present at low levels, and so are not totally suppressed.

Alternatively, bars or a central oval feature can trigger a global $m=2$ 
spiral pattern \citep{BT87}, or the spiral arms could arise 
due to manifold-driven trajectories in a barred potential \citep{Ath12}. 
It is interesting that the few LSB galaxies which show a well-defined though weak 
spiral pattern do have this starting at the end of a bar or an oval, as seen 
in F577-V1 (McGaugh et al. 1995), or in F568-1 \citep{Fu08}. Thus such a 
two-armed spiral pattern could have its origin due to the bar. The question 
then is what gave rise to the bar in the first place despite the halo
dominance in the LSB galaxies. 

A tidal encounter could trigger short-lived spiral 
features in the LSB galaxies, as it does in the HSB galaxies \citep[e.g.,][]{No87}, 
except that disk response would be lower due to the high $Q$ values and hence the spiral 
features would be weaker.

Some authors have argued that disks in LSBs should have higher density 
to explain the observed spiral features. We discuss below that the 
derivation of these results is problematic. \citet{Fu08} assumes $m=2$, and 
$X=2$ as seen for high amplification \citep{Too81} to be applicable, and 
then calculates the disk surface density and finds this to be high. We caution 
that this result is only be valid for those few LSBs which have an $m=2$ 
pattern and is not generally applicable. 
\citet{Sab11}, and \citet{SZ13} assume 
the disk in the LSBs to be in a marginal stability with $Q = 1$ and then use
this to obtain the stellar surface density. However, 
 the true $Q$ values for the LSBs are $\gg 1$ (see our 
Table~\ref{res-axi}). Thus artificially imposing a condition of $Q=1$ as done by \citet{Sab11} gives a spuriously 
high value of the disk surface density.

In summary, our stability analysis in this paper shows that strong spiral 
features are not likely to be seen in the LSBs due to the halo dominance. 
However, as argued above, the LSBs could still 
 support weak spiral structure due to the kinematic resonance 
between shear and epicyclic motion even without much help from disk self- 
gravity. This could explain the presence of occasional, weak 
spiral structure seen in some LSB galaxies.

\subsection{Generalization to other LSB galaxies}
We have presented results pertaining to a specific LSB galaxy, UGC 7321, 
since the various input parameters are known observationally for this galaxy.
Generalizing the results from this one case to all LSBs may seem a little 
far-fetched. However, we note that the two main physical properties that 
we find to be responsible for suppression of the instabilities in this galaxy 
are the low disk surface density and the dark matter halo that is dominant 
from inner regions. These two features are generally valid for all the LSBs 
\citep[e.g.,][]{deBMc96,BIMc97,deBMc97}. Hence, we expect our
conclusion about the suppression of gravitational instabilities to be valid in general
for all the LSB galaxies.

While for several LSBs, rotation curves \citep[e.g.,][]{deBMcvan96} and $HI$ surface density \citep[e.g.,][]{deBMcvan96,Pic97} have been obtained, the values of velocity dispersion are not known, these are needed so that the idea be applied to other LSBs as well.

\section{Conclusions}
\label{conclu1}
To summarize, we have studied the axisymmetric and non-axisymmetric local, linear  perturbations for an LSB galaxy, UGC 7321, treating the galactic disk as a system of stars-alone and then as a gravitationally coupled system of stars and gas embedded in a rigid  dark matter halo. We show that the galaxy
 is quite stable against both cases when the effect of the dark matter halo is incorporated while studying its dynamics. 
The inclusion of the massive halo results in a higher undisturbed rotational velocity and hence higher $Q$ values. The low disk surface density also contributes to a high $Q$ value. This results in the disk being stable against axisymmetric perturbations, and also it leads to a suppression of swing amplification of non-axisymmetric perturbations. This can explain the lack of star formation and the absence of strong spiral structure in this galaxy.
 Thus, the low disk surface density and the dominant
 dark matter halo together have a profound effect on stabilizing the disk against both axisymmetric and non-axisymmetric perturbations in 
LSB galaxies. 

{}

\thispagestyle{empty}

\chapter[Dwarf irregular galaxies with extended HI gas disks: Suppression of small-scale spiral structure by dark matter halo]{Dwarf irregular galaxies with extended HI gas disks: Suppression of small-scale spiral structure by dark matter halo {\footnote {Based on Ghosh \& Jog, 2018, New Astronomy, 63, 38}}}
\chaptermark{\it  Suppression of spiral arms by halo in dwarf irregular galaxies}
\vspace {2.5cm}

\section{Abstract}
Dwarf irregular galaxies with extended $HI$ disk distributions, such as DDO 154, allow measurement of rotation curves, hence
deduction of dark matter halo properties
to large radial distances, up to several times the optical radius.
 These galaxies contain a huge reservoir of dark matter halo, 
 which dominates over most of disk.
We study the effect of 
the dark matter halo on small-scale spiral features by
carrying out the local, non-axisymmetric perturbation analysis in the disks of five such late-type, gas-rich dwarf irregular galaxies, namely, DDO~154, NGC~3741, DDO~43, NGC~2366, and DDO~168 which host a dense and compact dark matter halo. 
We show that when the gas disk is treated alone, it allows a finite swing amplification; which would  result in 
small-scale spiral structure in the outer gas disk, but the addition of dark matter halo in the analysis results in a higher Toomre $Q$ parameter which prevents the amplification almost completely. This trend is also seen to be true in regions inside the optical radius because the dominant dark matter halo leads to higher Toomre $Q$ values than the gas-alone case.
This implies absence of strong small-scale spiral arms in  these galaxies, which is in agreement with observations. 
Hence despite being gas-rich, and in fact having gas as the main baryonic component, these galaxies cannot support small-scale spiral structure  which would otherwise have been expected in normal gas-rich galaxies.

\section{Introduction} 
\label{intro3}
Dwarf galaxies are  the most common type of galaxies in the Local group  \citep{Mat98,THT09}.
They exhibit a huge variety of properties, from gas-less dwarf elliptical galaxies to gas-rich irregular galaxies.
Dwarf irregular galaxies form a subset of dwarf galaxies, and can be said to lie at the tail end of the Hubble sequence. They are thus late-type,
gas-rich, bulge-less galaxies. While these galaxies are still rotationally dominated, the maximum rotation velocity is small ($< 60$ km s$^{-1}$) \citep{Io16} as compared to values for normal high surface brightness (hereafter HSB) galaxies, like our Milky Way.

In this paper, we focus on late-type, gas-rich, rotationally supported dwarf galaxies, 
which host an extended $HI$ gas disk. In extreme cases, it can extend up to several times the optical radius. 
For example, the $HI$ disk is extended up to 4 to 8 times the Holmberg radius  in case of DDO 154 \citep{CF88} or NGC 3741 \citep{BCK05,Gen07}  respectively; thus allowing one to model the dark matter halo parameters up to large radii in these galaxies.
{\emph{ Over most of the disk in these galaxies, the HI gas is the main constituent of the baryonic mass}.}
There is very little or no signature of molecular hydrogen (H$_2$) as traced by the $CO$ observations \citep{TaYo87,TKS87}, although it is also possible that the $CO$-to-H$_2$ conversion factor in such low metalicities is small, thus making $CO$ as a poor tracer for H$_2$ \citep{Ohta93,Buy06}. A study by \citet{McG00} showed that for these faint galaxies, the stellar mass alone can not account for the baryonic Tully-Fisher relation, instead one has to take account of the gas mass in order to satisfy the baryonic Tully-Fisher relation.

Various surveys on galaxy morphology have shown that the spiral structure in disk galaxies can be divided into two types: small-scale, patchy or flocculent structure; or the more regular, grand-design spiral structure \citep{Elm11}.

Surprisingly despite being gas-rich and in fact having gas as the dominant baryonic component,
the late-type dwarf irregular galaxies with extended $HI$ disk lack strong spiral structure \citep{Ash92,Tol06}. This is counter-intuitive in the sense that any gas-rich, normal HSB galaxy would generally be expected to produce more flocculent spiral structure \citep[e.g. see][]{Jog92}.
It is well-known that the spiral arms take part in the secular evolution of disk galaxies by scattering the stars off the galactic plane \citep{BW67} and via angular momentum transport \citep{LYKA72,SJ14}. Thus a lack of spiral structure would imply that the secular evolution of the galactic disk is hampered.

\citet{KoFre16} found a systematic trend that the dark matter halos in dwarf galaxies have smaller core radii and denser core density as the galaxies become progressively less luminous. Following this trend, we expect that the dark matter halo would be more important from smaller radii for low luminosity dwarf galaxies that we study here. 
 Thus these galaxies could be good candidates for probing the effect of dark matter halo on the dynamics and evolution of the baryonic disks at the far end of the Hubble sequence.

Several past studies have investigated the dynamical role of dark matter halo in disk galaxies. \citet{Mih97} showed that the dominant dark matter halo prevents the global non-axisymmetric bar mode in the low surface brightness (hereafter LSB) galaxies. \citet{DeSe98} showed that in barred galaxies the dynamical friction from a dense dark matter halo slows down drastically the rotation speed of the bar. A study by \citet{BJ13} showed that the superthin property of the superthin galaxies (a sub-class of LSB galaxies) is explained by a dense, compact halo that dominates from the innermost regions. Recent study by \citet{GJ14} showed that the dominant dark matter halo in LSB galaxies makes the disk rotationally more stable, and
 suppresses the swing amplification process which naturally 
explains why strong spiral structure is not observed in these galaxies.
Even in a typical HSB galaxy like the Milky Way, the disk by itself is close to being unstable to local axisymmetric features, thus the dark matter halo is crucial in stabilizing such disks \citep{Jog14}.

In many dwarf irregulars, the rotation curve rises all the way to the outermost point measured \citep{Bro92,Swa09}; also see the appendix of \citet{KaSa16}. For these galaxies, spiral structure is not expected since most of the disk displays a near solid body rotation, and hence there is no differential rotation. But the dwarf irregular galaxies with extended $HI$ disk where the rotation curve becomes flat and extends far beyond the central region of  solid body rotation, could host spiral structure since the disk does show differential rotation in these parts. Along with that, since these galaxies are extremely gas-rich, spiral structure would normally be expected for these systems. This is at odds with the observational fact that strong small-scale spiral features are absent in these galaxies (e.g., Ashman 1992). In this paper we investigate in these galaxies whether the dominant dark matter halo has any possible dynamical effect on suppression of spiral structure in these galaxies.

Here, we carry out the dynamical study of local non-axisymmetric perturbations in the $HI$-rich dwarf irregular galaxies for which the required quantities such as $HI$ surface density, rotation curve have been measured observationally and are available in the literature. We choose five galaxies, namely, DDO~154, NGC~3741, DDO~43, NGC~2366 and DDO~168, and it turns out that each of these has a dense and compact dark matter halo. Three of these (DDO~154, NGC~2366, IC~2574) are a part of the THINGS (The HI Nearby Galaxy Survey) survey by \citet{Wal08}, whereas DDO~43 and DDO~168 are a part of the LITTLE (Local Irregulars That Trace Luminosity Extremes) THINGS (The HI Nearby Galaxy Survey) survey by \citet{Hun12}. We then investigate the influence of dark matter halo on small-scale spiral structure as generated by swing amplification.

We find that for all the five galaxies, namely, DDO~154, NGC~3741, DDO~43, NGC~2366 and DDO~168, the dominant dark matter halo makes the Toomre $Q$ parameter values, calculated at several radii, very high ($\gg 1$); which in turn prevents the swing amplification process, and hence small-scale spiral features, almost completely.
We find that suppression of strong small-scale spiral structure by dominant dark matter halo turns out to be a generic result in the outer parts as well as regions inside the optical disks of the 
dwarf irregular galaxies with extended $HI$ disks.
 This result is due to the dense, compact dark matter halo and the low disk surface density which together lead to high Toomre $Q$ values.

\S~{\ref{formu_3}} gives the details of sample of dwarf galaxies selected and the formulation of the problem; \S~{\ref{res_3}} describes the input parameters for different galaxies and the results while \S~{\ref{dis_3}} and \S~{\ref{conclu_3}} contain the discussion and conclusions, respectively.

\section{Formulation of the problem}
\label{formu_3}

\subsection{Sample of dwarf galaxies}

We select five late-type dwarf irregular galaxies, namely, DDO~154, NGC~3741, DDO~43, NGC~2366 and DDO~168 in which we study the constraining influence of dominant dark halo on the formation of small-scale spiral arms. Our sample galaxies have the following characteristics:\\
 These have a large $HI$ disk (extending to several optical radii), and in the inner regions as well as in the outer regions the larger part of the baryonic contribution comes from $HI$ gas, and not stars. The galaxies are rotationally supported. We find that for these low luminosity galaxies, the corresponding dark matter halo is
dense and compact, i.e., the central density ($\rho_0$) is few times $10^{-2}$ M$_{\odot}$ pc$^{-2}$ and the core radius (R$_{\rm c}$) is less than two-three times the exponential disk scalelength (R$_{\rm d}$). Note that, the term `compact' does not imply anything about its extent.
 Thus for these dwarf galaxies, the dark matter halo dominates over most of the disk. In contrast, for HSB galaxies like the Milky Way and M~31, the core radius ($R_{\rm c}$) is of the order of 3--4 times the disk scalelength $R_{\rm d}$ \citep{BJ13}.

 Also as a counter-example, we consider IC~2574, a dwarf galaxy which has an extended $HI$ disk but the dark matter halo has low central density and is not compact (R$_{\rm c}>$ 3R$_{\rm d}$); a comparison with the above sample helps to bring out the  role of  a dominant dark matter halo on small-scale spiral features in $HI$-rich dwarf irregular galaxies.

\subsection{Non-axisymmetric perturbation : Swing amplification}
\label{swing_chap3}

 The formulation of local, non-axisymmetric linear perturbation analysis of a galactic disk is followed from \citet{GLB65}. For the sake of completeness here we only mention the relevant assumptions and equations, for details see \citet{GLB65}.

The galactic disk is taken to be infinitesimally thin which is embedded in a dark matter halo, concentric to the galactic disk. For the sake of simplicity, here we treat the dark matter halo as rigid and non-responsive, i.e., it remains inert to the perturbations by the galactic disk.

The perturbations are taken to be planar, and the gas disk is taken to be isothermal, characterized by the surface density $\Sigma$ and the one-dimensional velocity dispersion or the sound speed $c$.

First we introduce sheared coordinates (to take account of the shear introduced by the differential rotation of the disk) given by:\\
\begin{equation}
x'=x,\: y'=y-2Axt,\: z'=z, \: t'=t\\
\label{shear_chap3}
\end{equation}

Next we perform the linear perturbation analysis on the Euler's equations of motion, the continuity equation, and the Poisson equation, and a trial solution of the form exp[$i(k_x x' + k_yy')$] is introduced for the independent perturbed quantities such as the perturbed surface density $\delta \Sigma$.
We define $\tau$ as:\\
\begin{equation}
\tau \equiv 2At'-k_x/k_y \hspace{0.3 cm} \mbox{, for a wavenumber} \hspace{0.1 cm} k_y \ne 0\\
\label{def_tau_chap3}
\end{equation}
In the sheared coordinates, $\tau$ is a measure of time, and it becomes zero when the modes becomes radial, i.e., where $x$ is along the initial radial direction.

We define $\theta$, the dimensionless measure of the density perturbation as:\\
\begin{equation}
\theta = \delta \Sigma/ \Sigma\,,
\end{equation}
\noindent where $\Sigma$ denotes the unperturbed surface density and  $\delta \Sigma$ denotes the variation in surface density in the sheared frame whereas in the non-sheared galactocentric frame of reference, it denotes the density for a mode of wavenumber $k_y(1+\tau^2)^{1/2}$ that is sheared by an angle $\tan^{-1}\tau$ with respect to the radial position. It implies that higher the value of $\tau$, the more sheared will be the mode.

After some algebraic simplifications along with the usage of equations~(\ref{shear_chap3}) and (\ref{def_tau_chap3}), the perturbed equations of motion, the continuity equation, and the Poisson equation reduce to \citep[for details see][]{GLB65}
\begin{equation}
\left(\frac{d^2\theta}{d\tau^2}\right)-\left(\frac{d\theta}{d\tau}\right)\left(\frac{2\tau}{1+\tau^2}\right)+\theta\Bigg[\frac{\kappa^2}{4A^2}+\frac{2B/A}{1+\tau^2}\\
+\frac{k_y^2c^2}{4A^2}(1+\tau^2)-\Sigma\left(\frac{\pi G k_y}{2A^2}\right)(1+\tau^2)^{1/2}\Bigg]=0\,.
\label{swing-2comp}
\end{equation}
\noindent where $\kappa$ is the local epicyclic frequency and $A$, $B$ are the Oort constants (for definitions, see Chapter~1).

The four terms within the square brackets of equation (\ref{swing-2comp}) are due to the epicyclic motion, the unperturbed shear flow, the gas pressure, and the self-gravity, respectively.

 The systematic behavior of $\theta$ is as follows:

 When $\tau$ is large, the pressure term dominates over other terms, and hence the solution will be oscillatory in nature, but when $\tau$ is small the epicyclic motion term and the unperturbed shear flow dominate over the pressure term and they cancel each other completely for a flat rotation curve, thus resulting in setting up a kinematic resonance. In addition, if the self-gravity term dominates over the pressure term then the duration of kinematic resonance increases and the mode undergoes a swing amplification while evolving from radial position ($\tau =0$) to trailing position ($\tau > 0$) \citep[for details see][]{GLB65,Too81}.

Now we introduce three dimensionless parameters, namely, Toomre $Q$ parameter \citep{Too64} = $\kappa c / \pi G \Sigma$ where  $c$ is the velocity dispersion, $\eta$ (= $2A/\Omega$) which denotes the logarithmic shearing rate and $X$ = ($\lambda_y/\lambda_{\rm crit}$), where $\lambda_{\rm crit} (= 4\pi^2 G \Sigma/\kappa^2)$ is the critical wavelength for growth of instabilities in a one-fluid disk supported purely by rotation.

Putting these quantities in equation~(\ref{swing-2comp}), we finally get the evolution of $\theta$ with $\tau$ as
\begin{equation}
\left(\frac{d^2\theta}{d \tau^2}
\right)-\left(\frac{d\theta}{d\tau}\right)\left(\frac{2\tau}{1+\tau^2}
\right)+\theta\Big[{\xi^2}+\frac{2(\eta-2)}{\eta(1+\tau^2)}\\
+\frac{(1+\tau^2)Q^2\xi^2}{4X^2}-\frac{\xi^2}{X}(1+\tau^2)^{1/2}\Big]=0\,,
\label{swing-final}
\end{equation}
\noindent where ${\xi^2}={\kappa^2/4A^2}= 2(2-\eta)/\eta^2$.

For a given set of parameter values we solve equation~(\ref{swing-final}) numerically by fourth-order Runge-Kutta method while treating equation~(\ref{swing-final}) as two coupled, first--order linear differential equations in $\theta$ and $d\theta/d\tau$ \citep[for details see][]{Jog92}.

\section{Results}
\label{res_3}

In this section we present results of the swing amplification at different radii for the five galaxies considered here. First we give the sources from which we obtained the observed values of the required parameters such as the rotation curve, surface density etc.; as well as the halo model parameters. These values  are listed in Table~3.1 (for detailed description see text in \S~{\ref{input_chap3}}). For the sake of consistency, we choose the halo parameters which are obtained by fitting a pseudo-isothermal density profile as given in the cited references. 

 We clarify that for any individual galaxy, although various parameters for stellar, gas and dark matter halo are taken from different papers (as given below), the observational data-set used in those papers are the same. Therefore, these parameters are internally consistent for each individual galaxy. For example, \citet{deB08} gives the values of dark matter halo parameters and \citet{Ler08} gives the values of stellar parameters for DDO~154, although both of them use the same THINGS data.

\begin{table*}
\begin{minipage}{\textwidth}
   \caption{Input parameters for the sample galaxies}
\begin{tabular}{cccccccccc}
\hline
galaxy & RC ref.& $\Sigma_{\rm star}$ & $M_{\rm gas}$ & R$_{\rm d}$ & $R_{\rm HI}/ R_{\rm d}$ & $\rho_0$ & R$_{\rm c}$\\
& & (M$_{\odot}$ pc$^{-2}$) & ($\times$10$^8$ M$_{\odot}$)& (kpc) && (M$_{\odot}$ pc $^{-3}$) & (kpc)\\
\hline
DDO~154 &(1) & 5.7$^{a}$ & 3.6$^{e}$& 0.8$^{a}$ & 8.5$^{j}$ & 0.028$^{b}$ & 1.34$^{b}$\\
NGC~3741 &(2) & 2.7$^{d}$ & 1.7$^{c}$& 0.9$^{d}$ & 13.1$^{d}$ & 0.078$^{d}$ & 0.7$^{d}$ \\
DDO~43 & (4) & - & 2.3$^{h}$ & 0.43$^{f}$ & 9.1$^{m}$ & 0.033$^{h}$ & 0.94$^{h}$\\
NGC~2366 & (3) & 10.5$^{b}$ & 6.5$^{e}$& 0.5$^{b}$ & 13.6$^{j}$ & 0.035$^{b}$ & 1.36$^{b}$ \\
DDO~168 & (5) & 21.2$^{i}$ & 2.6$^{g}$& 0.66$^{h}$ & 11.7$^{k}$ & 0.039$^{h}$ & 2.8$^{h}$ & \\
$*$IC~2574 & (3) & 24.6$^{a}$ & 14.8$^{e}$& 2.1$^{a}$ & 4.4$^{j}$ & 0.004$^{b}$ & 7.23$^{b}$ \\
\hline
\end{tabular}
{ $\Sigma_{\rm star}$ denotes the central stellar surface density and $M_{\rm gas}$ denotes the total gas mass of the galaxy. $R_{\rm HI}$ and $R_{\rm d}$ denote the extent of the $HI$ disk and the disk scalelength, respectively. $\rho_0$ is the core density and $R_{\rm c}$ is the core radius of the dark matter halo.\\
$*$IC~2574 is used as an counter-example, as it has a non-dense and non-compact dark matter halo.\\
Rotation curve (RC) refs: (1)-\citet{deB08};(2)-\citet{BCK05}; (3)-\citet{Oh08}; (4)-\citet{Oh15}; (5)-\citet{John15}.\\
Other parameter refs.: (a)-\citet{Ler08}; (b)-\citet{deB08}; (c)-\citet{Beg13}; (d)-\citet{BCK05}; (e)-\citet{Wal08}; (f)-\citet{HE04};
 (g)-\citet{Hun12};(h)-\citet{Oh15};(i)-\citet{John15}; (j)-\citet{Bage11}; (k)-\citet{Bro92}; (l)-\citet{Oh08}; (m)-\citet{SHN05}.}
\end{minipage}
\end{table*}

\subsection{Input parameters}
\label{input_chap3}
{\emph{DDO~154}}: The rotation curve 
is taken from \citet{deB08}. The stellar parameters $\Sigma_0$ (central surface density) and exponential scalelength (R$_{\rm d}$) are taken from \citet{Ler08} whereas the dark matter halo parameters are taken from \citet{deB08}. We use the $HI$ surface density distribution and a constant value for the gas velocity dispersion (8 km s$^{-1}$) from the maps given in \citet{Wal08}. The extent of $HI$ disk is taken from \citet{Bage11}.

{\emph{NGC~3741}}: The rotation curve, stellar parameters and the dark matter halo parameters are taken from \citet{BCK05} whereas the gas mass and the gas velocity dispersion value are taken from \citet{Beg13}. The extent of $HI$ disk is taken from \citet{BCK05}.

{\emph{DDO~43}}: The rotation curve is taken from \citet{Oh15}. The disk scalelength is from \citet{HE04}. The $HI$ surface density, surface density and the total amount of $HI$ gas and the dark matter halo parameters are given by \citet{Oh15}. Note that, since \citet{Oh15} did not consider stellar component (as it is very small) and carried out the mass modeling from the rotation curve, therefore, for internal consistency, we do not consider the stellar component. We assumed a constant $HI$ velocity dispersion of 8 km s$^{-1}$, typical of what we found for other dwarf galaxies in this sample. The extent of $HI$ disk is taken from \citet{SHN05}.

{\emph{NGC~2366}}: The rotation curve and the stellar parameters ($\Sigma_0$, R$_{\rm d}$) are taken from \citet{Oh08} whereas the dark matter halo parameters are from \citet{deB08}. The $HI$ surface density distribution and a constant value for the gas velocity dispersion (9 km s$^{-1}$) are taken from the maps given in \citet{Wal08}. The extent of $HI$ disk is taken from \citet{Bage11}.

{\emph{DDO~168}}: The rotation curve, the component-wise decomposition, stellar mass and the average gas velocity dispersion are taken from \citet{John15} whereas the disk scalelength is taken from \citet{HE04}. The dark matter halo parameters are taken from \citet{Oh15}. The $HI$ surface density is taken from LITTLE THINGS by \citet{Hun12}. The extent of $HI$ disk is taken from \citet{Bro92}.

{\emph{IC~2574}}: The rotation curve is taken from \citet{Oh08}. We take the stellar parameters from \citet{Ler08} and the dark matter halo parameters from \citet{deB08}. The $HI$ surface density distribution and a constant value for the gas velocity dispersion (7 km s$^{-1}$) are taken from the maps given in \citet{Wal08}. The extent of $HI$ disk is taken from \citet{Bage11}.

There are other isolated, dwarf galaxies, e.g. DDO~170 \citep{LSV90}, DDO~46 \citep{Oh15} which could be good candidates for our study since these have dense and compact halo, however all the parameters are not known observationally; hence
this restricts us to consider the above sample of five galaxies.

\subsection{Effect of dark matter halo : suppression of swing amplification}
\label{res_swing_chap3}

 Before going into the details of the result we point out that for the radii considered here, the rotation curve is assumed to be flat and thus $\eta$ = 1 is used for all cases. Also we use $X$ = 1 for all cases. The assumption of flat rotation curve is reasonably well justified for most of the regions of interest in the galaxies we considered here. 

The dependence of our finding with different values of $\eta$ is studied later in this paper (see \S~{\ref{var_eta}} ).

 Since the disk scalelength is different for these galaxies, and the observational data-points for the rotation curve and surface densities of stellar and gas are available for different ranges of radii, therefore maintaining exact uniformity in selecting radii is 
not possible. 
We point out that, at the radii chosen, the larger contribution to the baryonic mass is due to $HI$ and not stars. Hence, the one-component treatment outlined in \S~{\ref{swing_chap3}} is valid.

Also note that we have used the stellar disk scalelength, R$_{\rm d}$, as a measure of spatial size as is the usual practice in dynamical studies of galaxies, even though the main baryonic component of the disk is $HI$ gas. Similarly, following the
usual practice, we have given $HI$ extent in terms of stellar disk scalelength, R$_{\rm d}$ (Table~3.1).

\subsubsection{DDO~154}
Keeping  the criteria as given in \S~{\ref{res_swing_chap3}}, we choose three radii, namely, 4 R$_{\rm d}$, 5 R$_{\rm d}$, and 6 R$_{\rm d}$. At these radii, the surface density values for $HI$ ($\Sigma_{HI}$) are 3.9, 3, and 2.1 M$_{\odot}$ pc$^{-2}$, respectively \citep{Wal08}. The Holmberg radius ($R_{\rm Ho}$) for this galaxy is 2.1 $R_{\rm d}$.
 
\begin{figure}
    \centering
    \begin{minipage}{.5\textwidth}
        \centering
        \includegraphics[height=2.5in,width=3.5in]{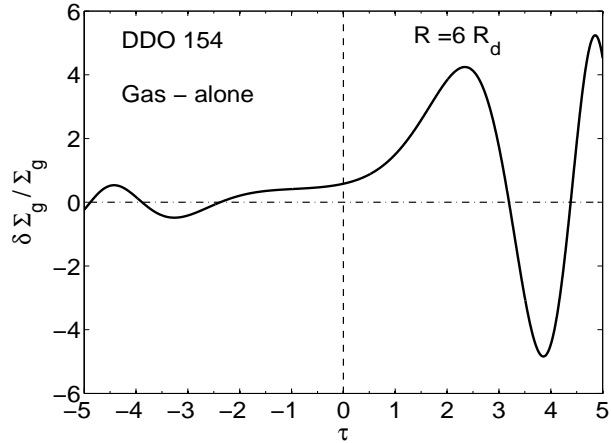}
       \vspace{0.2 cm}
	{\bf{(a)}}\\
    \end{minipage}
\begin{minipage}{.5\textwidth}
        \centering
        \includegraphics[height=2.5in,width=3.5in]{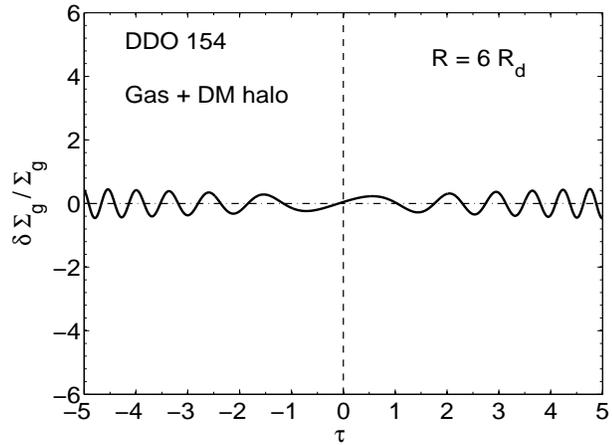}
        \vspace{0.2 cm}
	{\bf{(b)}}\\
    \end{minipage}
    \caption{DDO~154: Variation in $\delta \Sigma_{\rm g}/\Sigma_{\rm g}$, the ratio of the perturbed gas surface density to the unperturbed gas surface density plotted as a function of $\tau$, dimensionless measure of time in the sheared frame, at $R$ = 6 R$_{\rm d}$. (a)  For the gas-alone case ($Q = 1.3$) while (b) is for the gas plus dark matter halo case ($Q = 3.8$). While the gas-alone case shows finite amplification, the inclusion of the dark matter prevents the amplification, which would thus prevent the occurrence of strong small-scale spiral features.}
    \label{DDO154_first}
\end{figure}

Now we investigate the role of dominant dark matter halo using the following prescription:\\
 Using the rotation curve calculated from observed gas surface density values (as done in e.g. \citet{CF88}) and assuming a flat rotation curve (as mentioned in \S~{\ref{res_swing_chap3}}) we calculate $\kappa$ for gas-alone case and then use it to calculate the Toomre $Q$ value.
 Then we take the observed net rotation curve (also being flat, but with a higher value for the rotation velocity than for gas-alone case) which also includes the contribution of the dark matter halo \citep[see][]{deB08} to calculate $\kappa$ and the corresponding Toomre $Q$ value. Then for these two Toomre $Q$ values, we separately solve equation~(\ref{swing-final}) for these two cases. 
This procedure will tell us whether dark matter halo plays a significant role in suppressing  small-scale spiral features generated through the swing amplification mechanism. We choose 6 R$_{\rm d}$ for this detailed analysis and we find that $Q$ = 1.3 when the gas-alone contribution is taken and $Q$ = 3.8 when the net rotation curve is used for calculating $\kappa$. The resulting swing amplification is shown in Fig.~\ref{DDO154_first}.

From Fig.~\ref{DDO154_first} it is clear that when only  gas in the disk is considered, the system shows a finite swing amplification; which indicates that small-scale spiral structure is possible in the gas-alone system, but when we also include the dark matter halo in the system, the net $Q$ value is higher, and this almost completely suppresses the swing amplification, thus preventing the system from displaying small-scale spiral features.
For the other radii, namely, $R= 4R_{\rm d}$ and $R=5R_{\rm d}$, we got a similar behavior, i.e., the gas-alone system allows a finite growth of the non-axisymmetric perturbations, but the addition of dark matter halo prevents the growth of the perturbations. Therefore, for these two radii, we do not show any figure.

Thus the suppression of swing amplification and hence strong small-scale spiral arms by the dominant dark matter halo turns out to be a general result for the radii we considered for DDO~154. 
 This trend also continues well inside the optical radius (see \S~{\ref{var_toomre_q}}).

\subsubsection{NGC~3741}

We choose four radii, namely, 2.5 R$_{\rm d}$, 3 R$_{\rm d}$, 3.5 R$_{\rm d}$, and 4 R$_{\rm d}$. At these radii, the surface density values for $HI$ ($\Sigma_{HI}$) are 2.5, 1.5, 1, 0.5 M$_{\odot}$ pc$^{-2}$, respectively \citep{Beg13}. Note that the rotation curve of $HI$ disk is measured up to 8.3 R$_{\rm d}$ \citep[see fig. 1 in][]{BCK05}, but the $HI$ surface density data is limited to 4.5 $R_{\rm d}$ \citep[see fig. 5 in][]{Beg13}, thus restricting us to the above range chosen. The Holmberg radius ($R_{\rm Ho}$) for this galaxy is 1.6 $R_{\rm d}$.

\begin{figure}
    \centering
    \begin{minipage}{.5\textwidth}
        \centering
        \includegraphics[height=2.5in,width=3.5in]{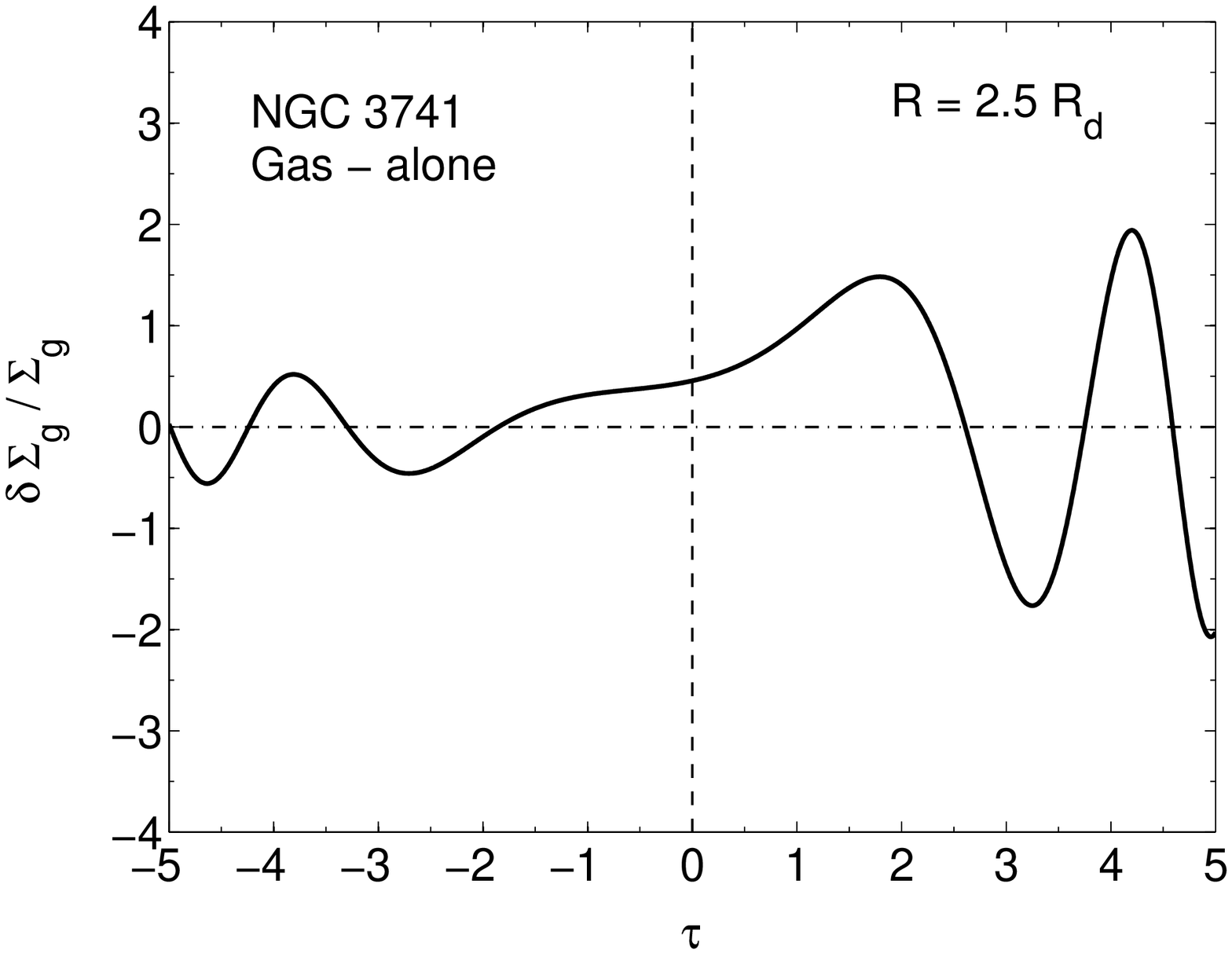}
       \vspace{0.2 cm}
	{\bf{(a)}}\\
    \end{minipage}
\begin{minipage}{.5\textwidth}
        \centering
        \includegraphics[height=2.5in,width=3.5in]{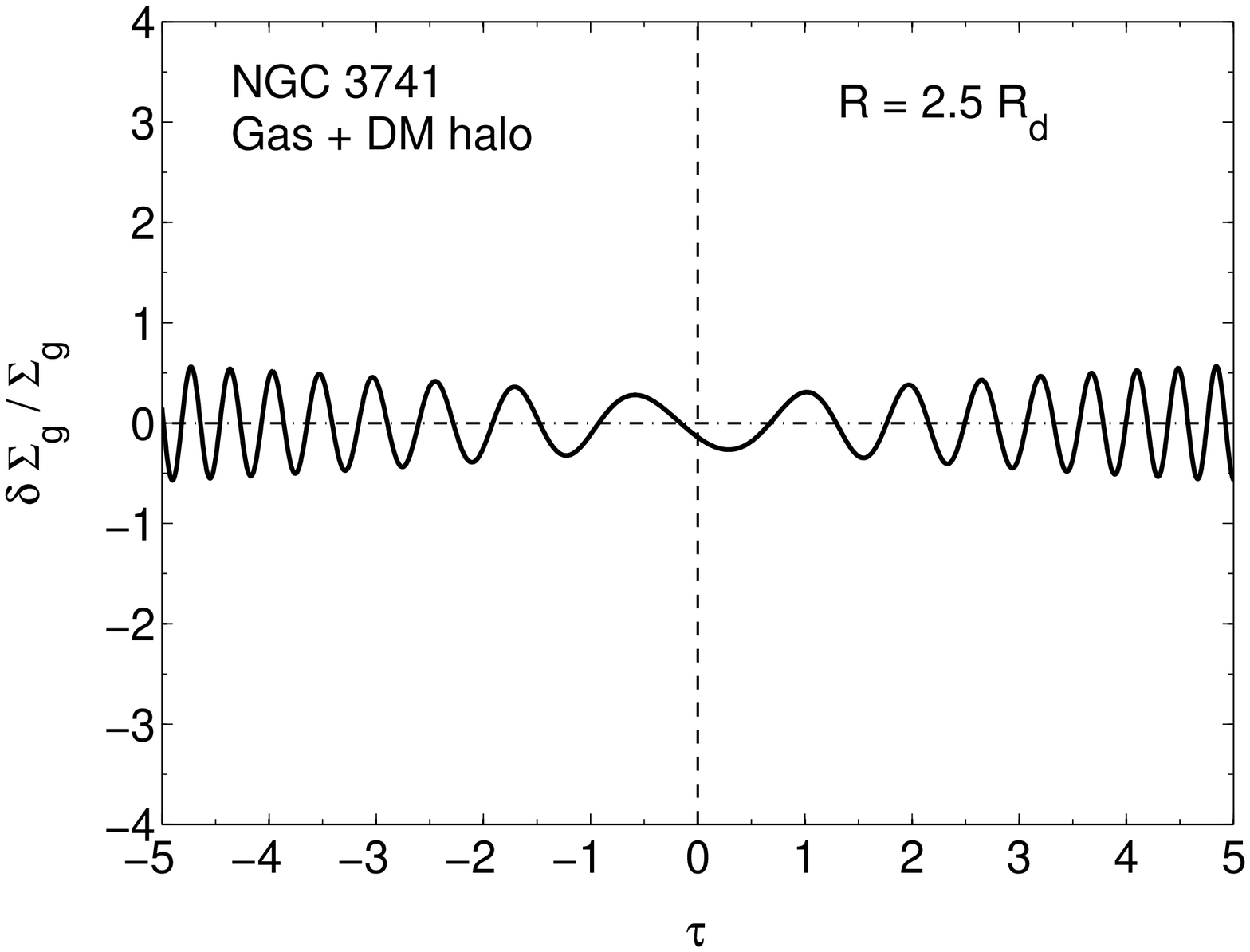}
        \vspace{0.2 cm}
	{\bf{(b)}}\\
    \end{minipage}
	\caption{NGC~3741: Variation in $\delta \Sigma_{\rm g}/\Sigma_{\rm g}$, the ratio of the perturbed gas surface density to the unperturbed gas surface density plotted as a function of $\tau$, dimensionless measure of time in the sheared frame, at $R$ = 2.5 R$_{\rm d}$. (a)  For the gas-alone case ($Q = 1.5$) while (b) is for the gas plus dark matter halo case ($Q = 5.3$). While the gas-alone case shows finite amplification, the inclusion of the dark matter halo prevents the amplification, which would thus suppress the occurrence of strong small-scale spiral features.}
    \label{fig-ngc3741}
\end{figure}

At $R=$ 2.5 R$_{\rm d}$, when the contribution of only gas to the rotation curve is taken into account, the Toomre $Q$ value is calculated to be 1.5, whereas by considering the net rotation curve we got the Toomre $Q$ value as 5.3. The resulting swing amplification is shown in Fig.~\ref{fig-ngc3741}.

From Fig.~\ref{fig-ngc3741} we see that gas-alone case allows a finite amplification in the perturbed gas surface density, but the addition of dark matter halo suppresses the amplification almost completely, thus indicating that small-scale spiral structure will be suppressed.  For other three radii, namely 3 R$_{\rm d}$, 3.5 R$_{\rm d}$, and 4 R$_{\rm d}$, when the net rotation curve (which represents the real galaxy) is used, the Toomre $Q$ values are found to be even higher, namely 8.3, 10.9 and 19.7, respectively. We checked that the values are too high to produce finite swing amplification. Therefore, dark matter halo is found to have a pivotal role in suppressing the small-scale spiral features at radii we considered for this galaxy. However, a weak spiral arm can be permitted in the disk (for details see discussion in \S~{\ref{occasional_spiral}}). This trend also continues well inside the optical radius (see \S~{\ref{var_toomre_q}}).

\subsubsection{DDO~43}

We choose three radii, namely, 6 R$_{\rm d}$, 7 R$_{\rm d}$, and 8 R$_{\rm d}$. At these radii, the surface density values for $HI$ ($\Sigma_{HI}$) are 5, 3, and 2.5 M$_{\odot}$ pc$^{-2}$, respectively \citep{Oh15}. The Holmberg radius ($R_{\rm Ho}$) for this galaxy is 4.1 $R_{\rm d}$.

\begin{figure}
    \centering
    \begin{minipage}{.5\textwidth}
        \centering
        \includegraphics[height=2.5in,width=3.5in]{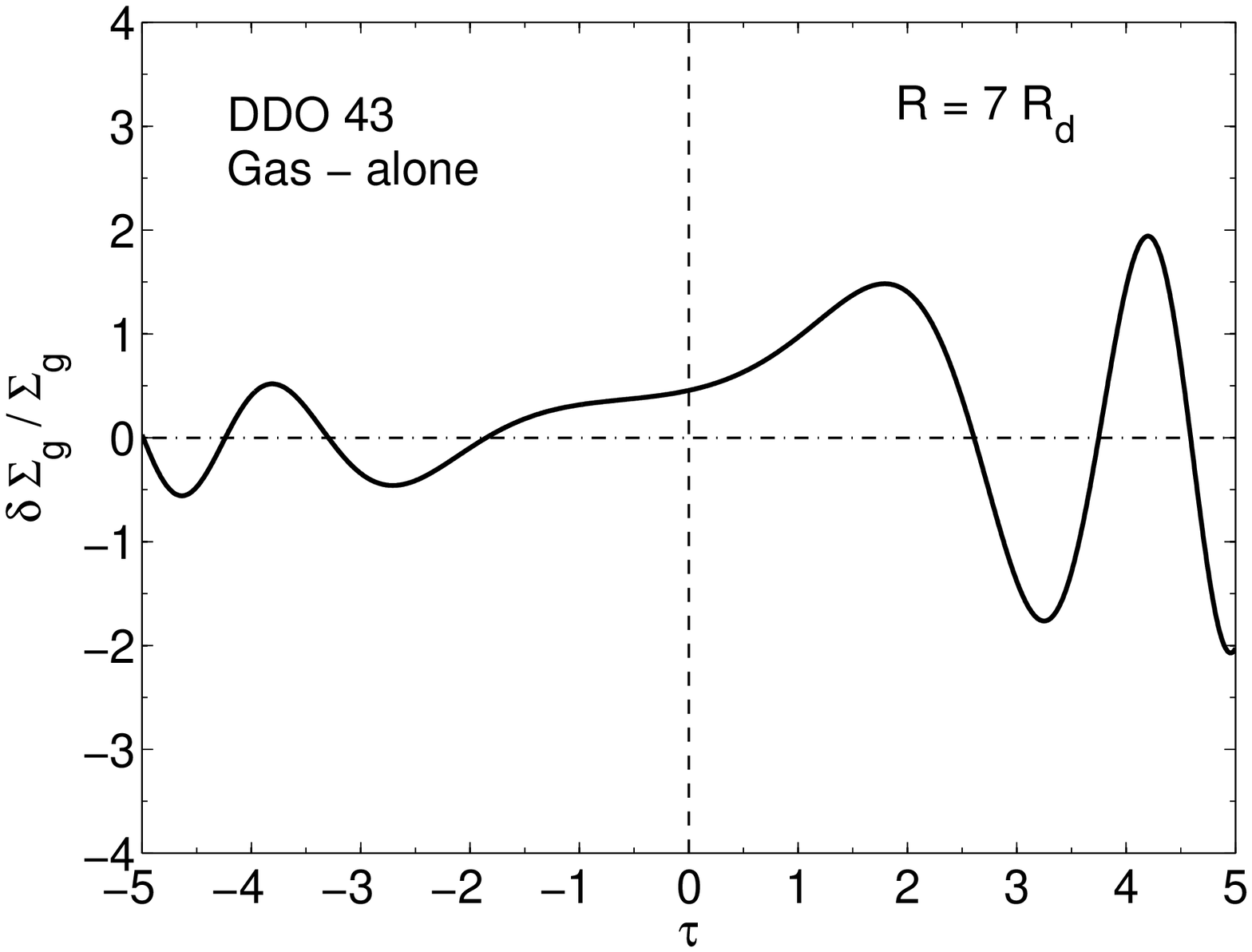}
       \vspace{0.2 cm}
	{\bf{(a)}}\\
    \end{minipage}
\begin{minipage}{.5\textwidth}
        \centering
        \includegraphics[height=2.5in,width=3.5in]{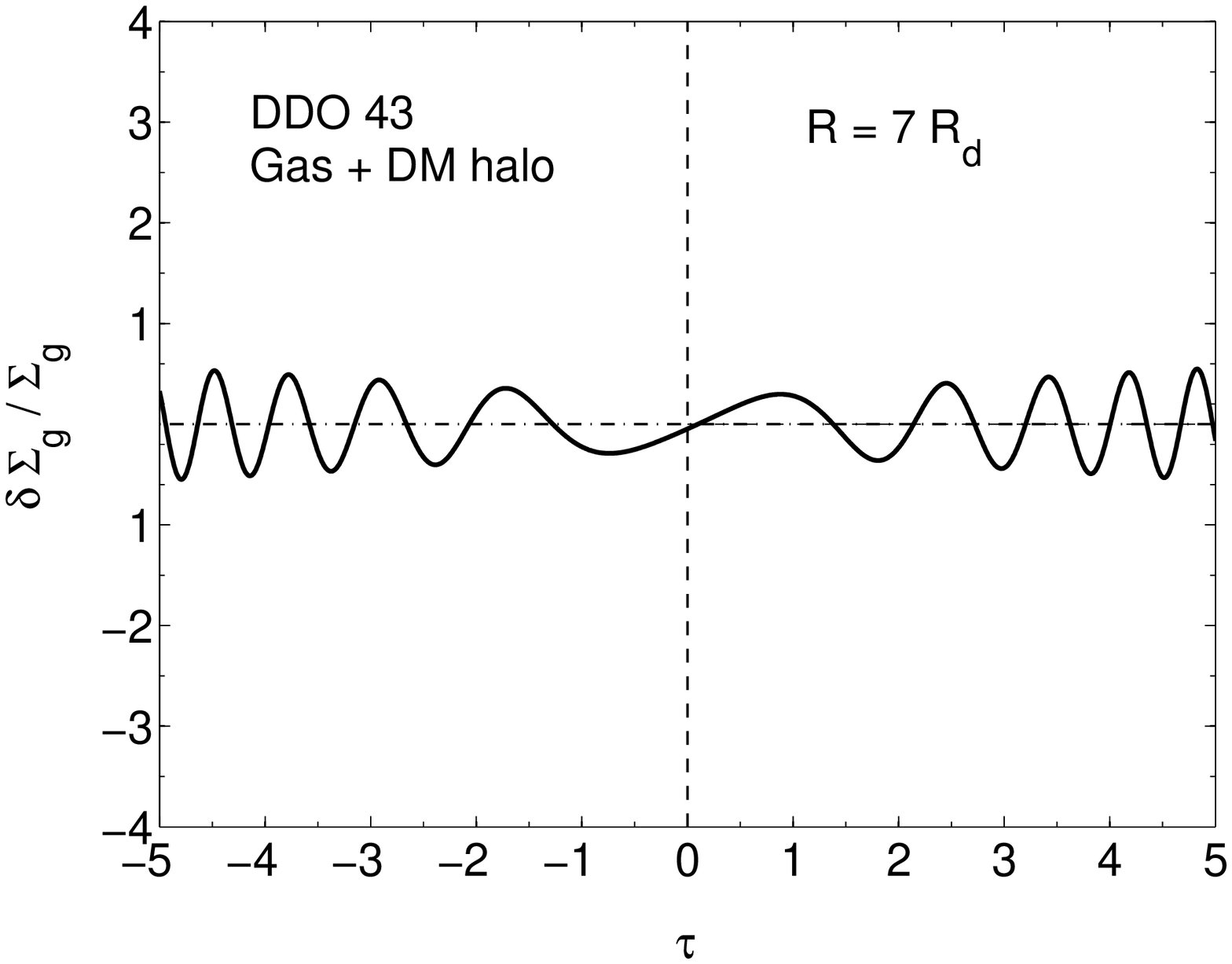}
        \vspace{0.2 cm}
	{\bf{(b)}}\\
    \end{minipage}
	\caption{DDO~43: Variation in $\delta \Sigma_{\rm g}/\Sigma_{\rm g}$, the ratio of the perturbed gas surface density to the unperturbed gas surface density plotted as a function of $\tau$, dimensionless measure of time in the sheared frame, at $R$ = 7 R$_{\rm d}$. (a)  For the gas-alone case ($Q = 1.5$) while (b) is for the gas plus dark matter halo case ($Q = 3.1$). While the gas-alone case shows finite amplification, the inclusion of dark matter halo prevents the amplification, which would thus suppress the occurrence of strong small-scale spiral features.}
    \label{fig-ddo43}
\end{figure}

 Following the same technique as mentioned in \S~{\ref{res_swing_chap3}}, we calculated the Toomre $Q$ parameter, first for the gas-alone case, and then for the gas plus dark matter halo case. For illustration purpose, we present the analysis for $R=7 R_{\rm d}$. At this radius, we solve equation~(\ref{swing-final}) by using the Toomre $Q$ parameter for both gas-alone and gas plus dark matter halo cases. This is shown in Fig.~\ref{fig-ddo43}. We find that, the gas-alone case allows finite swing amplification, but when the contribution of dark matter is taken into account, it completely damps the amplification that was earlier present in the gas disk.

For the other radii, namely, 6 R$_{\rm d}$ and 8 R$_{\rm d}$, we found the same trend as was seen for 7 R$_{\rm d}$, therefore, we do not present any figure for these cases. Hence, for this galaxy also, the dominant dark matter halo prevents the small-scale spiral structure at radii considered here. The role of dark matter on preventing the small-scale spiral arms in the regions well inside the optical disk is discussed in \S~{\ref{var_toomre_q}}.

\subsubsection{NGC~2366}

We choose four radii, namely, 6 R$_{\rm d}$, 7 R$_{\rm d}$, 8 R$_{\rm d}$ and 10 R$_{\rm d}$. At these radii, the surface density values for $HI$ ($\Sigma_{HI}$) are 7.6, 6.7, 4.9, and 2.8 M$_{\odot}$ pc$^{-2}$, respectively \citep{Wal08}. For all these radii we obtained the values of Toomre $Q$ parameter, and they are 2, 1.9, 2.4, and 3.4, respectively, by using the net observed rotation curve which includes the contributions of gas and dark matter halo. We did not find finite swing amplification at these four radii considered here. Thus dark matter turns out to be the key factor in preventing the small-scale spiral structure.

\subsubsection{DDO~168}
We choose three radii, namely, 3.5 R$_{\rm d}$, 4 R$_{\rm d}$, and 5 R$_{\rm d}$. The $HI$ surface density ($\Sigma_{HI}$) at these radii are 2.6, 1.3, and 1 M$_{\odot}$ pc$^{-2}$, respectively \citep{Hun12}. The Holmberg radius ($R_{\rm Ho}$) for this galaxy is 4.2 $R_{\rm d}$.

The values of Toomre $Q$ parameter for gas-alone case are calculated to be 3.7, 6.5, and 6.6, respectively. Then we calculated the values of Toomre $Q$ parameter for the gas plus dark matter halo case, and the values are 11.4, 20.4, and 19.8, respectively.

  We see that the Toomre $Q$ values are very high in both the gas-alone and gas plus dark matter halo cases as compared to the other galaxies. Note that, for this galaxy, the value of $\kappa$ is higher than the other cases and the $HI$ velocity dispersion (11 km s$^{-1}$) is also a bit higher (~7-8 km s$^{-1}$ being typical values) than the other galaxies. Also, the $HI$ surface density is lower than the other galaxies considered. Thus all factors comprise to produce  very high values of Toomre $Q$ parameter. We did not find any amplification at all at these radii. While the gas-alone case does not allow any swing amplification, the inclusion of dark matter in the system completely rules out any possibility for the system to host small-scale spiral structure at radii that we considered here.  
The role of dark matter and the low disk surface density on preventing the small-scale spiral features in the regions well inside the optical disk is discussed in \S~{\ref{var_toomre_q}}.

\subsection {Variation with $\eta$ -- departure from the assumption of flat rotation curve}
\label{var_eta}

So far we have considered only flat rotation curve which corresponds to $\eta = 1$. We note that for the galaxies considered here, sometimes the rotation curves are not strictly flat in the regions where we have carried out the analysis of swing amplification, e.g., see the rotation curve of DDO~43 in \citet{Oh15} which is not strictly flat, but is slowly rising at radii which we have considered here. We calculated the actual value of $\eta$ from the observed rotation curve of DDO~43 at $R=7 R_{\rm d}$, and the value of $\eta$ turned out to be $\sim$ 0.6. This indicates a departure from the flat rotation curve assumption.

 Here in this section, we study the dependence of our finding on the variation of the parameter $\eta$.

The parameter $\eta$ as defined earlier in \S~{\ref{swing_chap3}} can be expressed in detail as 
\begin{equation}
\eta = \frac{2A}{\Omega} = \frac{1}{\Omega}\left[\frac{v_{\rm c}}{R} - \frac{dv_{\rm c}}{dR}\right]
\end{equation}
\noindent where $v_{\rm c}$ denotes the circular velocity at radius $R$.

\begin{figure}
    \centering
    \begin{minipage}{.5\textwidth}
        \centering
        \includegraphics[height=2.5in,width=3.5in]{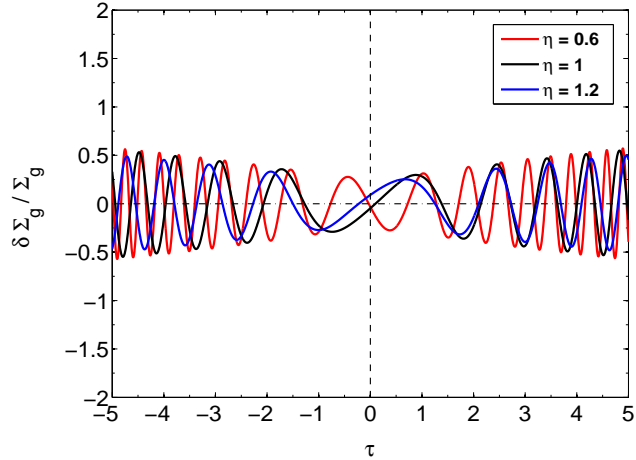}
    \end{minipage}
	\caption{DDO~43: The evolution of $\delta \Sigma_{\rm g}/\Sigma_{\rm g}$, the ratio of the perturbed gas surface density to the unperturbed gas surface density plotted as a function of $\tau$, dimensionless measure of time in the sheared frame at $R$ = 7 R$_{\rm d}$ ($Q$ = 3.1), with $\eta = 1.2$ and $0.6$, along with the standard case of $\eta = 1$. The suppression of small-scale spiral features by dominant dark matter halo remains unchanged with the variation of $\eta$ (for details see text).}
    \label{fig-variation_eta}
\end{figure}

Therefore, $\eta >1$ and $\eta < 1$ would imply a falling rotation curve and a rising rotation curve, respectively. For illustration purpose we choose $R=7 R_{\rm d}$ of DDO~43 where the Toomre $Q$ value for gas plus dark matter halo system is found to be 3.1. Then we carry out the swing amplification analysis with $\eta =0.6,0.8, 1.2$. The results for swing amplifications for $\eta = 1.2$ and $\eta = 0.6$ , along with $\eta =1$ are shown in Fig~\ref{fig-variation_eta}.

From Fig.~\ref{fig-variation_eta} it is clear that for $\eta = 1.2$ and $\eta = 0.6$, the corresponding solutions obtained from equation (\ref{swing-final}) remains indistinguishable from that obtained for $\eta = 1$. Therefore, even if the rotation curve is either slightly falling or rising, the main finding of this paper remains unchanged. In other words, if for a radius $R$ and for $\eta = 1$, the dominant dark matter halo is found to suppress the small-scale spiral features, the same will also hold for $\eta > 1$ and $\eta < 1$.

The general trend of change in swing amplification for different $\eta$ values is as follows:\\
For $\eta$ $< 1$ (i.e. for rising rotation curve), the pressure term stops the growth of the perturbations at an early epoch and hence would result in more open spiral structure, and for $\eta >1$ (i.e. for falling rotation curve) the effect will be opposite to the case for $\eta <1$ \citep[e.g., see][]{GLB65,Jog92}.

Here, the Toomre $Q$ value is so high that the system will not be able to support any swing amplification, and hence the issue of relative openness of the spiral structure (as generated by swing amplification) will appear for this work.  Thus, variation of $\eta$ has little impact on the main finding of this paper. The high Toomre Q values effectively damp swing amplification for any of the $\eta$ values considered here.

\subsection { Suppression of spiral structure at all radii-- a generic trend ?}
\label{var_toomre_q}

 So far, we showed the effect of dark matter halo on suppression of the strong small-scale spiral features at some radii for each galaxy considered. 
The set of radii we chose for the calculation of swing amplification lie all outside the optical disk of the respective galaxies. 
Hence, the question remains whether this effect of dark matter halo on prevention of strong small-scale spiral structure is strictly valid only in the outer disks or this trend also holds true for regions well inside the optical disk? 

\begin{figure*}
\begin{multicols}{2}
    \includegraphics[width=\linewidth]{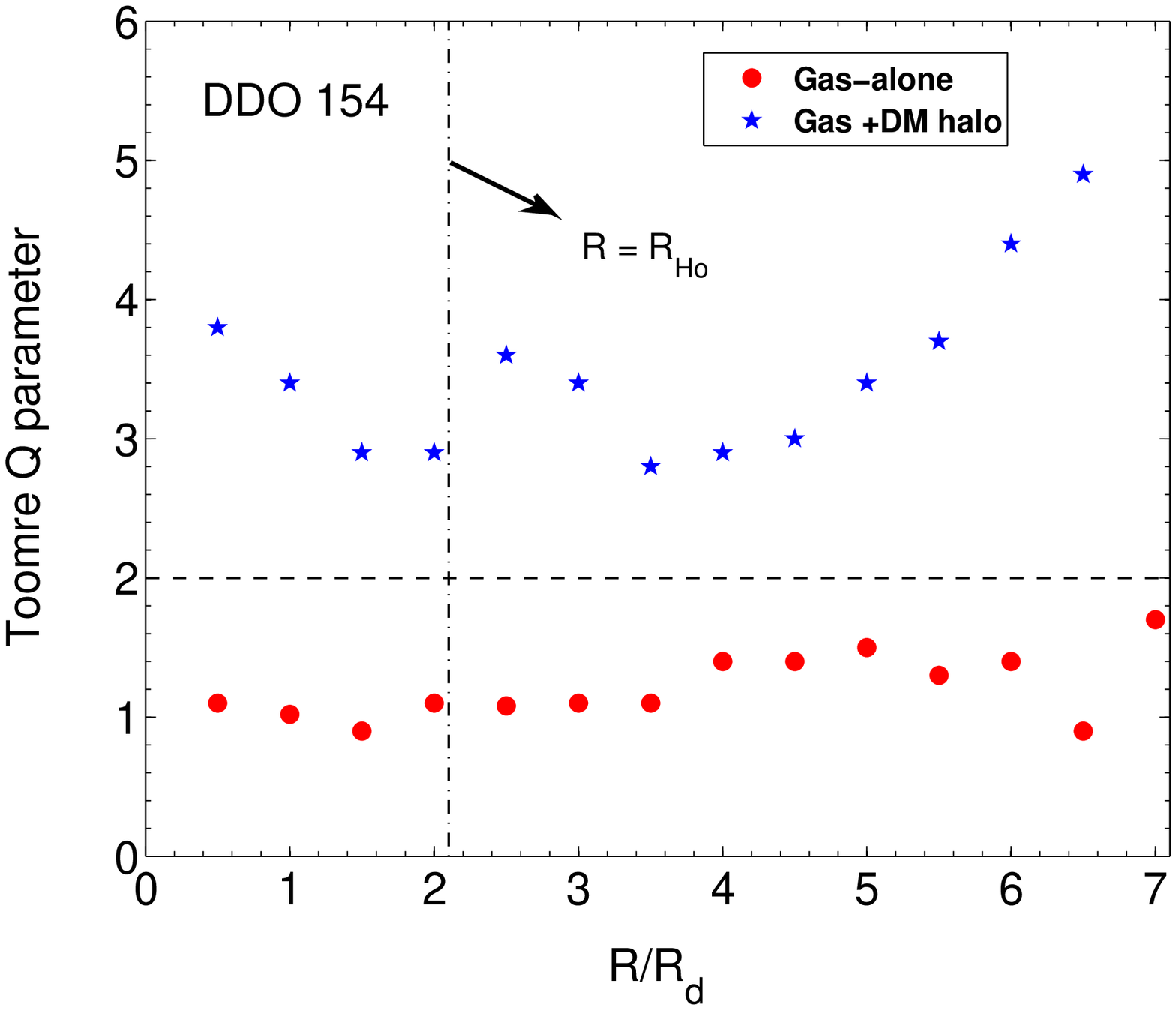}\par 
    \includegraphics[width=\linewidth]{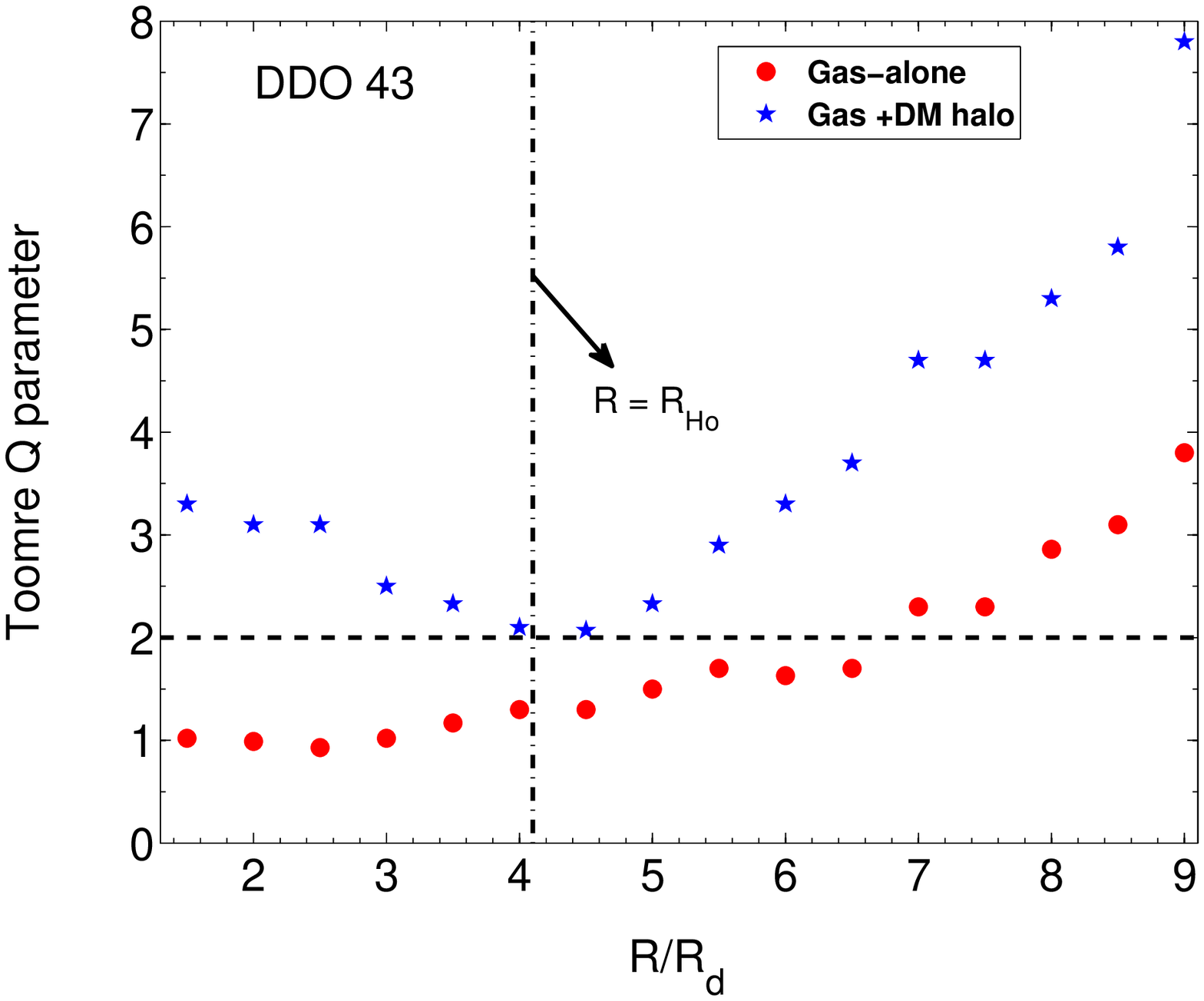}\par 
    \end{multicols}
\begin{multicols}{2}
    \includegraphics[width=\linewidth]{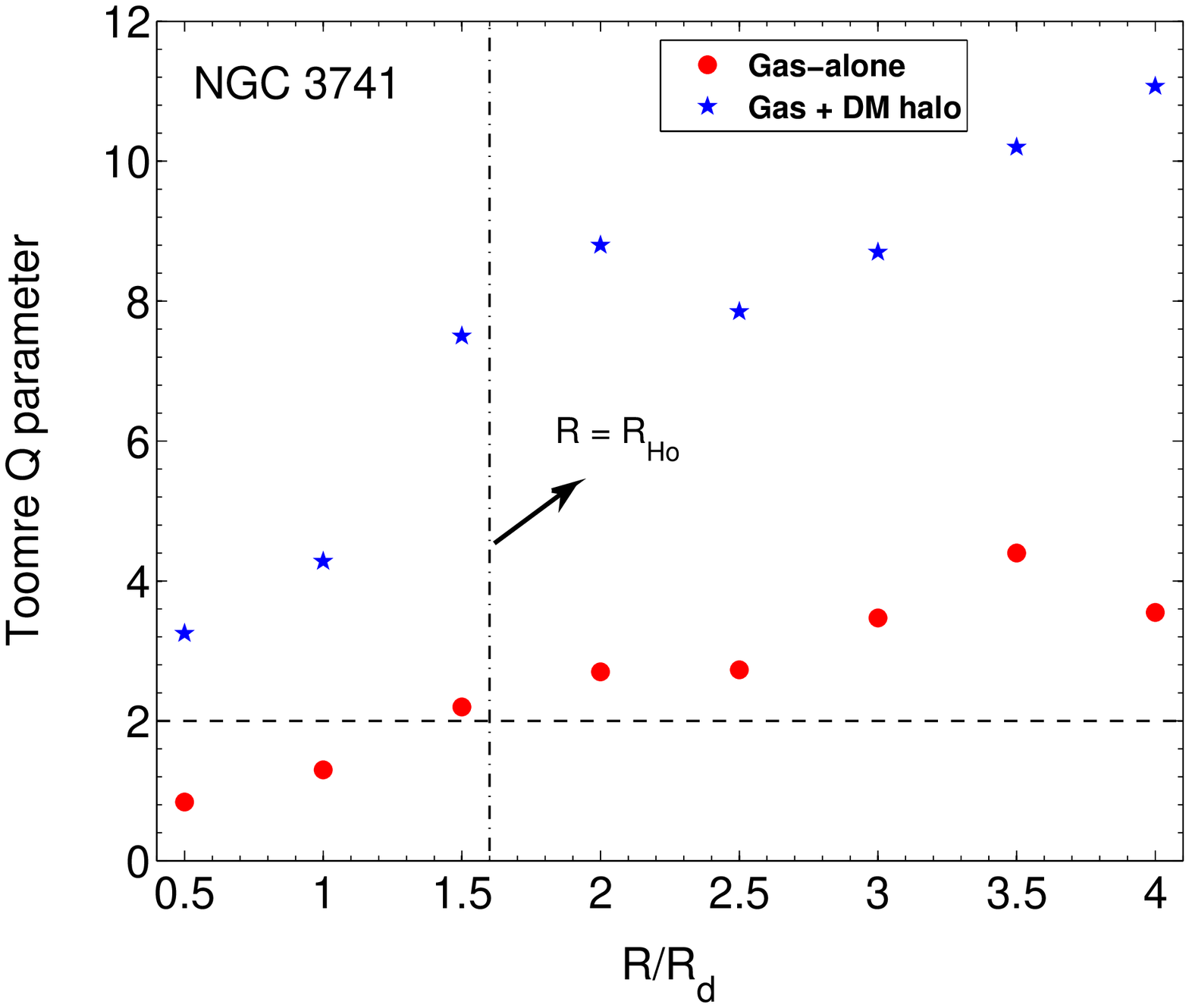}\par
    \includegraphics[width=\linewidth]{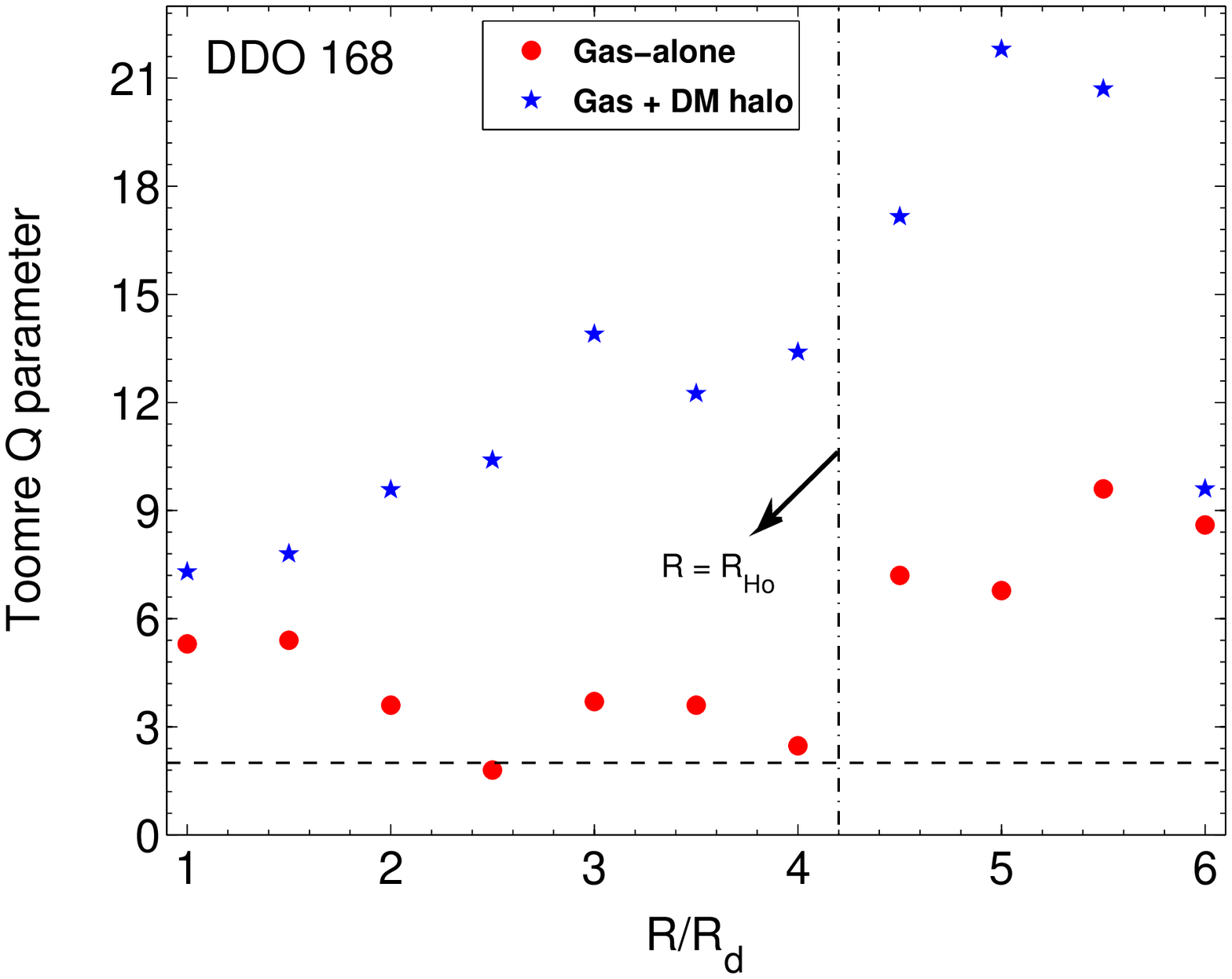}\par
\end{multicols}
\caption{Toomre $Q$ parameter for gas-alone and gas plus dark matter halo case, plotted as a function of radius for galaxies DDO~154, DDO~43, NGC~3741, and DDO~168.
The vertical line shown in each plot indicates the Holmberg radius ($R_{\rm Ho}$) which denotes the optical extent
for each galaxy. The horizontal dashed  line indicates $Q =2$ for each case which is taken to separate the region which allows swing amplification ($Q <2$) from region where it is suppressed ($Q>2$). The vertical dashed line denotes the Holmberg radius in each case.
Thus, this figure shows that the dark matter suppresses swing amplification at all radii, even inside of the optical radius in each case.}
\label{fig-q_plots}
\end{figure*}

To address this question, in this section we calculated the Toomre~$Q$ parameter for gas-alone case ($\kappa$ for only the gas component) and gas plus dark matter halo ($\kappa$ from the observed rotation curve) for a wide range of radii, ranging from the very  inner regions to the outermost possible region. This is shown in Figure~(\ref{fig-q_plots}).

 We point out that, here we relaxed the assumption of flat rotation (used in Section 3.2 to calculate $\kappa$), and calculated the $\kappa$ values from the observed rotation curve using the standard definition $\kappa^2 = -4B(A-B)$, where A, B are Oort constants. 
When calculating the radial variation of $Q$, it is necessary to take account of $\kappa$ as dictated by the observed rotation curve. 
This difference in calculating $\kappa$ will give the Toomre $Q$ values slightly   different in these two cases. However, we checked that this difference is small, and will not alter the findings presented in the earlier subsections. 

The dashed horizontal line in each case denoted $Q=2$, which is taken to indicate the threshold so that $Q < 2$ permits swing amplification while it is damped for $Q > 2$, as is typically done \citep[e.g.][]{Too81}. However we caution that $Q = 2$ is an indicator but not a rigorous cut-off : 
 \citet{Too81} showed that for disks with {\it flat rotation curve}, $1\le X \le 3$ and $Q \le 2$ are the necessary and sufficient condition for swing amplification factors of more than a few \citep[also see][]{BT87}. However, we point out that the rotation curves of these dwarf galaxies are not strictly flat, therefore the above condition will not be exactly valid. 

 The vertical dashed line denotes the Holmberg radius, R$_{\rm Ho}$ in each case. We show next that the dark matter damps swing amplification even at radii inside of the optical radius in most cases.

 We note that even in the very inner regions of the most of our sample galaxies, the gas surface density dominates the baryonic component.

Here, we summarize the main results that we found from the study of the variation of Toomre $Q$ parameter as a function of radius, for each galaxy.\\

{\it{DDO~154 :}} The Toomre $Q$ parameter for the gas-alone case turned out to be less than $Q =2$, for all radii shown in Figure~(\ref{fig-q_plots}), thus indicating that the gas disk is able to support the small-scale spiral features even within the optical disk. But the addition of dark matter produces a larger value of Toomre $Q$ parameter ($Q \sim 4$ being the typical values), and hence the system will no longer be able to produce the strong small-scale spiral features generated via swing amplification mechanism at all radii, even inside of the optical radius, R$_{\rm Ho}$.

{\it{DDO~43 :}} Here also for the gas-alone case, the values of Toomre $Q$ parameter are less than $Q = 2$ up to $R = 6.5 R_{\rm d}$ which also includes the optical disk and hence the system can display small-scale spiral structure. The addition of dark matter in the calculation increases the values of Toomre $Q$ parameter (see Figure~(\ref{fig-q_plots})), thus the suppressing effect of dark matter halo on 
local spiral structure is seen at all radii.

{\it{NGC~3741 : }}The Toomre $Q$ parameter for the gas-alone case rises gradually to $Q \sim 4$ in the outermost point considered here. The addition of dark matter makes the values of Toomre $Q$ parameter as high as $11$ (a factor of $\sim$ 3 increment in Toomre $Q$ value) in the outer part (see Figure~(\ref{fig-q_plots})). As indicated by the Toomre $Q$ for gas-alone case in the regions outside the optical disk, the gas disk still may display some weak small-scale spiral structure, the addition of dark matter rules out the possibility of having {\it small-scale} spiral features (but also see the \S~{\ref{occasional_spiral}}). Again, the suppressing effect of dark matter halo on 
local spiral structure is seen at all radii, even inside of the optical radius, R$_{\rm Ho}$.

{\it{DDO~168 : }} The Toomre $Q$ parameter is found to be greater than $Q =2$ for all radii. The addition of dark matter makes the Toomre $Q$ parameter as high as $Q \sim 21$ in some radii. Hence, even though some weak small-scale spiral structure is possible to exist in the gas disk, the dominant dark matter completely rules out the possibility of having small-scale structure. Also, we note that the low disk surface density is another simultaneous possible reason (along with dominant dark matter halo) for not supporting any strong small-scale spiral features in the disk, as indicated by higher values of Toomre $Q$ parameter for gas-alone case.

Thus, we have shown that dark matter halo plays a vital role in preventing the strong small-scale spiral features in the outer disk as well as regions inside the optical disk. Also, for some cases, we found that the low disk surface density plays an important role along with the dark matter halo to prevent the small-scale spiral structure (e.g see cases of DDO~168 and NGC~3741 in Figure~(\ref{fig-q_plots})).

\subsection{A counter-example : IC~2574}

\begin{figure}
    \centering
    \begin{minipage}{.5\textwidth}
        \centering
        \includegraphics[height=2.5in,width=3.5in]{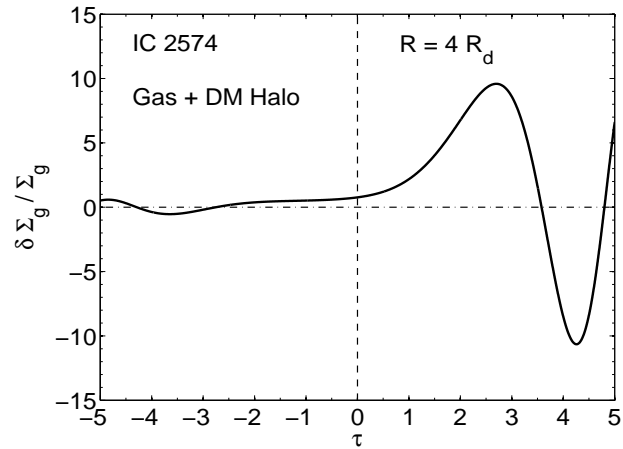}
    \end{minipage}
	\caption{IC~2574: variation in $\theta_{\rm g}$ = $\delta \Sigma_{\rm g}/\Sigma_{\rm g}$, the ratio of the perturbed gas surface density to the unperturbed gas surface density plotted as a function of $\tau$, dimensionless measure of time in the sheared frame at $R$ = 4 R$_{\rm d}$ ($Q$ = 1.2). The Toomre $Q$ value is calculated by taking the contributions of both gas and dark matter halo. The disk still allows a finite amplification, thus resulting in small-scale spiral features.}
    \label{fig-ic2574}
\end{figure}

 From Table~3.1, it is clear that the dark matter halo parameters for IC~2574 are quite different from the rest of the sample galaxies, i.e., it has a dark matter halo which is not dense and not compact. We point out that in the inner part ($R$ $\le$ 5 kpc) of this galaxy, baryons (stars and gas) are able to produce the observed rotation curve and only in the outer parts dark matter halo contribution takes over \citep[for details see fig.21 in][]{Oh08}. This is in a sharp contrast with the other galaxies in our sample where the dark matter halo is found to dominate from the innermost region. Therefore it is interesting to investigate the role of the dark matter halo on swing amplification process in such a different situation. We choose three radii, namely, 3 R$_{\rm d}$, 3.5 R$_{\rm d}$, and 4 R$_{\rm d}$. At these radii, the surface density values for $HI$ ($\Sigma_{HI}$) are 8, 6.5, and 4.9 M$_{\odot}$ pc$^{-2}$, respectively \citep{Wal08}. The Toomre $Q$ values for the gas plus dark matter halo case at radii 3 R$_{\rm d}$ and 3.5 R$_{\rm d}$ turn out to be less than one, indicating that the gas disk is unstable against the local axisymmetric perturbation and also is likely to host strong small-scale spiral structure. We note that the maximum rotation velocity is similar to those other galaxies, but due to a larger value of R$_{\rm d}$, positions of 3R$_{\rm d}$ and 3.5 R$_{\rm d}$ are further out as compared to the other galaxies, implying a lower value of $\kappa$ (as for a flat rotation curve $\kappa$ inversely proportional to the radius $R$); combinations of all these yield such a low value for Toomre $Q$. At $R$ = 4 R$_{\rm d}$, $Q$ is found to be 1.2, and it allows the system to have a finite swing amplification. This is shown is Figure~\ref{fig-ic2574}.

Thus for IC~2574,  where the dark matter halo does not dominate  from the innermost radii, we find that 
the galaxy is likely to host small-scale spiral features. This could explain the spiral structure seen in IC~2574 which becomes visible when long exposure time is used \citep{dede64}.

\section{Discussion}
\label{dis_3}
\subsection{Occasional presence of spiral arms}
\label{occasional_spiral}

Our present analysis clearly shows that the dominant dark matter halo prevents almost completely the small-scale spiral structure by making the swing amplification process inefficient. However we note that one of our sample galaxies, namely, NGC~3741 shows a 
spiral arm in the disk, 
albeit it is faint and so in view of the results in our paper, its origin
remains a puzzle.

We point out that the swing amplification process stops being effective when $Q$ is greater than  2.5 and value of $X$ being greater than 3 \citep{Too81,SC84} and outside this range of parameter values the amplification will not be very strong \citep{BT87}. However it does not necessarily nullify the possibility of having occasional spiral features, as say triggered
by tidal encounters  \citep{ToTo72,BT87}.  The disk still can support occasional small-scale features although the higher values of $Q$ indicate that the response of the disk will not be very strong, thus making the spiral features weak \citep[for a detailed discussion, see \S~4.1 in][]{GJ14}. 

Some dwarf galaxies, which are not included in our sample, also show spiral structure. For example, the rotation curve decomposition of UGC~2259 shows that the behavior is typical of HSB galaxies \citep[see fig.9 in][]{CSV88}. It shows a grand-design spiral structure, although  faint: it is interesting to see that galaxies with such low luminosity (M$_B$ $\sim$ -16) are still susceptible to spiral instabilities \citep[for detailed discussion see \S~5 in][]{Ash92}.

 Also we caution the reader that here we are concerned about only late-type dwarf irregular galaxies 
which are rotationally supported.
Some of the early-type dwarf galaxies do show some spiral features \citep[e.g., see][]{Jer00,Gra03,Lis09}, but those are largely pressure-supported, and not rotationally-supported. Hence, they are dynamically different from the ones we consider in this paper.

Secondly note that two of our sample galaxies host a bar, e.g, DDO~154 and NGC~3741. 
However, the measured pattern speed of the bar is low, as in NGC~3741 \citep{Ban13}. This could be because of the dynamical friction due the dark matter halo which slows down the rotation speed of the bar significantly, as  proposed by \citet{DeSe98}. Bars can excite the transient spiral arms in the disk and it can heat up the disk as well, as was shown for HSB galaxies \citep{STT10}. The present paper shows that 
even when a bar exists, it seems unable to trigger strong spiral arms in our sample galaxies, perhaps these are suppressed by the the dominant, dense and compact dark matter halo.

\subsection{Other issues}

Here we mention a few other points regarding this work.

First of all, galaxies with highly extended $HI$ distributions which are mainly studied here are not too common \citep{BCK05}, however so far no systematic study seems to have been done in the literature to see how common such galaxies are. Such a study ( Faint Irregular Galaxies GMRT survey 2 -- FIGGS2) has now been started by \citet{Pat16}.
Therefore, an important question is: whether our finding is valid only for those dwarf irregular galaxies with extremely large $HI$ disks, or can it also be applicable to more typical dwarf irregular galaxies with moderately extended $HI$ disks?

In order to address this issue, it should first be noted that the size of $HI$ disk present in a galaxy sensitively depends on the limiting $HI$ column density, as the latter sets the limit up to which one can trace the $HI$ extent. 
For example, in NGC~3741, the measurement were done to the limiting $HI$ column density of 10$^{19}$ cm$^{2}$, while for DDO~154, NGC~2366 and IC~2574 taken from the THINGS sample \citep{Wal08}, the limiting value is typically 4 $\times$ 10$^{19}$ cm$^{-2}$. For LITTLE THINGS (from which DDO~168, DDO~43 are taken), in some of their samples (e.g., DDO~126, DDO~155) $HI$ column density is measured as low as a few times 10$^{17}$ cm$^{-2}$\citep[see fig.~5 in][]{Hun12}. For a detailed discussion of the size of the $HI$ disk see section 12.2 in \citet{GioHay88}. This choice affects the value of the outermost radius to which $HI$ is detected (see Table~3.1).

Further, in typical HSB galaxies, the optical radius is about 4--5 times the disk scalelength \citep{BM98}. This follows from the fact that the central luminosity is nearly constant and the outer
limit is set by the limiting magnitude of 26.5 mag arcsec$ ^{-2}$ \citep[see e.g.][]{vase81}; also for a detailed discussion see \citet{NJ03}. However in the dwarf irregular galaxies, the ratio can vary from 1 to 5 \citep{Swa99}. The reason is that the central value could be much lower while the outer limiting magnitude set by the sky brightness remains the same.
\citet{Swa99} showed that the ratio of the radial $HI$ extent to the optical size in dwarf irregular galaxies is $\sim 1.8$ similar to the ratio of $\sim 1.5$ for larger galaxies. We checked from the references used in Table~3.1 that, the ratio $R_{ HI}/ R_{\rm Ho}$ for DDO~154 and NGC~3741 is 4 and 8.3, respectively. For the other galaxies, namely, NGC~2366, DDO~43, and DDO~168, this ratio is 2.3, 2.2, and 2.8, respectively.

However, even the dwarfs that do not have very extended $HI$ distributions,  have a high $M/L$ ratio of $\sim 10$ \citep[e.g. see][]{CCF00}. Some of these dwarf galaxies do have a dense, compact halo (e.g., UGCA 442) where the dark matter halo dominates the rotation curve even at the inner radii.
 Thus the main result from this paper, namely, the dominant dark matter halo suppressing the generation of strong, small-scale spiral structure via swing amplification, is also valid for a much larger number of dwarf galaxies than the prototypical galaxies like DDO 154 and NGC 3741 that we have highlighted in our work.
However,
 the radial distance over which the dark matter halo dominates is not so large unlike the sample we have studied, so the cushioning influence
of the dark matter halo in preventing spiral structure, and resultant galaxy evolution, would not be as strong as in the case of sample galaxies studied in our work.

 We note that there are many other dwarf galaxies which have a lower density, non-compact halo (with $R_c/R_d > 3$), such as
the counter-example IC 2574 considered here, or HO II \citep[see e.g.][]{BJ13}. The dark matter halo in such galaxies is likely to play little role in the spiral structure formation in these.

Secondly, in any late-type galaxy stars and gas co-exist, and hence a more realistic approach would be to model the galactic disk as a gravitationally coupled two-fluid system and then do the non-axisymmetric perturbation analysis \citep[as done in][]{Jog92}. For the radii that we have considered here, $HI$ surface density greatly dominates over stars, e.g., for DDO~154, at $R$ = 6 R$_{\rm d}$, the stellar surface density is 0.015 M$_{\odot}$ pc$^{-2}$ as compared to a $HI$ surface density of 2.1 M$_{\odot}$ pc$^{-2}$. Therefore the system effectively reduces to a one-component system. This justifies our use of one-component formalism in \S~{\ref{swing_chap3}}.

\section{Conclusion}
\label{conclu_3}
In summary, we have carried out a 
 local, non-axisymmetric perturbation analysis for a sample of five late-type
dwarf irregular galaxies with extended $HI$ disks, and which have a dense and compact dark matter halo. 
  We show that when the gas disk is treated as an isolated system (with no dark matter included), the disk allows the growth of non-axisymmetric perturbations via swing amplification, but  the addition of a dominant dark matter halo in the analysis results in a higher rotational velocity and hence a higher Toomre $Q$ parameter which prevents the amplification almost completely. We further calculated the values of Toomre $Q$ parameter for a wide range of radii for these galaxies and showed that even in the regions inside the optical radius, the dominant dark matter halo yields  higher Toomre $Q$ values than the gas-alone case.
This naturally explains why these galaxies do not generally show strong small-scale spiral features despite being gas-rich.

This suppression of spiral features inhibits further dynamical evolution, that would otherwise have been expected via angular momentum transport due to spiral features (see \S~{\ref{intro3}}). This could be of interest in understanding early galaxy evolution, since in the hierarchical scenario of galaxy formation the smaller galaxies form first and then merge to form larger galaxies.
The  dwarf irregular galaxies, being  small, low-metalicity, gas-rich systems; are believed to be the present-day analogs of the high-redshift unevolved galaxies. Hence understanding the dynamical evolution of dwarf irregular galaxies could possibly give some insight in to the early evolution of high-redshift galaxies.

 \citet{GJ14} showed that for the LSBs, where the stars dominate the baryonic disk, the dense and compact halo that dominates the disk at all radii suppresses the strong small-scale spiral features almost completely. 
{\emph {Thus, whether the larger baryonic contribution is in the form of stars or in the low velocity dispersion component $HI$ gas, in the region where the galaxy is dominated by dark matter halo, 
the disk will not be able to host small-scale spiral structure. }}

\thispagestyle{empty}

\chapter[Role of gas in supporting grand spiral  structure]{Role of gas in supporting grand spiral  structure {\footnote {based on Ghosh \& Jog, 2015, MNRAS, 451, 5868}}}
\chaptermark{\it Role of gas in grand spiral structure}
\vspace {2.5cm}

\section{Abstract}
The density wave theory for the grand two-armed spiral pattern in galaxies 
is successful in explaining several observed features.
 However, the long-term persistence of this spiral structure is a serious problem  since the group transport would destroy it within about a billion years as shown 
in a classic paper by Toomre.
In this paper, we include the low velocity dispersion component, namely gas, on an equal footing with stars in the formulation of the density wave theory, and obtain the dispersion relation for this coupled system.   
We show that the inclusion of gas makes the group transport slower by a factor of few, thus allowing the pattern to persist longer -- for several billion years. Though still less than the Hubble time,
this helps in making the spiral structure more long-lived. 
Further we show that addition of gas is essential to get a stable wave for the observed pattern speed for the Galaxy,
 which otherwise is not possible for a one-component stellar disk.

\section{Introduction} 
\label{intro4}

The grand two-armed spiral pattern as seen in M~81 or M~51 makes a striking visual impression. Although these spiral patterns  have been studied for over five decades, their origin and persistence are still not fully understood.

It was realized early on that a material spiral feature would get wound up in a few rotation periods, due to the differential rotation in a galactic disk. Since spiral features are commonly seen, it was
proposed by several authors starting from B. Lindblad, and others including \citet{LS64,LS66}, that at 
least the grand spiral patterns seen in spiral galaxies are density waves governed mainly by gravity. In this theory, the spiral pattern is claimed to be stationary which gets around the winding problem.
Thus the spiral pattern is claimed to last for times much longer than rotational time-scales. Further, \citet{LYS69} successfully interpreted some of the observable features of spiral galaxies  by this theory. For a good exposition of the density wave theory, see \citet{Roh77}.

However, several questions have been raised about the validity of this theory, such as whether or not these density waves
 are truly stationary \citep{Too69}, and whether galaxies
 indeed admit spiral waves as self-consistent modes of oscillation \citep{LBO67}, see \citet{Pas04} for a review of this topic.
Over the years, many additional aspects have been studied that underline the complexity of the spiral structure, see e.g. the reviews by \citet{Too77,Sel03,DB14}. Bars have been suggested as a possible mechanisms responsible for the origin of density wave \citep{SR97,Ath12}. Numerical simulations \citep{SH76,CG85} have shown how even a weak barred potential can trigger a spiral perturbation in the gaseous component.
A tidal encounter generally leads to a global $m=2$ spiral pattern \citep{ZRR93,SDH11,DB14}. Many galaxies
also show flocculent spiral arms which can be explained as being  transient, material spiral features that arise due to swing amplification of non-axisymmetric perturbations, as originally proposed  by \citet{GLB65}, also see \citet{Too81}.
In many cases both density waves as well as transient, material spiral arms can co-exist which makes the analysis and application to a particular galaxy more complicated. A galaxy may show more than one pattern speed for $m=2$ \citep[e.g.,][]{Ger11}.
Thus, the subject of density wave theory as applied to galaxies is complex.

In this paper, we will focus on  one specific issue, namely the effect of gas on the existence and the group transport of spiral density waves. 
In a classic paper, \citet{Too69} showed that any packet of such waves moves 
radially, and towards  increasingly shorter wavelengths, with a group velocity that is sufficient to destroy the wave packet itself within a few galactic revolutions. This was a major setback to the persistence of `stationary' density waves as proposed by \citet{LS66}. 
Since it is known observationally that regular, grand spiral features are common, 
Toomre argued that the presence of a mechanism to regenerate those grand features is required, such as due to tidal forcing, for the density wave picture to be saved. \citet{MR87} addressed the question of persistence of the grand spiral structure by means of a wave packet with vanishing group velocity.

A typical spiral galaxy also contains a low velocity dispersion component, namely gas, in addition to stars. For a more realistic and complete treatment of galactic dynamics, the dynamical effect of gas needs to be taken into account. The addition of gas is shown to make the disk significantly more unstable as shown for local, axisymmetric perturbations \citep{JS84a,JS84b,BR88,Jog96}. In the non-axisymmetric case,  gas  has a strong effect on the resulting swing amplification \citep{Jog92}, and it is shown to result in broader arms as observed
\citep{Sch76}.

The gas fraction by mass is measured to vary from $\sim 4$ per cent for Sa-type galaxies to $\sim 25$ per cent for Scd-type galaxies \citep{YS91,BM98}. A study 
using the deep Spitzer survey data \citep{Elm11} revealed the trend that early type galaxies tend to have multiple arms and grand design spiral pattern, while the late type galaxies show  mainly flocculent spiral features.
Thus gas fraction may be correlated to the type of spiral structure.
Though the response of gas component to spiral density wave has been studied theoretically \citep[e.g.,][]{Rob69} as well as observationally for galaxies like M~51 \citep{Ran93}, 
the role of gas in the frame work of density wave theory has so far not drawn the attention it deserves.

In this paper we consider gas along with stars in the formulation of the density wave theory, and obtain the dispersion relation for this coupled system.
Although the response of interstellar gas to the stellar density wave has been well-studied, the role of gas in the group transport has not been considered so far. 
 We find that the addition of gas lowers the group velocity thus allowing the grand design spiral pattern to persist longer.
Further, we show that addition of gas permits waves to be real for the observed pattern speed, which otherwise cannot be realized for a one-component stellar disk.
Thus, the dynamical effect of gas helps spiral arms to exist and persist longer.

\S~\ref{formu4}  contains the details of the formulation of the problem, \S~\ref{res4} presents 
the results. \S~\ref{dis4} and \S~\ref{con4} contain the discussion and conclusion respectively.

\section{Formulation of the Problem}
\label{formu4}

\subsection{Dispersion Relation for a two-component disk}
Here, we treat a galactic disk as a gravitationally coupled two-component (stars plus gas) system, where the stars are taken to be collisionless  and characterized by a surface density $\Sigma_{\rm 0s}$ and a one-dimensional velocity dispersion, $\sigma_{\rm s}$, and gas as a fluid characterized by surface density $\Sigma_{\rm 0g}$ and a one-dimensional velocity dispersion or the sound speed, $\sigma_{\rm g}$.
For simplicity of calculation, the galactic disk is assumed to be infinitesimally thin and pressure acts only in the disk plane i.e. in this paper we are interested in gravitational instabilities in the disk plane ($z=0$) only. We use cylindrical co-ordinates $(R, \phi, z$).

We derive the dispersion relation for such a joint system in the WKB (Wentzel -- Kramers -- Brillouin) limit or the tightly wound case, following the procedure as in \citet{BT87}.
The small density perturbations are taken to be of type exp $ [i(\omega t - m \phi + kR) ]$ where $\omega$ is the frequency, and $k$ is the wavenumber. This simple modal approach assumes a constant pattern speed $\Omega_{\rm p} = \omega/ m$. However, in a realistic case, the quantities $\omega$ and $k$ could vary gradually with 
radial location and time, as centered around a mode \citep[as in][]{Too69}. 
Here, the same dispersion relation as obtained for the modes is taken to be valid to describe the behavior of these general disturbances that can be used to study wave packets with gradually varying properties \citep[e.g.,][]{Wit60,Lit65}.

The dispersion relation is obtained to be
\begin{equation}
\frac{2\pi G \Sigma_{\rm 0s} |k|{\mathcal F}\Big(\frac{\omega-m \Omega}{\kappa},\frac{k^2\sigma^2_{\rm s}}{\kappa^2}\Big)}{\kappa^2-(\omega-m\Omega)^2}+\frac{2\pi G \Sigma_{\rm 0g} |k|}{\kappa^2-(\omega-m\Omega)^2+\sigma^2_{\rm g} k^2} = 1\\.
\label{cont-comp}
\end{equation}
Here the function ${\mathcal F}$ is the reduction factor
with an expression as given in \citet{BT87}, also see~\S~\ref{appen4}. 
 This factor  physically takes account of the reduction in self-gravity of the galactic disk due to the velocity dispersion of stars.

\citet{Raf01} had  obtained the dispersion relation for a system comprised of $n$ distinct collisionless systems along with gas for the axisymmetric case ($m=0$), (see equation (22) in that paper). Thus the dispersion relation for our case
is given as a special case corresponding to $n=1$.

Next we define
\begin{equation}
\begin{split}
\alpha_{\rm s} = \kappa^2-2\pi G\Sigma_{\rm 0s}|k| {\mathcal F}\Big(\frac{\omega-m \Omega}{\kappa},\frac{k^2\sigma^2_{\rm s}}{\kappa^2}\Big)\\
\alpha_{\rm g} = \kappa^2-2\pi G\Sigma_{\rm 0g}|k|+k^2{\sigma}^2_{\rm g}\\
\beta_{\rm s} = 2\pi G\Sigma_{\rm 0s}|k|{\mathcal F}\Big(\frac{\omega-m \Omega}{\kappa},\frac{k^2\sigma^2_{\rm s}}{\kappa^2}\Big)\\
\beta_{\rm g} = 2\pi G\Sigma_{\rm 0g}|k| \\
\end{split}
\end{equation}
Then equation (\ref{cont-comp}) reduces to
\begin{equation}
(\omega-m\Omega)^4-(\alpha_{\rm s}+\alpha_{\rm g})(\omega-m\Omega)^2+(\alpha_{\rm s}\alpha_{\rm g}-\beta_{\rm s}\beta_{\rm g})=0\\.
\end{equation}
This is a quadratic equation in $(\omega-m\Omega)^2$. Solving it we get
\begin{equation}
(\omega-m\Omega)^2=\frac{1}{2}\left[(\alpha_{\rm s}+\alpha_{\rm g})\pm \left\{(\alpha_{\rm s}+\alpha_{\rm g})^2-4(\alpha_{\rm s}\alpha_{\rm g}-\beta_{\rm s}\beta_{\rm g})\right\}^{1/2}\right]\\\,.
\end{equation}
The  additive root for $(\omega-m\Omega)^2$ always lead to a positive quantity, hence it indicates always oscillatory perturbations under all conditions \citep[same as for axisymmetric case; see][]{JS84a}. In order to study the stability of the system and its further consequences, we therefore consider only the negative root which is
\begin{equation}
(\omega-m\Omega)^2=\frac{1}{2}\left[(\alpha_{\rm s}+\alpha_{\rm g})- \left\{(\alpha_{\rm s}+\alpha_{\rm g})^2-4(\alpha_{\rm s}\alpha_{\rm g}-\beta_{\rm s}\beta_{\rm g})\right\}^{1/2}\right]\\\,.
\label{disp-main}
\end{equation}
Next define two dimensionless quantities, $s$, the dimensionless frequency, and $x$, the dimensionless wavenumber as
\begin{equation}
s=({\omega-m\Omega})/{\kappa} =   {m(\Omega_{\rm p} - \Omega})/{\kappa}, \: \:  x = k / k_{\rm crit}\,,
\end{equation}
\noindent where $k_{\rm crit} = \kappa^2 / 2 \pi G  (\Sigma_{\rm 0s} + \Sigma_{\rm 0g})$. For one-component case, say with $\Sigma_{\rm 0g}=0$, this is the largest stable wavenumber for a pressureless stellar disk.

Dividing both sides of equation (\ref{disp-main}) by $\kappa^2$ and using the above two dimensionless quantities, we get the dimensionless form of the above dispersion relation as
\begin{equation}
s^2=\frac{1}{2}\left[(\alpha'_{\rm s}+\alpha'_{\rm g})-\left \{(\alpha'_{\rm s}+\alpha'_{\rm g})^2-4(\alpha'_{\rm s}\alpha'_{\rm g}-\beta'_{\rm s}\beta'_{\rm g})\right \}^{1/2}\right]\\,
\label{disp-twocom}
\end{equation}
where
\begin{equation}
\begin{split}
\alpha'_{\rm s}=1-(1-\epsilon)|x|{\mathcal F}(s,\xi)\\
\alpha'_{\rm g}=1-\epsilon |x|+\frac{1}{4}Q^2_{\rm g}\epsilon^2x^2\\
\beta'_{\rm s}=(1-\epsilon)|x|{\mathcal F}(s,\xi) \\
\beta'_{\rm g} = \epsilon |x|\\
\end{split}
\end{equation}
where, $\chi$= ${k^2\sigma^2_{\rm s}}/{\kappa^2}$ = $0.286 Q_{\rm s}^2 (1-\epsilon)^2 x^2$.  

The three dimensionless parameters $Q_{\rm S}$, $Q_{\rm g}$ and $\epsilon$ are, respectively the Toomre $Q$ factors for stars as a collisionless system  $Q_{\rm s}$(=$\kappa \sigma_{\rm s} /(3.36 G \Sigma_{\rm 0s})$), and for gas $Q_{\rm g}$ = ($\kappa \sigma_{\rm g} /(\pi G \Sigma_{\rm 0g})$)
and;  $\epsilon$ =${\Sigma_{\rm 0g}}/( {\Sigma_{\rm 0s}+\Sigma_{\rm 0g}})$ the gas mass fraction in the disk, respectively.

Similarly, the one-component analog of this dispersion relation is \citep{BT87}
\begin{equation}
s^2= 1-|x|{\mathcal F}\Big(\frac{\omega-m \Omega}{\kappa},\frac{k^2\sigma^2_{\rm s}}{\kappa^2}\Big)\,.
\label{disp-onecom}
\end{equation}

\subsection {Group velocity for a two-component disk}

It is well known that information from a disturbance, generated at a given radius, propagates radially with its group velocity \citep[e.g.,][]{Wit99}.
For an inhomogeneous medium, the group velocity at a fixed radius is defined as \citep{Wit60,Lit65}
\begin{equation}
c_{\rm g}(R)= \frac{\partial \omega(k,R)}{\partial k}\\.
\end{equation}

Here, we study effect of gas on the radial group velocity
of a wavepacket in a gravitationally coupled star-gas system. 
From the dispersion relation for a collisionless case, it can be seen that evaluating an analytic expression for the group velocity is cumbersome due to the implicit form of the reduction factor. This is particularly so for the two-component case.
However, interestingly, the group velocity can be easily estimated graphically from the slope of $s$ versus $x$ (see Figure~\ref{gas-effect} in ~\S~\ref{res4}), and
 is given as \citep{Too69,BT87}

\begin{equation}
c_g(R)= sgn(ks) (k/k_{\rm crit}) \frac{ds}{dx}\\,
\label{grp-vel}
\end{equation}

\noindent where $sgn(ks) = \pm 1$ depending on whether $ks >0$ or is $< 0$.
This approach was developed by \citet{Too69}, we apply it here for the frequency $s$ obtained above for the two-component case (equation~(\ref{disp-twocom})).

\section{Results} 
\label{res4}

We next investigate how the addition of gas affects the group velocity and hence the radial group transport of a wavepacket. 
Driven by purely theoretical interest, here we carried out the analysis at a fixed radius in the disk varying the gas surface density from $5$ per cent to $20$ per cent of the stellar surface density. For a real galaxy, at a particular radius one would expect to get a unique ratio of gas to stellar surface density, nevertheless our approach would 
throw some light on what is likely to happen at those radii for which the observed gas fraction matches with the values considered here in the analysis.
We will compare our results with those of \citet{Too69}, the latter obtained for the stars-alone case, with $Q_{\rm s}$ = 1. For a two-component case, when $Q_{\rm s}$ = 1, the system will
 become unstable \citep{JS84a}, and the group velocity concept is not applicable in that case.
Hence, we consider a slightly higher $Q_{\rm s}$ = 1.3 since that leaves the two-component system stable even for higher gas fraction of $20$ per cent \citep{Jog96}.
For given $Q_{\rm s}$ and $\epsilon$ values, and assuming a certain value for the ratio of dispersions $\sigma_{\rm s}$ to $\sigma_{\rm g}$ (taken to be = 3.5 as observed in the solar neighborhood, see \citet{NJ02}), one can obtain $Q_{\rm g} = (0.306 Q_{\rm s})(1-\epsilon) / \epsilon$. Using these as input parameters, the values of $s$ versus $x$ can be obtained for different gas fraction values.

\begin{figure}
\centering
\includegraphics[height=3.0in,width=4.0in]{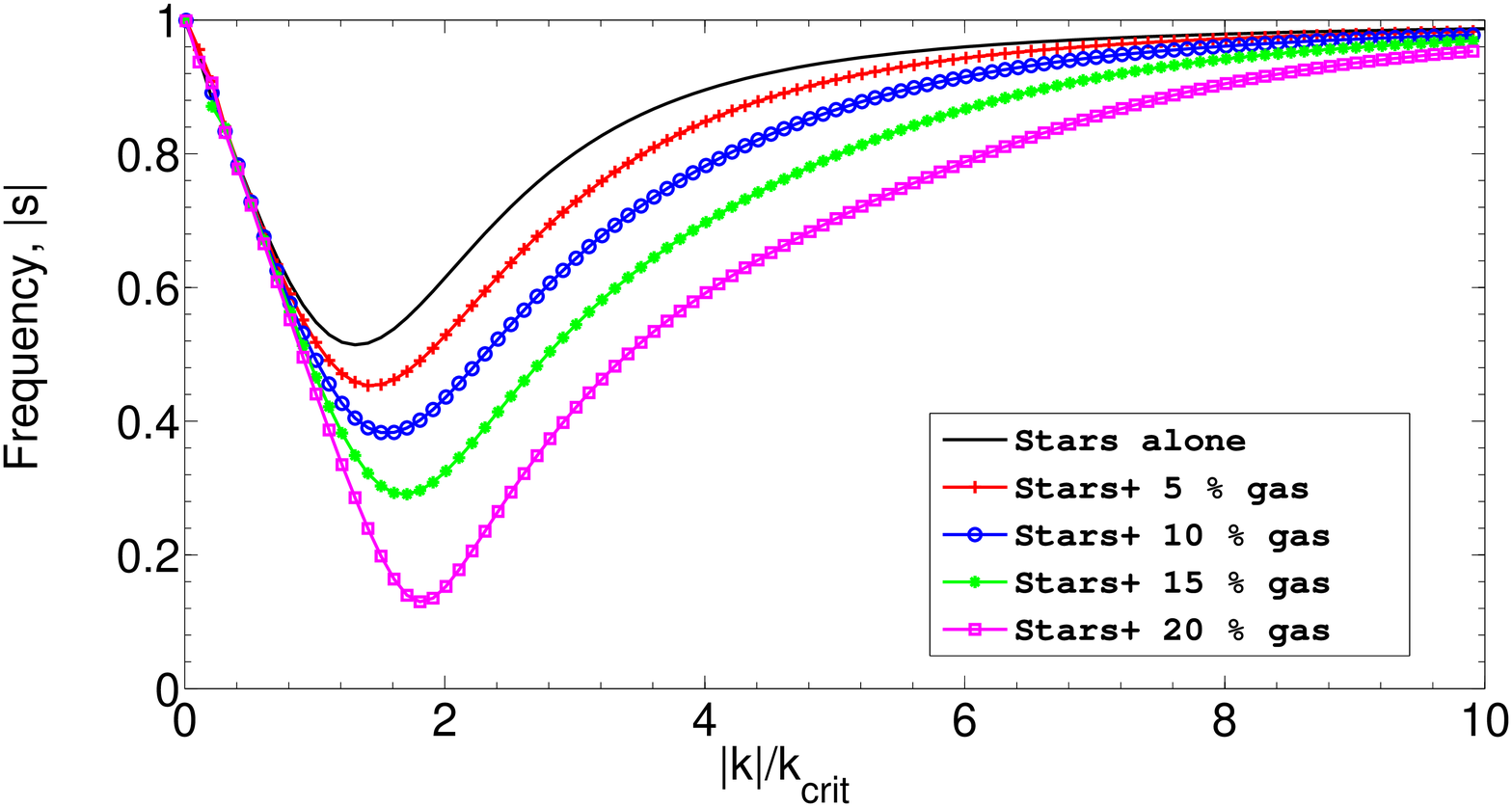}
\caption{Dispersion relation(equations (\ref{disp-twocom}) \& (\ref{disp-onecom})), plotted in its dimensionless form. 
Since the dispersion relation is symmetric with respect to both $s$ and $x$, only their absolute
values are shown in this figure. As the fraction of gas is increased, the system gets closer to being unstable.}
\label{gas-effect}
\end{figure}

In Figure~\ref{gas-effect}, we plot the dispersion relation in its dimensionless form 
for different gas fractions (equation~(\ref{disp-twocom})), including for the stars-alone case (equation~(\ref{disp-onecom})).

\noindent Two points are clear from this figure.
First, as more gas is added, the plot for dispersion relation progressively occupies more of the forbidden region between corotation ($s=0$) and inner Lindblad resonance (ILR, $s =-1$), implying that the system becomes more and more prone to being unstable, along the same line as seen for the axisymmetric case \citep{JS84a}. 
Secondly, a careful inspection also reveals that in the relatively short branch (high wavenumbers) of the dispersion relation, the plots become moderately flatter as more gas is added to the system.

While the first feature is somewhat expected and relatively well-understood, the second point is particularly interesting. As the group velocity is directly related to the slope of these curves (see~\S~\ref{formu4}), these curves give a clear hint  that when gas is included, the radial group velocity of a wave packet whose central wavenumber falls in the region, is likely to decrease, though moderately.

To investigate this issue further, we estimated the value of group velocity at $R= 8$ kpc for different gas fractions.
The values of $\kappa$ and $\Omega$ are calculated from a flat rotation curve with a constant radial velocity of 220 km s$^{-1}$ at $R=8$ kpc near the solar neighborhood in the Galaxy, obtained from the standard mass model by \citet{MCS98}. 
These are: $\kappa$ = 38.9 km s$^{-1}$ kpc$^{-1} $, and $\Omega$ = 27.5 km s$^{-1}$ kpc$^{-1}$. The pattern speed $\Omega_{\rm p}$ for $m=2$ is taken to be 12.5 km s$^{-1}$ kpc$^{-1}$ as in \citet{Too69}, so as to ensure that any the variation in results is due to the inclusion of gas. The value of the slope is obtained graphically from the curve of $s$ versus $x$ (Figure~\ref{gas-effect}) and is calculated at a point $x$ where the line $s$ = constant (for a particular pattern speed) intersects the curve. The value of $x$ corresponding to the higher $x$ or $k$ value is chosen rather than the lower $x$ value since the tight-winding approximation is better satisfied at the shorter wave branch. This gives the value of the group velocity (equation~(\ref{grp-vel})), see Table~\ref{group-res} for the results.

\begin{table*}
\centering
   \caption{Group velocity for various gas fraction values}
\begin{tabular}{lllll}
\hline
$\Sigma_{\rm 0g}/\Sigma_{\rm 0s}$   & Slope & Group velocity  & Time to travel \\%
&  &  (km s$^{-1}$)& 10 kpc  (Gyr) \\
\hline
0.0 &  0.14 & 8.0 & 1.2\\
0.05 &  0.11 & 6.0  & 1.6\\
0.10 &  0.10 & 5.5 & 1.7\\
0.15 &  0.08 & 4.4 & 2.2 \\
0.20 &  0.07 & 4.0 & 2.4 \\
\hline
\end{tabular}
\label{group-res}
\end{table*} 

As the gas fraction increases from $0$ to $20$ per cent,  the group velocity is reduced by a factor of 2. Consequently, the time taken by a wavepacket to travel a distance of 10 kpc is about two times longer. Even these rough estimates tell us that 
the addition of gas helps the density pattern to persist for a relatively longer time-scale. This is the main finding of this work.
We stress that the quantitative result is not robust as it depends on the slope of the curve measured graphically. For example, for a slightly higher pattern speed of 14 km s$^{-1}$ kpc$^{-1}$, the group velocity has a range of 11--4.4 km s$^{-1}$ as the gas fraction is increased from $0$ to $20$ per cent, hence the time taken for a wavepacket to travel the same distance of 10 kpc increases almost by a factor of 3. Thus the results in Table~\ref{group-res} give a typical
sense of increase in the lifetime of a spiral pattern on including gas in the picture.

The modern observations of the pattern speed for the Milky Way show higher values, that lie in a range between 17 and 28 km s$^{-1}$ kpc$^{-1}$ \citep[e.g.,][]{Ger11,Sie12,Jun15}, with a typical value of $\Omega_{\rm p}$ = 18 km s$^{-1}$ kpc$^{-1}$. For most of this range of values for $\Omega_{\rm p}$, the value of $s$ in the middle range of the Galaxy is $< 0.5$. For this value of $s$, the dispersion relation (equation~(\ref{disp-onecom})) does not admit a real solution for $Q_{\rm s}$ = 1.3.
Instead its solution  has an imaginary wavenumber, and hence it decays exponentially with radius on scales of $\sim$ a few $\lambda_{\rm crit}$ \citep[Chapter 6.2 in][]{BT87}. 
For the solar neighborhood, with $(\Sigma_{\rm 0s}+\Sigma_{\rm 0g})$ = 52 M$_{\odot}$ pc$^{-2}$ \citep{NJ02}, the value of $\lambda_{\rm crit}$ is $\sim$ 6 kpc. 
This aspect does not seem to be recognized or at least mentioned in the papers which give the observational determination of the pattern speed.
Thus, a one-component stellar disk is not adequate to describe the observed pattern, and therefore taking account of gas is essential in order for the  density waves to be stable as shown next.

For the recent observed value of the pattern speed of 18 km s$^{-1}$ kpc$^{-1}$, the value of the frequency is obtained as $s=0.5$. For this value of $s$,
the wavenumber for the $s$ versus $x$ curve for  the stars-alone case would be imaginary. Note, however, that in contrast,
for gas fractions of $10$--$15$ per cent, the curve $s$ versus $x$ does have real solutions for the wavenumber and the slope at these wavenumbers 
remains  nearly constant $\sim 0.2 $. The corresponding group velocity is  11 km s$^{-1}$,
comparable to but higher than the values in Table~\ref{group-res}. Thus the time to travel 10 kpc is $\sim 10^ 9$ yrs.
Thus,  interestingly, we find that for the observed pattern speed in the Galaxy, the waves for stars-alone case have an imaginary wavenumber, hence are evanescent. For a stable wave solution one needs to take account of gas.  
Moreover, addition of gas allows somewhat higher pattern speeds which are consistent with observations, to be valid in a galaxy. Since the pattern speed decides the location of resonance points, the inclusion of gas thus allows the corotation to be shifted to an inner radius.

\section{Discussion}
\label{dis4}
\subsection{Stars: collisionless versus fluid approach}

In this paper, we have treated stars as a collisionless system involving no pressure term, as required for a correct treatment of the group velocity.
There are some problems in dynamics where the result does not depend critically on whether a fluid or a collisionless representation is used for stars. For example, 
in the local stability analysis of axisymmetric perturbations in a galactic disk leading to the Toomre $Q$ criterion, results obtained from both approaches 
match quite satisfactorily. The Toomre
$Q$ factors \citep{Too64} for stars and gas differ by only $7$ per cent, with a factor 3.36 replacing $\pi$ in the denominator for a collisionless case \citep[e.g.,][]{BT87}.

\begin{figure}
\centering
\includegraphics[height=3.0in,width=4.0in]{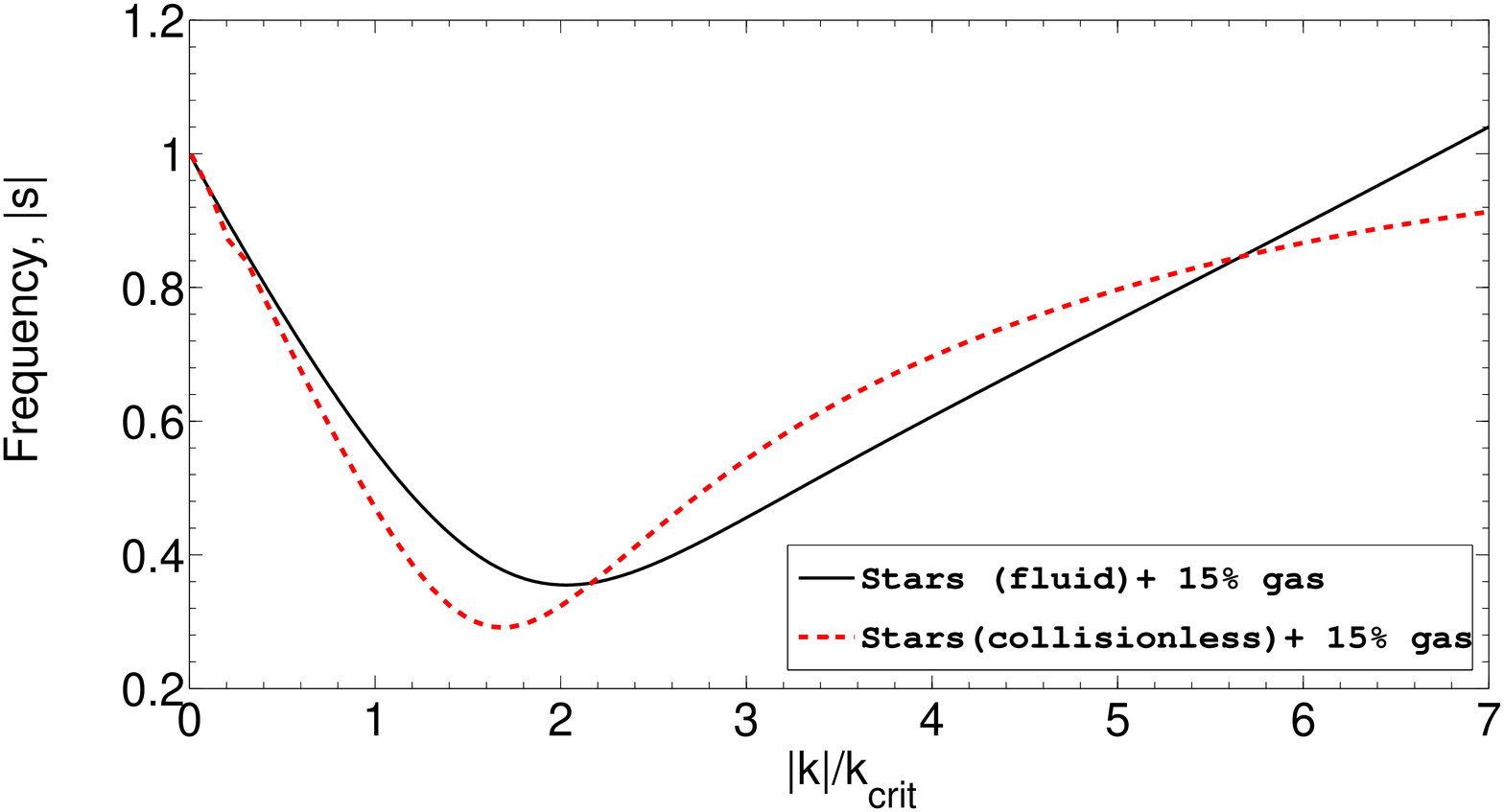}
\caption{Dispersion relation(equations~(\ref{disp-twocom}) \& (\ref{disp-onecom})), plotted in its dimensionless form,  
at $R = 8$ kpc for a disk with 15 per cent gas, for
two cases: first treating stars as a collisionless system and then treating stars as a fluid. The two approaches agree at low $x$ values, but differ at large $x$ values.}
\label{com-flucol}
\end{figure}

However, it is known that the two approaches differ substantially with the system being more bouncy at high wavenumbers in the fluid case as can be seen by the plot of the dispersion relation in the two cases (e.g., Fig. 6.14 in \citet{BT87}; also see \citet{Raf01}). We plot the dispersion relation with 15 per cent  gas for two cases: first stars as a collisionless system (equation (\ref{disp-twocom}) above; and then treating stars as a fluid (with the dispersion relation as in \citet{JS84a}), see Figure~\ref{com-flucol}.

From Figure~\ref{com-flucol}, it is  evident that for the case where the stars are treated as a fluid, the curve for the dispersion relation is steeper than the case where stars are treated as collisionless system and the $s$ values exceed 1 beyond $x=7$. In contrast, in the collisionless treatment, the $s$ values are always less than 1 and saturate to 1 at $x$ $\gg 7$. Since the technique adopted here to calculate group velocity is strongly dependent on slope of these curves, consequently these two different approaches, using the same gas fraction, are bound to give different estimates for the group velocity.

\subsection{Other Issues:}

We would like to caution that, from our result it may appear that with the inclusion of gas by more than 20 per cent or so which is the extreme case considered for the present work, may still improve the persistence of the spiral features. But in a real gas-rich galaxy, other processes  like swing amplification generating local flocculent spiral features do take place simultaneously. In particular, for gas rich galaxies it is probably the swing amplification which supersede the grand spiral structure, as evident from the abundance of more flocculent spiral features found in gas-rich Sc-type galaxies. 

Another point to note is that the analysis presented here is based on a linear perturbation theory so that the non-linear effects have been neglected. But these non-linear effects could have far-reaching consequences. 
A recent study using high-resolution $N$-body simulations \citep{Don13} showed that the nonlinear effects can significantly modify the process of origin and persistence of floccculent spiral features, produced through swing amplification. So, it is worth checking the significance of nonlinearity in the frame work of density wave theory with gas treated on equal footing with stars.

\section{Conclusion:}
\label{con4}

In summary, we have studied the effect of gas by treating a galactic disk as a two-component gravitationally coupled stars plus gas
system in the framework of density wave theory. The resulting frequency versus wavenumber curve is flatter which leads to a
lower group velocity.  
The idea of group velocity as an indicator of transport of information is 
routinely studied in other branches of physics, such as quantum mechanics, its usefulness is underutilized in astrophysics. In the case of 
galactic disks, this idea was first applied by \citet{Too69} to show that the density waves cannot last for more than $\sim$ 10$^9$ yrs. 
We have shown that taking account of gas lowers the group velocity by a factor of few, which allows the 
density waves to last longer, to about few $10^9$ yr. This helps persistence of the spiral structure for a longer time-scale in a gas-rich galaxy,
irrespective of the mechanism, such as tidal interaction, that  gives rise to the grand two-armed spiral structure.

The second important result from this work is that we show that for the observed pattern speed of $\sim 18$ km s$^{-1}$ kpc$^{-1}$ for the Galaxy \citep{Sie12}, 
the solution gives an evanescent wave (i.e., a wave with an imaginary wavenumber) for a one-component stellar disk, hence the wave would disperse radially. 
We show that it is the inclusion of gas that makes it possible to have a stable wave for the measured pattern speed.  
The addition of gas thus allows somewhat higher pattern speeds to be valid in a galaxy
which otherwise cannot be realized for a one-component stellar disk. Since the pattern speed decides the location of resonance points, the addition of gas thus allows the corotation to be shifted to an inner radius.
This could have implications for the angular transport properties due to the spiral pattern.

\medskip
\section{Appendix}
\label{appen4}
\subsection{Calculation of the Reduction Factor}

The dispersion relation for collisionless stellar disk in the tight-winding limit is (see equation~(\ref{disp-onecom})):
\begin{equation}
s^2= 1-|x|{\mathcal F}(s, \chi)\,,
\label{disp-supp}
\end{equation}
\noindent where the reduction factor ${\mathcal F}(s, \chi)$ is given by \citep{BT87}:\\
\begin{equation}
{\mathcal F}(s, \chi)=\frac{2}{\chi}exp(-\chi)(1-s^2)\sum_{n=1}^\infty\frac{I_n(\chi)}{1-\frac{s^2}{n^2}}\,,
\label{eqn-term}
\end{equation}
\noindent where $I_n (\chi)$ is a modified Bessel function of order $n$.
Now the explicit dependence of reduction factor  ${\mathcal F}(s, \chi)$ on $s$ makes the dispersion relation (equation~(\ref{disp-supp})) an implicit relation which then has to be solved in a self-consistent manner. For obtaining the solution numerically, one has to truncate the infinite sum after a finite terms in such a way so that the solution is not affected by the truncation process. In other words, if the solution with $k+1$ terms of the series matches well (within the pre-defined tolerance limit) with the solution having $k$ (not to confuse with wavevector $k$) terms of the series, one can safely truncate the series after $k$ terms.

In some special cases, for example while studying the $m=1$ slow modes in Keplerian disks, \citet{Tre01} showed that 
the reduction factor in the dispersion relation simplifies to one containing only the first term of the infinite sum, hence becomes an explicit relation in $s$ and $x$ ($|k|/k_{\rm crit}$)(see equation (12) there). But we caution that, in general, such a simplification is not always valid and one has to incorporate more terms of the series to get the actual solution.

Here we illustrate how incorporating further terms of the infinite sum will affect the dispersion relation for a collisionless stellar disk for $Q =1.3$.

\begin{figure}
\centering
\includegraphics[height=3.0in,width=4.0in]{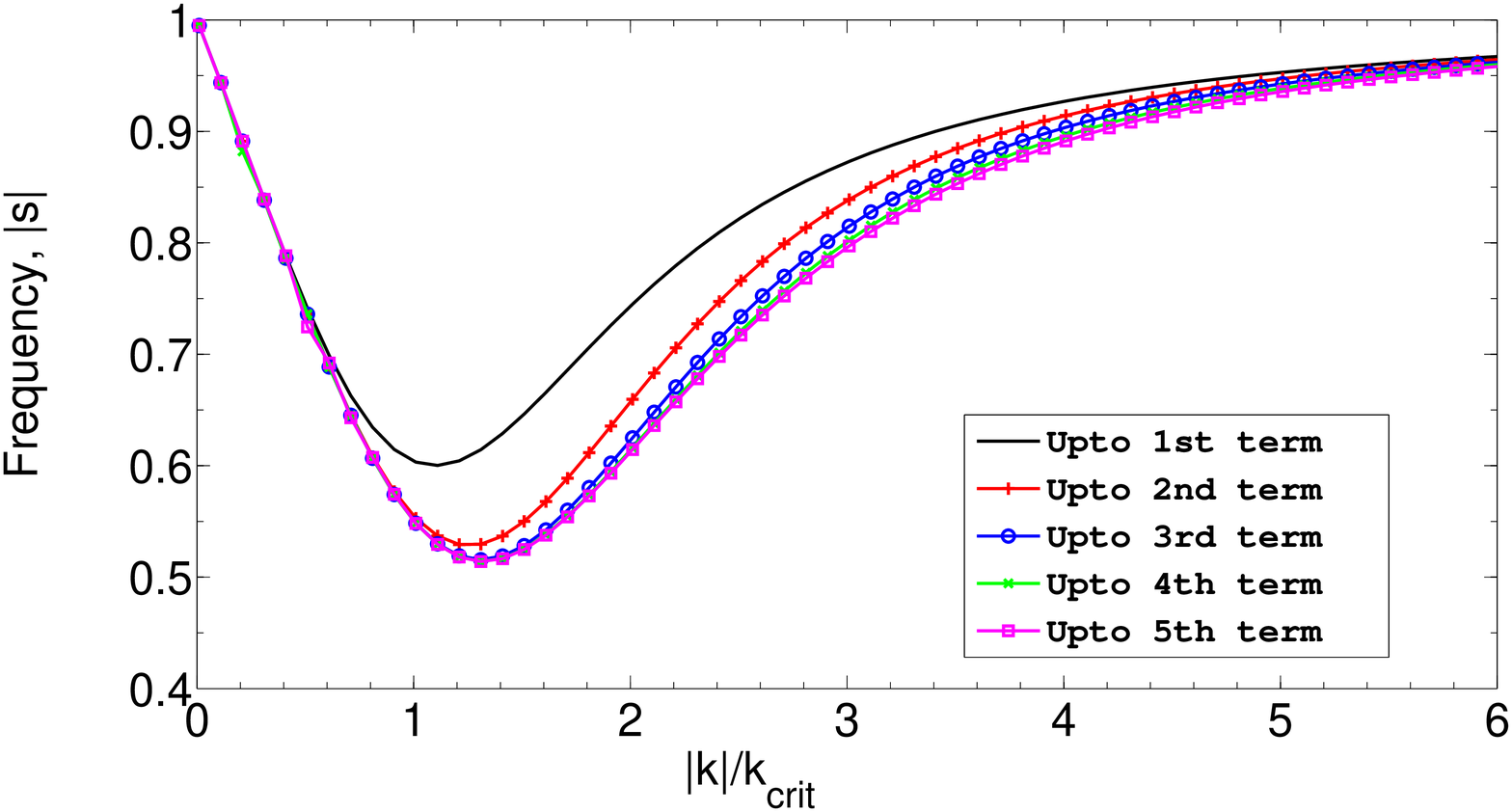}
\caption{Variation of dispersion relation for collisionless stellar disk with $Q =1.3$ while incorporating subsequent terms of the infinite series in the expression for the reduction factor (equation~(\ref{eqn-term})). 
The solution with first four terms matches well with that having first five terms to within 1 per cent, so the series can safely be truncated after the first four terms.}
\label{eff-term}
\end{figure}

From Figure~\ref{eff-term}, it is  seen  that with the inclusion of subsequent terms, the solution changes significantly. Allowing more terms of that infinite sum into the solution makes the forbidden region between corotation($s =0$) and Inner Lindblad resonance(ILR, $s=-1$) reduce up to a certain limit, after that the effect gets saturated. This happens for
$n=4$. Hence in our calculation (~\S~\ref{formu4}), we include terms up to $n=4$ in the reduction factor ${\mathcal F}$ as suggested by this figure. The result shown in Figure~\ref{eff-term} agrees with that in \citet{Too69}.

\subsection{WKB dispersion relation for a two-fluid galactic disk}
For the sake of completeness, the derivation of the WKB dispersion relation for two-fluid galactic disk is given next.

The disk taken as infinitesimally thin so that the pressure acts only in the mid-plane ($z=0$). Both the stars and gas are treated as fluid, each characterized by the surface density $\Sigma_i$ and the one-dimensional velocity dispersion or sound speed $c_i$. We use the galactocentric cylindrical coordinates ($R$, $\phi$, $z$) throughout in this section. Also, we use the subscripts $i=s, g$ for stars and gas, respectively.

In this cylindrical coordinates, the radial and azimuthal components of the Euler equation of motion are

\begin{equation}
\frac{\partial v_{R_i}}{\partial t}+v_{R_i}\frac{\partial v_{R_i}}{\partial R}+\frac{v_{\phi_i}}{R}\frac{\partial v_{R_i}}{\partial \phi}-\frac{v^2_{\phi_i}}{R}=-\frac{1}{\Sigma_{0_i}}\frac{\partial p_i}{\partial R}-\frac{\partial \Phi_{\rm tot}}{\partial R}
\label{intro-Euler_vr}
\end{equation}

\noindent and,

\begin{equation}
\frac{\partial v_{\phi_i}}{\partial t}+v_{R_i}\frac{\partial v_{\phi_i}}{\partial R}+\frac{v_{\phi_i}}{R}\frac{\partial v_{\phi_i}}{\partial \phi}+\frac{v_{R_i} v_{\phi_i}}{R}=-\frac{1}{\Sigma_{0_i} R}\frac{\partial p_i}{\partial \phi}-\frac{1}{R}\frac{\partial \Phi_{\rm tot}}{\partial \phi}\,,
\label{intro-Euler_vphi}
\end{equation}

\noindent where $\Phi_{\rm tot}$ denotes the joint gravitational potential due to both stars and gas.

Following \citet{BT87}, we choose a simple equation of state given as\\

\begin{equation}
p_{i}=K \Sigma_i^{\gamma}\,.
\end{equation}

The corresponding specific enthalpy ($h$) is given as
\begin{equation}
h_{i}=\frac{\gamma}{\gamma-1} \Sigma_i^{\gamma-1}
\end{equation}

With this definition of specific enthalpy, equations~(\ref{intro-Euler_vr}) and (\ref{intro-Euler_vphi}) reduce to
\begin{equation}
\frac{\partial v_{R_i}}{\partial t}+v_{R_i}\frac{\partial v_{R_i}}{\partial R}+\frac{v_{\phi_i}}{R}\frac{\partial v_{R_i}}{\partial \phi}-\frac{v^2_{\phi_i}}{R}=-\frac{\partial (\Phi_{\rm tot} + h_i)}{\partial R}
\end{equation}

\noindent and,

\begin{equation}
\frac{\partial v_{\phi_i}}{\partial t}+v_{R_i}\frac{\partial v_{\phi_i}}{\partial R}+\frac{v_{\phi_i}}{R}\frac{\partial v_{\phi_i}}{\partial \phi}+\frac{v_{R_i} v_{\phi_i}}{R}=-\frac{1}{R}\frac{\partial (\Phi_{\rm tot}+ h_i)}{\partial \phi}\,.
\end{equation}

The spiral wave is treated as only a small perturbation on a steady-state axisymmetric disk \citep[for details see][]{BT87} which allows one to carry out linear perturbation analysis in the Euler equations of motion. The unperturbed part of a quantity will be accompanied by a subscript `0' and the perturbed part of any quantity will be accompanied by a subscript `1'.

The unperturbed Euler equation of motion gives

\begin{equation}
v_{\phi_0}=\sqrt{R\frac{d \Phi_{\rm 0, tot}}{dR}} = R \Omega(R)\,,
\end{equation}
\noindent where $\Omega(R)$ denotes the circular frequency at any radius $R$.

Now we carry out the linear perturbation analysis, and write different quantities sin the following way:\\
$v_{Ri}= v_{R_{1i}}$, $v_{\phi i}= v_{\phi_{0i}}+v_{\phi_{1i}}$, $\Sigma_{i}= \Sigma_{0i}+\Sigma_{1i}$, $\Phi_{\rm tot}=\Phi_{\rm 0, tot}+\Phi_{\rm 1, tot}$ etc., where the perturbed quantities are very small as compared to the unperturbed quantities.

Then the perturbed equations of motion reduce to
\begin{equation}
\frac{\partial v_{R_{i1}}}{\partial t}+\Omega\frac{\partial v_{R_{i1}}}{\partial \phi}-2\Omega v_{\phi_{i1}}=-\frac{\partial}{\partial R}\left(\Phi_{\rm 1, tot}+h_{i1}\right)\\
\label{intro-per_vr}
\end{equation}
\noindent and,

\begin{equation}
\frac{\partial v_{\phi_{i1}}}{\partial t}+\left[\frac{d(\Omega R)}{dR}+R\right]v_{R_{i1}}+\Omega \frac{v_{\phi_{i1}}}{\partial \phi}=-\frac{1}{R}\frac{\partial}{\partial \phi}\left(\Phi_{\rm 1, tot}+h_{i1}\right)\\\,.
\label{intro-per_vphi}
\end{equation}

We take the trial solution as
\begin{equation}
\chi_1=Re[\chi_a e^{i(m\phi-\omega t)}]\\\,,
\label{intro-trial_sol}
\end{equation}
\noindent where $\chi_1$ denotes any perturbed dynamical quantity of interest, and $m\ge 0$ is any integral values which indicates that the perturbation has a $m$-fold symmetry (here $m=2$ for spiral arms).

Putting the trial solutions in equations~(\ref{intro-per_vr}) and (~\ref{intro-per_vphi}) and after some algebraic simplification we get \citep[for details see][]{BT87}

\begin{equation}
v_{R_{ai}}(R)=\frac{i}{\Delta}\left[(\omega-m\Omega)\frac{d}{dR}(\Phi_{a, \rm tot}+h_{ai})-\frac{2m\Omega}{R}(\Phi_{a, \rm tot}+h_{ai})\right]\\
\label{intro-vr_ori}
\end{equation}

\begin{equation}
v_{\phi_{ai}}(R)=-\frac{1}{\Delta}\left[2B(\omega-m\Omega)\frac{d}{dR}(\Phi_{a, \rm tot}+h_{ai})+\frac{m(\omega-m\Omega)}{R}(\Phi_{a, \rm tot}+h_{ai})\right]\,,
\label{intro-vphi_ori}
\end{equation}
\noindent where $\kappa$ is the local epicyclic frequency, and $\Delta$ is defined as
\begin{equation}
\Delta \equiv \kappa^2-(\omega-m\Omega)^2\,.
\end{equation}

Similarly, the linearized equation of state is 

\begin{equation}
h_{ai}=\gamma K \Sigma_{0i}^{\gamma-2}\Sigma_{ai}\,,
\end{equation}
\noindent and the linearized continuity equation in the cylindrical coordinates is
\begin{equation}
\frac{\partial \Sigma_{1i}}{\partial t}+\Omega\frac{\partial \Sigma_{1i}}{\partial \phi}+\frac{1}{R}\frac{d}{dR}\left(R v_{R_{1i}}\Sigma_{0i}\right)+\frac{\Sigma_{0i}}{R}\frac{v_{\phi_{1i}}}{\partial \phi}=0\\\,.
\end{equation}

Putting the trial in this equation, the perturbed continuity equation reduces to
\begin{equation}
-i(\omega-m\Omega)\Sigma_{ai}+\frac{1}{R}\frac{d}{dR}(R v_{R_{ai}}\Sigma_{0i})+\frac{im\Sigma_{0i}}{R}v_{\phi_{ai}}=0\,.
\label{intro-con_ori}
\end{equation}

Now we invoke the tight winding approximation or the WKB approximation to get an {\it local}, analytical expression for the dispersion relation for this two-fluid galactic disk.

The potential of such a tightly wound density wave can be written as \citep[for details see][]{BT87}
\begin{equation}
\Phi_{a}(R)= F(R) exp \left[\int^R k dR\right]
\end{equation}
\noindent where $F(R)$ is a slowly varying function of $R$ and $|kR| \gg 1$.

Neglecting the terms higher than $O(|kR|^{-1})$ and after some algebraic simplifications, equations~(\ref{intro-vr_ori}), (\ref{intro-vphi_ori}), and (\ref{intro-con_ori}) reduce to

\begin{equation}
\begin{split}
v_{R_{ai}}=-\frac{(\omega-m\Omega)}{\Delta}k(\Phi_{a, \rm tot}+h_{ai})\\
v_{\phi_{ai}}=-\frac{2iB}{\Delta}k(\Phi_{a, \rm tot}+h_{ai})\\
\end{split}
\end{equation}
\noindent and,
\begin{equation}
-(\omega-m\Omega)\Sigma_{ai}+k\Sigma_{0i}v_{R_{ai}}=0\,.
\end{equation}

Eliminating $v_{R_{ai}}$ and $h_{ai}$ from these equations we get

\begin{equation}
\Sigma_{ai}(R)= {\mathcal P}_i(k, R, \omega) \Sigma_{ai}(R)\,,
\end{equation}
\noindent where the polarization function ${\mathcal P}_i$ which relates the response density $\Sigma_{ai}$ to the total density and the form is given as

\begin{equation}
{\mathcal P}_i(k, R, \omega) = \frac{2 \pi G \Sigma_{0i}|k|}{\kappa^2-(\omega-m\Omega)^2+c^2_{i}k^2}\,.
\end{equation}

In order to obtain a {\it self-consistent} density wave, the response surface density should be equal to the total surface density. Thus, we get,
\begin{equation}
\frac{2 \pi G \Sigma_{\rm 0s}|k|}{\kappa^2-(\omega-m\Omega)^2+c^2_{\rm s}k^2}+\frac{2 \pi G \Sigma_{\rm 0g}|k|}{\kappa^2-(\omega-m\Omega)^2+c^2_{\rm g}k^2}=1\,.
\end{equation}

After some algebraic simplification, we get,
\begin{equation}
(\omega-m\Omega)^4-(\omega-m\Omega)^2(\alpha_{\rm s}+\alpha_{\rm g})+(\alpha_{\rm s}\alpha_{\rm g}-\beta_{\rm s}\beta_{\rm g})=0
\label{intro-def_disp}
\end{equation}
\noindent where,
\begin{equation}
\begin{split}
\alpha_{i}=\kappa^2-2\pi G \Sigma_{0i}|k|+c^2_{i}k^2\\
\beta_{i}=2 \pi G \Sigma_{0i}|k|\\
\end{split}
\label{intro-def_alpha}
\end{equation}
\noindent and, $i =s, g$ for stars and gas, rexpectively.

Solving equation~(\ref{intro-def_disp}) we get,
\begin{equation}
(\omega-m\Omega)^2=\frac{1}{2}\left[(\alpha_{\rm s}+\alpha_{\rm g})\pm \left\{(\alpha_{\rm s}+\alpha_{\rm g})^2-4(\alpha_{\rm s}\alpha_{\rm g}-\beta_{\rm s}\beta_{\rm g})\right \}^{1/2}\right]\,.
\end{equation}

Now the additive root for $(\omega-m\Omega)^2$ will always be positive, and hence it indicates the oscillatory perturbations under all conditions; thereby leaving the other root of $(\omega-m\Omega)^2$ meaningful for studying the stability of the two-fluid gravitationally coupled system \citep[for details see][]{JS84}. Hence, the WKB dispersion relation for a gravitationally coupled two-fluid system is 

\begin{equation}
(\omega-m\Omega)^2=\frac{1}{2}\Big[(\alpha_{\rm s}+\alpha_{\rm g})- \left\{(\alpha_{\rm s}+\alpha_{\rm g})^2-4(\alpha_{\rm s}\alpha_{\rm g}-\beta_{\rm s}\beta_{\rm g})\right\}^{1/2}\Big]\,,
\label{intro-disp_2flu}
\end{equation}
\noindent where the quantities $\alpha_{i}$ and $\beta_{i}$ are defined in equation~(\ref{intro-def_alpha}), and $i= s, g$ are for stars and gas, respectively.

By putting $m=0$ in equation~(\ref{intro-disp_2flu}), we recover the dispersion relation for the axisymmetric case \citep[see equation~(20) in][]{JS84}, as expected.

\subsection{WKB dispersion relation for a two-component disk}
For the sake of completeness, the derivation of the WKB dispersion relation for two-component galactic disk is given next.

Here, we treat the stellar component as a {\it collisionless} system, characterized by the surface density $\Sigma_{\rm 0s}$ and the one-dimensional velocity dispersion $\sigma_{\rm s}$, and the gas as {\it fluid}, characterized by the surface density $\Sigma_{\rm 0g}$ and the sound speed $c_{\rm g}$.

 Much of the formulation will be similar to what it is done in the earlier section, except there will be some modification due to the different modeling of the stellar component as fluid and collisionless system. 

We consider the infinitesimally thin galactic disk, and the pressure acts only in the mid-plane ($z=0$). We use the galactocentric cylindrical coordinate system ($R$, $\phi$, $z$) throughout in this section.

The basic outline for the formulation is the following:\\
Equations of motion are used to compute the perturbation in surface density $\Sigma_{1i}$ which arises due to the perturbation in the potential $\Phi_{1,\rm tot}$, and then in order to get a {\it self-consistent} density wave we equate the perturbation in surface density to the total density. Also for the stellar component, instead of the continuity equation, the Jean's equation is used for the collisionless stellar component \citep[for details see][]{BT87}. However, the main difficulty is in computing the quantity $\bar{v}_{R_{1\rm s}}$, the perturbation in the mean velocity of the stars, introduced by the perturbation in the potential $\Phi_{1,\rm tot}$, as also mentioned in \citet{BT87}.

For a cold stellar disk (where the unperturbed orbits are circular), the quantity $\bar{v}_{R_{1 \rm s}}$ can be written as
\begin{equation} 
\bar{v}_{R_{1\rm s}}=-\frac{\omega-m\Omega}{\Delta}k\Phi_{a, \rm tot}\\
\label{intro-cold_vr}
\end{equation}
\noindent because a cold stellar disk essentially is  dynamically equivalent to a fluid disk with zero pressure \citep[see discussion in][]{BT87}. However, note that if the stellar disk is not cold enough so that the typical epicyclic amplitude is not much smaller than the wavelength $2\pi/k$, the expression in equation~(\ref{intro-cold_vr}) will no longer remain accurate. In that case, stars coming from a distance of twice the epicyclic amplitude and which are passing through a certain point ($R$, $\phi$) at any given time can be sampled entirely different parts of the spiral potential. This effect causes an effective reduction in the response of the disk to the spiral perturbations \citep[for detailed discussion see][]{BT87}. Therefore, including the effect of this reduction we get,

\begin{equation} 
\bar{v}_{R_{1\rm s}}=-\frac{\omega-m\Omega}{\Delta}k\Phi_{a, \rm tot}{\mathcal F}\\\,,
\label{intro-reduc_vr}
\end{equation}
\noindent where ${\mathcal F} \le 1$ is the reduction factor. The form for ${\mathcal F}(s, \chi)$ for a razor-thin disk whose stellar equilibrium state can be described by the Schwarzchild distribution function is given by \citep{BT87}:
 \begin{equation}
{\mathcal F}(s, \chi)=\frac{2}{\chi}\exp(-\chi)(1-s^2)\sum_{n=1}^\infty\frac{I_n(\chi)}{1- s^2/n^2}\,,
\label{intro-reduc_factor}
\end{equation}
\noindent where  $s(=({\omega-m\Omega})/{\kappa})$  and  $\chi (=k^2\sigma^2_{\rm s}/\kappa^2)$ are the dimensionless frequencies, and $\sigma_{\rm s}$ is the velocity dispersion of the stellar component. $I_n$ is the modified Bessel function of first kind.

After putting the trial solution as given in equation~(\ref{intro-trial_sol}), the perturbed Jean's equation reduces to
\begin{equation}
-(\omega-m\Omega)\Sigma_{a \rm s}+k \Sigma_{\rm 0s} {\bar v}_{R_{a _{\rm s}}}=0
\end{equation}

Following the same way as what is done in the earlier section, we eliminate ${\bar v}_{R_{a\rm s}}$, and $v_{R_{a\rm g}}$ to get the polarization functions for both the stellar and gas components which are  given as
\begin{equation}
{\mathcal P}_{\rm s}(k, R, \omega) = \frac{2 \pi G \Sigma_{0\rm s}|k| {\mathcal F}}{\kappa^2-(\omega-m\Omega)^2}\\
\label{intro-pola_star}
\end{equation}
\noindent and,
\begin{equation}
{\mathcal P}_{\rm g}(k, R, \omega) = \frac{2 \pi G \Sigma_{0\rm s}|k|}{\kappa^2-(\omega-m\Omega)^2+c^2_{\rm g}k^2}\\
\label{intro-pola_gas}
\end{equation}

To get a self-consistent density wave, the sum of these two polarization function should be equal to one. Therefore, we get,
\begin{equation}
\frac{2 \pi G \Sigma_{0\rm s}|k| {\mathcal F}}{\kappa^2-(\omega-m\Omega)^2}+ \frac{2 \pi G \Sigma_{0\rm s}|k|}{\kappa^2-(\omega-m\Omega)^2+c^2_{\rm g}k^2} = 1\\\,.
\end{equation}

After some algebraic simplifications we get,

\begin{equation}
(\omega-m\Omega)^2=\frac{1}{2}\left[(\alpha'_{\rm s}+\alpha'_{\rm g})\pm \left \{(\alpha'_{\rm s}+\alpha'_{\rm g})^2-4(\alpha'_{\rm s}\alpha'_{\rm g}-\beta'_{\rm s}\beta'_{\rm g})\right \}^{1/2}\right]\,,
\end{equation}
\noindent where,

\begin{equation}
\begin{split}
\alpha'_{\rm s}= \kappa^2-2 \pi G \Sigma_{\rm 0s}|k|{\mathcal F}\\
\alpha'_{\rm g}= \kappa^2-2 \pi G \Sigma_{\rm 0g}|k|+c^2_{\rm g}k^2\\
\beta'_{\rm s}= 2 \pi G \Sigma_{\rm 0s}|k|{\mathcal F}\\
\beta'_{\rm g}= 2 \pi G \Sigma_{\rm 0g}|k|\,.
\end{split}
\end{equation}

Here also, we neglect the additive root as it remains always positive, and hence indicates the oscillatory behavior of the perturbation in all conditions. Therefore, the dispersion relation for a gravitationally coupled two-component (stars plus gas) system in the tight-winding limit is

\begin{equation}
(\omega-m\Omega)^2=\frac{1}{2}\left[(\alpha'_{\rm s}+\alpha'_{\rm g}) - \left \{(\alpha'_{\rm s}+\alpha'_{\rm g})^2-4(\alpha'_{\rm s}\alpha'_{\rm g}-\beta'_{\rm s}\beta'_{\rm g})\right \}^{1/2}\right]\,.
\end{equation}

\thispagestyle{empty}

\chapter[Dynamical effect of gas on spiral pattern speed in galaxies]{Dynamical effect of gas on spiral pattern speed in galaxies\footnote{based on Ghosh \& Jog, 2016, MNRAS, 459, 4057}}
\chaptermark{\it Effect of gas on spiral pattern speed}
\vspace {2.5cm}

\section{Abstract}
In the density wave theory of spiral structure, the grand-design two-armed spiral pattern is taken to rotate rigidly in a  galactic disk with a constant, definite pattern speed. The observational measurement of the pattern speed of the spiral arms, though difficult, has been achieved in a few galaxies such as NGC 6946, NGC 2997, and M 51 which we consider here. We examine whether the theoretical dispersion relation permits a real solution for wavenumber corresponding to a stable wave, for the observed rotation curve and the pattern speed values. We find that the disk when treated to consist of stars alone, as is usually done in literature, does not generally support a stable
density wave for the observed pattern speed. Instead the inclusion of the low velocity dispersion component, namely, gas, is essential to 
obtain a stable density wave. Further, we obtain a theoretical range of allowed pattern speeds that correspond to a stable 
 density wave at a certain radius, and check that for the three galaxies considered,  
the observed pattern speeds fall in the respective prescribed range. The inclusion of even a small amount ($\sim$ ~ 15 per cent) of gas by mass fraction in the galactic disk is shown to to have a significant dynamical effect on the dispersion relation and hence on the pattern speed that is likely to be seen in a real, gas-rich spiral galaxy.

\section{Introduction} 
\label{intro5}

Various surveys on galaxy morphology have revealed that the spiral structure is mainly of two types: flocculent structure and  grand-design spiral structure \citep{Elm11}. According to the density wave theory, the grand-design spiral pattern is a density wave that  rotates rigidly in a galactic disk, and is maintained by the disk gravity  \citep{LS64,LS66}, also see  \citet{DOBA14} for a recent review.

The determination of pattern speed is of particular interest in galactic dynamics as it sets the location of resonance points where angular momentum transport is believed to occur, thus it has direct implications for the secular evolution of the galactic disk \citep{LYKA72}. Over the years, several techniques have been devised to measure the pattern speed of these spiral structures. A common technique that is used is based on the assumed knowledge of the location of the resonance points \citep{BuCs96}, which  involves understanding the behaviour of stars and gas at these resonance points. Thus, estimation of the resonance points from the observed surface brightness profiles, will give the value of pattern speed \citep{PD97}. 
Another approach relies on information regarding the sign reversal of the radial streaming motion across the corotation. This can be done by studying a strip covering the kinematical minor-axis, where the radial component of the velocity has a non-zero projection along the line of sight. This method requires the position angle of the strip to be known accurately, which is not easy to check observationally because warps, often present in the outer parts of a galaxy, twist the location of the position angles.
A related method, first proposed by \citet{CAN93}, employs the change of sign of radial streaming motion across corotation and also takes account of the geometric phase values. This technique has been applied successfully first to NGC~4321 \citep{Sem95}. Also, the pattern speed can be estimated from the observed azimuthal age-gradient of the young stellar complexes which are  seen to be associated with the spiral arms \citep{GP98}. This technique has been applied to NGC~2997 to measure its pattern speed \citep{GD09}. Other popular method is the Tremaine-Weinberg method (hereafter, the TW method) which requires no specific dynamic model and predicts the value of pattern speed from kinematic measurements only \citep{TW84}. In the past, the TW method has been used to deduce the pattern speed of bars \citep{MK95,COR03,MA06} and spiral structures \citep{Fat07}. Each of the above methods involves different possible sources of errors or inaccuracies \citep[see e.g. the discussion in][]{Jun15}.

Spiral galaxies also contain a certain amount of interstellar gas whose fraction varies with their Hubble type \citep[e.g
][]{YoSc91,BM98}. The role of gas has been studied in various contexts in galactic dynamics, and it has been shown that the low velocity dispersion component, namely, gas has a significant effect on stability against both local axisymmetric \citep{JS84a,JS84b,BR88,Jog96,Raf01} and non-axisymmetric \citep{Jog92} perturbations.

 The longevity of a density wave was questioned by \citet{Too69} who showed that a wavepacket of density waves would propagate radially with a group velocity $c_{\rm g}(R)$ = $\partial \omega(k, R) / \partial k$, where $\omega$ and $k$ are the frequency and the wavenumber, respectively, and $R$ is the radius; and the dependence of $\omega$ on $k$ is determined by the corresponding dispersion relation. This results in winding up of the wavepacket in a time-scale of about $10^9$ yr. A recent work by \citet{GJ15} showed that the inclusion of gas in the disk makes the group transport slower by a factor of few, thus allowing the pattern to persist for a longer time-scale. They also showed that for the observed pattern speed of 18 km s$^{-1}$ kpc$^{-1}$ \citep{Sie12} and for assumed values of Toomre $Q$-parameters for our Galaxy, the disk when modeled as a stars-alone case, does not give a stable wave solution. Instead, one needs to invoke gas in order to get a stable density wave solution for the observed pattern speed.

In this paper, we study this in more detail and show this to be a general result: for this we study three external galaxies, NGC~6946, NGC~2997, and M~51 (NGC~5194)
for which the observational values of the pattern speeds for the spiral structure and the rotation curves are available in the literature. We first treat the galactic disk as comprised only of stars and from the dispersion relation we obtain the lowest possible value of the dimensionless frequency (see \S~{\ref{method-chap5}} for details) for which stars-alone will allow a stable wave. We next include gas on an equal footing with stars and follow the same procedure except for a two-component dispersion relation. We find that, at a radius equal to two disk scalelengths and for an assumed set of Toomre $Q$-parameter values, the stars-alone cases for NGC~6946 and NGC~2997 do not support a stable wave while for M~51, the stars-alone case marginally supports a stable density wave for the observed pattern speed. One has to include a gas fraction appropriate for the galaxies considered here, to get a stable density wave corresponding to the observed pattern speed.
As a check, we also varied the parameters considered, covering a reasonable range of values, and confirmed the validity of this finding. Also, based on the calculations, we derive a range of allowed pattern speeds that yields a stable density wave at a given radius of a galaxy.  We apply this method to these three galaxies, and find that the observed pattern speed values indeed fall in this prescribed range.

\S~\ref{formu5} describes the formulation of the problem while \S~\ref{res5} presents the results. \S~\ref{dis5} and \S~\ref{con5} contain the discussion and conclusion, respectively.

\section{Formulation of the problem}
\label{formu5}

We treat the galactic disk as a gravitationally coupled two-component (star plus gas) system, where the stars are treated as a collisionless system and characterized by the surface density $\Sigma_{\rm 0s}$ and the one-dimensional velocity dispersion $\sigma_{\rm s}$. The gas is treated as a fluid, characterized by the surface density $\Sigma_{\rm 0g}$ and a one-dimensional velocity dispersion or sound speed $\sigma_{\rm g}$.
Note that in any real galaxy, gas is seen to be present in both atomic ($HI$) and molecular ($H_2$) hydrogen form, having different surface density \citep{YoSc91,BB12} and velocity dispersion profiles \citep{Tam09}. For the sake of simplicity, here we treat gas as a single-component and study its effect on the dispersion relation.
 The galactic disk is taken to be infinitesimally thin and the pressure acts only in the disk plane. In other words, we are interested in gravitational perturbations in the disk plane only. We use the cylindrical co-ordinates (R, $\phi$, z).\\
\subsection{Dispersion relation in the WKB limit}
Consider the above system being perturbed by linear perturbations of the type  exp[${i (kr - \omega t)}$], where $k$ is the wavenumber and $\omega$ is the frequency of the perturbation.
For such a system, the dispersion relation in the WKB (Wentzel - Kramers - Brillouin) limit or the tightly wound case is \citep{GJ15}
\begin{equation}
\frac{2\pi G \Sigma_{\rm 0s} |k|\mathcal{F}\Big(\frac{\omega-m \Omega}{\kappa},\frac{k^2\sigma^2_{\rm s}}{\kappa^2}\Big)}{\kappa^2-(\omega-m\Omega)^2}+\frac{2\pi G \Sigma_{\rm 0g} |k|}{\kappa^2-(\omega-m\Omega)^2+\sigma^2_{\rm g}k^2} = 1\\,
\label{disp-raw}
\end{equation}
\noindent where $\mathcal{F}$ is the reduction factor which physically takes into account the reduction in the self-gravity of the disk arising due to the velocity dispersion of stars. The functional form of $\mathcal{F}$ is as given in \citet{BT87}. Here, $\Omega$ is the angular frequency and $\kappa$ is the local epicyclic frequency at a given radius.

After some algebraic simplifications \citep [for details see][]{GJ15}, the dispersion relation (equation~\ref{disp-raw}) reduces to
\begin{equation}
(\omega-m\Omega)^2=\frac{1}{2}\left \{(\alpha_{\rm s}+\alpha_{\rm g})- \left[(\alpha_{\rm s}+\alpha_{\rm g})^2-4(\alpha_{\rm s}\alpha_{\rm g}-\beta_{\rm s}\beta_{\rm g})\right]^{1/2}\right \}
\label{disp-2component}
\end{equation}
where,
\begin{equation}
\begin{split}
\alpha_{\rm s} = \kappa^2-2\pi G\Sigma_{\rm 0s}|k| {\mathcal F}\Big(\frac{\omega-m \Omega}{\kappa},\frac{k^2\sigma^2_{\rm s}}{\kappa^2}\Big)\\
\alpha_{\rm g} = \kappa^2-2\pi G\Sigma_{\rm 0g}|k|+k^2{\sigma}^2_{\rm g}\\
\beta_{\rm s} = 2\pi G\Sigma_{\rm 0s}|k|{\mathcal F}\Big(\frac{\omega-m \Omega}{\kappa},\frac{k^2\sigma^2_{\rm s}}{\kappa^2}\Big)\\
\beta_{\rm g} = 2\pi G\Sigma_{\rm 0g}|k| \,.
\end{split}
\end{equation}
Now, we define two dimensionless quantities, $s$, the dimensionless frequency and $x$, the dimensionless wavenumber as\\
\begin{equation}
s=({\omega-m\Omega})/{\kappa} =   {m(\Omega_p - \Omega})/{\kappa}, \: \:  x = k / k_{\rm crit}\\,
\label{dim-quan}
\end{equation}
\noindent where $k_{\rm crit} = \kappa^2 / 2 \pi G  (\Sigma_{\rm 0s} + \Sigma_{\rm 0g})$.
Substituting equation (\ref{dim-quan}) in equation (\ref{disp-2component}), we get the dimensionless form of the dispersion relation as \citep[see equations (7) \& (8) in][]{GJ15}:
\begin{equation}
s^2=\frac{1}{2}\left[(\alpha'_{\rm s}+\alpha'_{\rm g})-\left \{(\alpha'_{\rm s}+\alpha'_{\rm g})^2-4(\alpha'_{\rm s}\alpha'_{\rm g}-\beta'_{\rm s}\beta'_{\rm g})\right \}^{1/2}\right]
\label{stargas-disp}
\end{equation}
where,
\begin{equation}
\begin{split}
\alpha'_{\rm s}=1-(1-\epsilon)|x|{\mathcal F}(s,\chi) \\
\alpha'_{\rm g}=1-\epsilon |x|+\frac{1}{4}Q^2_{\rm g}\epsilon^2x^2 \\
\beta'_{\rm s}=(1-\epsilon)|x|{\mathcal F}(s,\chi) \\
\beta'_{\rm g} = \epsilon |x|\\
\end{split}
\end{equation}
and, $\chi$= ${k^2\sigma^2_{\rm s}}/{\kappa^2}$ = $0.286 Q_{\rm s}^2 (1-\epsilon)^2 x^2$.  
The three dimensionless parameters $Q_{\rm s}$, $Q_{\rm g}$, and $\epsilon$ are respectively the Toomre $Q$ factors for stars as a collisionless system  $Q_{\rm s}$(=$\kappa \sigma_{\rm s} /(3.36 G \Sigma_{\rm 0s})$), and for gas $Q_{\rm g}$ = ($\kappa \sigma_{\rm g} /(\pi G \Sigma_{\rm 0g})$) \citep{Too64}, 
and $\epsilon$ =${\Sigma_{\rm 0g}}/( {\Sigma_{\rm 0s}+\Sigma_{\rm 0g}})$ is the gas mass fraction in the disk.

Similarly, the one-component analogue of this dispersion relation is \citep{BT87}
\begin{equation}
s^2= 1-|x| {\mathcal F} \Big(\frac{\omega-m \Omega}{\kappa},\frac{k^2\sigma^2_{\rm s}}{\kappa^2}\Big)\\.
\label{onefluid-disp}
\end{equation}
The dispersion relations (equation (\ref{onefluid-disp}) for stars-alone and equation (\ref{stargas-disp}) for the two-component case) provide the information about how the dimensionless frequency $s$ varies locally with respect to the dimensionless wavenumber $x$.  Note that the equations (\ref{stargas-disp}) and (\ref{onefluid-disp}) are symmetric with respect to both $s$ and $x$, hence we consider only their absolute values throughout this paper. It is clear that the absolute value of $s$ ($|s|$) ranges from 0 to 1, where $s=0$ yields the position of corotation (hereafter CR) and $|s| =1$ gives the positions of Lindblad resonances \citep{BT87}.

 It has been shown that for any $Q_{\rm s}> 1$, there exists a zone between Lindblad resonance and corotation, known as the forbidden region where the dispersion relation (equation (\ref{onefluid-disp})) has no real solution for $|k|$. Hence, the corresponding density wave solutions, that fall in that range, will be evanescent, i.e. they will have complex wavenumber $k$, and they will either decay or grow exponentially \citep{BT87}. A similar result holds when a two-component system has real $|s|$ solution.

 A wavepacket, starting from the long-wave branch of the dispersion relation ($|x| < 1$), travels inward with a negative group velocity and gets reflected from the edge of the forbidden region near the CR, and then start travelling radially outward in the short-wave branch ($|x|> 1$) with a positive group velocity (see Figure~\ref{fig-explain}) before finally being absorbed at a large wavenumber by a process similar to Landau damping \citep{BT87}.

\subsection {Method}
\label{method-chap5}

Note that, in the density wave theory, the pattern speed of the grand-design two-armed spiral structure, $\Omega_{\rm p}$ (= $\omega/m$; $m = 2$ here for spirals) is treated as a free parameter \citep[e.g.][]{LS66}, and its value is obtained from observations. Thus, for a given observed value of $\Omega_ {\rm p}$ and from the observed rotation curve, the dimensionless frequency  $|s|$ ($=m|\Omega_p- \Omega|/\kappa$) (equation (\ref{dim-quan})), has a definite value at a certain radius $R$, say $|s|_{\rm obs}$. Therefore, for getting a stable density wave, the observationally found $|s|_{\rm obs}$ value should fall in the allowed range of $|s|$, derived from the dispersion relation, both obtained at the same radius. In other words, the horizontal line $|s|$ = $|s|_{\rm obs}$ should cut the plot of $|s|$ versus $|x|$ at that radius, to yield a stable wave solution (i.e. a real solution for $|k|$), otherwise it will give an evanescent density wave solution.

\begin{figure}
\centering
\includegraphics[height=3.0in,width=4.0in]{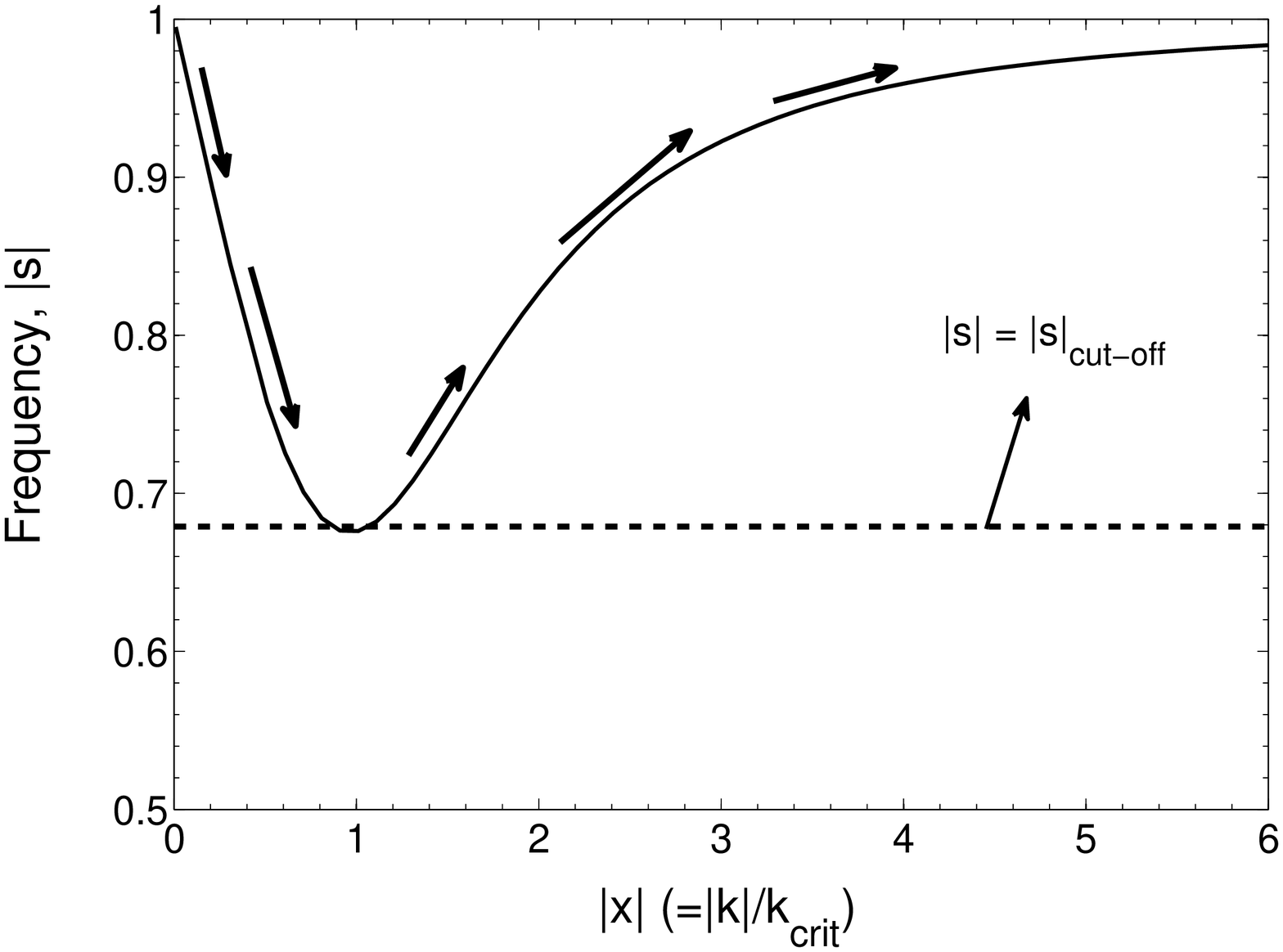}
\caption{The dimensionless frequency $|s|$ versus the dimensionless wavenumber $|x|$ from the dispersion relation for the stars-alone case (equation (\ref{onefluid-disp})), plotted for $Q_{\rm s} =1.7$. Since the dispersion relation is symmetric in both $s$ and $x$, only their absolute values are shown here. The arrows denote the direction of propagation of a typical density wavepacket. The horizontal line $|s|$ = $|s|_{\rm cut-off}$ indicates the lowest possible value of $|s|$, for which one can get a stable wave solution.}
\label{fig-explain}
\end{figure}

We define $|s|_{\rm cut-off}$ as the lowest possible value of $|s|$ for which one is able to get a stable wave solution from the dispersion relation  at a given radius $R$. We set $|s|_{\rm cut-off}$ as the lowest $|s|$-value, where the plot of the dispersion relation turns around (equation (\ref{stargas-disp}) or equation (\ref{onefluid-disp}), whichever is applicable). In other words, $|s|_{\rm cut-off}$ indicates the edge of the forbidden region. A typical example of how $|s|_{\rm cut-off}$ is obtained from a dispersion relation for a stars-alone case, is shown in Figure~{\ref{fig-explain}}. For the sake of illustration, we assume $Q_{\rm s}$ = 1.7, as observed in the solar neighbourhood \citep{BT87}.

Thus, if the following inequality holds for a certain radius $R$: 
\begin{equation}
|s|_{\rm obs} \ge |s|_{\rm cut-off}
\label{stable-cond}
\end{equation}
then one can say that the observed pattern speed will give a stable density wave, and the dispersion relation will have real solution for $|k|$.
Throughout this paper, $|s|_{\rm cut-off}$ will be used for making quantitative statements
regarding the existence of stable solutions.

\section{Results}
\label{res5}

\subsection{Input parameters}
We consider three galaxies for which the observational values for the input parameters, namely the pattern speed and the rotation curve, are available in the literature.
\subsubsection{NGC~6946}
This is a barred grand-design spiral galaxy with the Hubble type Scd. It has an angle of inclination $38^0$ \citep{CAR90,Boo07} and it is located at a distance of 5.5 Mpc \citep{KEN03}. The pattern speed of the main $m = 2$ gravitational perturbation i.e. the large oval and the two prominent spiral arms, is measured to be $22_{-1}^{+4}$ km s$^{-1}$ kpc$^{-1}$ by \citet{Fat07}, using the TW method. They also derived the locations of different resonance points (see table 2 there). The $HI$ rotation curve is taken from \citet{Boo07}. The exponential disk scalelength is measured as 1.9 arcmin or 3.3 kpc, where 1 arcmin = 1.75 kpc \citep{CAR90}.

\subsubsection{NGC~2997}
This is a grand-design spiral galaxy of the Hubble type Sc \citep{MM81}. The pattern speed of the grand-design spiral structure is measured
to be  16 km s$^{-1}$ kpc$^{-1}$ by \citet{GD09}, by using the measurement of azimuthal age-gradient of newly formed stars. 
The rotation curve is taken from \citet{PET78}. The exponential disk scalelength is measured as 4.0 kpc \citep{GBP99}

\subsubsection{M~51 (NGC~5194)}
This is a face-on spiral galaxy of the Hubble type Sc and is located at a distance of 9.6 Mpc \citep{SanTamm75}, in close interaction with NGC~5195. The pattern speed of the spiral structure is measured to be 38 km s$^{-1}$ kpc$^{-1}$ by \citet{Zimm04}, using $CO$ as a tracer and by applying the TW method. The rotation curve for M~51 is found to be steeply rising up to $\sim$ $R = 25$ arcsec and then for R $>$ $25$ arcsec it saturates to 210 km s$^{-1}$ \citep{RAND93}, where 1 arcsec = 46.5 pc, assuming a distance of 9.6 Mpc \citep{SanTamm75}. The observed rotation curve and the observed pattern speed together place the CR at a distance of 5.5 kpc. The exponential disk scalelength is measured in various wavelengths and is found to vary from 4.36 kpc in $B$ band to 3.77 kpc in $R$ band \citep{Beck96}. We took a mean value of 4.0 kpc for the present purpose.

\subsection{Stars-alone case}

We first investigate whether a disk, consisting only of stars, can support a stable density wave for the observed values of pattern speed in different galaxies. To do this, first we obtained $|s|_{\rm obs}$ 
at a radius of 2$R_{\rm d}$, $R_{\rm d}$ being the exponential disk scalelength, in each of these three galaxies.
The choice of 2R$_{\rm d}$ is made because the spiral structure is typically seen in the middle part of an optical disk whose size is $\sim$ 4-5 R$_{\rm d}$ \citep[e.g.][]{BM98}. We found that the $|s|_{\rm obs}$ values for NGC~6946, NGC~2997, and M~51 are 0.38, 0.44, and 0.63, respectively.

\begin{figure}
\centering
\includegraphics[height=3.0in,width=4.0in]{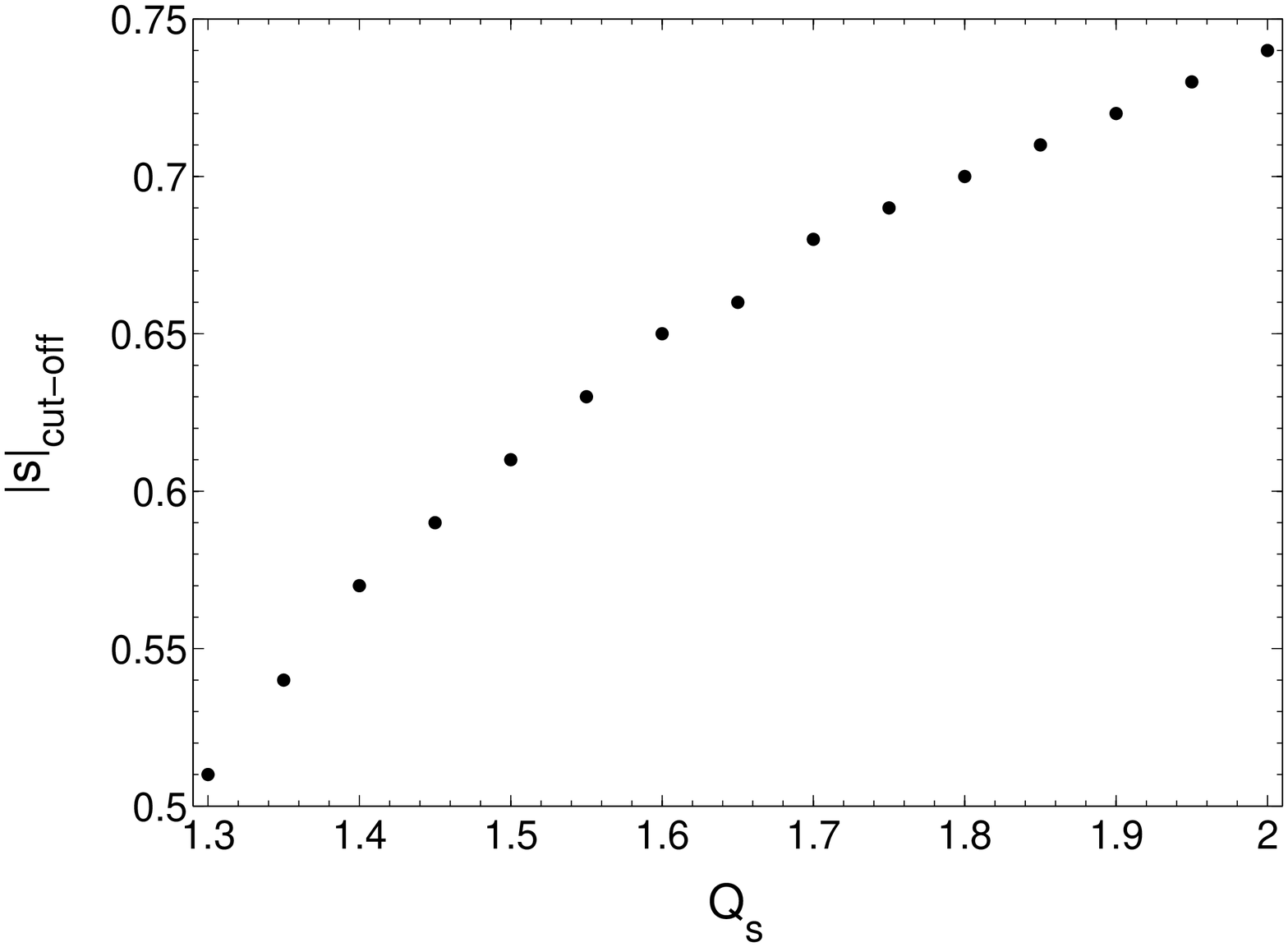}
\caption{$|s|_{\rm cut-off}$, the lowest values of the dimensionless frequency for which a stable density wave solution is possible, 
plotted as a function of $Q_s$, obtained from the one-component dispersion relation (equation~\ref{onefluid-disp}). The resulting $|s|_{\rm cut-off}$ value shows a steady increase with $Q_{\rm s}$, thereby implying an increase in the forbidden region.}
\label{fig-1cutoff}
\end{figure}

Now to check whether the inequality given in equation (\ref{stable-cond}) is satisfied or not, we need to calculate the $|s|_{\rm cut-off}$ value from the dispersion relation  (equation (\ref{onefluid-disp})) for these galaxies. Note that the calculation of $|s|_{\rm cut-off}$ from the dispersion relation requires the knowledge of $Q_{\rm s}$, and the values of $Q_{\rm s}$ at different radii for  a galaxy are not known observationally for most galaxies. We assume $Q_{\rm s}$ to be constant with $R$ for simplicity, but in reality its values will vary \citep[e.g., see][]{GJ14,Jog14}. Further, for a theoretical study, we varied the value of this radially constant $Q_{\rm s}$ from $1.3$ to $2.0$, while $1.7$ is the typical value in the solar neighbourhood \citep{BT87}. The plot of the resulting $|s|_{\rm cut-off}$ versus $Q_{\rm s}$ is shown in Figure~\ref{fig-1cutoff}. 

From Figure~{\ref{fig-1cutoff}}, we see that the $|s|_{\rm cut-off}$ value increases steadily with the increase of $Q_{\rm s}$, thereby implying a steady increase in the forbidden region around the CR as the $Q_{\rm s}$ value increases. 

On comparing the $|s|_{\rm obs}$ values that we obtained for both NGC~6946 and NGC~2997, with Figure~{\ref{fig-1cutoff}}, we find that the inequality given by equation (\ref{stable-cond}) is not satisfied for any value of $Q_{\rm s}$, considered here. Thus, the dispersion relation for a purely stellar disk does not yield a real solution for $|k|$ and hence it does not yield a stable density wave corresponding to the observed pattern speeds of both these galaxies. For M~51, the stars-alone case barely supports a stable density wave when $Q_{\rm s} =$ 1.5, but if it is set to a value larger than 1.5, then the stars-alone case no longer supports a stable density wave for the observed pattern speed.

\subsection{Stars plus gas case}
\label{res-stargas}

\begin{figure}
\centering
\includegraphics[height=3.0in,width=4.0in]{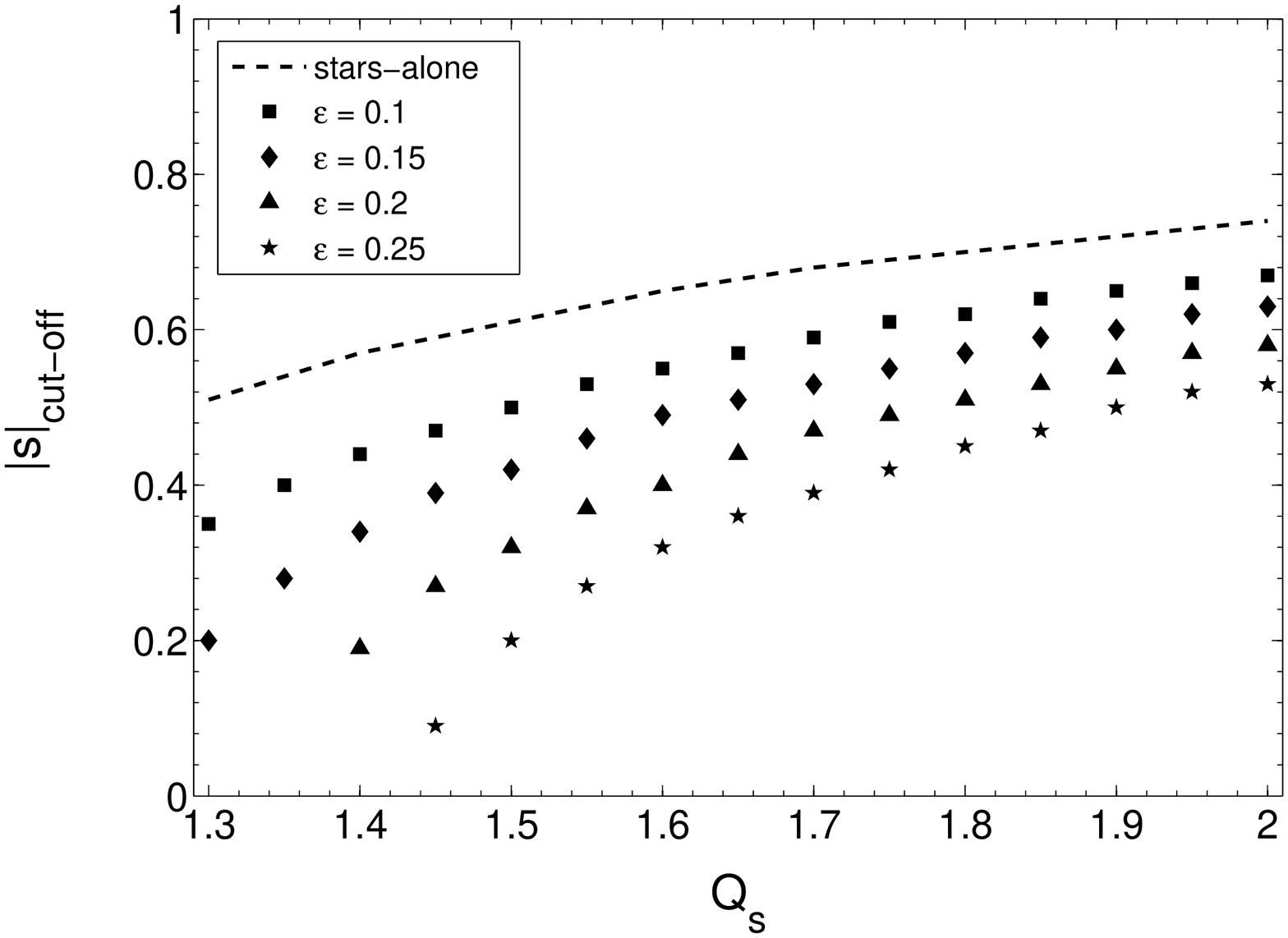}
\caption{$|s|_{\rm cut-off}$, the lowest values of the dimensionless frequency for which a stable density wave solution is possible for the two-component case, 
plotted as a function of $Q_{\rm s}$, for different gas fractions ($\epsilon$) and $Q_{\rm g}$ = 1.5. The labels used for different gas fractions are as indicated in the legend. With increasing gas fraction, the $|s|_{\rm cut-off}$ decreases steadily for the whole range of $Q_{\rm s}$ considered here. Thus the extent of the forbidden region decreases with increasing gas fraction.}
\label{cutoff-2flu}
\end{figure}

We next include the interstellar gas, which has a low velocity dispersion as compared to the stars, in the system, and study whether the inclusion of gas helps in getting a stable wave solution for the observed pattern speeds.
First, we investigated how the values of $|s|_{\rm cut-off}$ change with different values for the three parameters $Q_{\rm s}$, $Q_{\rm g}$, and $\epsilon$. We varied $Q_{\rm s}$ from 1.3 to 2.0 and $\epsilon$ from 0.1 to 0.25. In each case we fix $Q_{\rm g}$, and then compute $|s|_{\rm cut-off}$ from equation (\ref{stargas-disp}), as a function of $Q_{\rm s}$, and repeat this procedure for different values of $\epsilon$. For comparison, we replotted the $|s|_{\rm cut-off}$ values for the stars-alone case, as a function of $Q_{\rm s}$. The result for $Q_{\rm g} = 1.5$ is shown in Figure~\ref{cutoff-2flu}. From Figure~\ref{cutoff-2flu}, it is clear that with the inclusion of more gas (i.e. higher $\epsilon$ value), the $|s|_{\rm cut-off}$ value steadily decreases. This holds true for the above-mentioned range of $Q_{\rm s}$ values, at a given $Q_{\rm g}$. In other words, a larger value of gas fraction ($\epsilon$) helps to decrease the forbidden region around the CR, and thus allowing a higher range of permitted pattern speeds (for details see \S~\ref{range-predict}). For the other values of $Q_{\rm g}$ in the range of 1.4--1.8, we got a similar trend, hence we do not produce them here.

\begin{figure}
   \centering
    \begin{minipage}{.49\textwidth}
        \centering
        \includegraphics[height=2.5in,width=3.in]{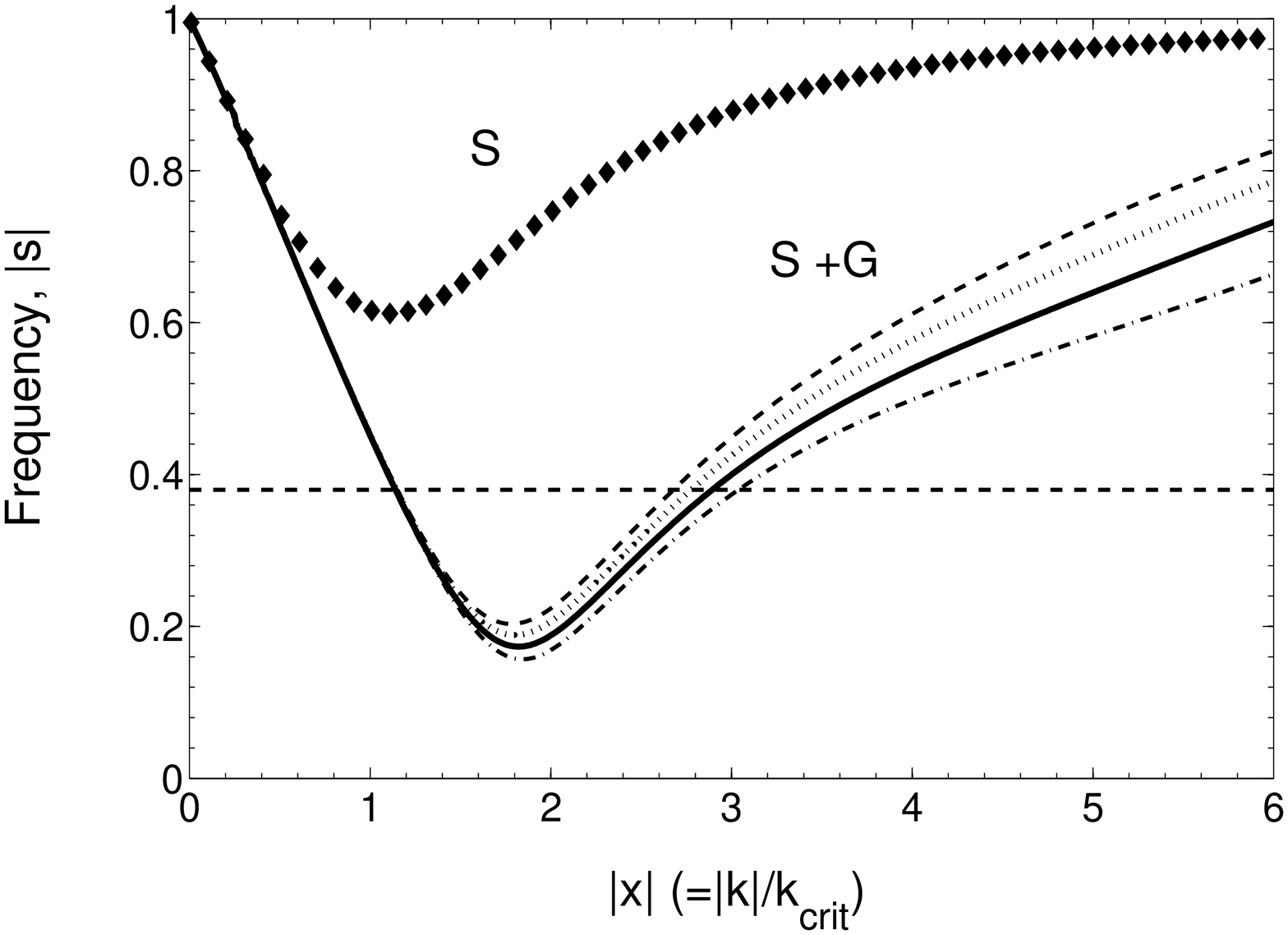}
       \vspace{0.2 cm}
	{\bf{(a)}}\\
    \end{minipage}
	\medskip
    \begin{minipage}{.49\textwidth}
        \centering
        \includegraphics[height=2.5in,width=3.in]{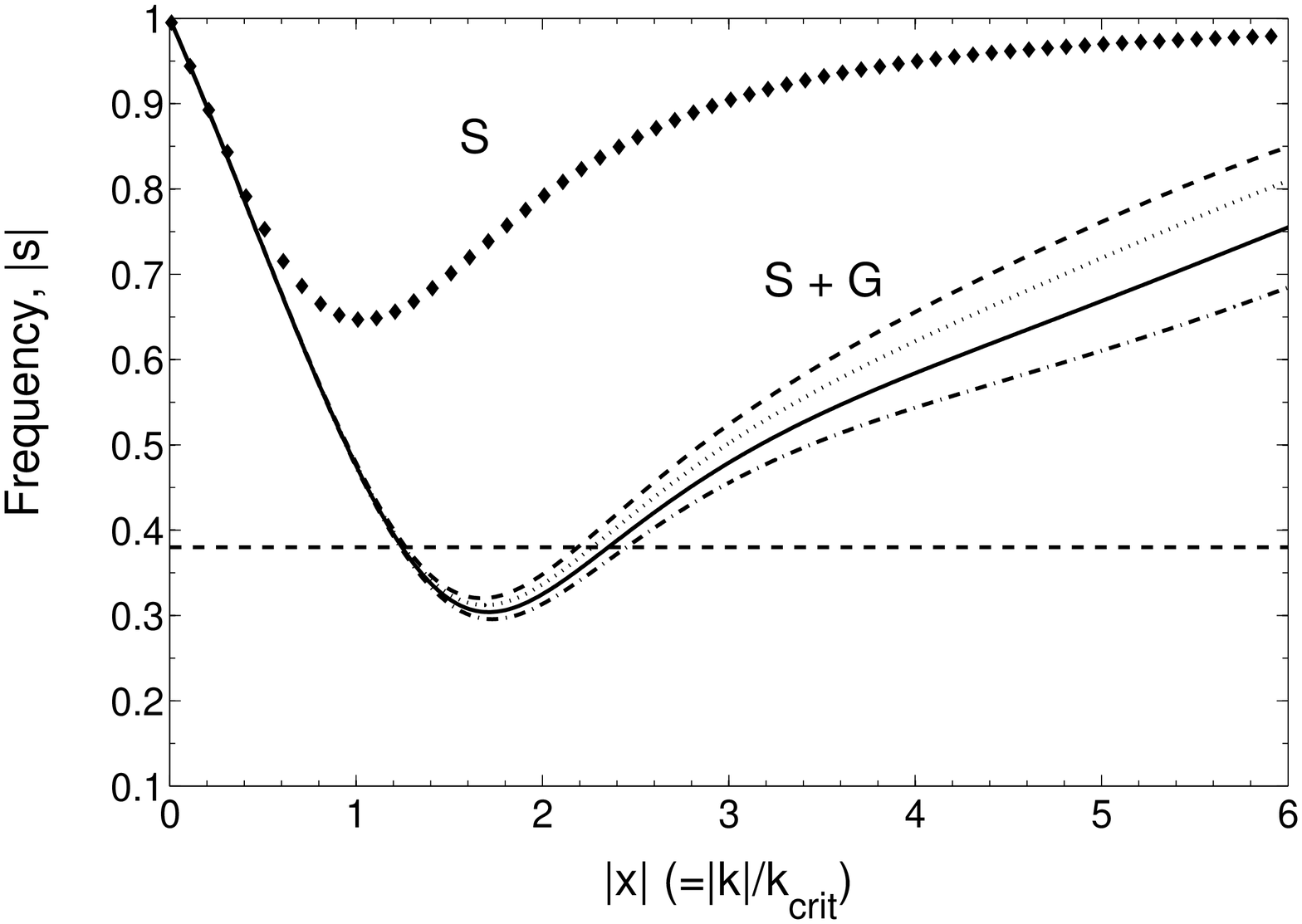}
        \vspace{0.2 cm}
	{\bf{(b)}}\\
    \end{minipage}
	\medskip
\begin{minipage}{.55\textwidth}
        \centering
        \includegraphics[height=2.5in,width=3.in]{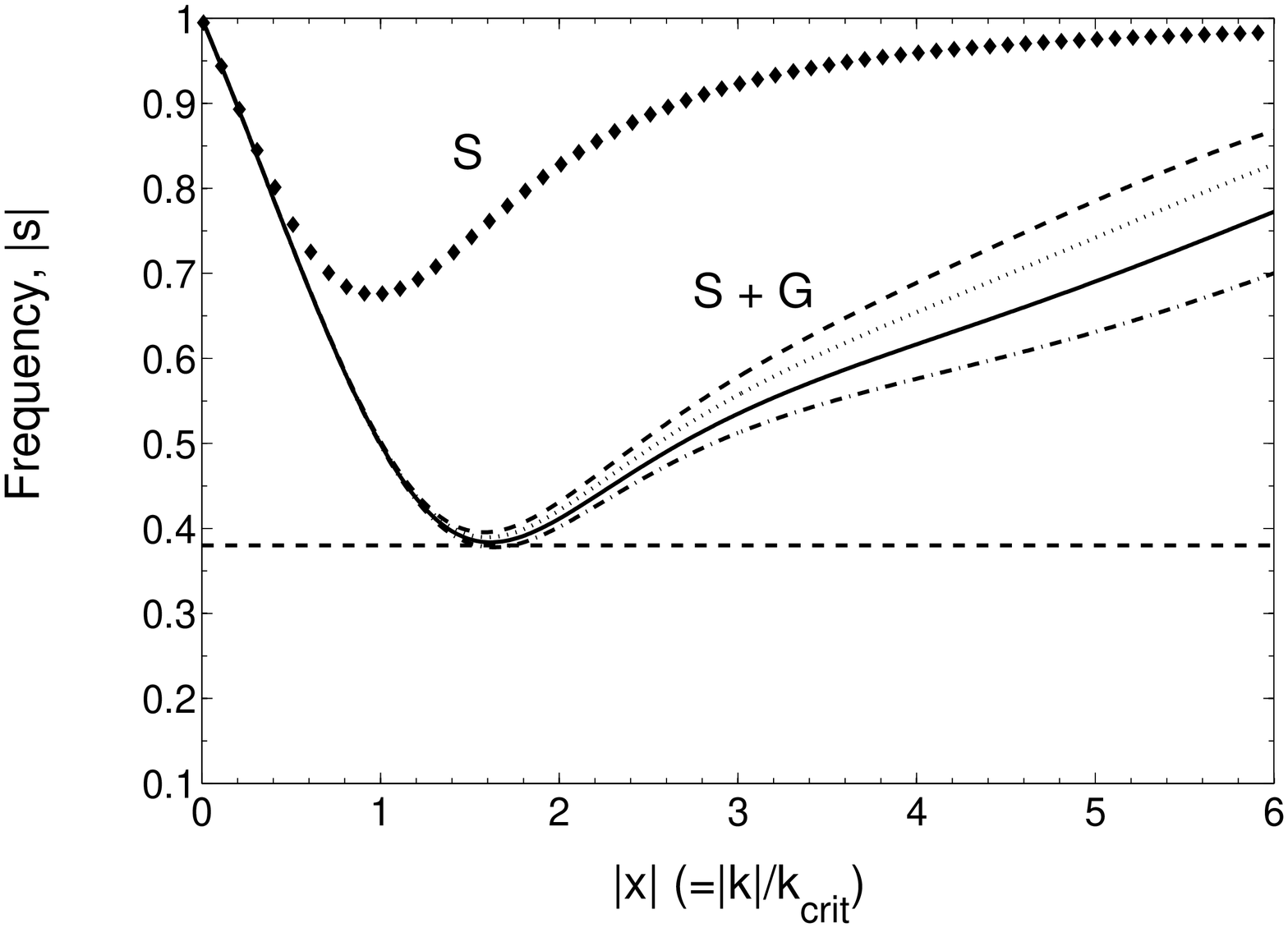}
        \vspace{0.2 cm}\\
	\hspace{-.2cm} {\bf{(c)}}\\
    \end{minipage}
    \caption{{\it{NGC~6946}} : Dispersion relations for stars-alone (S) and stars plus gas (S + G) cases, plotted in a dimensionless form, for a range of $Q_{\rm s}$ and $Q_{\rm g}$ values, at R = 2R$_{\rm d}$. Panel (a) for $Q_{\rm s}=1.5$, panel (b) for $Q_{\rm s}=1.6$, and panel (c) for $Q_{\rm s}=1.7$. In each panel, $Q_{\rm g}$ is taken to be 1.4, 1.5, 1.6, and 1.7, successively. The corresponding dispersion relations are shown from bottom to top. The horizontal line indicates the value $|s|_{\rm obs}$, derived from the observed pattern speed and the rotation curve. Here in all the cases, the two-component case allows a real solution for $|k|$ and thus a stable density wave solution for the observed value of the pattern speed, but this is not true for the stars-alone case.}
    \label{fig-ngc6946}
\end{figure}

\begin{figure}
    \centering
    \begin{minipage}{.49\textwidth}
        \centering
        \includegraphics[height=2.5in,width=3.in]{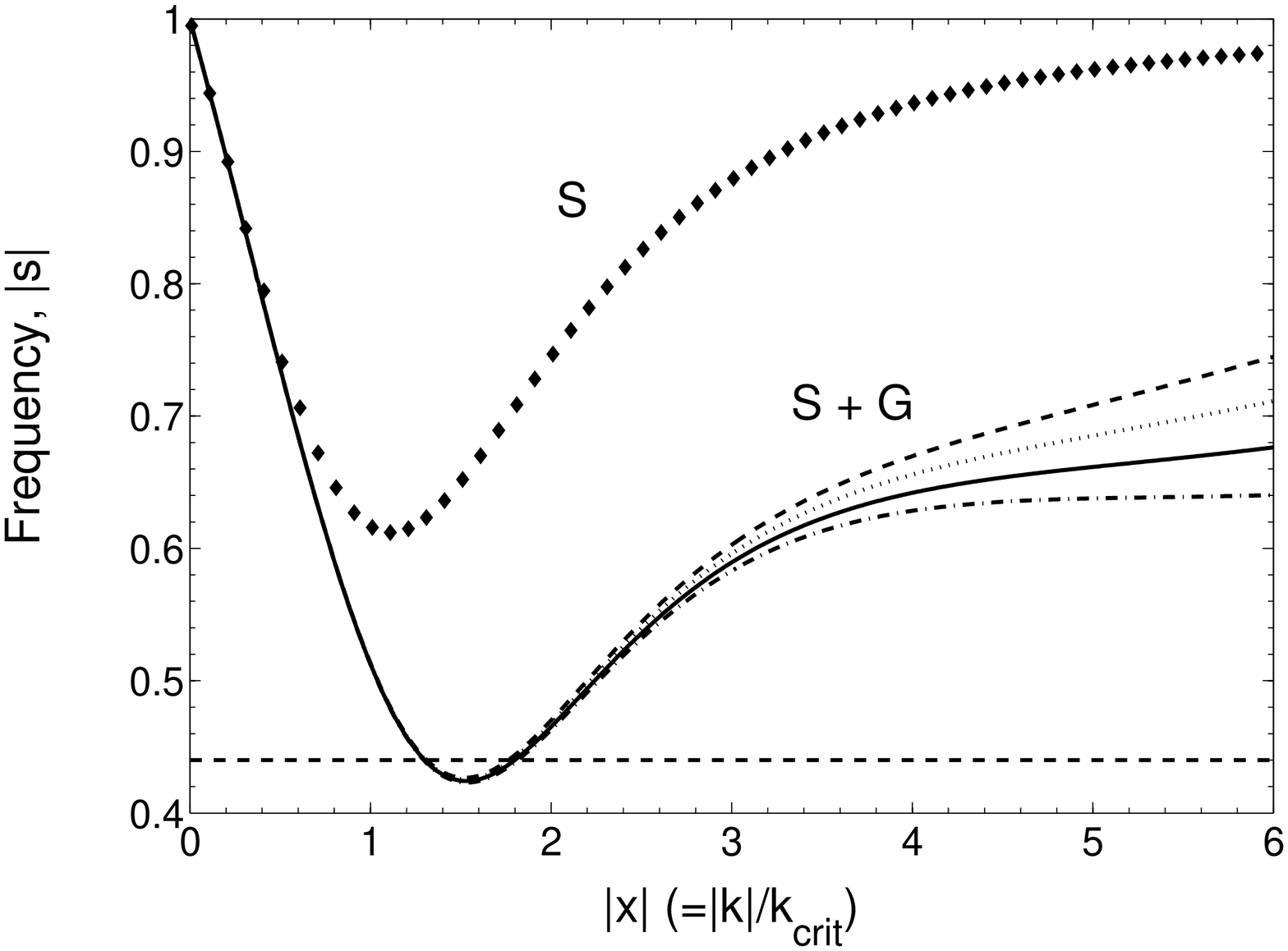}
       \vspace{0.2 cm}
	{\bf{(a)}}\\
    \end{minipage}
	\medskip
    \begin{minipage}{.49\textwidth}
        \centering
        \includegraphics[height=2.5in,width=3.in]{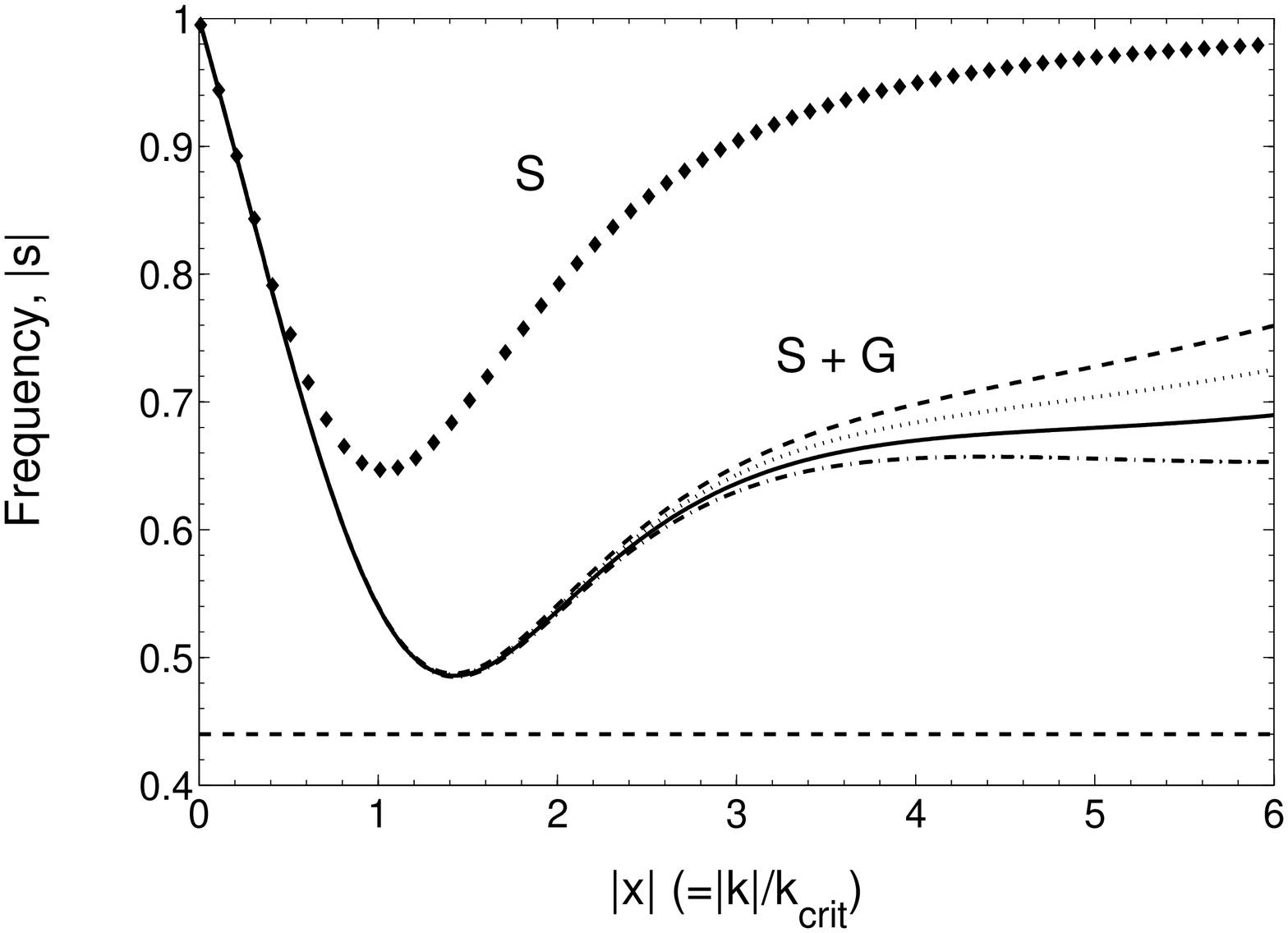}
        \vspace{0.2 cm}
	{\bf{(b)}}\\
    \end{minipage}
	\medskip
\begin{minipage}{.55\textwidth}
        \centering
        \includegraphics[height=2.5in,width=3.in]{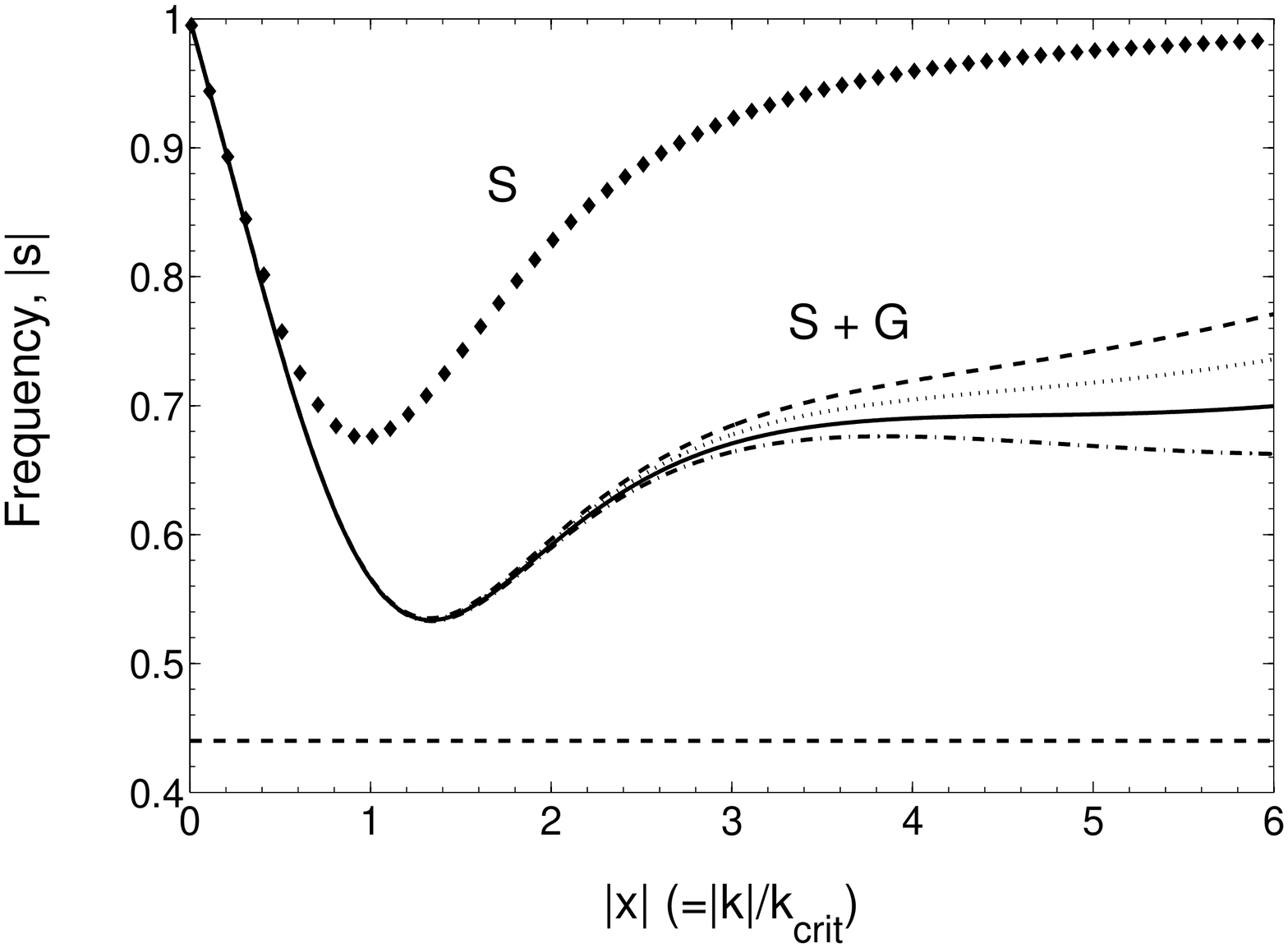}
        \vspace{0.2 cm}\\
	\hspace{-.2cm} {\bf{(c)}}\\
    \end{minipage}
    \caption{{\it{NGC~2997}} : Dispersion relations for stars-alone (S) and stars plus gas (S + G) cases, plotted in a dimensionless form, for a range of $Q_{\rm s}$ and $Q_{\rm g}$ values, at R = 2R$_{\rm d}$. Panel (a) for $Q_{\rm s}=1.5$, panel (b) for $Q_{\rm s}=1.6$, and panel (c) for $Q_{\rm s}=1.7$. In each panel, $Q_{\rm g}$ is taken to be 1.4, 1.5, 1.6, and 1.7, successively. The corresponding dispersion relations are shown from bottom to top. The horizontal line indicates the value $|s|_{\rm obs}$, derived from the observed pattern speed and the rotation curve. Here for $Q_{\rm s} = 1.5$, the two-component case, but not the stars-alone case, allows a stable density wave solution while for other values of $Q_{\rm s}$, none of the stars-alone and two-component case allows a stable density wave for the observed value of the pattern speed.}
\label{fig-ngc2997}
\end{figure}

\begin{figure}
    \centering
    \begin{minipage}{.49\textwidth}
        \centering
        \includegraphics[height=2.5in,width=3.in]{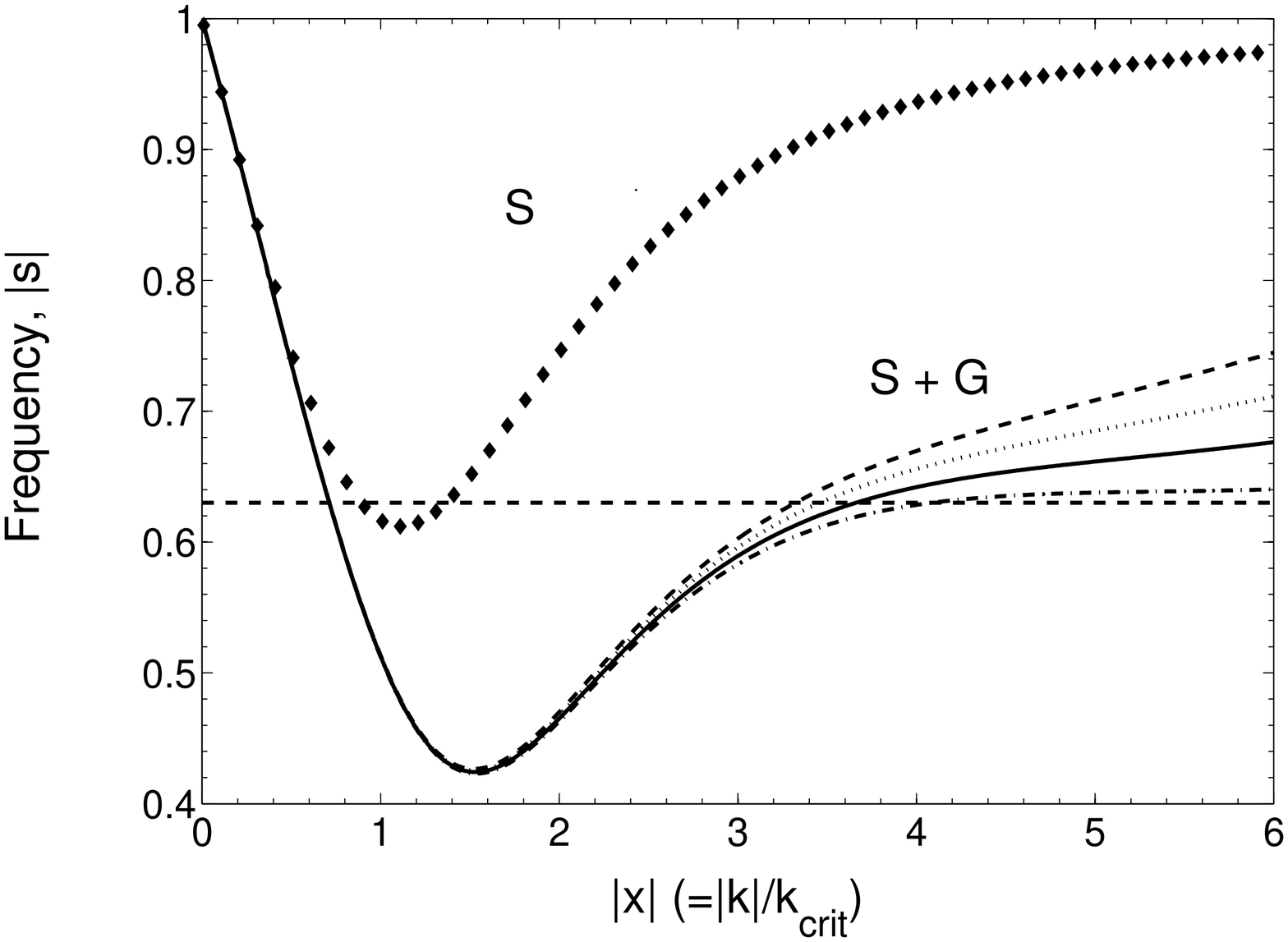}
       \vspace{0.2 cm}
	{\bf{(a)}}\\
    \end{minipage}
	\medskip
    \begin{minipage}{.49\textwidth}
        \centering
        \includegraphics[height=2.5in,width=3.in]{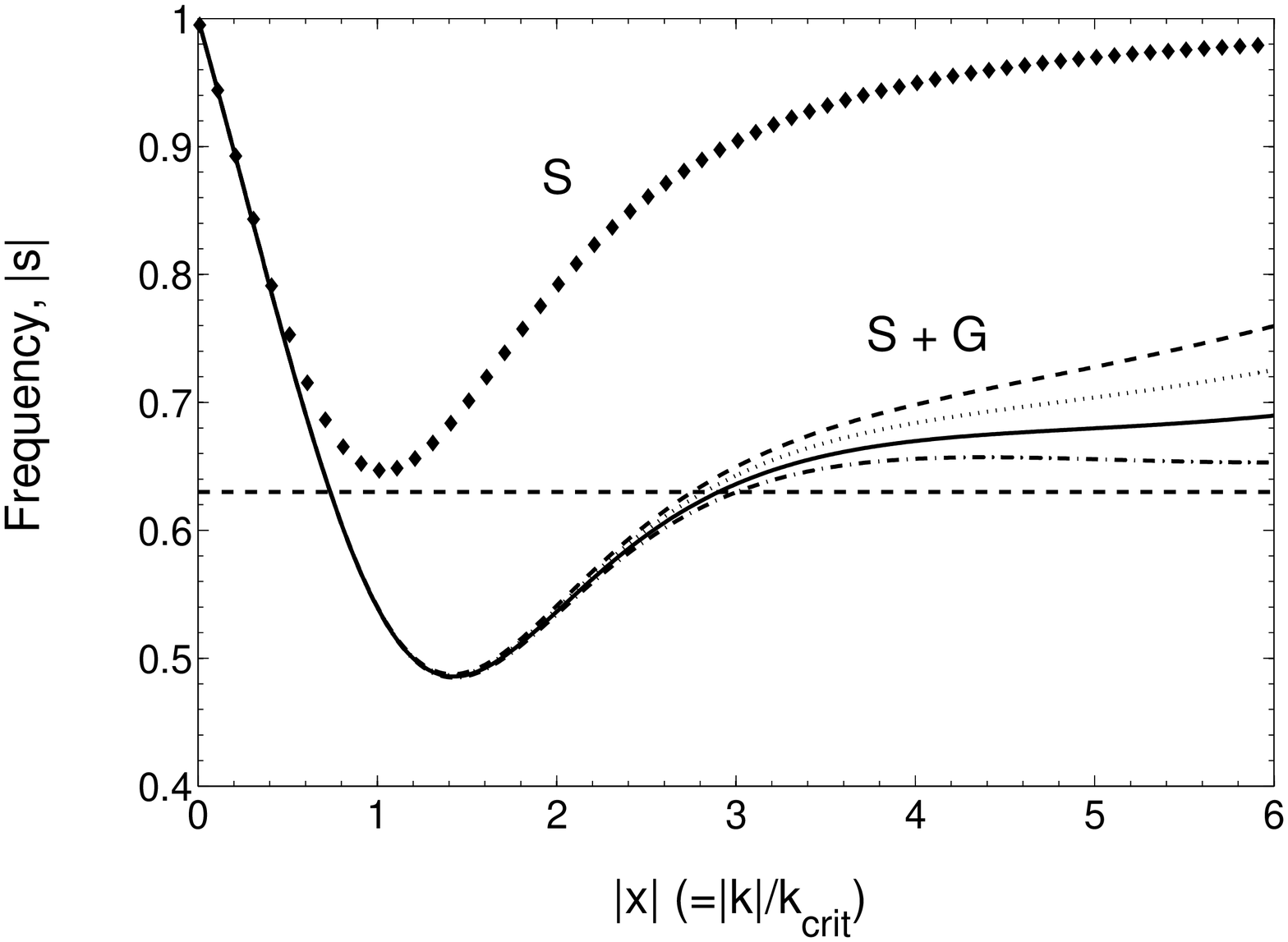}
        \vspace{0.2 cm}
	{\bf{(b)}}\\
    \end{minipage}
	\medskip
\begin{minipage}{.55\textwidth}
        \centering
        \includegraphics[height=2.5in,width=3.in]{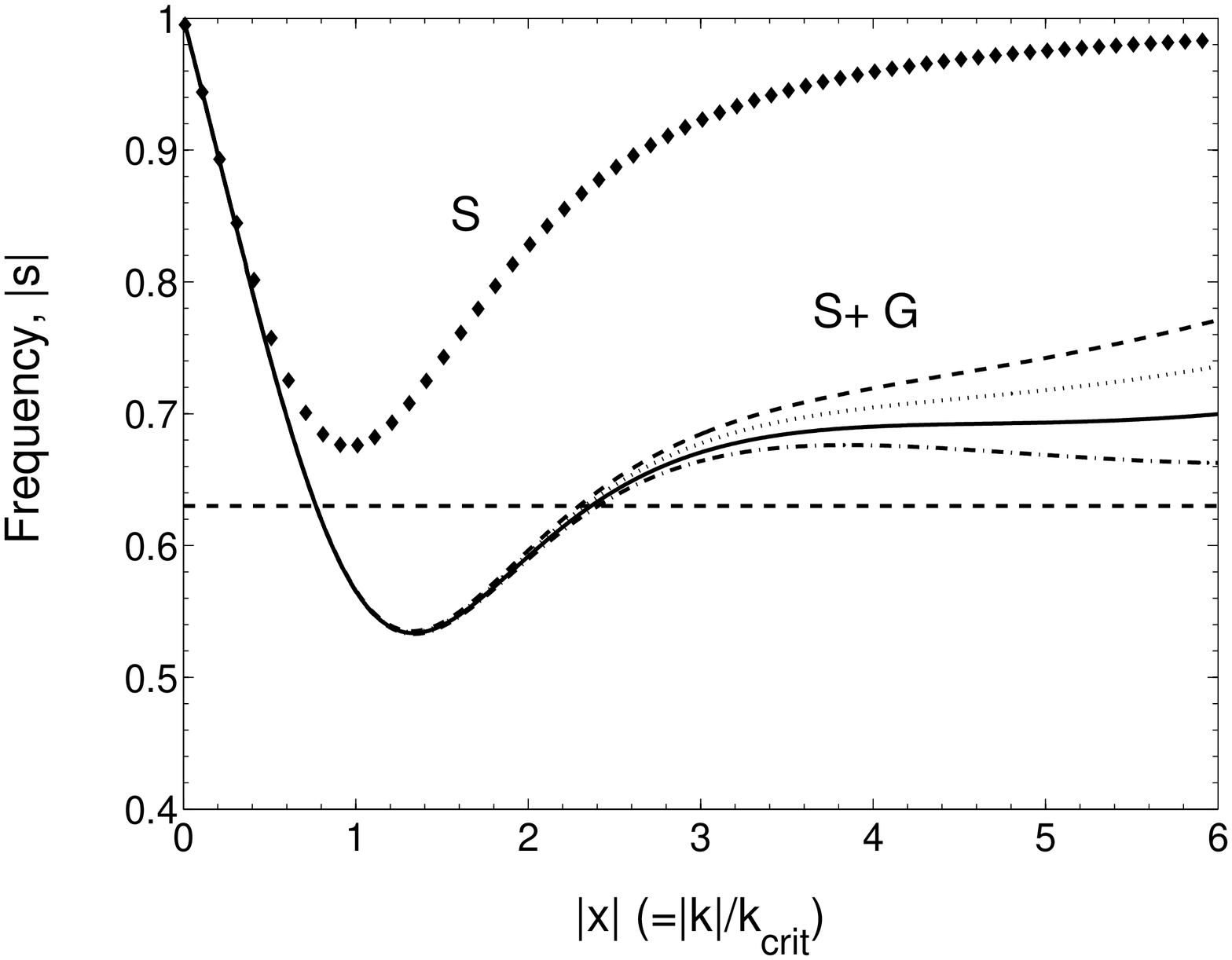}
        \vspace{0.2 cm}\\
	\hspace{-.2cm} {\bf{(c)}}\\
    \end{minipage}
    \caption{{\it{M~51}} : Dispersion relations for stars-alone (S) and stars plus gas (S + G) cases, plotted in a dimensionless form, for a range of $Q_{\rm s}$ and $Q_{\rm g}$ values, at R = 2R$_{\rm d}$. Panel (a) for $Q_{\rm s}=1.5$, panel (b) for $Q_{\rm s}=1.6$, and panel (c) for $Q_{\rm s}=1.7$. In each panel, $Q_{\rm g}$ is taken to be 1.4, 1.5, 1.6, and 1.7, successively. The corresponding dispersion relations are shown from bottom to top. The horizontal line indicates the value $|s|_{\rm obs}$, derived from the observed pattern speed and the rotation curve. Here for $Q_{\rm s}=1.5$, both the two-component case and the stars-alone case, allow a stable density wave solution, but for larger values of $Q_{\rm s}$, only the two-component case allows a stable density wave for the observed value of pattern speed.}
    \label{fig-m51}
\end{figure}

Now, we investigate the dynamical effect of gas on the grand-design spiral structures for the three specific galaxies, chosen for this work. The value of $|s|_{\rm cut-off}$ is obtained from the dispersion relation for a two-component system (equation~(\ref{stargas-disp})) for a set of values for the three
dimensionless input parameters ($Q_{\rm s}$, $Q_{\rm g}$, $\epsilon$). The values used are $Q_{\rm s} = 1.5$ and $Q_{\rm g} = 1.5$ for the stars plus gas case, and
$Q_{\rm s} = 1.5$ for the stars-alone case, with $\epsilon=0.25$ for NGC 6946 and $\epsilon=0.15$ for NGC 2997 and M~51, as typical for their Hubble types \citep[see figure~5,][]{YoSc91}. Then, we checked whether or not the value of $|s|_{\rm cut-off}$ obtained theoretically from the dispersion relation for the above input parameters, and the value of $|s|_{\rm obs}$ obtained from observations satisfy the inequality in equation (\ref{stable-cond}). 
The results for NGC~6946, NGC~2997, and M~51 are shown in Figure~\ref{fig-ngc6946}, Figure~{\ref{fig-ngc2997}}, and in Figure~{\ref{fig-m51}}, respectively.

Since, the above values of $Q_{\rm s}$ and $Q_{\rm g}$ are chosen in a somewhat ad hoc way, hence for each galaxy, we next study the variation in the dispersion relation for a reasonable range of $Q_{\rm s}$ and $Q_{\rm g}$ values, $Q_{\rm s}$ = 1.5, 1.6, 1.7 and $Q_{\rm g}$ = 1.4, 1.5, 1.6 and 1.7. The typical gas fraction value, chosen as per the Hubble type of any individual galaxy \citep[for details see][]{YoSc91} is kept constant. We found that, for a fixed $\epsilon$, there is a strong variation in the behaviour of the dispersion relation for different $Q_{\rm s}$ values, as compared to the different $Q_{\rm g}$ values.
A higher gas fraction ($\epsilon$) would be expected to change the results substantially, as suggested by \citet[see figure~3 there]{JS84a}, but here we have kept $\epsilon$ constant as typical for a given Hubble type.

Figure~{\ref{fig-ngc6946}}, and Figure~{\ref{fig-m51}}  show that for NGC~6946 and M~51, the $|s|_{\rm obs}$ value lies above the $|s|_{\rm cut-off}$ value for the star-gas system, thus the inequality given by equation (\ref{stable-cond}) is satisfied, and this is true for the whole range of parameter space considered here. Thus, the stars plus gas case allows a stable solution. For NGC~2997, the inequality given by equation (\ref{stable-cond}) is satisfied only for $Q_{\rm s} = 1.5$. For a $Q_{\rm s}$ value higher than 1.5, even the addition of gas in the two-component system no longer admits a stable density wave (see Figure~\ref{fig-ngc2997}). In contrast, for $Q_{\rm s} < 1.5$, the two component system admits a stable density wave solution. This can be seen from the result that as $Q_{\rm s}$ decreases, the $|s|_{\rm cut-off}$ for the stars plus gas also decreases, so that the forbidden region is smaller (see Figure~{\ref{cutoff-2flu}}).

 Hence, the galactic disk when treated as a gravitationally coupled stars plus gas system,
with the observed gas fraction,  allows a stable wave solution for the observed pattern speed, for most of the parameter ranges considered here. This is the main finding from this paper.

Interestingly, we see that for both NGC~6946 and NGC~2997 (Figure~\ref{fig-ngc6946} \& Figure~\ref{fig-ngc2997} (a)), the observed value of $|s|_{\rm obs}$ is close to the $|s|_{\rm cut-off}$ value which is obtained theoretically from the dispersion relation for the stars plus gas case (equation ({\ref{stargas-disp})), but not the  $|s|_{\rm cut-off}$ value obtained for the stars-alone case (equation (\ref{onefluid-disp})). Note that the curve for the dispersion relation for stars-alone case lies above that for the stars plus gas case, so if the pattern speed were such that the corresponding observed $|s|_{\rm obs}$ were to be  greater than $|s|_{\rm cut-off}$ value for the stars-lone case it would also be greater than the cut-off value for the two-component case.
 Thus, if a pattern speed value gives a stable density wave solution for a one-component case, it will also give so for the two-component case. Also the observed value of $|s|_{\rm obs}$ lies close to  the two-component $|s|_{\rm cut-off}$ value (but see M~51), and
this means that a galaxy seems to `prefer' to have a pattern speed that is indicated by the stars-plus-gas case for the observed gas fraction.

This can be explained as follows.

A joint two-component system is more unstable to the growth of perturbations or is closer to being unstable 
than the stars-alone case \citep{JS84a}, and this results in a lower cut-off $|s|_{\rm cut-off}$
for the two component case than the stars-alone case. If the galactic disk were subjected to perturbations having a range of values for pattern speeds (say as arising
due to a tidal interaction), then the perturbation most likely to be amplified in a galaxy is the one for which the dimensionless frequency $|s|_{\rm cut-off}$
has the lowest value, as this would correspond to the fastest growing perturbation.

This indicates that the inclusion of even 15 to 25 per cent gas fraction by mass has a non-trivial effect on the determination of the pattern speed that a galaxy is likely to have. 
The value of the pattern speed, along with the rotation curve, sets the locations of the Lindblad resonances in the disk. These are important in determining the secular evolution of galactic disk via angular momentum transport \citep{LYKA72}. 
Thus, our work shows that the gas 
is important in setting the observed pattern speed in a galaxy and hence it plays a crucial role in future dynamical evolution of a galactic disk.

We caution that the assumption of tight-winding (or, WKB limit), which played the crucial role in deriving the dispersion relations in equation~({\ref{stargas-disp}}) and equation~({\ref{onefluid-disp}}), is suspect for the long-wave branch ($|x| < 1$). This is important especially for NGC~2997, where the real solution ($|x|$), i.e. where the line $|s|=|s|_{\rm obs}$ cuts the dispersion relation, is less than 2 (see Figure~{\ref{fig-ngc2997}}). To verify the validity of the tight-winding limit in a more quantitative manner for all three galaxies considered here, we calculated the quantity $X$, at a radius $R$ equals to 2R$_{\rm d}$, where $X$ is defined as follows

\begin{equation}
X=\frac{\kappa^2 R}{2\pi G m \Sigma_{\rm tot}}\\,
\label{x-factor}
\end{equation} 
$\Sigma_{\rm tot}$ denotes the sum of the surface densities of the stellar and the gaseous components and $m$ (= 2, here) denotes the number of spiral arms. For the gas surface density tending to zero, it reduces to the usual $X$, defined for the one-component stellar case, as expected \citep[for details see][]{BT87}. The quantity $X$ denotes the cotangent of the pitch angle of waves of the critical wavenumber $k_{\rm crit}$ and $X \gg 1$ implies that the spiral arm is  tightly wound, and hence the tight-winding approximation holds good \citep{BT87}. We followed the following prescription for calculating the quantity $X$.\\
For a given gas fraction $\epsilon$ and for an observed gas surface density at a given radius $R$, we can obtain the total surface density at that radius. Further from the observed rotation curve, the epicyclic frequency $\kappa$ at $R=2R_{\rm d}$ is obtained. The total gas surface density for NGC~6946 is taken from \citet{Cros07} and for M~51, it is taken from \citet{Schu07}. A similar distribution of the total gas surface density for NGC~2997, as a function of radius, is not available in the literature, to the best of our knowledge. So for this case, we used the following technique to have an estimate of gas surface density. The $HI$ observations for NGC~2997 gives a total $HI$ mass of 4.2$\times$ 10$^9$ M$_\odot$, which extends over an area having mean diameter of 31.2 kpc \citep{Kod11}. This, in turn, gives a mean $HI$ surface density of 5.5 M$_\odot$ pc$^{-2}$ for NGC~2997. Now assuming a constant value of 0.5 for the ratio of mass in $HI$ to mass in $H_2$ \citep[for details see][]{YoSc91}, we get the mean surface density of $H_2$. Then this total mean surface density is used in equation (\ref{x-factor}) to derive the quantity $X$. The results for value of $X$ for the three galaxies are given in Table~{\ref{table-xvalues}}.

\begin{table}
\centering
\caption{Values for $X$ for three galaxies, calculated at $R=2R_{\rm d}$}
\begin{tabular}{ccccccc}
\hline
Galaxy & $R$ & $\kappa$ &$\Sigma_{\rm g}$ & $\epsilon$ & $\Sigma_{\rm tot}$ & $X$\\
name & (kpc) & (km s$^{-1}$  & (M$_{\odot}$  && M$_{\odot}$ \\
&&kpc$^{-1}$) & pc $^{-2}$) && pc $^{-2}$)\\
\hline
NGC~6946 & 6.6 &42.4 & 22.4 & 0.25 & 89.6 & 2.3 \\
NGC~2997 & 8.0 & 32.7 & 8.3 & 0.15 & 55.3 & 2.7\\
M~51 & 8.0 & 37.1 & 10 & 0.15 & 66.7 & 2.9  \\
\hline
\end{tabular}
\label{table-xvalues}
\end{table}
From Table~{\ref{table-xvalues}}, we see that the values of $X$ is greater than 1, thus the tight-winding approximation is satisfied, though not by a comfortable margin.

\subsection {A general constraint on pattern speeds}
\label{range-predict}
For a collisionless disk, the density wave can exist only in the regions where
\begin{equation}
\Omega-\kappa/2 \le \Omega_{\rm p} \le \Omega+\kappa/2\\
\label{range-col}
\end{equation}
\noindent holds, and the equality holds only at the resonance points.

Here in this section, we use our technique based on the calculation of $|s|_{\rm cut-off}$ to constrain the range of allowed pattern speed for the $m=2$, grand-design spiral structure, which will give a stable density wave. The prescription is as follows.

First, we consider $Q_{\rm s}$ values in the range from $1.3$ to $2.0$ and $Q_{\rm g}$ also in the range from $1.4$ to $1.8$. The gas-fraction, $\epsilon$, is varied from 5 to 25 per cent of the total disk mass, depending on the Hubble-type of a particular galaxy. Then for a fixed value of $\epsilon$, we obtained the value of $|s|_{\rm cut-off}$ for the whole range of $Q_{\rm s}$ and $Q_{\rm g}$ considered here. We found that for the gas-fraction of 5 per cent, the gas does not contribute much towards getting a stable density wave, but as the gas-fraction increases steadily, the effect of gas becomes more prominent. However, galaxies having $\epsilon$ $>$ 25 per cent mainly show a flocculent spiral structure and do not generally host a grand-design spiral structure \citep{Elm11}. Considering the above points, we have restricted the value of $\epsilon$ from 10 to 25 per cent.

Now suppose, at a certain radius $R$ for a particular galaxy, one knows the value of $\epsilon$, and one can make a reasonable choice for the values of $Q_{\rm s}$ and $Q_{\rm g}$. Then from these parameters, we can  obtain the value of $|s|_{\rm cut-off}$, call it $\alpha$. 
Then, to get a stable wave, the pattern speed ($\Omega_{\rm p}$) has to satisfy equation (\ref{stable-cond}) with $|s|_{\rm obs}$ being replaced by $|s|$.
This in turn gives
\begin{equation}
s \ge \alpha \hspace{0.5 cm}\mbox{or} \hspace{0.5 cm}s\le -\alpha\,
\end{equation}
\noindent i.e. 
\begin{equation}
\Omega_{\rm p} \ge \Omega+\frac{\alpha \kappa}{2} \hspace{0.3 cm} \mbox{or} \hspace{0.3 cm} \Omega_{\rm p} \le \Omega-\frac{\alpha \kappa}{2}
\label{range-allowed}
\end{equation}
depending upon whether one is outside the CR or inside the CR.
Now combining inequalities as given in the equations (\ref{range-col}) and (\ref{range-allowed}), we can say that when one is inside the CR then the allowed range of pattern speed would be ($\Omega- \kappa/2$, $\Omega-\alpha \kappa/2$) and when one is outside the CR then the allowed range would be ($\Omega+\alpha \kappa/2$, $\Omega+\kappa/2$).

We applied this technique to the three galaxies considered here, at $R=2R_d$, using $Q_{\rm s}=1.5$ and $Q_{\rm g}=1.5$ that were used as a set of parameters in Figure~{\ref{fig-ngc6946}}, Figure~{\ref{fig-ngc2997}} and Figure~{\ref{fig-m51}}. For NGC~6946 and NGC~2997, $R=2R_{\rm d} $ lies inside the CR, and for M~51, $R=2R_{\rm d} $ lies outside the CR. Hence following the above prescription, the predicted range of allowed pattern speeds that give a stable density wave, would be ($\Omega- \kappa/2$, $\Omega-\alpha \kappa/2$) for NGC~6946 and NGC~2997, and for M~51, predicted range of allowed pattern speeds would be ($\Omega + \alpha\kappa/2$, $\Omega + \kappa/2$). The results are summarized in Table~{\ref{table-range}}. We check that the observed values of the pattern speed, 22 km s$^{-1}$ kpc$^{-1}$ for NGC 6946, 16 km s$^{-1}$ kpc$^{-1}$ for NGC 2997, and 38 km s$^{-1}$ kpc$^{-1}$ for M~51, do indeed lie within the respective range of allowed pattern speeds obtained theoretically.

\begin{table*}
\centering
\caption{Range of allowed pattern speeds for NGC~6946, NGC~2997 \& M~51 at $R=2R_{\rm d}$}
\begin{tabular}{cccccccc}
\hline
Galaxy & $\Omega$ & $\kappa$ & $\epsilon$ & observed  & $\alpha$ & Lower  &  Upper   \\
name & (km   & (km   & (gas  & $\Omega_{\rm p}$ && bound on $\Omega_{\rm p}$   & bound on $\Omega_{\rm p}$    \\
& s$^{-1}$ & s$^{-1}$ && (km  s$^{-1}$ && (km  s$^{-1}$ & (km s$^{-1}$\\
& kpc$^{-1}$) & kpc$^{-1}$) & fraction) &  kpc$^{-1}$)&& kpc$^{-1}$) &  kpc$^{-1}$) \\
\hline
NGC~6946 & 30.0 & 42.4 & 0.25 & 22 & 0.17 & 8.8 & 26.4 \\
NGC~2997 & 23.1 & 32.7 & 0.15 & 16 & 0.42 & 6.8 & 16.2 \\
M~51 & 26.2 & 37.1 & 0.15 &  38 &0.42 & 34.0 & 44.8 \\
\hline
\end{tabular}
\label{table-range}
\end{table*}

 We then varied $Q_{\rm s}$ from 1.3 to 2.0 and $Q_{\rm g}$ from 1.4 to 1.8 to see what ranges of these two parameter will allow the observed pattern speed of three galaxies, to fall in the range predicted from our theoretical work, and thus give a stable density wave solution. These are summarized in Table~{\ref{table-qrange}}.

\begin{table}
\centering
\caption{Ranges of $Q_{\rm s}$ and $Q_{\rm g}$ for three galaxies,
that give a stable wave solution}
\begin{tabular}{ccccc}
\hline
Galaxy & $\epsilon$ & $R (=2R_{\rm d})$ & Range in & Range in \\
name & &(kpc) & $Q_{\rm s}$ & $Q_{\rm g}$ \\

\hline
\vspace {0.3 cm}
NGC~6946 & 0.25 & 6.6 &1.3 - 1.7 & 1.4 - 1.8 \\
NGC~2997 & 0.15 & 8.0  &1.3 - 1.5 & 1.4 - 1.8 \\
M~51 & 0.15 & 8.0 &1.3 - 1.9 & 1.4 - 1.8 \\
\hline
\end{tabular}
\label{table-qrange}
\end{table}

We note that a lower value of $Q_{\rm s}$ or $Q_{\rm g}$ than their respective minima chosen here will give an even smaller forbidden region, hence the observed pattern speed would be also permitted for $Q_{\rm s} < 1.3$ and $Q_{\rm g} < 1.4$.

It is interesting to note that for NGC~6946, the allowed range for the pattern speed for a stable
wave is obtained to be 
8.8--17.1 km s$^{-1}$ kpc$^{-1}$ for stars-alone case, while the observed pattern speed 22 km s$^{-1}$ kpc$^{-1}$ clearly lies outside this allowed range.
Similarly, for NGC 2997, the allowed range for the stars-alone case is calculated to be 6.8--13.5 km $^{-1}$ kpc$^{-1}$, while again the observed pattern speed
16 km s$^{-1}$ kpc$^{-1}$ lies outside this permitted range. Thus, real galaxies seem to have pattern speed values that lie beyond the values required for
a stable solution for the stars-alone case. The presence of gas pushes the galaxy to adopt a pattern speed which is higher than the stars-alone case. Thus,
the inclusion of gas will have effect in setting the pattern speed and thus in turn will effect the secular evolution of the disk  galaxy (see \S~\ref{res-stargas}).

We also illustrate the effect of gas on pattern speeds in another way. Suppose we were to artificially decrease the gas fraction in NGC 6946 to be 15 per cent, then the allowed range for the pattern speeds is obtained to be 8.8--20.8 km $^{-1}$ kpc$^{-1}$ -- this does not cover the observed pattern speed of 22 km s$^{-1}$ kpc$^{-1}$. Thus, the actual value of gas mass fraction ($\epsilon$) also matters in setting the value of pattern speed. The real galaxies studied seem to have pattern speeds that are close to the 
upper value in the allowed range, or in other words when  $|s|_{\rm obs}$ is just higher than $|s|_{\rm cut-off}$ (see the discussion in \S~\ref{res-stargas}).

\section {Discussion}
\label{dis5}

Here, we mention a few points regarding this current work. First, the shape of the dispersion relation and hence the value of $|s|_{\rm cut-off}$ is largely dependent on the values of the parameters ($Q_{\rm s}$,  $Q_{\rm g}$, $\epsilon$) chosen.
 Note that, for most of the galaxies, the observed radial profiles for $Q_{\rm s}$ and $Q_{\rm g}$ are not available. Consequently, one has to rely on educative guesses for these two $Q$-values, as we have done here. If the actual observed values for $Q_{\rm s}$ and $Q_{\rm g}$ are known, the findings of this paper can be put in a more accurate way.

Secondly, the density wave corresponding to the observed pattern speed may not be long lasting. In fact, some of the past $N$- body simulations of spiral galaxies have showed that the spiral arms fade out quickly in time \citep{DOBA14}. However, when a stellar disk is represented with sufficiently higher number of particles ($\sim$ orders of 10$^8$), the spiral structure in the simulations lasts for a time-scale of about a Hubble time, because in these cases the Poisson noise of the system is minimized \citep{Fuji11,Don13}. But, in general, $N$-body simulations of spiral galaxies show that an individual spiral arm is transient and gets wound up, which is at odds with a classical long-lived quasi-stationary density wave scenario \citep{Sel11,Gra12,Ba13}.
 The spiral structures could be more complex, for example,  there is evidence that some galaxies could show either a long-lived spiral pattern or a short-lived pattern, or a galaxy could exhibit both types of patterns, e.g. see the discussion in \citet{Sie12}.

 Thirdly, all the results presented in this work, are based on the assumption of the existence of a quasi-stationary density wave that rotates in the galactic disk with a constant pattern speed. However, observational studies by \citet{Foy11} for a sample of 12 nearby late-type star-forming galaxies and by \citet{Fer12} for NGC~4321 did not find any angular off-set in age as predicted by the classical density wave theory. This shows that the density wave theory does not seem to be applicable for these galaxies.

Finally, note that, the results presented here are based on a local calculation, whereas the density wave extends over a large range of radii in the galactic disk. 
The results indicate that gas may have a significant effect on the spiral arms for the large-scale as well. A global modal analysis for a gravitationally coupled two-component (star plus gas) galactic disk to study the effect of the gas on large-scale spiral arms will be taken up in a future study.

\section{Conclusion}
\label{con5}

In summary, we have shown that the inclusion of gas is essential to get a stable density wave solution corresponding to the observed pattern speed in a spiral galaxy.
We use a two-component (stars plus gas) dispersion relation for a galactic disk, and assume reasonable gas fraction and Toomre $Q$ values for stars and gas. Then, we check whether the theoretical dispersion relation permits a real $|k|$ solution (or a stable wave) for the observed rotation curve and the pattern speed.
 Three galaxies of different Hubble type, NGC 6946, NGC 2997, and M~51, are chosen for which the pattern speed values of the grand-design spiral structure and the rotation curves are known from observations. We show that for both NGC~6946 and NGC~2997, at a radius of two disk scalelengths, the stars-alone case is not able to produce a stable density wave corresponding to the observed pattern speeds, while for M~51, stars-alone case barely supports a stable density wave. One has to include the observed gas fraction in the study, in order to get a stable density wave for these pattern speeds.
 This is the main result from this work. 
 Since these galaxies are typical representatives of their Hubble type, we expect that this finding will hold true for other grand-design spiral galaxies as well.

Based on the technique used here, we obtain a theoretical range of allowed pattern speeds that give a stable density wave at a certain radius of a galaxy. We apply this technique to the three galaxies considered here, and find that the observed pattern speeds of these three galaxies indeed fall in the respective prescribed range.
 
We show that the inclusion of even 15 to 25 per cent gas by mass fraction has a significant dynamical effect on the dispersion relation and hence on the pattern speed which is likely to be seen in a real, gas-rich galaxy. The resulting allowed range of pattern speeds is higher in presence of gas.
The value of pattern speed affects the angular momentum transport. Thus, the gas plays a crucial role in the secular evolution of a galaxy. The effect of gas on the large-scale spiral structures will be investigated in a future study.

\section{Appendix}
\subsection{Application to the Galaxy}

Earlier in this chapter, using different observational inputs we have shown for three late-type disk galaxies (NGC~6946, NGC~2997, M~51) that the inclusion of interstellar gas is needed in order to obtain a stable density wave corresponding to the observed values of the pattern speed for the spiral arms. Here we check this for our Galaxy also for which the pattern speed is measured and in addition to that other parameters such as the rotation curve and the amount of interstellar gas present in the Galaxy are relatively well-known.

In the section, we apply the technique for studying the effect of the interstellar gas on getting a stable density wave for the observed pattern speed for the Galaxy. The rotation curve and the disk scalelength are taken from \citet{MCS98}. The pattern speed for the spiral arm in the Galaxy is measured to be 18 km s$^{-1}$ kpc$^{-1}$ \citep{Sie12}. We perform this exercise at $R$ = 8 kpc, near the solar neighborhood, which is equal to 2.5 times of the disk scalelength. We took $Q_{\rm s}$ = 1.7, $Q_{\rm g}$ = 1.5, and $\epsilon$ = 0.2 for this chosen radius. The resulting effect of the interstellar gas is shown in Figure~\ref{fig-galaxy}.

Since, the above values of $Q_{\rm s}$ and $Q_{\rm g}$ are chosen in a somewhat ad hoc way, hence for the Galaxy, we next study the variation in the dispersion relation for a reasonable range of $Q_{\rm s}$ and $Q_{\rm g}$ values, $Q_{\rm s}$ = 1.5, 1.6, 1.7 and $Q_{\rm g}$ = 1.4, 1.5, 1.6, and 1.7, while keeping the gas fraction ($\epsilon$) fixed at 0.2. The result is summarized in Figure~\ref{fig-parameter_galaxy}.

\begin{figure}
    \centering
    \begin{minipage}{.49\textwidth}
        \centering
        \includegraphics[height=2.5in,width=3.in]{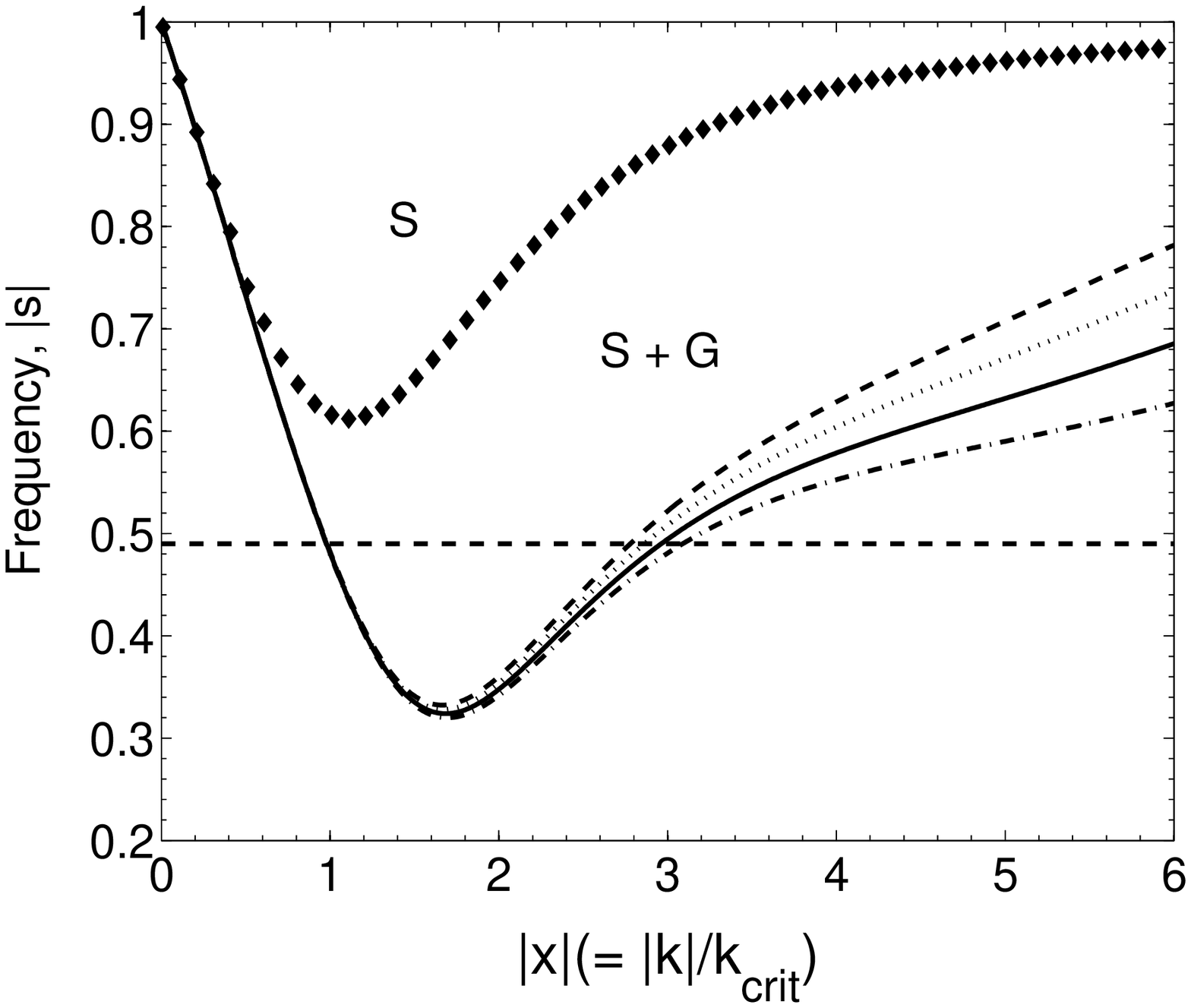}
       \vspace{0.2 cm}
	{\bf{(a)}}\\
    \end{minipage}
	\medskip
    \begin{minipage}{.49\textwidth}
        \centering
        \includegraphics[height=2.5in,width=3.in]{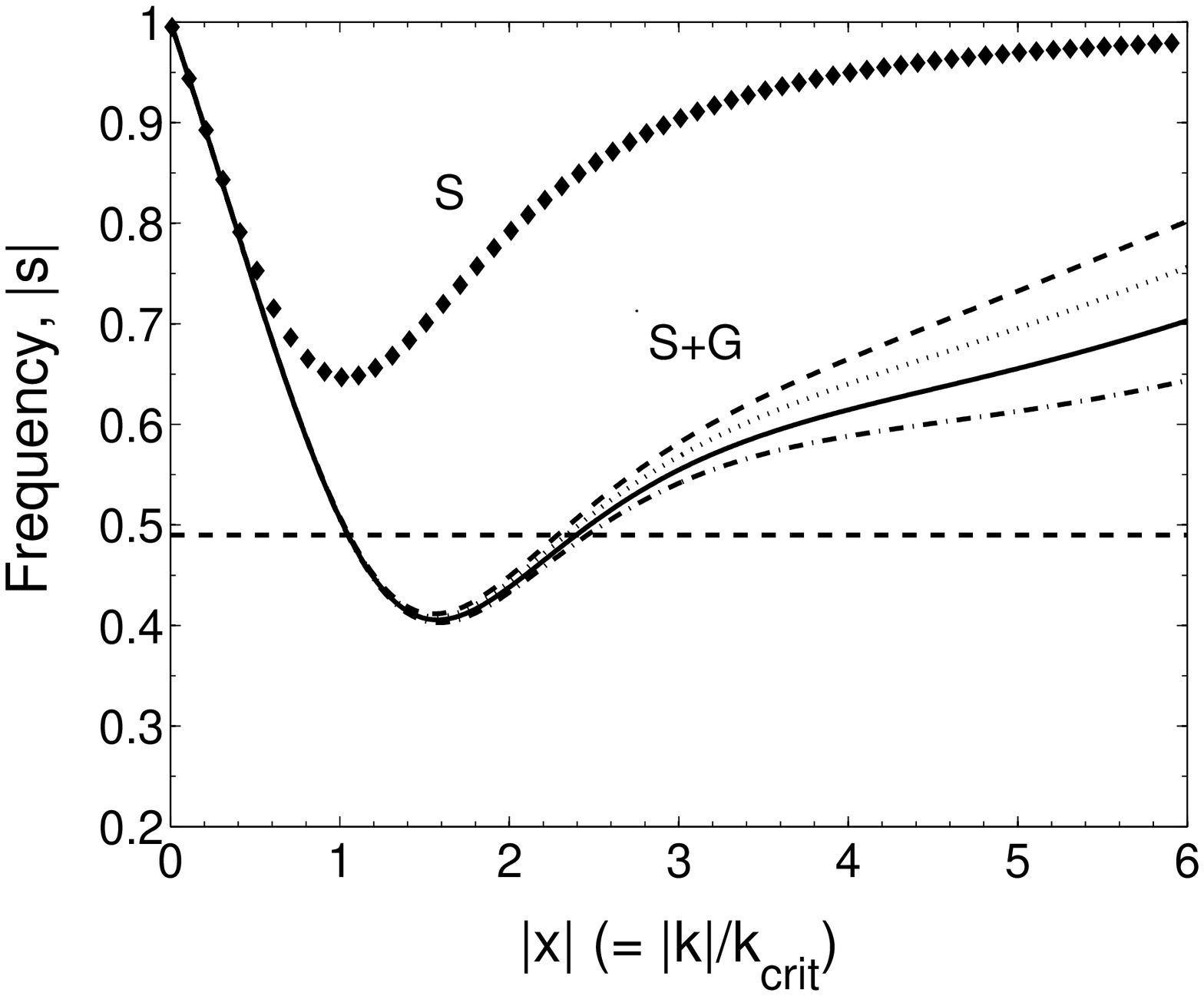}
        \vspace{0.2 cm}
	{\bf{(b)}}\\
    \end{minipage}
	\medskip
\begin{minipage}{.55\textwidth}
        \centering
        \includegraphics[height=2.5in,width=3.in]{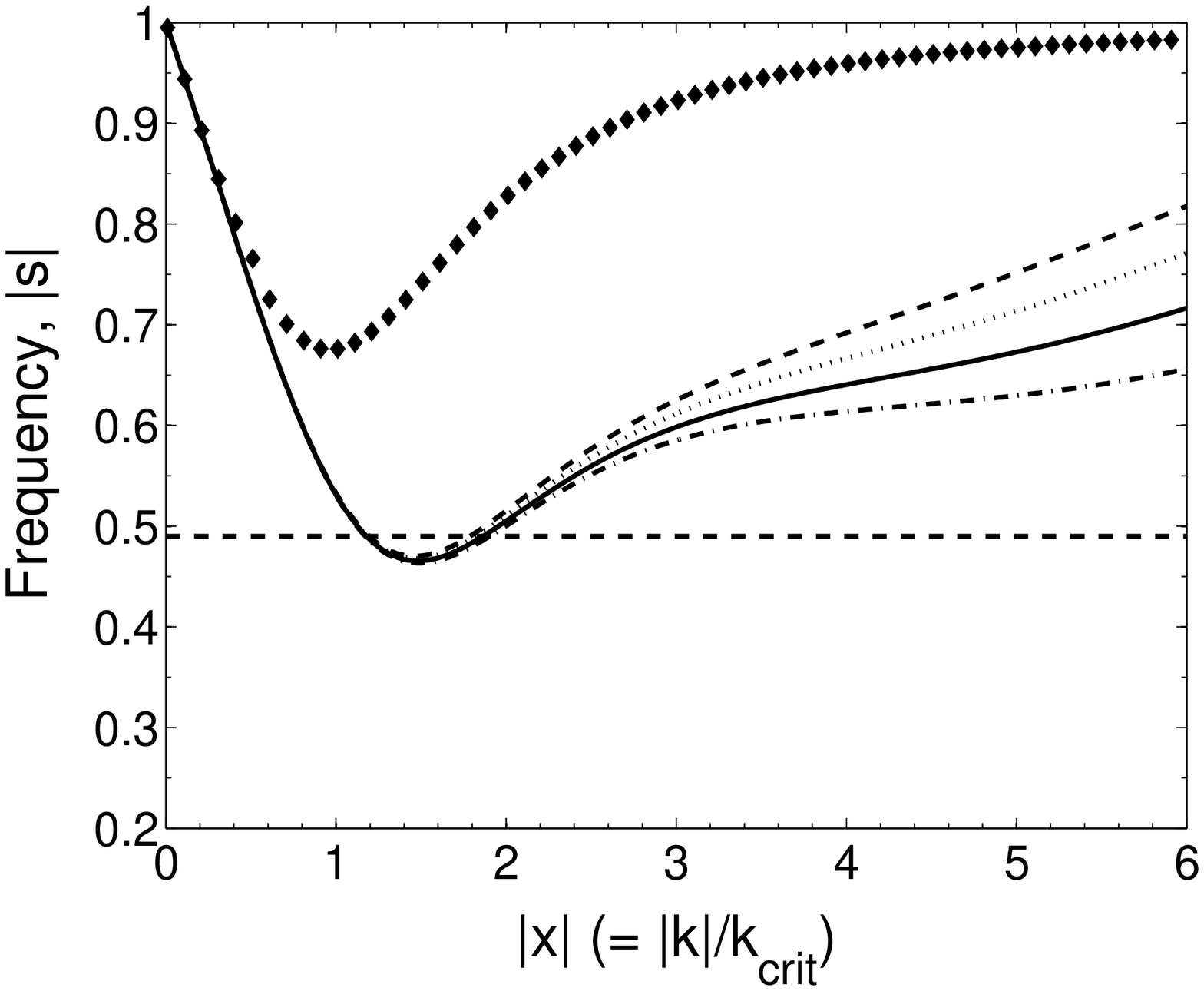}
        \vspace{0.2 cm}\\
	\hspace{-.2cm} {\bf{(c)}}\\
    \end{minipage}
    \caption{{\it{The Galaxy}} : Dispersion relations for stars-alone (S) and stars plus gas (S + G) cases, plotted in a dimensionless form, for a range of $Q_{\rm s}$ and $Q_{\rm g}$ values, at R = 2.5 R$_{\rm d}$. Panel (a) for $Q_{\rm s}=1.5$, panel (b) for $Q_{\rm s}=1.6$, and panel (c) for $Q_{\rm s}=1.7$. In each panel, $Q_{\rm g}$ is taken to be 1.4, 1.5, 1.6, and 1.7, successively. The corresponding dispersion relations are shown from bottom to top. The horizontal line indicates the value $|s|_{\rm obs}$, derived from the observed pattern speed and the rotation curve. Here for the whole ranges of $Q_{\rm s}$ and $Q_{\rm g}$, the stars plus gas system allows a stable density wave corresponding to the observed pattern speed, but in none of these cases the stars-alone system allows a stable wave solution.}
    \label{fig-parameter_galaxy}
\end{figure}

We next calculated the quantity $X$ using the input parameters used in plotting the dispersion relation in Figure~\ref{fig-parameter_galaxy}. We found that at $R$ = 8 kpc, $X$ = 4.2, indicating that the WKB approximation is satisfied here also, though not by a very comfortable margin.

Next we derive a range of pattern speeds that would allow the system to sustain a stable density wave. Here also we employ the same set of input parameters ($Q_{\rm s}$ = 1.7, $Q_{\rm g}$ = 1.5, $\epsilon$ = 0.2) to derive the value of $\alpha$ theoretically from the dispersion relation. Note that, for this case, $R$= 2.5 $R_{\rm d}$ lies within the CR, therefore the allowed range of pattern speeds would be ($\Omega-\kappa/2$, $\Omega-\alpha\kappa/2$). The results are given in Table~\ref{table-parameter_galaxy}.

\begin{table*}
\centering
\caption{Range of allowed pattern speeds for the Galaxy at $R=2.5 R_{\rm d}$}
\begin{tabular}{cccccccc}
\hline
Galaxy & $\Omega$ & $\kappa$ & $\epsilon$ & observed  & $\alpha$ & Lower  &  Upper   \\
name & (km   & (km   & (gas  & $\Omega_{\rm p}$ && bound on $\Omega_{\rm p}$   & bound on $\Omega_{\rm p}$    \\
& s$^{-1}$ & s$^{-1}$ && (km  s$^{-1}$ && (km  s$^{-1}$ & (km s$^{-1}$\\
& kpc$^{-1}$) & kpc$^{-1}$) & fraction) &  kpc$^{-1}$)&& kpc$^{-1}$) &  kpc$^{-1}$) \\
\hline
The Galaxy & 27.5 & 38.9 & 0.2 & 18 & 0.46 & 8.1 & 18.6\\
\hline
\end{tabular}
\label{table-parameter_galaxy}
\end{table*}

It is clear from Table~\ref{table-parameter_galaxy} is that the observed pattern speed falls in this theoretically predicted range. Thus, the results for our Galaxy confirm the results obtained in the earlier part of this chapter for other galaxies, namely, the observed gas fraction needs to be included in order to have a stable density wave corresponding to the observed pattern speed.

\thispagestyle{empty}

\chapter[Effect of dark matter halo on global spiral modes in galaxies]{Effect of dark matter halo on global spiral modes in galaxies \footnote{Ghosh, Saini \& Jog, 2016, MNRAS, 456, 943}}
\chaptermark{\it Effect of dark matter halo on global spiral modes}
\vspace {2.5cm}

\section{Abstract}
Low surface brightness (LSB) galaxies form a major class of galaxies, and are 
characterized by low disk surface density and low star formation rate. These 
are known to be dominated by dark matter halo from the innermost regions. Here, 
we study the role of dark matter halo on the grand-design, $m=2$, spiral modes 
in a galactic disk by carrying out a global mode analysis in the WKB approximation. 
The Bohr--Sommerfeld quantization rule is used to determine how many discrete global 
spiral modes are permitted. First, a typical superthin, LSB galaxy UGC 7321 is 
studied by taking only the galactic disk, modeled as fluid; and then the disk 
embedded in a dark matter halo. We find that  both cases permit the existence of 
global spiral modes.  This is in contrast to earlier results where the 
inclusion of dark matter halo was shown to nearly fully suppress local, 
swing-amplified spiral features. Although technically global modes are permitted in the fluid model
as shown here, we argue that due to lack of tidal interactions, these are not 
triggered in LSB galaxies. For comparison, we carried out a similar analysis for 
the Galaxy, for which the dark matter halo does not dominate in the inner regions. 
We show that here too the dark matter halo has little effect, hence the disk 
embedded in a halo is also able to support global modes. The derived pattern speed 
of the global mode agrees fairly well with the observed value for the Galaxy.

\section{Introduction} 
Various surveys on galaxy morphology have revealed that the spiral arms, making a spectacular visual impression, are mainly of two types, namely, grand-design spiral arms and flocculent arms. It is also found that the fractional abundance of these two types varies with the Hubble type \citep{Elm11}. The origin and maintenance of these features has been extensively studied over the past five decades, though many aspects are still not fully understood.
In general, the grand-design spiral features are explained as density waves, governed mainly by gravity \citep{LS64, LS66}, whereas the flocculent spiral features are material arms, caused by swing amplification \citep{GB65, Tom81}.

A major class of galaxies are the so-called low surface brightness (LSB) galaxies, which are characterized by low star formation rate \citep{IB97}, and low disk surface density \citep{dBM96,dBM01}.
LSB galaxies do not show grand-design spiral structure. These can show spiral structure but it is fragmentary,  extremely faint, and difficult to trace \citep{Sch90, Mcg95,Sch11}. The LSBs are dark matter dominated starting right from the very inner radii
\citep{Bot97,dBM97,dBM01,Com02,Ban10}. This aspect of dark matter dominance is different from that seen in case of normal spiral galaxies. For example, in the LSBs, the dark matter constitutes about 90 percent of the total mass within the optical disk, whereas for the `normal' or High Surface Brightness (HSB) galaxies, such as the Milky Way, the contribution of stellar mass and dark matter halo mass is comparable \citep[e.g.][]{dBM01, Jog12} within the optical disk. In the Milky Way and other HSB galaxies, the dark matter dominates only in regions way outside the optical disk. Consequently, the LSB galaxies naturally offer a good place for probing the effect of the dark matter halo on the dynamics  of a galactic disk. We caution that we only consider the small size LSBs which are more common, and do not discuss giant LSBs like Malin1. The latter sub-class have massive disks and  show fairly strong spiral arms as in UGC 6614 \citep{Das13}.

In the literature, several studies have demonstrated that the dark matter halo has a profound effect on various dynamical properties of LSBs. For example, early numerical simulation by \citet{Mih97} showed that the dominant dark matter halo and the low surface density make the LSB galaxies stable against the growth of global non-axisymmetric modes such as bars. Further, there exists a subclass of LSBs, namely, superthin galaxies. A study by \citet{BJ13} showed that the superthin property is explained by a dense, compact halo that dominates from the innermost regions.
 
A recent work by \citet{GJ14} showed that a dark matter halo that is dominant from the innermost regions  makes the galactic disk stable against both local, axisymmetric and non-axisymmetric perturbations, by making the Toomre $Q$ parameter very high ($>$ 3). This suppresses the swing amplification process, thereby explaining the observed lack of star formation and lack of strong small-scale spiral features in LSB galaxies. But the role of dark matter has not been studied so far in the context of grand-design spiral structure. Motivated by the results for small-scale spiral structures, here we plan to study the effect
 of dark matter halo in the context of global spiral modes.

The idea that spiral patterns could be long-lived density waves was first proposed by \citet{LS64}. In this hypothesis the spiral pattern is a rigidly rotating density wave through which differentially-rotating stars and gas can flow. The pattern is maintained by the self-gravity and pressure of the density wave self-consistently. The pattern retains its shape over long periods of time without suffering the winding problem. Investigation of self-sustaining density waves in a disk is essentially a problem of wave-mechanics in a differentially rotating, self-gravitating disk. However, the long-range nature of gravitational interaction makes studying them difficult due to the non-local gravitational interactions between different parts of the perturbed disk. 

A very useful approach for studying the dynamics of disks is the tight-winding approximation. This approximation makes the gravitational effects of the density perturbation local. For a wave-like, non-axisymmetric, small amplitude perturbation, the temporal oscillation frequency $\omega$ can be obtained as a function of the radial wavevector $k$, the radial coordinate $R$, and the azimuthal wavenumber $m$, known as WKB (Wentzel - Kramers - Brillouin) dispersion relation \citep[see, e.g., Chapter~6 of][for derivation and applications]{BT87}. The same local dispersion relation can also be used to construct global standing-wave like solutions through the Bohr-Sommerfeld quantization condition. In the past, this approach has been used by \citet{BR88}, \citet{Shu90}; and more recently by \citet{Tre01}, \citet{Sai09}, \citet{Gul12} for $m=1$ modes. Here, we are using this alternative approach to investigate the role of dark matter on the existence of large-scale spiral patterns in a disk galaxy.

In this paper, we investigate the number of allowed discrete global mode(s)---the term is used as a proxy for the grand spiral arms---within the WKB approximation, using the Bohr--Sommerfeld quantization condition. We treat the stellar disk as a fluid, which allows a considerably simple analytical dispersion relation. This assumption is good so long as one is away from the resonance points (although see \S~\ref{res-ugcflu}). However, we remove the artifacts, intrinsically associated with the fluid approximation, where necessary (for details see \S~\ref{sec-qcond}). Our analysis is carried out for UGC 7321, a superthin, LSB galaxy, first for a stellar disk alone and then for a disk embedded in a dark matter halo. We find that in both the disk-alone and disk plus halo models, global spiral modes are present i.e. based on this linear calculation and under fluid approximation, dark matter does not produce any significant change for global spiral modes. Similarly, we followed the same course of modal analysis for  our Galaxy, a typical `normal', HSB galaxy. For our Galaxy, dark matter halo is shown to have negligible effect on global modes. This trend for our Galaxy follows our expectation since dark matter halo is not a dominant component in the inner/optical region of the Galaxy. For this work  the observed profiles of surface density and stellar velocity dispersion for both UGC 7321 and the Galaxy are used, so as to make our calculation more realistic.

In \S~\ref{formu6}  we present the model used, and the details of model input parameters; \S~\ref{sec:WKBflu} presents the details of the WKB approximation and \S~\ref{res6} describes the results while \S~\ref{dis6} and \S~\ref{con6} contain the discussion and conclusions, respectively.

\section{Formulation of the Problem}
\label{formu6}

We treat the galactic stellar disk as a fluid, characterized by an exponential surface density $\Sigma_{\rm s}$, and the one-dimensional velocity dispersion $v_{\rm s}$ for pressure support, which is treated as the fluid sound speed. To keep our analysis simple the disk is taken to be infinitesimally thin. The disk is embedded in a spherically symmetric dark matter halo with a pseudo-isothermal radial profile. We have used the cylindrical coordinates ($R$, $\phi$, $z$) below for both disk as well as the spherical halo. The dynamics of the disk is calculated first under the gravity of disk only (referred to as disk-alone case) and then under the combined gravity of the halo and the disk (referred to as disk plus dark matter case). The halo is assumed to be gravitationally inert, i.e.,  it is assumed to be non-responsive to the gravitational force of the perturbations in the disk.  

\subsection{Model of disk and halo}
\label{sec:model}
For a galactic disk embedded in a dark matter halo concentric to the galactic disk, the net angular speed, $\Omega$, and the net epicyclic frequency, $\kappa$ (used later in the dispersion relation), are obtained by adding the disk and the halo contribution in quadrature as follows:
\begin{equation}
\kappa^2= \kappa^2_{\rm disk}+\kappa^2_{\rm DM}\,; \quad \Omega^2= \Omega^2_{\rm disk}+\Omega^2_{\rm DM}\,.
\label{kappa-contri}
\end{equation}
The frequencies can be computed from the standard expressions in terms of the gravitational  potentials of the disk and the halo. For an exponential disk, the potential $\Phi(R, 0)$ in the equatorial plane is given by \citep[][equation (2.168)]{BT87}
\begin{equation}
\Phi(R, 0)= -\pi G \Sigma_0R [I_0(y)K_1(y) - I_1(y)K_0(y)]\,,
\end{equation}
where $\Sigma_0$ is the disk central surface density, $y$ is the dimensionless quantity, defined as $y = R/2R_{\rm d}$, where $R$ is the galactocentric radius, $R_{\rm d}$ is the disk scalelength. $I_n$ and $K_n$ $(n= 0, 1)$ are the modified Bessel functions of first and second kind, respectively. In terms of the disk potential,  $\kappa^2_{\rm disk}$ and $\Omega^2_{\rm disk}$ are given as
\begin{equation}
\begin{split}
 \kappa^2_{\rm disk} = \frac{\pi G \Sigma_0}{R_{\rm d}}  \Big[4I_0(y) & K_0(y)  -2I_1(y) K_1(y) +\\
   & 2y\left(I_1(y) K_0(y) - I_0(y) K_1(y) \right)\Big]
\end{split}
\end{equation}
and
\begin{equation}
\Omega^2_{\rm disk}= \frac{\pi G \Sigma_0}{R_{\rm d}}\Big[I_0(y) K_0(y) - I_1(y) K_1(y) \Big]\,.
\end{equation}

For both the galaxies considered in this paper, UGC 7321 and the Galaxy, a pseudo-isothermal profile has been shown to yield a good fit for the DM halo \citep{Mer98, Ban10}. The gravitational potential of a pseudo-isothermal halo is given by
\begin{equation}
\begin{split}
\Phi_{\rm DM}=4 \pi G \rho_0 R^2_{\rm c} \Bigg[\frac{1}{2}\log (R^2_{\rm c}+R^2+z^2)+
\quad\Bigg(\frac{R_{\rm c}}{(R^2+z^2)^{1/2}}\Bigg)\times\\
 \tan^{-1}\Bigg(\frac{(R^2+z^2)^{1/2}}{R_{\rm c}}\Bigg)-1\Bigg]\,,
\end{split}
\end{equation}
where $R_{\rm c}$ is the core radius and $\rho_0$ is the core density. The corresponding  $\kappa^2_{\rm DM}$ and $\Omega^2_{\rm DM}$, in the mid-plane ($z$ = 0), are given as
\begin{equation}
\begin{split}
\kappa^2_{\rm DM}=4 \pi G \rho_0 \Bigg[\frac{2}{1+(R/R_{\rm c})^2}+\Big(\frac{R_{\rm c}}{R}\Big)^2\frac{1}{1+(R/R_{\rm c})^2}-
\Big(\frac{R_{\rm c}}{R}\Big)^3\tan^{-1}\Big(\frac{R}{R_{\rm c}}\Big)\Bigg]\,,
\end{split}
\end{equation}
and
\begin{equation}
\begin{split}
\Omega^2_{\rm DM}=4 \pi G \rho_0\Bigg[\frac{1}{1+(R/R_{\rm c})^2}+\Big(\frac{R_{\rm c}}{R}\Big)^2\frac{1}{1+(R/R_{\rm c})^2}
-\Big(\frac{R_{\rm c}}{R}\Big)^3 \tan^{-1}\Big(\frac{R}{R_{\rm c}}\Big)\Bigg]\,.
\end{split}
\end{equation}

\subsection{Model parameters}
To calculate the rotation and epicyclic frequencies needed for the WKB analysis below, we have used the following set of parameters.
\subsubsection{UGC 7321}
For the stellar disk and halo parameters, the values we used are obtained either observationally or by modeling (Banerjee et al. 2010). The stellar disk is an exponential disk with central surface density of 50.2 M$_\odot$ pc$^{-2}$ and a disk scalelength of 2.1 kpc. For the dark matter halo, a pseudo-isothermal profile with core density of 0.057 M$_\odot$ pc$^{-3}$ and core radius of 2.5 kpc yields good fit for the observed rotation curve and the vertical gas scaleheight distribution \citep{Ban10}.

For the one-dimensional stellar velocity dispersion in the radial direction, the profile is taken as $v_{\rm s}$ =$v_{\rm s0} \exp(-R/2R_{\rm d})$. The observed central velocity dispersion along $z$ direction is equal to 14.3 km s$^{-1}$ \citep{Ban10}. For the solar neighborhood, it is observationally found that, the ratio of velocity dispersion in the $z$ direction to that of radial direction is $\sim$ 0.5 \citep[e.g.,][]{BT87}. Here we assume the same conversion factor for all radii in this galaxy.

\subsubsection{The Galaxy}
The stellar disk parameters for the Galaxy are taken from the standard mass model by \citet{Mer98} which gives an exponential radial profile for the disk surface density with a disk scalelength of 3.2 kpc and a central surface density of 640.9 M$_\odot$ pc$^{-2}$. For the dark matter halo, a pseudo-isothermal profile is used with a core radius of 5.0 kpc and a core density of 0.035 M$_\odot$ pc$^{-3}$ \citep{Mer98}.

The one-dimensional stellar velocity dispersion is observed to be of  the form:
$v_{\rm s} = v_{s0}\exp(-R/8.7)$, where $v_{\rm s0} = 95$ km s$^{-1}$ \citep{LF89}.

\section{WKB analysis}
\label{sec:WKBflu}

For small non-axisymmetric perturbations of the form 
\begin{equation}
X(R,\phi, t) = X_{\rm a} \exp \left [{\rm i} \left(\int^R k(R')dR' -m\phi + \omega t \right) \right]\,,
\end{equation}
where $X$ is a generic dynamical quantity, and assuming the tight-winding approximation, which formally requires $|kR| \gg 1$, the Euler equations reduce to the standard dispersion relation \citep{BT87}
\begin{equation}
(\omega - m\Omega)^2=\kappa^2-2\pi G \Sigma_{\rm s} |k|+v^2_{\rm s} k^2\,,
\label{disp-flu_model_equation}
\end{equation}
where $\kappa$ and $\Omega$ are as defined in the \S~\ref{sec:model}, with the parameters given in the previous subsection. We are interested in spiral patterns corresponding to $m =2$. A slight rearrangement of equation (\ref{disp-flu_model_equation}) yields
\begin{equation}
4(\Omega_{\rm p}-\Omega)^2=\kappa^2-2\pi G \Sigma_{\rm s}|k|+v^2_{\rm s} k^2\,,
\end{equation}
where $\Omega_{\rm p}= \omega/2$ is the pattern speed at which the spiral pattern rotates rigidly. For a given pattern speed $\Omega_{\rm p}$, this dispersion relation can be thought of as an implicit relation between the phase space variables $k$ and $R$ (the resulting graph is also known as the propagation diagram).
 In general this relation is multi-valued, therefore, it is convenient to obtain its two branches explicitly for computational purposes. Expressing the relation in terms of the wavevector $k$ as a function of $R$ we get
\begin{equation}
|k_{\pm}|(R)=[2\pi G \Sigma_{\rm s} \pm \Delta^{1/2}]/2v^2_{\rm s}\,,
\label{wavevector}
\end{equation}
where 
\begin{equation}
\Delta = [(2\pi G\Sigma_{\rm s})^2-4v^2_{\rm s}\{\kappa^2-4(\Omega_{\rm p}-\Omega)^2\}] \,.
\end{equation}
Note that the dependence on the wavevector is through its absolute value $|k|$; and therefore, these two branches in general correspond to four possible values of $k$ at any given radius $R$. However, if $\Delta^{1/2} > 2\pi G \Sigma_{\rm s}$, then only the `+' branch is physical, and the total number of solutions for $k$ reduces to two. Waves cannot exist in regions where $\Delta < 0$, called the forbidden region, where there are no real solutions for $k$. From equation~(\ref{wavevector}) it can be seen that when $\Delta=0$, $k_{+}=k_{-}$, the two branches merge at that point, and we say that the wave reflects from trailing to leading branch and vice versa.
 The Lindblad resonances occur when $\Omega_{\rm p} = \Omega \pm \kappa/2$. The dispersion relation then gives $k=0$ for the $k_{-}$ branch. The $k_{+}$ branch, however, can extend beyond the Lindblad radius for gas. This generic behavior
is evident in the constant $\Omega_{\rm p}$ plots in Figure~\ref{fig:flumodel_UGC7321} where the plot is for the LSB galaxy UGC 7321, with and without the dark matter halo.

A constant $\Omega_{\rm p}$ (or equivalently constant $\omega$) contour represents the trajectory of a wavepacket which travels at a radial velocity given by the corresponding group velocity $v_{\rm g} = d\omega/dk$  \citep{Tom69}.

Figure~\ref{fig:flumodel_UGC7321} exhibits three different types of contours. Their physical significance in terms of propagation of a wavepacket is as follows.

\begin{itemize}
\item{{\it Type A}:} These are the `W' shaped contours in the top panel of Figure~\ref{fig:flumodel_UGC7321}. A wavepacket may start at a large radius (to the left of the curve) on the short leading branch and then travel inwards with a negative radial group velocity. It first reflects from the edge of the forbidden region at the lowest point of the curve, moves radially outwards and reaches the outer Lindblad radius (at $k=0$) from where it is reflected back into the long trailing branch, and gets reflected again from the edge of the forbidden region. The wavevector of perturbations comprising the wavepacket then continues to increase steadily. In collisionless disk, the fate of such wavepackets would be to eventually dissipate at Lindblad resonance by a process similar to Landau damping \citep{BT87}. 

\item{{\it Type B}:} These are the oval shaped closed curves and they always occur in pairs. A wavepacket would move in and out indefinitely by being reflected at the edge of the two forbidden regions, from short branch to relatively long branch and vice versa, this being true for both the leading as well as the trailing branches. Also note that the wavevector remains non-zero all along the closed curve.

\item{{\it Type C}:} Unlike {\it Type B} contours, this type of closed contours are not degenerate. They involve reflection of a wavepacket from the leading to the trailing branch, and vice versa, by the edge of the forbidden region as well as the inner and outer Lindblad radius.
\end{itemize}

\subsection{Global standing waves: the quantization condition}
\label{sec-qcond}
In contrast to the behavior of wavepackets, the global modes of a disk are characterized by a single frequency $\omega$ (or $\Omega_{\rm p}$) and extend over a large region of the disk. These essentially non-local perturbations are characterized by a wavefunction that extends across the disk, and which is expected to behave in a manner consistent with the boundary conditions across the disk. The density perturbation across the disk for these standing waves has a specific pattern shape (given by the wavefunction) and pattern speed $(\Omega_{\rm p} = \omega/2)$. 

 In principle, the pattern speed $\Omega_{\rm p}$ could be either positive or negative, implying a prograde or retrograde motion for the pattern with respect to the galactic rotation. However, we find that for the two galaxies considered by us, $\Omega - \kappa/2 > 0$; and collisionless disks do not support wave-like solutions for negative pattern speeds under this condition, although fluid disks do allow them. We have verified that the resulting propagation diagrams for negative $\Omega_{\rm p}$ are not closed, therefore, in the rest of the paper we do not consider negative $\Omega_{\rm p}$. All the global modes described below are thus prograde.

In the previous section, we saw that the fate of {\it Type A} wavepackets is to eventually disperse, and unless the boundary condition at large radius is reflecting them back, they will not give rise to long-lived standing waves. However, the contours of {\it Type B} and {\it C} seem to have the correct behavior to give rise to global standing waves due to their closed shape. Interestingly, note that for contours of {\it Type~C}, the wave goes right past the Lindblad resonances on either side before getting reflected from the forbidden region. This behavior is peculiar to fluid disks. For a collisionless disk the waves can exist only in regions where $\Omega - \kappa/2 \le \Omega_p \le \Omega + \kappa/2$. In Figure~\ref{fig2} we plot the line $\Omega = \Omega_{\rm p}$ with $\Omega - \kappa/2$,  $\Omega$, and $\Omega + \kappa/2$, where we can see that the contours of {\it Type~B} do obey this condition, but not the contours of {\it Type~C}. 

To elaborate, in the fluid model a wave can partially transmit through the Inner Lindblad Resonance (ILR), and consequently the wave cycle, operating between the co-rotation (CR) and ILR to maintain a standing wave, will get hampered. In other words, to maintain a wave cycle between CR and ILR, one has to make sure that the inequality
\begin{equation}
 R_{\rm ILR} < R_{-} < R_{+} < R_{\rm CR}\,,
\label{ineq}
\end{equation} 
where $R_{\pm}$ are the boundaries of the forbidden region, holds for each mode, otherwise for that mode ILR would be exposed. Since the inner regions of galactic disks are dominated by stars, only the closed contours for which the above mentioned inequality holds can be realized in a real galaxy \citep[see page 205][for a detailed discussion]{Ber00}. Therefore, in our analysis the contours of {\it Type~C} are considered as unphysical, and hence are discarded.

To construct the discrete global density waves from the closed contours of {\it Type B}, it is useful to recall that in quantum mechanics the WKB approximation is used to calculate the eigenvalues of a Hamiltonian in the following manner: for a one dimensional system the phase plot (in the $p$--$q$ plane, where $q$ and $p$ are generalized coordinates and generalized momenta respectively) of constant energy for a bound state forms a closed loop.  The total area inside the loop $\oint pdq$ can then be shown to be some integral multiple of $\hbar$ \citep[see, e.g.,][]{LL65}. In our case this Bohr--Sommerfeld quantization rule essentially reduces to constructing the constant $\omega$ curves in the $k$--$R$ plane. Provided that closed loops exist, they can be quantized by calculating the $\oint kdR$ and equating it to integral multiples of $\pi$, apart from some extra factors of $\pi/2$ that arise due to the boundary condition at the turning points. The quantization condition ensures that only discrete values of $\omega$ are allowed, which are called the eigenfrequencies \citep[e.g.][]{BR88}.

For the closed contours of {\it Type~B} to represent standing waves they must satisfy a quantization condition. The appropriate WKB quantization rule is given by \citet{Tre01}
\begin{equation}
2 \int_{R_-}^{R_+} [k_{+}(R)-k_{-}(R)] dR = 2\pi \left(n-\frac{1}{2}\right)\,,
\label{flumodel_qcond}
\end{equation}
where $n=1,2,3,\cdots$. Although the WKB approximation requires $|kR| \gg 1$, its validity is often sufficiently accurate even for $|kR| \simeq 1$. \begin{figure}
\centering
\includegraphics[height=3.0in,width=4.0in]{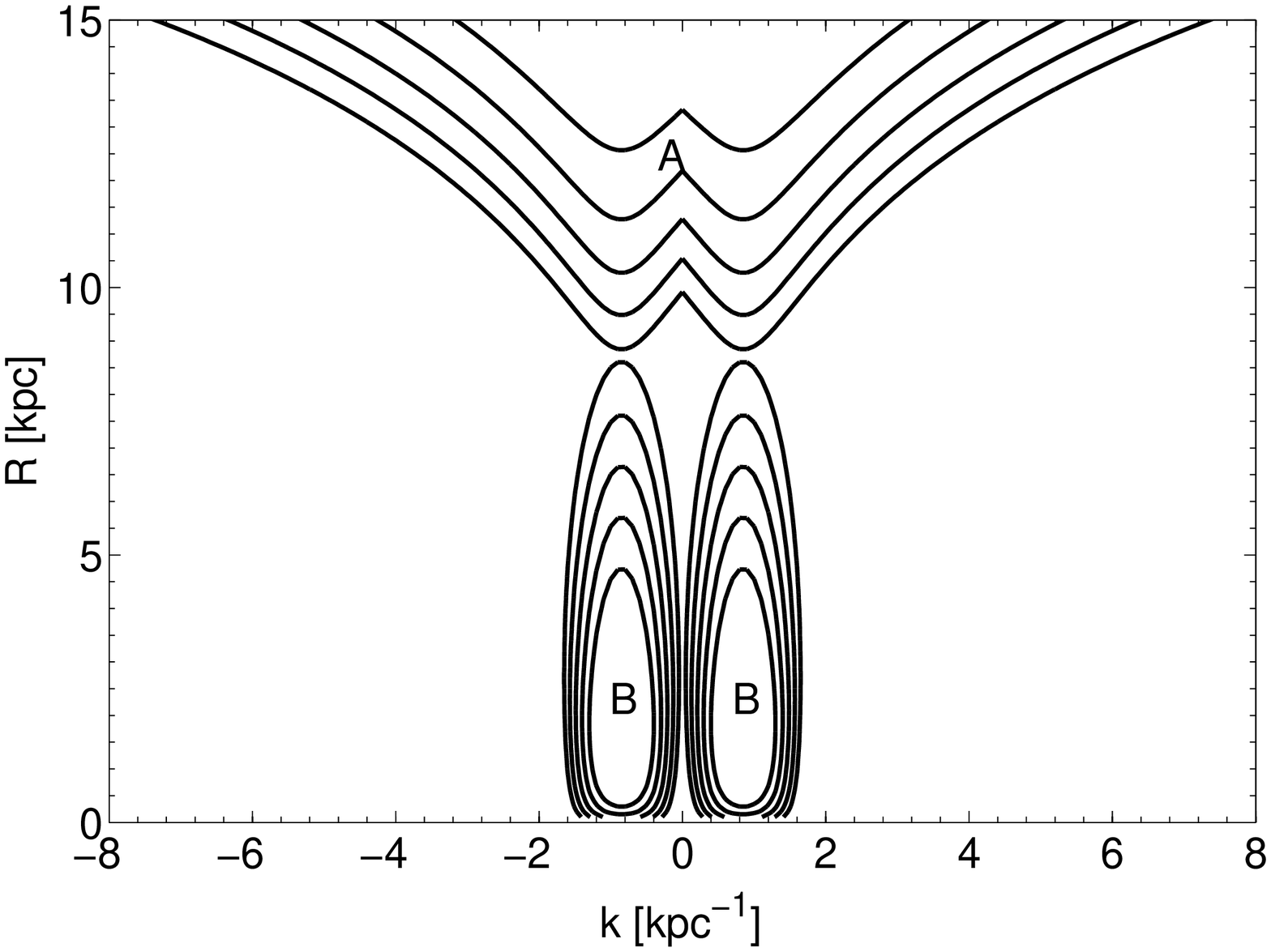}
\medskip
\includegraphics[height=3.0in,width=4.0in]{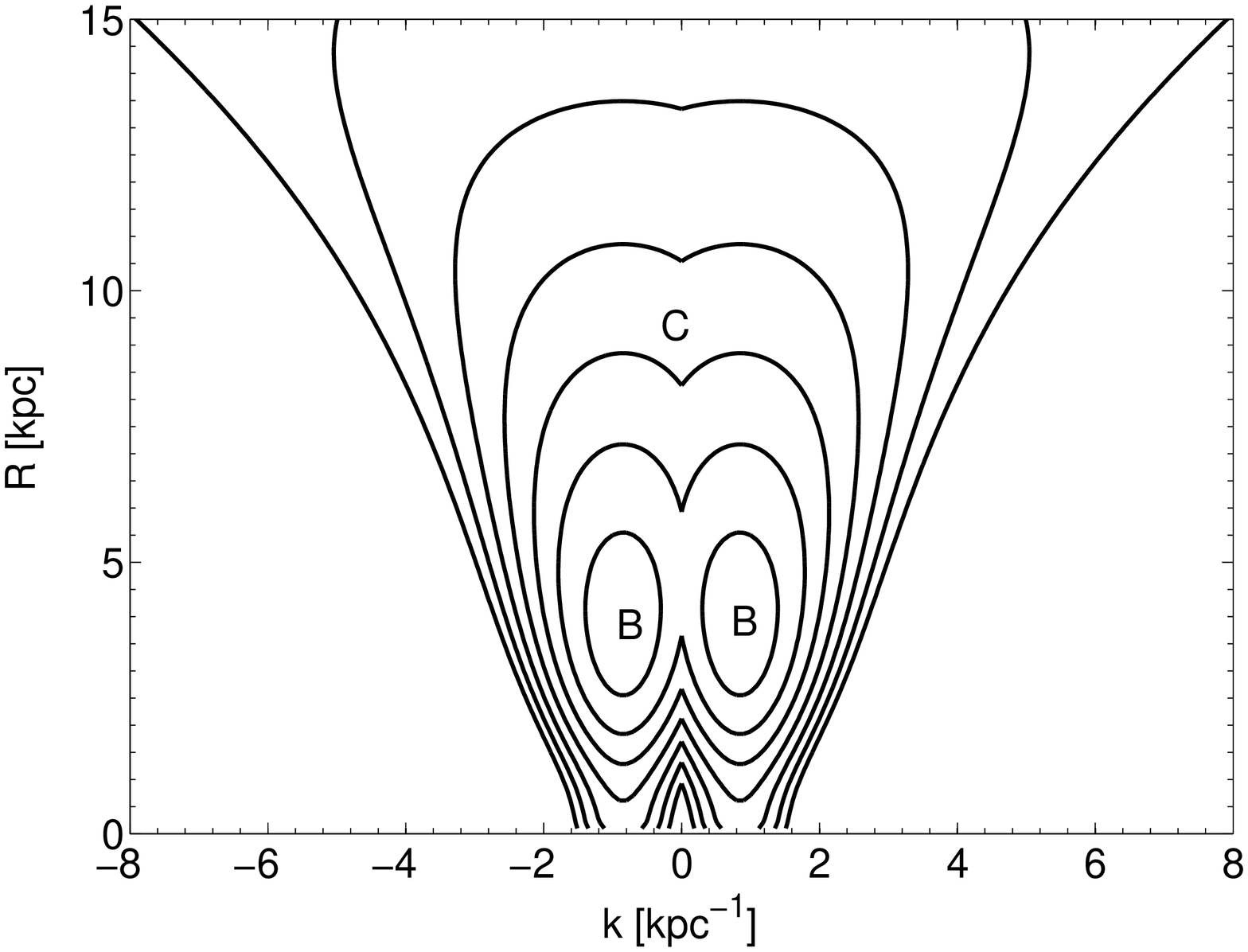}
\caption{Propagation diagrams (contours of constant $\Omega_{\rm p}$) plotted for different pattern speeds, by using the input parameters of LSB galaxy UGC 7321. The top panel shows contours for the disk-alone case where the range of $\Omega_{\rm p}$ varies from 2.3 km s$^{-1}$ kpc$^{-1}$ to 4.3 km s$^{-1}$ kpc$^{-1}$  and the bottom panel shows the contours for the disk plus dark matter halo case where the range of $\Omega_{\rm p}$ varies from 1.5 km s$^{-1}$ kpc$^{-1}$ to 4.1 km s$^{-1}$ kpc$^{-1}$. The contours are plotted at intervals of 0.2 for both the panels. Different types of contours present here are marked as A, B and C. $\Omega_{\rm p}$ value increases from the outer to the inner contours.}
\label{fig:flumodel_UGC7321}
\end{figure}

\begin{figure}
\centering
\includegraphics[height=3.0in,width=4.0in]{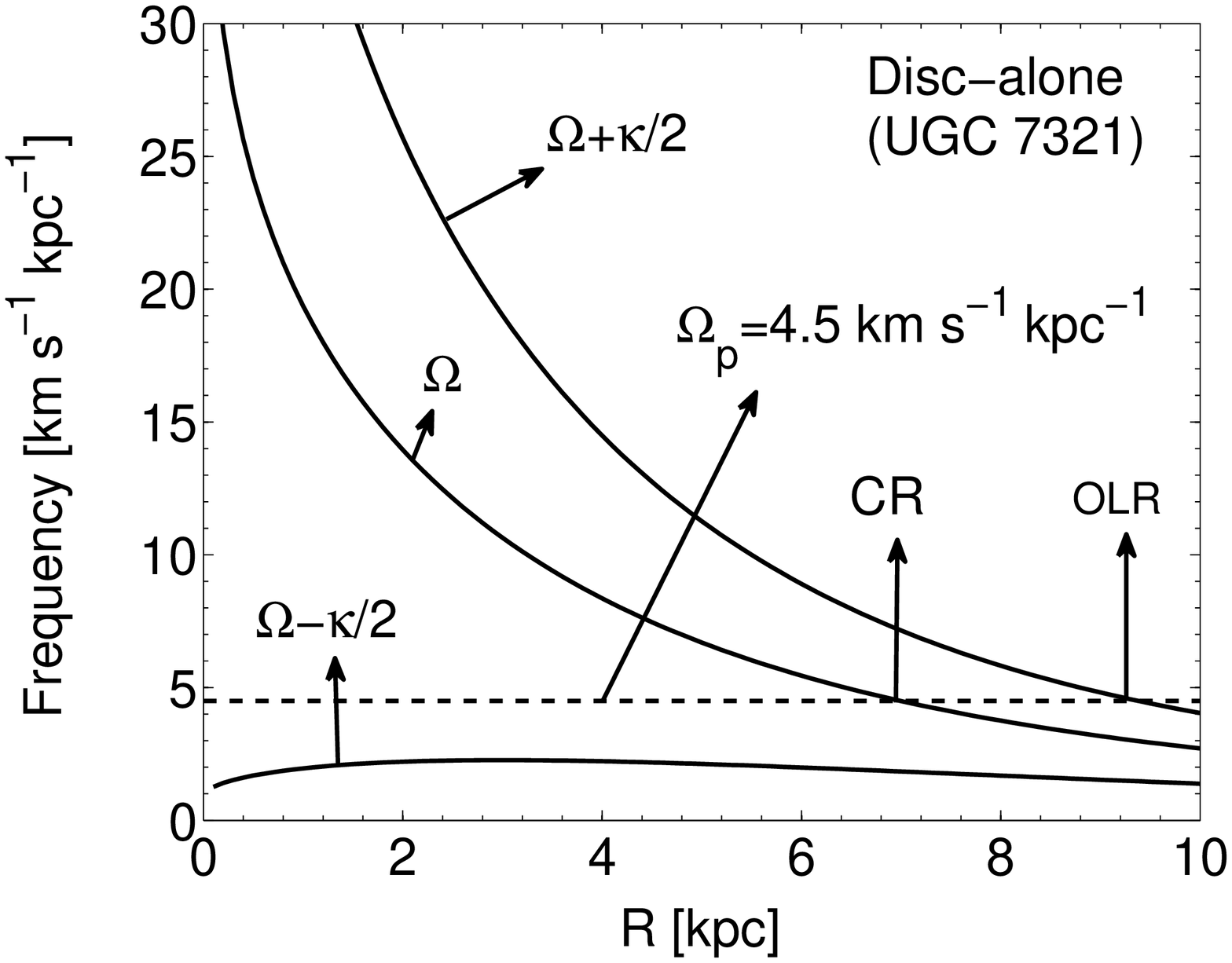}
\medskip
\includegraphics[height=3.0in,width=4.0in]{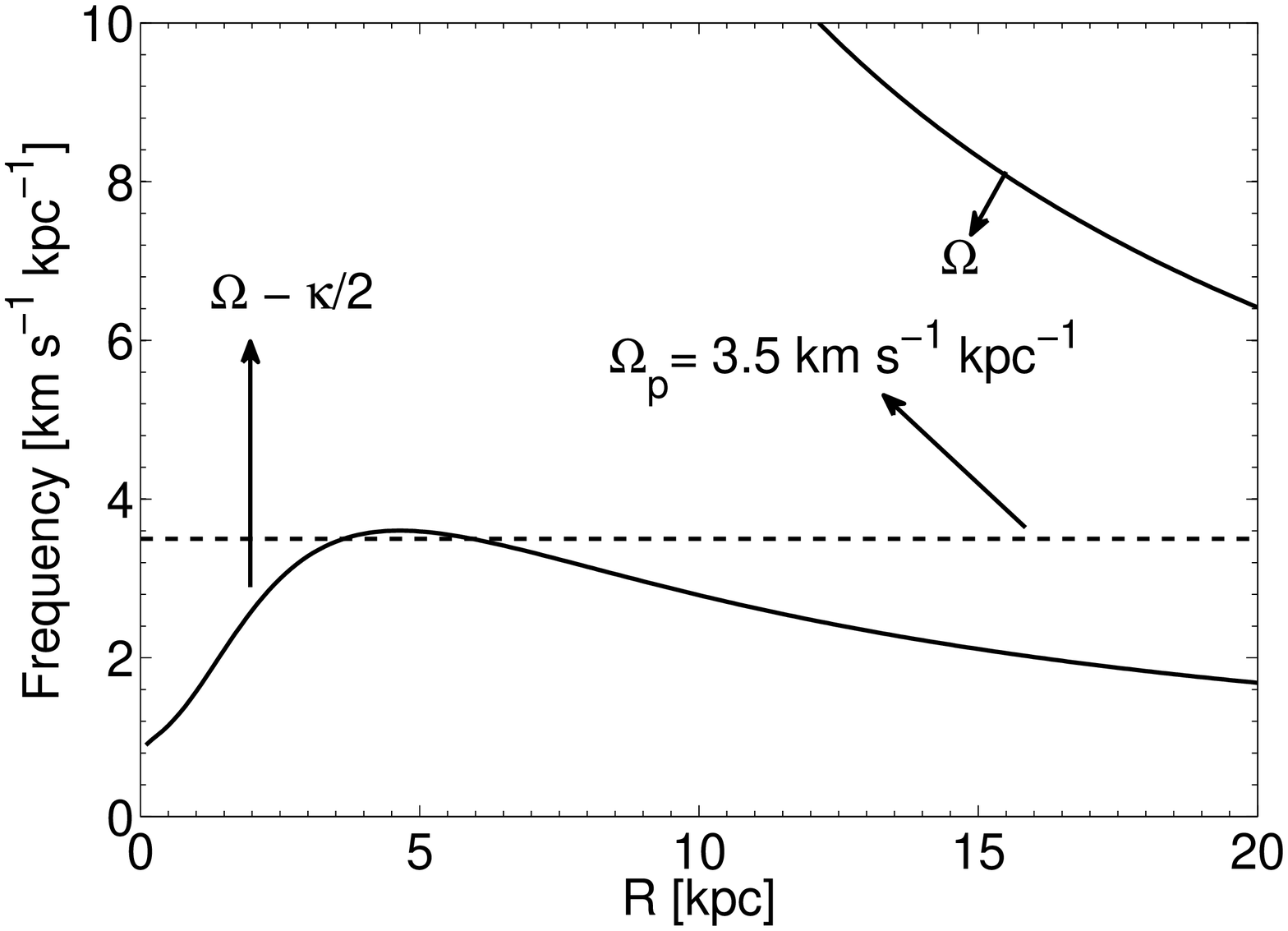}
\caption{ The $\Omega$, $\Omega-\kappa/2$ \& $\Omega+\kappa/2$ (in units of km s$^{-1}$ kpc$^{-1}$) curves are shown as a function of radius for UGC 7321. Top panel shows the disk-alone case, where the pattern speed ($\Omega_{\rm p}$) of 4.5 km s$^{-1}$ kpc$^{-1}$ gives a closed contour of {\it{Type B}} and the bottom panel showing disk plus dark matter halo case, where the pattern speed ($\Omega_{\rm p}$) of 3.5 km s$^{-1}$ kpc$^{-1}$ gives a closed contour of {\it{Type C}}. 
Note that, for {\it Type C} contour, the shown $\Omega_{\rm p}$ lies below $\Omega-\kappa/2$ for a certain range of $R$, thus it does not satisfy the inequality mentioned in the text (see \S~\ref{sec-qcond})
 }
\label{fig2}
\end{figure}

\section{Results}
\label{res6}

\subsection{UGC 7321}
\label{res-ugcflu}
The results for the contours of {\it Type B} for both disk-alone and disk plus halo cases, using the input parameters of UGC 7321 for different pattern speeds, are displayed in Figure~\ref{fig3}.

\begin{figure}
\centering
\includegraphics[height=3.0in,width=4.0in]{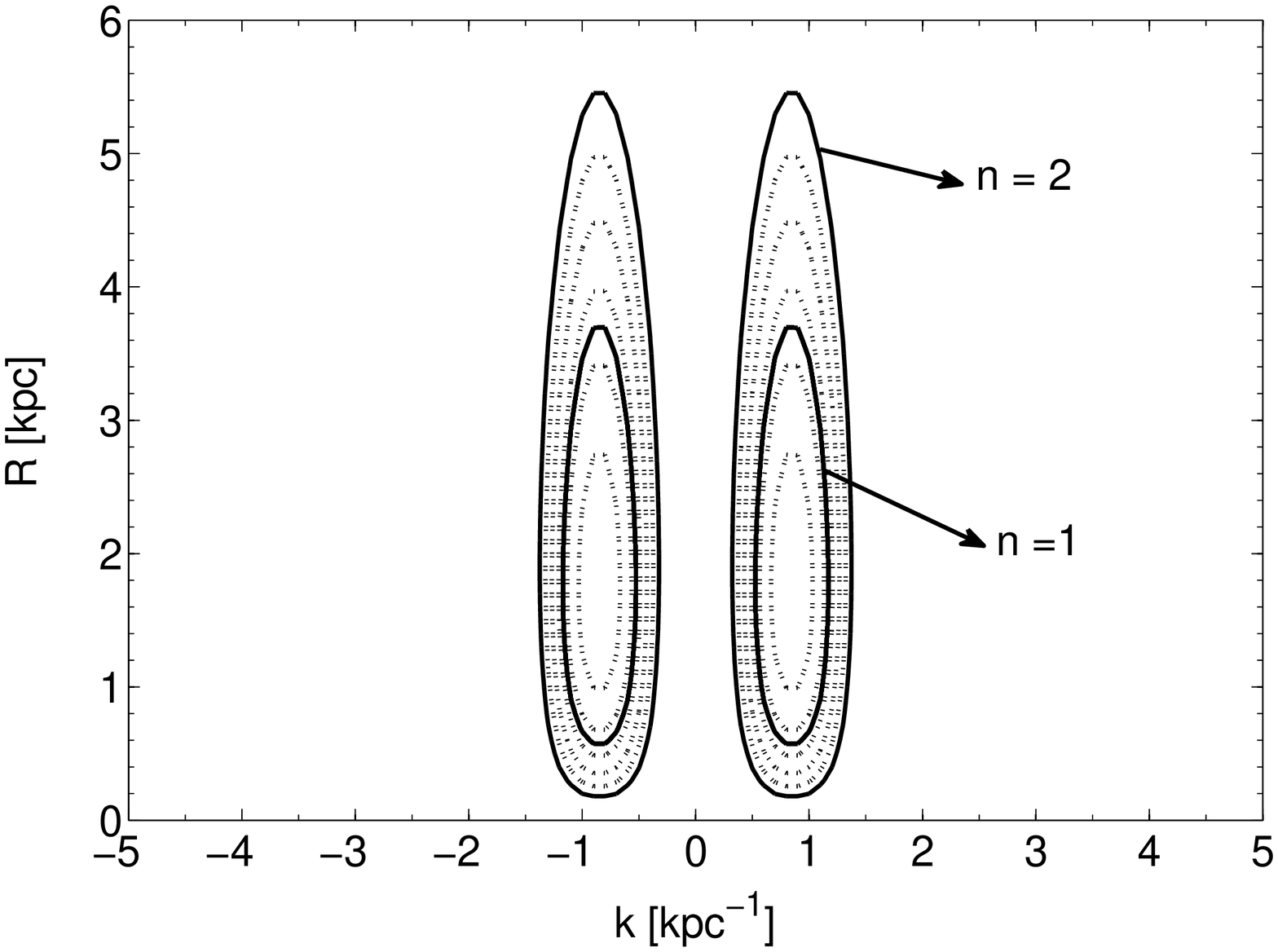}
\medskip
\includegraphics[height=3.0in,width=4.0in]{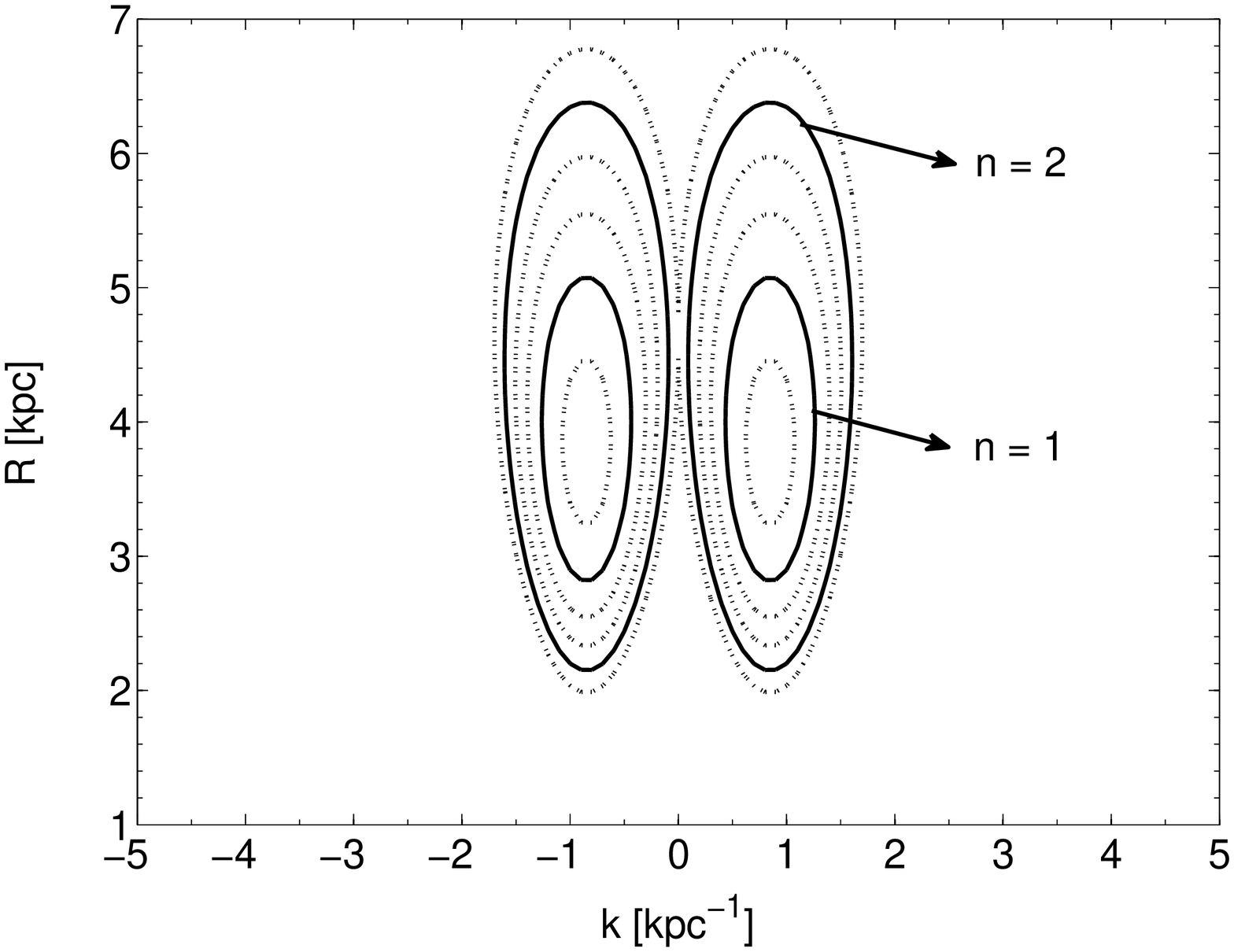}
\caption{Propagation diagrams (contours of constant $\Omega_{\rm p}$) for those pattern speeds which give closed loops of {\it{Type B}}. The input parameters used are for the LSB galaxy, UGC 7321. The top panel shows contours for disk-alone case where the range of $\Omega_{\rm p}$ varies from 3.8 km s$^{-1}$ kpc$^{-1}$ to 4.8 km s$^{-1}$ kpc$^{-1}$, at intervals of 0.2 and the bottom panel shows the contours for disk plus dark matter halo case where the range of $\Omega_{\rm p}$ varies from 3.6 km s$^{-1}$ kpc$^{-1}$ to 4.1 km s$^{-1}$ kpc$^{-1}$, at intervals of 0.1. The closed loops that correspond to the global modes for different models, are indicated by solid lines. $\Omega_{\rm p}$ value increases from the outer to the inner contours.}
\label{fig3}
\end{figure}
We then applied the quantization condition (equation~(\ref{flumodel_qcond})) to these contours to determine the quantum number $n$. The resulting values of $n$ for different pattern speeds that correspond to permitted global modes for both cases, are summarized in Table~\ref{tab-resugcflu}.
\begin{table*}
\centering
\caption{ Results for global modes for UGC~7321}
\begin{tabular}{ccccc}
\hline
$\Omega_{\rm p}$  &  $R_- $  & $R_+$ & $R_{\rm CR}$ & $n$ \\
(km s $^{-1}$ kpc$^{-1}$)&  (kpc) & (kpc) & (kpc) & \\
\hline
disk-alone case :\\
4.5  &   0.5  & 3.8 & 7& 1\\
3.8  &  0.08  & 5.8  & 8& 2\\
\hline
disk plus halo case :\\
4.0  &   2.8  & 5.1 & 33 & 1\\
3.7  &  2.1  & 6.4  & 36.5 & 2\\
\hline
\end{tabular}
\label{tab-resugcflu}
\end{table*} 

Note that in the statistical majority of galaxies, grand-design spiral structure should be quasi-stationary; and this can be possible if the dynamics of the disk is dominated by a single mode or by a small number of modes. Our finding, summarized in Table~\ref{tab-resugcflu}, agrees quite well with this point. Note that higher values of $n$ correspond to loops enclosing larger area, and for such loops $k_{-} \simeq 0$. Therefore, the smaller values of $n$ are more likely to satisfy the WKB approximation than the larger values of $n$.  

 However, we caution the reader that the modes that we have obtained are for a fluid disk (not collisionless), under the WKB approximation; and therefore the specific values may change if either of these assumptions is relaxed. Table~\ref{tab-resugcflu} shows that inclusion of dark matter halo has a negligible effect on the global spiral modes. The only change that we can observe from the Table~\ref{tab-resugcflu} is that the inclusion of halo appreciably increases the extent of the forbidden region (determined from the difference of $R_{\rm CR}$ and $R_{+}$) and it only changes the specific values of pattern speed which correspond to different values of $n$. Consequently,  the position of CR for different modes also changes. \citet{GJ14} showed that inclusion of dark matter halo supresses the local, transient, swing-amplified spiral features almost completely and therefore it is somewhat puzzling why dark matter halo fails to make any impression on global modes permitted in a galactic disk, modeled as a fluid.

We argue that there are several reasons why global modes, though technically permitted, may not materialize in a LSB galaxy. First,  a galaxy encounter is known to be a possible mechanism to excite the global spiral modes in a galactic disk. For example, spiral structures of M~51 and many other grand-design spirals with a companion have been modeled successfully \citep{TOMTOM72,BT87}. Since LSBs are observed to be isolated \citep{MO94}  or are located at the edges of voids \citep{ROSEM09}, they are less likely to undergo tidal encounters as compared to the HSB galaxies. Therefore, even if theoretically the modes are
permitted for the disk case, as well as the disk plus dark matter halo case for UGC 7321, in reality the triggering mechanism for exciting the global spiral modes may not operate. Secondly, the resulting pattern speed for the permitted global modes is small (Table~\ref{tab-resugcflu}). To excite these by interactions, a distant encounter is necessary which in turn will lead to a smaller amplitude. 
 Thirdly, even if the above permitted modes were to be triggered, the growth rate of the modes will depend on the surface density of the disk which is very low for the LSB galaxies, hence the amplitude of such modes is likely to be very small.
 Finally, we would like to mention that from the linear calculation done in this paper, we cannot fully comment on the effect of dark matter halo. 
A full $N$-body treatment which includes the above non-linear effects would be necessary in getting a comprehensive picture of effect of dark matter halo on global spiral modes.

Note that in contrast, the lack of nearby neighbors does not prevent local modes being triggered in the disks of LSB galaxies \citep [studied by][]{GJ14}, since local, transient swing-amplified features can be generated by internal triggers such as a gas cloud or star formation \citep{Sellcal84,Tom90}.
  In this paper, we have treated the stellar disk as a fluid for mathematical simplicity. At large values of $|k|$, the behavior of the dispersion relation for a collisionless system is very different from that for a fluid system \citep{Raf01,GJ15}. In a future work, we will follow up this problem treating the galactic disk as a collisionless system.

\subsection {The Galaxy}
\label{res-galflu}
For comparison, we carried out a similar modal analysis for the Galaxy, for which the dark matter halo is known to be not a dominant component in the inner regions of the disk. The closed contours present in the propagation diagram for different pattern speeds for the Galaxy in both disk-alone and disk plus dark matter halo cases are shown in Figure~\ref{fig-resgalflu}.\\

\begin{figure}
\centering
\includegraphics[height=3.0in,width=4.0in]{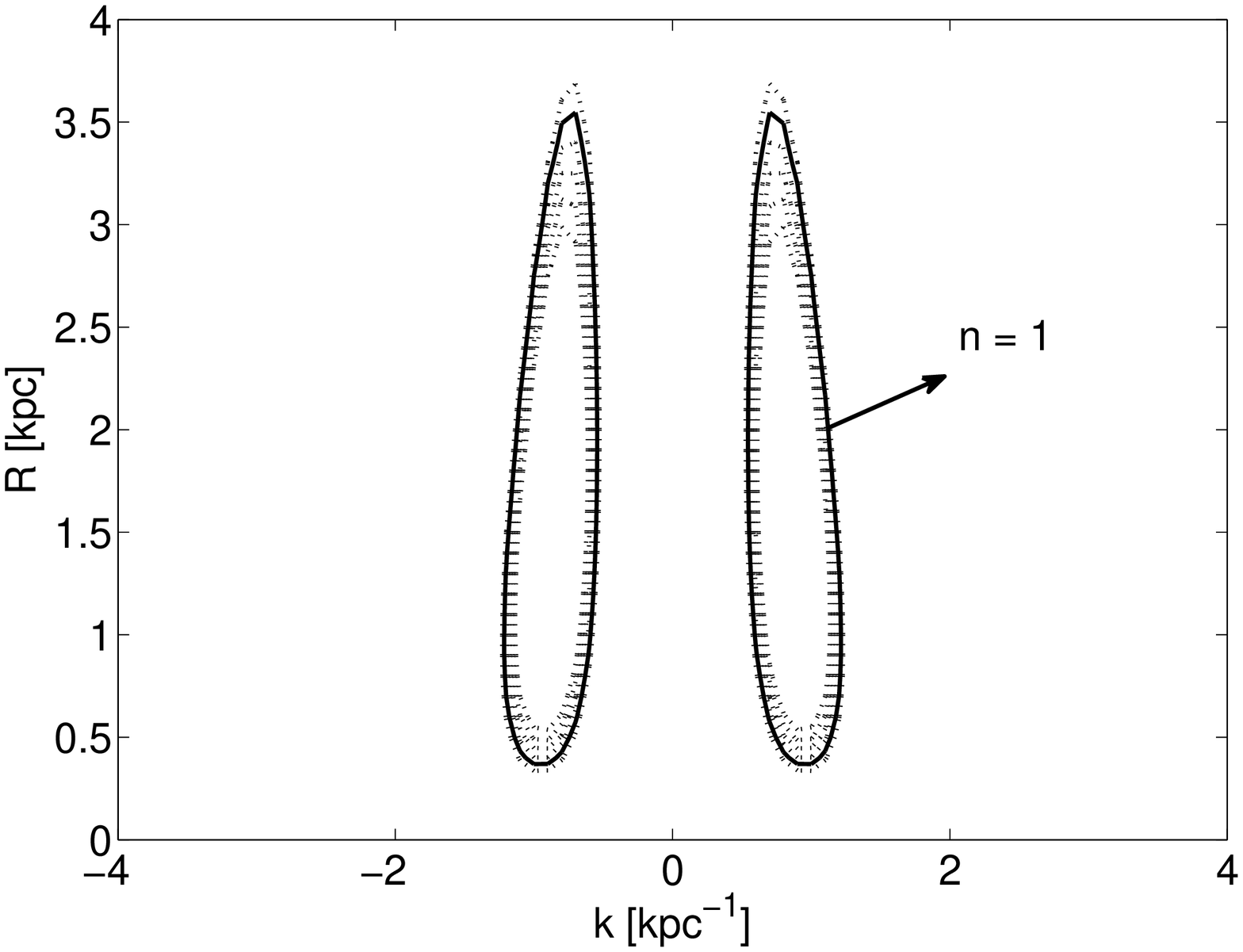}
\medskip
\includegraphics[height=3.0in,width=4.0in]{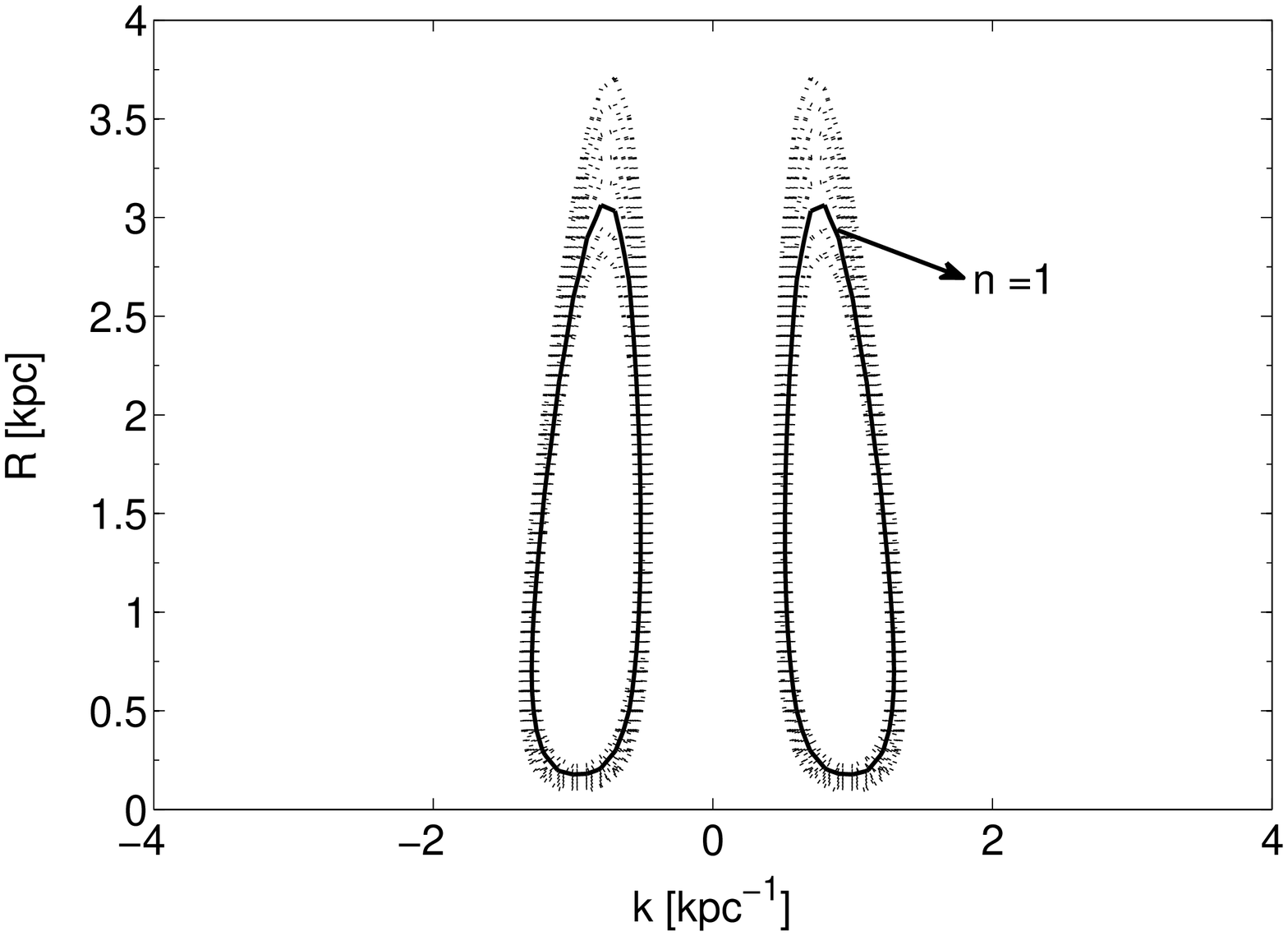}
\caption{Propagation diagrams (contours of constant $\Omega_{\rm p}$) for those pattern speeds which give closed loops of {\it{Type B}}. The input parameters used are for The Galaxy. The top panel shows contours for disk-alone case where the range of $\Omega_{\rm p}$ varies from 17 km s$^{-1}$ kpc$^{-1}$ to 18 km s$^{-1}$ kpc$^{-1}$, at intervals of 0.2 and the bottom panel shows the contours for disk plus dark matter halo case where the range of $\Omega_{\rm p}$ varies from 14 km s$^{-1}$ kpc$^{-1}$ to 15.4 km s$^{-1}$ kpc$^{-1}$, at intervals of 0.1. The closed loops that correspond to the global modes for different models, are indicated by solid lines. $\Omega_{\rm p}$ value increases from the outer to the inner contours.}
\label{fig-resgalflu}
\end{figure}
 For a quantitative verification, following the same procedure as taken for UGC 7321, we calculated the possible modes that could be present in both models. These are summarized in Table~\ref{tab-resgalflu}, which shows that there is only a small quantitative change in the overall spiral mode properties when we include the dark matter halo in the system. But, note that, unlike the case of UGC 7321, on inclusion of dark matter halo, the extent of forbidden region has not increased appreciably. This is in the line of our expectation because within the solar radius, the dark matter halo contributes 50 per cent or less to the rotation curve \citep{KG91,Sack97}.

\begin{table*}
\centering
\caption{Results for global modes for the Galaxy}
\begin{tabular}{ccccc}
\hline
$\Omega_{\rm p}$  &  $R_- $  & $R_+$ & $R_{\rm CR}$ & $n$ \\
(km s $^{-1}$ kpc$^{-1}$)&  (kpc) & (kpc) & (kpc) & \\
\hline
disk-alone case :\\
17.2 & 0.3  & 3.6 & 8 & 1\\
disk plus halo case:\\
15.2  & 0.2 & 3.0 & 13 & 1 \\
\hline
\end{tabular}
\label{tab-resgalflu}
\end{table*} 
 
\section{Discussion}
\label{dis6}

\subsection{Effect of Bulge}

At this point note that, for the Galaxy, the radial extent of the closed loops of ${\it Type ~ B}$ is in the very inner region (see Figure~\ref{fig-resgalflu}) where the bulge component dominates, and the spiral structures are not believed to be present at those radii. So far in the models of both galaxies, we have not included bulge. Normally, in the early type spirals, bulge dominates in the inner regions, and the late-type (e.g. Scd type) galaxies have no significant bulge \citep[]{BM98}. Consequently, the exclusion of the bulge will lead to a significant underestimation of the rotation curve in the inner parts of the early-type galaxies.

In this section, we have included the bulge component to the existing mass models, namely, disk-alone and disk plus dark matter cases for our Galaxy. On the other hand, UGC 7321 has no discernible bulge \citep{MAT99,MAT03}. It is a typical bulgeless galaxy \citep{Kau09}. So, we redid the global modal analysis, this time taking into account of the bulge component, for the Galaxy only. We adopt a Plummer--Kuzmin bulge model to derive the bulge contribution for the Galaxy.

In the spherical coordinates, the density profile of the bulge is given by the following formula (Binney \& Tremaine 1987):
\begin{equation}
\rho_{\rm bulge}= \frac{3 M_{\rm b}}{4 \pi R^3_{\rm b}}\Bigg(1+\frac{R^2}{R^2_{\rm b}}\Bigg)^{-5/2}\,,
\end{equation}
where $R_{\rm b}$ is  the  bulge  scalelength  and $M_{\rm b}$ is  the  total  bulge mass.\\
The corresponding potential in the cylindrical coordinates ($R$, $\phi$, $z$) is given as
\begin{equation}
\Phi_{\rm bulge} (R, z)=-\frac{G M_{\rm b}}{R_{\rm b}}{\Bigg(1+\frac{R^2+z^2}{R^2_{\rm b}}\Bigg)^{-1/2}}\,.
\end{equation}
Corresponding rotational frequency $\Omega_{\rm bulge}$ and the epicyclic frequency $\kappa_{\rm bulge}$ in the mid-plane ($z$ = 0), are
\begin{equation}
\Omega^2_{\rm bulge}=\frac{G M_{\rm b}}{R^3_{\rm b}}\Bigg(1+\frac{R^2}{R^2_{\rm b}}\Bigg)^{-3/2}\,
\end{equation}
and,
\begin{equation}
\kappa^2_{\rm bulge}=\frac{G M_{\rm b}}{R^3_{\rm b}}\Bigg[4\Bigg(1+\frac{R^2}{R^2_{\rm b}}\Bigg)^{-3/2}\\
-3\Bigg(\frac{R}{R_{\rm b}}\Bigg)^2\Bigg(1+\frac{R^2}{R^2_{\rm b}}\Bigg)^{-5/2}\Bigg]\,.
\end{equation}
Therefore, while taking the bulge component into account, these terms $\kappa^2_{\rm bulge}$ and $\Omega^2_{\rm bulge}$ will be added in R. H. S. of equation (\ref{kappa-contri}).
For the parameters of the bulge component, we have used a $R_{\rm b}$ of 2.5 kpc and a $M_{\rm b}$ of 2.8 $\times$ $10^{10}$ $M_{\odot}$ \citep{BLUM95}.

We did a global mode analysis, similar to that in \S~{\ref{res-galflu}}, for the disk plus bulge case, and then for the disk plus bulge plus dark matter halo case. As explained earlier, we considered only closed loops of $\it{Type ~ B}$ for both cases. Then the relevant quantization condition is applied to obtain the principle quantum number $n$. The results are summarized in Table~\ref{tab-galbulge}.

\begin{table*}
\centering
\caption{Results for global modes for the Galaxy (including bulge)}
\begin{tabular}{ccccc}
\hline
$\Omega_{\rm p}$  &  $R_- $  & $R_+$ & $R_{\rm CR}$ & $n$ \\
(km s $^{-1}$ kpc$^{-1}$)&  (kpc) & (kpc) & (kpc) & \\
\hline
disk plus bulge case :\\
18.3 & 1.8  & 4.5 & 9.7 & 1\\
15.1 & 1.1  & 6.3 & 11.3 & 2\\
\hline
disk plus bulge plus halo case:\\
17.1 & 1.6  & 4.2 & 13.3 & 1\\
14.0 & 0.9  & 5.8 & 16 & 2\\
\hline
\end{tabular}
\label{tab-galbulge}
\end{table*} 
In both models where bulge is included, two modes are present while models without bulge gave only one mode (see Table~\ref{tab-resgalflu} and \ref{tab-galbulge}). Also, the resulting corotation ($R_{\rm CR}$) changes quite a bit when the bulge component is added. More importantly, the extent of global spiral arms (indicated by $R_{-}$ and $R_{+}$) show an extended range and closer to the range where arms are observed \citep{BM98}. Note that the resulting pattern speed for $n=1$ is close to the observed value for the Galaxy \citep{Sie12}.
\subsection {Other issues}
In this subsection, we would like to mention  a few other issues regarding this work. First, in this work, we have not considered the low dispersion component in the disk, namely gas. In reality, a spiral galaxy contains a finite amount of gas. The role of gas has been studied in several dynamical issues, e.g. in the stability of local axisymmetric perturbations \citep{JS84a,JS84b, BR88,Raf01}, local non-axisymmetric perturbations \citep{Jog92}, and in the radial group transport \citep{GJ15} etc. Here, in this paper, our main aim was to address the role of only dark matter halo on the grand-design spiral structure, hence we excluded gas from this formalism. This ensures that whatever change we see in these models, they are purely due to the dark matter halo. Besides, the superthin LSB galaxy UGC 7321 considered here has small gas content as compared to that of normal HSB galaxies \citep{UM03}. 
A full investigation of role of DM halo in a system with gas and stars coupled via gravitation will be followed in a future work.

Another point to note is that, this calculation essentially is based on linear perturbation theory, i.e. all the non-linear effects are neglected. Interestingly a recent study by \citet{Don13} using high resolution $N$-body simulations shows that the nonlinear effect can significantly modify the origin and persistence scenario of flocculent spiral features. So, it is worth checking the nonlinear effect in this context with gas treated on an equal footing with stars.

\section{Conclusions}
\label{con6}
We have studied the existence of global modes in spiral galaxies in the WKB limit using the Bohr--Sommerfeld quantization condition. This approach has been used for the first time to study the effect of dark matter halo on the grand-design spiral structure in a galactic disk. Using the input parameters of a typical superthin, LSB galaxy  UGC 7321, we found that in both disk-alone and disk plus halo cases, global modes are permitted. While the small-scale spiral structure is suppressed by the dark matter halo \citep{GJ14}, the halo does not have a significant effect on the global spiral modes. We argue that the tidal interactions causing global spiral modes are less likely to occur in LSBs, as compared to the HSB counterparts. Thus even though the model for UGC 7321 including dark matter halo permits the existence of global spiral modes, in reality they are not likely to materialize, since these galaxies are isolated and hence may not experience tidal forces due to galaxy encounters.
Also, we carried out a similar analysis for our Galaxy. We found that, for our Galaxy, inclusion of the dark matter halo does not affect the existence of global spiral modes.
 Note, however, that these results are obtained while treating the galactic disk as a fluid. These results may get modified when the galactic disk is treated as a collisionless system. This problem of investigating the role of dark matter halo on large-scale spiral arms in a collisionless galactic disk will be followed up in a future paper.

\thispagestyle{empty}

\chapter[Effect of dark matter halo on global spiral modes in a collisionless\\
 galactic disk]{Effect of dark matter halo on global spiral modes in a collisionless galactic disk \footnote{Ghosh, Saini \& Jog, 2017, New Astronomy, 54, 72}}
\chaptermark{\it Effect of dark matter halo on global modes in collisionless disk}
\vspace {2.5cm}

\section{Abstract}
 Low surface brightness (LSB) galaxies are dominated by dark matter halo from the innermost radii; hence they are ideal candidates to investigate the influence of dark matter on different dynamical aspects of spiral galaxies. Here, we study the effect of dark matter halo on grand-design, $m=2$, spiral modes in a galactic disk, treated as a collisionless system, by carrying out a global modal analysis within the WKB approximation. First, we study a superthin, LSB galaxy UGC~7321 and show that it does  not support discrete global spiral modes when modeled as a disk-alone system or as a disk plus dark matter system. Even a moderate increase in the stellar central surface density does not yield any global spiral modes. This naturally explains the observed lack of strong large-scale spiral structure in LSBs. An earlier work \citep{GSJ16} where the galactic disk was treated as a fluid system for simplicity had shown that the dominant halo could not arrest global   modes. We found that this difference arises due to the different dispersion relation used in the two cases and which plays a crucial role in the search for global spiral modes. 
Thus the correct treatment of stars as a collisionless system as done here results in the suppression of global spiral modes, in agreement with the observations. We performed a similar modal analysis for the Galaxy, and found that the dark matter halo has a negligible effect on large-scale spiral structure.

\section{Introduction}  
Low Surface Brightness (LSB) galaxies are characterized by low star formation rate \citep{IB97} and low disk surface density 
\citep{dBM96,dBM01}. 
The spiral structure in LSBs is often incipient or fragmentary and usually faint and difficult to trace \citep{Sch90,Mcg95,Sch11}, and they generally do not host any strong large-scale spiral structure, the kind we see in case of normal high surface brightness (HSB) galaxies like our Milky way. We note that, we are interested only in small LSBs, which are more abundant, and do not include the giant LSBs like Malin~1. Some of the giant LSBs show fairly strong, large-scale spiral structure as in UGC~6614 \citep{Das13} and in the inner regions they are dynamically similar to their High Surface Brightness (HSB) counterparts \citep{Lelli10}.  \citet{Fu02} has applied the technique of density-wave theory to put constraint on the decomposition of the rotation curve in LSBs. However we note that in the sample considered by \citet{Fu02} contains giant LSB (e.g., UGC~6614) which often show large-scale spiral structure \citep[e.g., see][]{Das13}.

 The LSBs are dark matter dominated from the very inner regions \citep{Bot97,dBM97,dBM01}. Within the optical disk, the dark matter constitutes about $90$ per cent of the total mass of LSBs, whereas for the HSBs the contributions of the dark matter halo mass and stellar mass are comparable \citep{dBM01,Jog12}. Thus, the LSBs constitute a natural laboratory to study the effect of dark matter halo on different aspects of galactic dynamics.

Several past studies have shown the effect of dominant dark matter halo in the suppression of global non-axisymmetric bar modes \citep{Mih97}, in making the galactic disks superthin \citep{BJ13} and in prohibiting the swing amplification mechanism from operating, thus explaining the lack of small-scale spiral structure as noted observationally \citep{GJ14}. 

According to the density wave theory, the grand-design spiral arms are the high density regions of a rigidly rotating spiral density wave, with a well defined pattern speed, that are self-consistently generated by the combined gravity of the unperturbed disk and the density wave \citep{LS64,LS66}. For a recent review on this see \citet{DOBA14}.

 In a recent work \citet{GSJ16} (hereafter Paper~1) investigated the role of a dominant dark matter halo on the global spiral modes within the framework of the density wave theory by treating the  galactic disk as a fluid. Using the input parameters of a superthin LSB galaxy UGC~7321 and the Galaxy, they found that for UGC~7321, the dark matter halo has a negligible effect on arresting the global spiral modes when the disk is modeled as a fluid. This is in contrast to the results for small-scale spiral features where the dark matter was shown to suppress the small-scale, swing-amplified spiral structures almost completely \citep{GJ14}. \citet{GSJ16} (Paper 1) argued that that since  LSBs are relatively isolated, tidal interactions are less likely to occur compared to those for the HSB galaxies.  Thus even though the global spiral modes are formally permitted in the fluid disk plus dark matter halo model, it was argued that it is the lack of tidal interaction that makes it difficult for the global spiral structure to develop in these galaxies.

In this paper we address the effect of dark matter halo on global $m=2$ modes by modeling the galactic disk more realistically as a $\emph{collisionless}$ system. A tidal encounter is likely to give rise to global modes, as has been seen in simulations, as in M51 \citep[see e.g.,][]{TT72}.
 
 We use the dispersion relation for a $\emph{collisionless}$ disk to construct global standing-wave like solutions by invoking the Bohr-Sommerfeld quantization condition (for details see Paper~1). Note that fluid disks allow wavelike solutions at small wavelengths, since fluid pressure provides the restoring force; but collisionless disks suppress wavelike modes for very small wavelengths \citep[e.g. see][]{BT87}.

The \S~{\ref{formu7}} contains formulation of the problem and the input parameters. In \S~\ref{wkb-col} we present the WKB analysis and the relevant quantization rule. \S~\ref{res7} and \S~\ref{dis7} contain the results and discussion, respectively while \S~\ref{con7} contains the  conclusions.

\section{Formulation of the problem}
\label{formu7}

We model the galactic disk as a collisionless system characterized by an exponential surface density $\Sigma_{s}$, and one-dimensional velocity dispersion $\sigma_{\rm s}$. For simplicity, the galactic disk is taken to be infinitesimally thin. In other words, we are interested in perturbations that are confined to the mid-plane ($z=0$). The dark matter halo is assumed to be non-responsive to the gravitational perturbations of the disk. We have used cylindrical coordinates $(R, \phi, z)$ in our analysis.
\subsection{Details of models}
In this subsection, we describe the models that we have used for the study of effect of dark matter halo on the global spiral modes.

The dynamics of the disk is calculated first only under the gravity of the disk (referred to as disk-alone case) and then under the joint gravity of disk and the dark matter halo (referred to as disk plus halo case).
We took an exponential stellar disk with central surface density $\Sigma_0$  and the disk scalelength $R_{\rm d}$, which is embedded in a concentric dark matter halo whose density follows a pseudo-isothermal profile characterized by core density $\rho_0$ and core radius $R_{\rm c}$.

The net angular frequency, $\Omega$ and the net epicyclic frequency, $\kappa$ for a galactic disk embedded in a dark matter halo, concentric to the galactic disk, are given as:\\
\begin{equation}
\kappa^2= \kappa^2_{\rm disk}+\kappa^2_{\rm DM}\,; \quad \Omega^2= \Omega^2_{\rm disk}+\Omega^2_{\rm DM}\,.
\end{equation}

The expressions for $\kappa^2_{\rm disk}$, $\Omega^2_{\rm disk}$ in the mid-plane ($z=0$) for an exponential disk and $\kappa^2_{\rm DM}$, $\Omega^2_{\rm DM}$ for a pseudo-isothermal halo in the mid-plane ($z=0$) have been calculated earlier (see Paper 1 for details).

In the early-type galaxies, the bulge component dominates in the inner regions, and hence exclusion of the bulge component from the models for early-type galaxies will underestimate the rotation curve in the inner regions. In Paper 1, it was shown that for the Galaxy, inclusion of bulge yielded more realistic results (for details see \S~5.1 in Paper 1).
Also note that UGC~7321 has no discernible bulge \citep{MAT99,MAT03}. Therefore, only for our Galaxy we have included bulge in both the disk-alone and disk plus dark matter halo models. We adopt a Plummer--Kuzmin bulge model for our Galaxy which is characterized by total bulge mass $M_{\rm b}$ and bulge scalelength $R_{\rm b}$.

The expressions for $\kappa^2_{\rm bulge}$, $\Omega^2_{\rm bulge}$ in the mid-plane ($z=0$) for such a bulge having a Plummer-Kuzmin profile are given in Paper 1. Therefore, for the Galaxy, these $\kappa^2_{\rm bulge}$ and $\Omega^2_{\rm bulge}$ terms will be added in quadrature to the corresponding terms due to disk and dark matter halo.

\subsection{Input parameters}
The input parameters for different components of UGC~7321 \citep{Ban10} and the Galaxy \citep{Mer98,BLUM95} are summarized in Table~\ref{table-galinputs}.

\begin{table*}
\centering
\caption{Input parameters for UGC~7321 and the Galaxy}
\begin{tabular}{ccccccc}
\hline
Galaxy & $\Sigma_0$   &  $R_{\rm d}$  & $\rho_0$  & $R_{\rm c}$  & $M_{\rm b}$  & $R_{\rm b}$  \\
& (M$_{\odot}$ pc$^{-2}$) & (kpc) & (M$_{\odot}$ pc$^{-3}$) & (kpc) &  ($\times$ 10$^{10}$~M$_{\odot}$) & (kpc)\\
\hline
UGC~7321  & 50.2 & 2.1 & 0.057  & 2.5  & - & -\\
The Galaxy & 640.9 & 3.2 & 0.035 & 5 & 2.8 & 2.5\\
\hline
\end{tabular}
\label{table-galinputs}
\end{table*} 

For UGC~7321, the stellar velocity dispersion in the radial direction is taken to be: $\sigma_{\rm s}$ =$(\sigma_{\rm s0})_R \exp(-R/2R_{\rm d})$, where $(\sigma_{\rm s0})_R$ is the central velocity dispersion in the radial direction. The observed central velocity dispersion in the $z$ direction ($(\sigma_{\rm s0})_z$) is 14.3 km sec$^{-1}$ \citep{Ban10}. In the solar neighborhood, it is observationally found that $(\sigma_{\rm s})_z$/$(\sigma_{\rm s})_R$ $\sim$ 0.5 \citep[e.g.,][]{BT87}. Here we assume the same conversion factor for all radii.

For the Galaxy, the observed stellar velocity dispersion in the radial direction is : $\sigma_{\rm s}$ =$(\sigma_{\rm s0})_R \exp(-R/8.7)$, where $(\sigma_{\rm s0})_R$ = 95 km sec$^{-1}$ \citep{LF89}.

\section{WKB analysis}
\label{wkb-col}

The dispersion relation for an infinitesimally thin galactic disk, modeled as a collisionless system, in the WKB limit, is given by \citep{BT87}:\\
\begin{equation}
(\omega - m\Omega)^2=\kappa^2-2\pi G \Sigma_{\rm s} |k|{\mathcal F}(s, \chi),
\label{disp-equation}
\end{equation}
where  $s(=({\omega-m\Omega})/{\kappa})$  and  $\chi (=k^2\sigma^2_{\rm s}/\kappa^2)$ are the dimensionless frequencies. ${\mathcal F}(s, \chi)$ is the reduction factor which physically takes into account the reduction in self-gravity due to the velocity dispersion of stars. 
The form for ${\mathcal F}(s, \chi)$ for a razor-thin disk whose stellar equilibrium state can be described by the Schwarzchild distribution function is given by \citep{BT87}: 
\begin{equation}
{\mathcal F}(s, \chi)=\frac{2}{\chi}\exp(-\chi)(1-s^2)\sum_{n=1}^\infty\frac{I_n(\chi)}{1- s^2/n^2}\,.
\label{reduc-factor}
\end{equation}
Since we are interested in $m=2$ grand spiral modes, a rearrangement of equation (\ref{disp-equation}) yields\\
\begin{equation}
4(\Omega_{\rm p}-\Omega)^2=\kappa^2-2\pi G \Sigma_{\rm s}|k| {\mathcal F}(s, \chi)\,,
\label{disp-mod}
\end{equation}
where $\Omega_{\rm p} $ (= $\omega /m$) is the pattern speed with which the spiral arm  rotates rigidly in the galactic disk.

For a given $\Omega_{\rm p}$, equation~(\ref{disp-mod}) is an implicit relation between the phase space variables $k$ and $R$. A contour of constant $\Omega_{\rm p}$ denotes the path of a wavepacket that propagates with a group velocity $v_{\rm g}$ = $d\omega/dk$ \citep{Tom69}.

The simpler analytic dispersion relation for the fluid disk allows one to express $k$ explicitly as a function of $R$ (for details see \S~3 of Paper 1), but for a collisionless system, the dispersion relation (equation \ref {disp-equation}) is transcendental in nature, therefore it is impossible to analytically express the wavevector $k$ as an explicit function of $R$. However, it is straightforward to obtain this relation numerically.

\begin{figure}
\centering
\includegraphics[height=3.0in,width=4.0in]{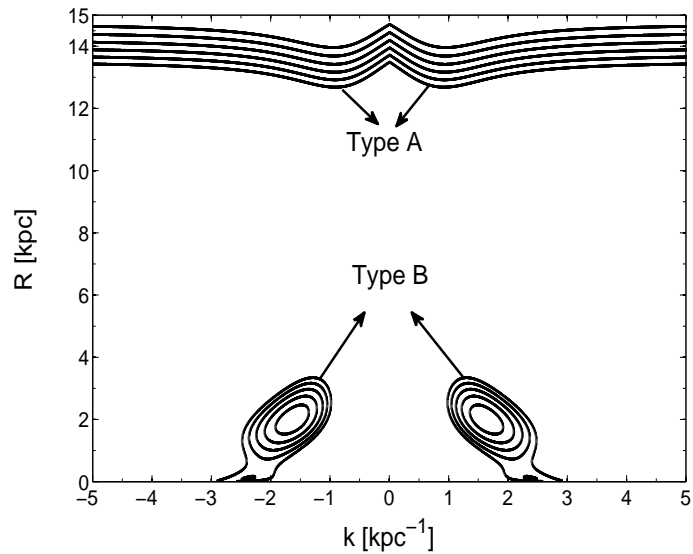}
\caption{Propagation diagrams (contours of constant $\Omega_{\rm p}$) for different pattern speeds. The input parameters used are for the Galaxy, where the Galaxy is modeled as a collisionless disk plus dark matter halo. Different types of contours present here are marked as A and B. The range of $\Omega_{\rm p}$ varies from 25.8 km s$^{-1}$ kpc$^{-1}$ to 28.7 km s$^{-1}$ kpc$^{-1}$, with a spacing of 0.5 km s$^{-1}$ kpc$^{-1}$. $\Omega_{\rm p}$ value increases from the outer to the inner contours patterns.}
\label{fig:UGC7321}
\end{figure}

Figure~\ref{fig:UGC7321} shows the typical contours that are present in different models considered in this work. We refer the reader to Paper 1 for a physical interpretation of these contours in terms of propagation of wavepackets.
In Paper 1, another type of contour (called {\it {Type C}}) was also present, but such contours occurred due to the fluid treatment of the disk, and consequently we did not consider them in our analysis (for details see \S~3.1 of Paper 1). As expected, these contours are absent when the disk is treated as collisionless, thus justifying our omitting them in our previous analysis.
Since global spiral modes are basically standing waves in a differentially rotating galactic disk \citep[e.g. see][]{Ber00}, therefore, in order to obtain a standing wave, the wave has to be reflected/refracted back into the wave cycle by the reflecting/refracting barrier. Although contours of {\it {Type A}} allow wavelike solutions, the wavevector $k$ can become quite large. Such waves will be dissipated at $k$ corresponding to the epicyclic length scale, and thus do not constitute valid standing waves. Hence, only a closed contour has the correct behaviour to represent a standing wave in a galactic disk. Hence, from now on, we will consider contours only of {\it {Type B}} and discard the contours of {\it {Type A}}.

The global spiral modes can be constructed from the WKB dispersion relation by using the Bohr-Sommerfeld quantization condition (for details see \S~3.1 in Paper 1).

The appropriate WKB quantization rule for the closed contours of {\it Type~B} is given by \citep{Tre01}
\begin{equation}
2 \int_{R_-}^{R_+} [k_{+}(R)-k_{-}(R)] dR = 2\pi \left (n-\frac{1}{2} \right)\,,
\label{qcond}
\end{equation}
where $n=1,2,3,\cdots$; and 
 $k_{+}$, $k_{-}$ (where $k_{-} \le k_{+}$ ) are the two solutions of equation~(\ref{disp-mod}), and the equality sign holds only at the turning points ($R_{\pm}$).

\section {Results}
\label{res7}

In principle, the pattern speed $\Omega_{\rm p}$ could be either positive or negative, implying prograde or retrograde motion of the spiral pattern,  respectively. For a negative pattern speed we can write $\Omega_{\rm p} = -|\Omega_{\rm p}|$, and rewrite equation (\ref{disp-mod}) in the form 
\begin{equation}
4(|\Omega_{\rm p}|+\Omega-\kappa/2)(|\Omega_{\rm p}|+\Omega+\kappa/2)=-2\pi G \Sigma_{\rm s}|k|{\mathcal F}(s, \chi)\,.
\label {disp-negative}
\end{equation}
For both the galaxies considered here, the quantities $\Omega-\kappa/2$ and $\Omega+\kappa/2$ are positive, hence the left hand side of equation (\ref{disp-negative}) is always positive while the right hand side equation is always negative. Consequently, there is no possible solution of equation (\ref{disp-negative}) for negative pattern speeds. This implies 
that no global spiral modes can have negative pattern speeds,  or in other words, all  discrete global spiral modes present in different models of this paper are prograde. Therefore, all the figures displayed in this paper are for prograde spiral patterns.

Before proceeding to investigate the global modes for galaxies UGC~7321 and the Galaxy, we note that \citet{GJ14,Jog14} have calculated the Toomre Q-value \citep{Too64} for these two galaxies and have shown it to be always $>$ 1 for the input parameters used in this paper. Therefore the galactic disks of the two galaxies considered by us are stable against local axisymmetric perturbations.

\subsection{UGC~7321}

Starting from the dispersion relation (equation \ref{disp-mod}), we obtained contours of {\it{Type B}}, first for the disk-alone case, and then for the disk plus dark matter halo case.

\subsubsection{ Disk-alone case}

 For the disk-alone case we found closed contours of {\it{Type B}}, but on applying the quantization condition (equation {\ref {qcond}}), we found no discrete global modes, i.e., the area enclosed by the closed loops is not sufficient to satisfy equation~({\ref {qcond}}) for any integral value of $n$ (see Figure~\ref{diskalone-ugc7321}).

 \begin{figure}
\centering
\includegraphics[height=3.0in,width=4.0in]{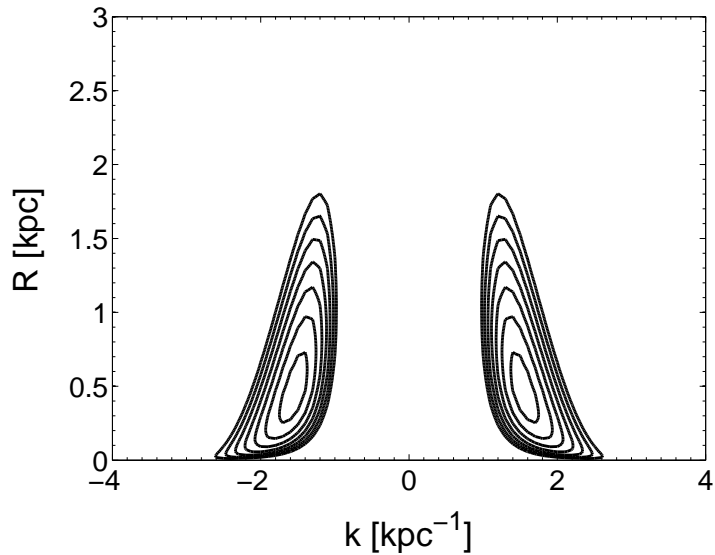}
\caption{ Propagation diagrams (contours of constant $\Omega_{\rm p}$) for the disk-alone case, corresponding to those pattern speeds which give closed loops of {\it{Type B}}. The input parameters used are for UGC~7321. The range of $\Omega_{\rm p}$ varies from 7.1 km s$^{-1}$ kpc$^{-1}$ to 7.7 km s$^{-1}$ kpc$^{-1}$, with a spacing of 0.1 km s$^{-1}$ kpc$^{-1}$. $\Omega_{\rm p}$ value increases from the outer to the inner contours patterns.}
\label{diskalone-ugc7321}
\end{figure}

Note that the closed contours in Figure~\ref{diskalone-ugc7321} do not satisfy the condition under which WKB approximation is valid, i.e.,  $|kR| \gg 1$. In fact, some parts of the contours have $|kR| < 1$. The standard use of the WKB approximation is in quantum mechanics where it is used to compute the energy spectrum of bound systems. It is well known that at the classical turning points the WKB condition fails since $k = 0$ at the turning points. However, despite this deficiency, the approximation furnishes useful qualitative results. Note, moreover, that in Figure~\ref{diskalone-ugc7321} the WKB approximation does not fail so extremely. However, the failure of WKB  also implies the failure of the tight-winding approximation that makes it possible to relate the gravitational potential to the \emph{local} perturbed density. Typically, the local approximation suffices to obtain useful qualitative information even in this case, even though the numerical results are less robust. For example, in a recent paper \citet{JT12} have calculated the slow-modes of collisionless  near-Keplerian disks. They have computed eigenvalues using exact numerical methods as well as through the WKB approximation. Although their contours in some parts also do not satisfy $|kR| > 1$, they find that their results are in good qualitative agreement with the frequencies calculated using the exact method. However, note that this concern is not valid for our main result for UGC~7321 below. 

\subsubsection{Disk plus halo case}
We find that for the disk plus dark matter halo case, the closed loops of {\it{Type B}} are \emph{not present} at all, and hence no discrete global modes exist in this case. This is in contrast to Paper 1 where, based on the fluid model of the disk, we found two eigenmodes for the disk plus halo case.  Since stable modes do not excite by themselves, we had argued that they would be absent in UGC 7321 due to its isolation and its consequent lack of tidal interactions with other galaxies. It is useful to note that 
it is possible to excite stable modes by tidal encounters as shown by \citet{JT12}. The difference in results can be ascribed to the difference in the treatment of the disk as a collisionless system here as opposed to a fluid  description in Paper 1. Thus, the correct treatment of stars as a collisionless system has directly led to the result that global modes are not permitted.

\subsubsection{Other LSB galaxies}
To investigate the role of dark halo on the existence of global modes in LSBs in general, we constructed models where we set the central surface density to  two and three times the value used in the original model for UGC~2371 (see Table~\ref{table-galinputs}), and modified the radial velocity dispersion profile in such a way that the resulting Toomre $Q$-parameter remains the same as that in the original model. The net rotation curve changes less than 20 per cent of that in the original model. Though somewhat contrived, the parameters for these additional cases were devised to check the effect of disk surface density on the existence of global modes. We did not find any global spiral modes in these modified models as well. Since a dominant dark matter halo is a characteristic feature of LSBs \citep {Bot97,dBM97}, our results indicate that low disk surface density and dark matter halo that is dominant from the innermost regions in LSBs together explain the suppression of local, swing-amplified spiral features \citep{GJ14} in these galaxies, as well as suppression of global $m=2$ spiral modes as shown here. 

In our analysis, we have used the input parameters of the LSB galaxy UGC~7321, which is an edge-on galaxy (angle of inclination 88$^{\circ}$, \citet{MA00}). However, we note that the generic features of typical LSBs that play the dominant role in the dynamics of disks in LSBs, i.e., low surface density and dark matter dominance from the innermost radius, are both observed in UGC~7321. Due to the edge on nature of this galaxy it is difficult to conclude if large scale spiral features are absent in this galaxy, although the absence of dust lanes is suggestive of this fact. In any case, for completeness of analysis we have also considered synthetic models where the disk mass was artificially enhanced by a factor of few. Therefore, the main result of this paper that the dominant dark matter halo suppresses the large-scale spiral features in LSBs is likely to hold for all LSB galaxies. 

\subsection{The Galaxy}

As a check, we carried out the same modal analysis to examine the effect of the dark matter halo on the global spiral modes for the Galaxy. The possible closed contours of {\it{Type B}} for both disk-alone and disk plus dark matter halo are shown in Figure~\ref{fig-galaxy}.

\begin{figure}
\centering
\includegraphics[height=3.0in,width=4.0in]{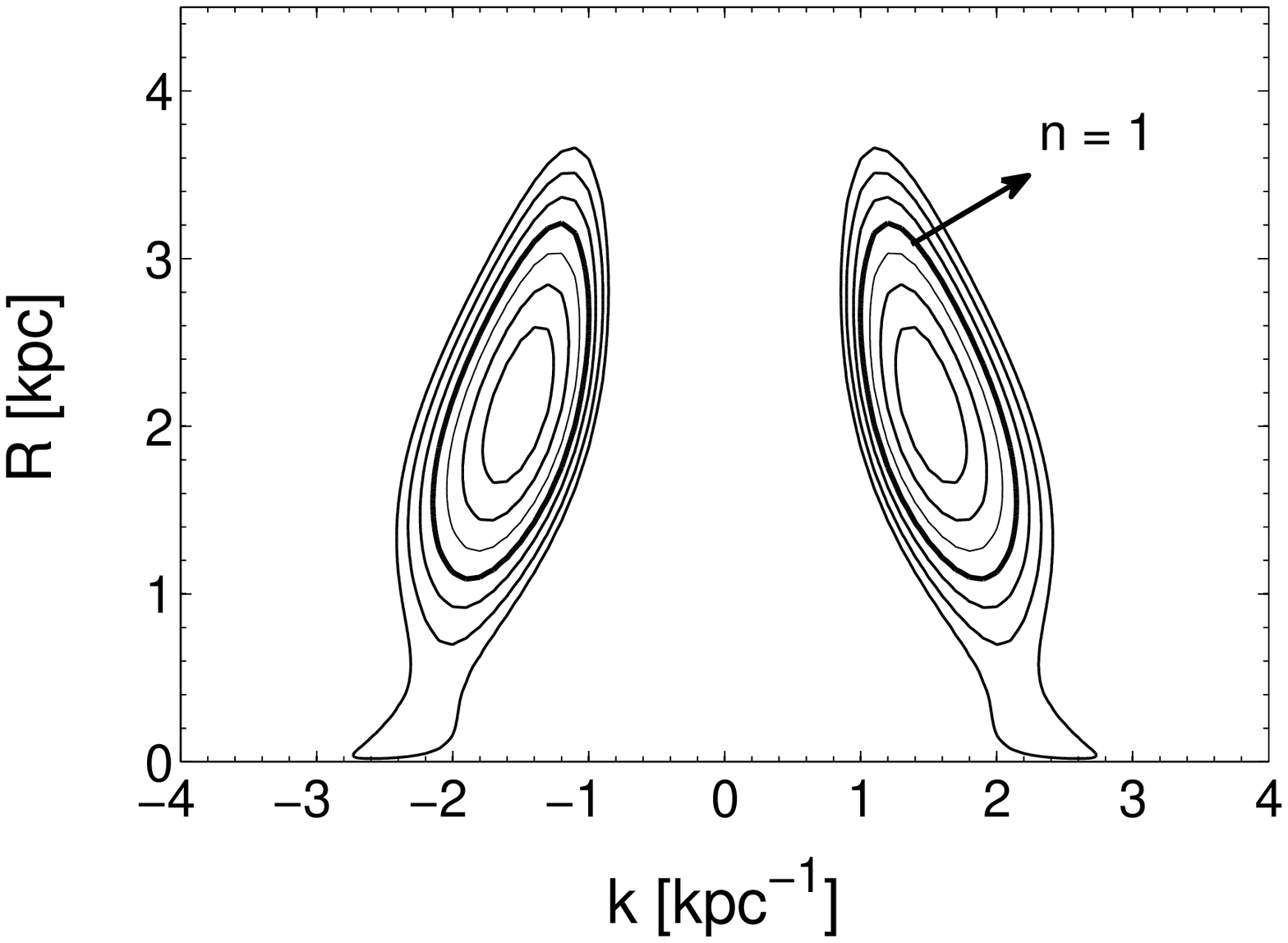}
\medskip
\includegraphics[height=3.0in,width=4.0in]{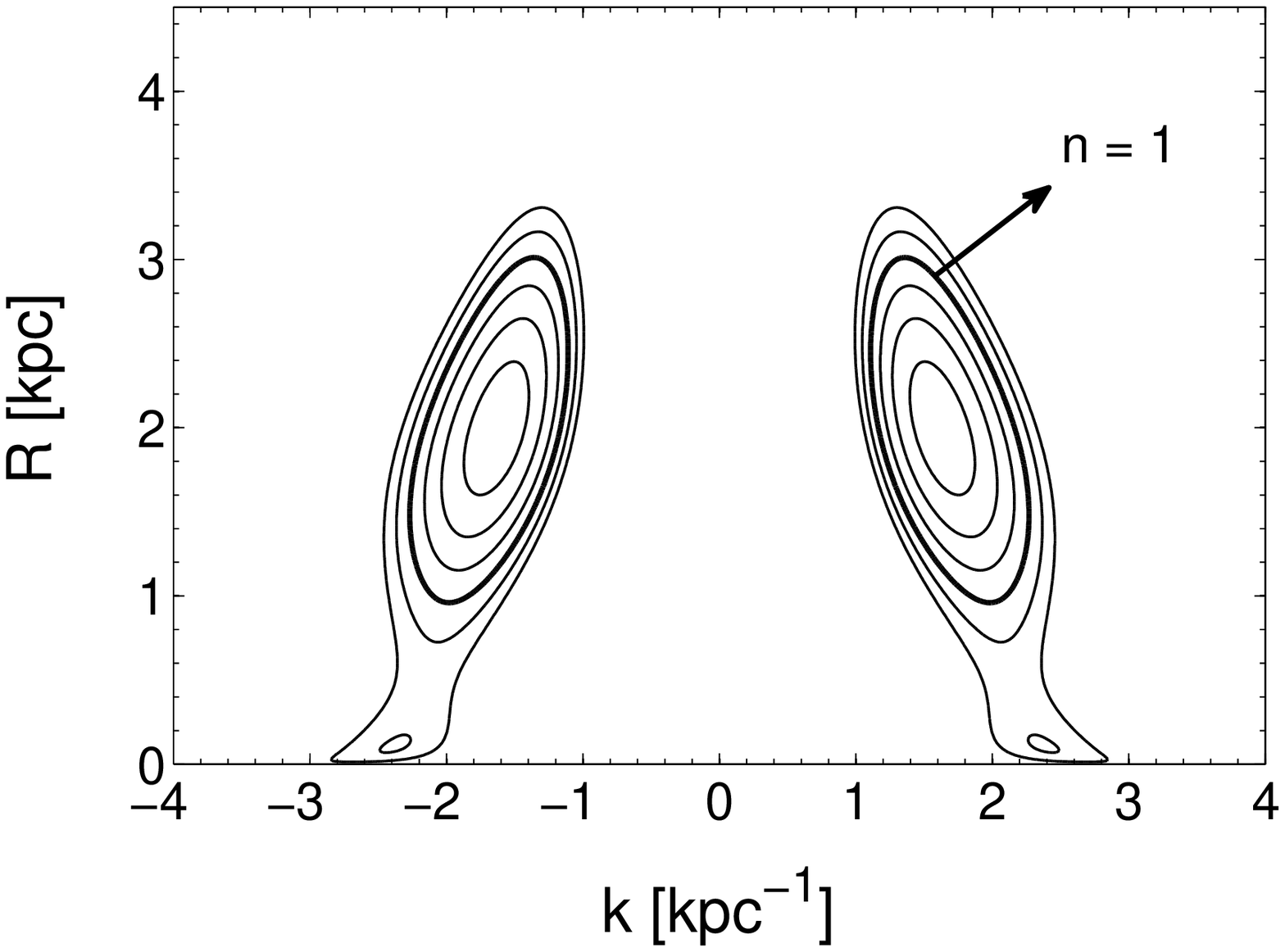}
\caption{ Propagation diagrams (contours of constant $\Omega_{\rm p}$) for those pattern speeds which give closed loops of {\it{Type B}}. The input parameters used are for the Galaxy. The top panel shows contours for disk-alone case (including the bulge) where the range of $\Omega_{\rm p}$ varies from 26.1 km s$^{-1}$ kpc$^{-1}$ to 28.8 km s$^{-1}$ kpc$^{-1}$, at intervals of 0.4 and the bottom panel shows the contours for disk plus dark matter halo case (including the bulge) where the range of $\Omega_{\rm p}$ varies from 25.9 km s$^{-1}$ kpc$^{-1}$ to 28.1 km s$^{-1}$ kpc$^{-1}$, at intervals of 0.4. The closed loops that correspond to the global modes for different models, are indicated by solid lines. $\Omega_{\rm p}$ value increases from the outer to the inner contours.}
\label{fig-galaxy}
\end{figure}

Next, we applied the quantization condition to seek the discrete global mode(s), and the results are summarized in Table~\ref{galaxycase}. Also the positions of different resonance points for the two modes in Table~\ref{galaxycase} are presented Figure~\ref{fig-galaxyreso}.\\

\begin{table*}
\centering
\caption{Results for global modes for the Galaxy}
\begin{tabular}{cccccc}
\hline
Case& $\Omega_{\rm p}$  &  $R_- $  & $R_+$ & $R_{\rm CR}$ & $n$ \\
&(km s $^{-1}$ kpc$^{-1}$) &  (kpc) & (kpc) & (kpc) & \\
\hline
Disk-alone case (including bulge)& 27.3 & 1.1  & 3.3 & 8.3 & 1\\
Disk plus halo case (including bulge) & 26.7  & 0.9 & 3.0 & 8.6 & 1 \\
\hline
\end{tabular}
\label{galaxycase}
\end{table*}

\begin{figure}
\centering
\includegraphics[height=3.0in,width=4.0in]{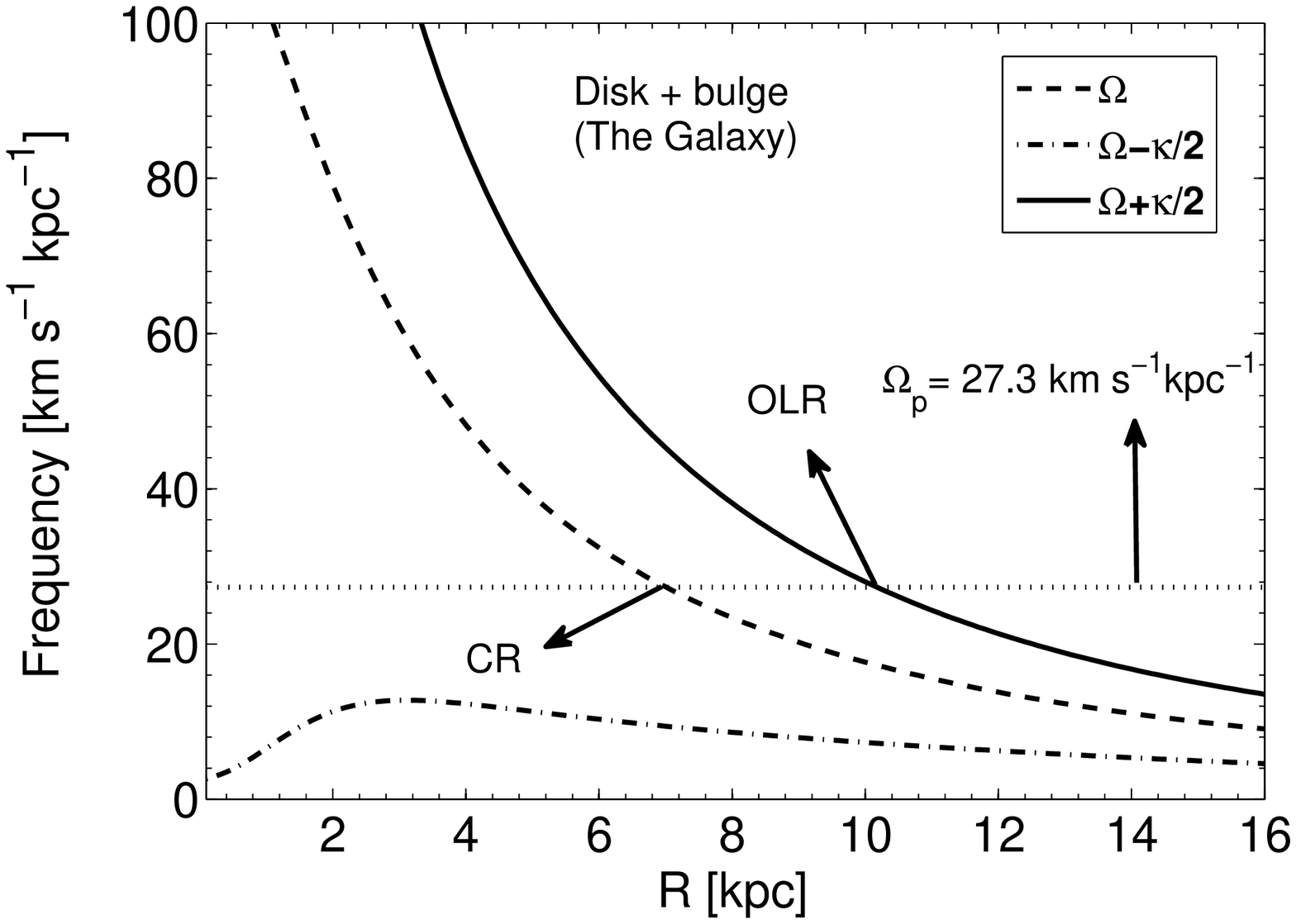}
\medskip
\includegraphics[height=3.0in,width=4.0in]{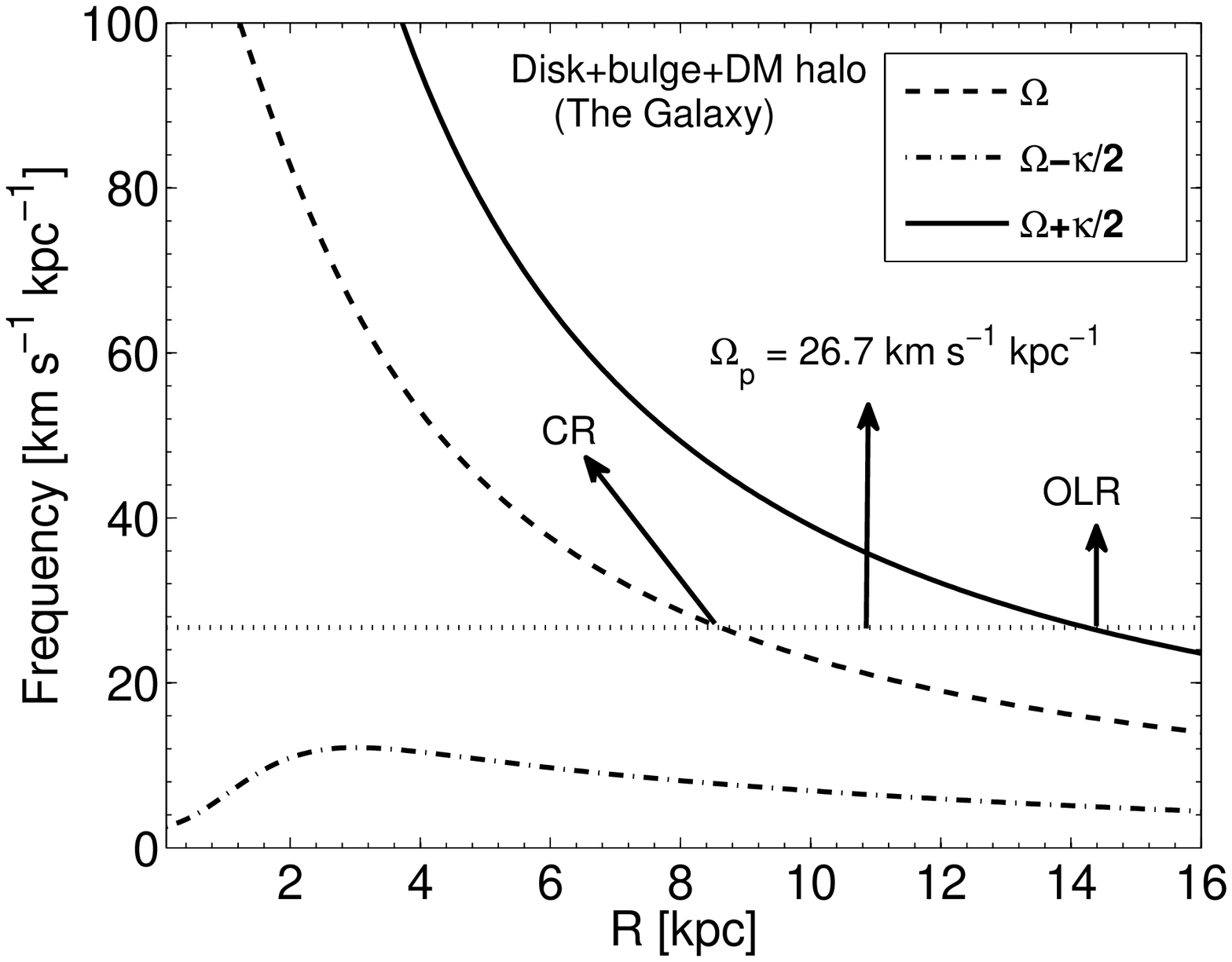}
\caption{Positions of different resonance points for the two modes in Table~\ref{galaxycase} are presented here. Comparison with Figure~\ref{fig-galaxy} shows that the modes do not extend up to the resonance points, consistent with the fact that $k$ is not equal to zero anywhere for modes in Figure~\ref{fig-galaxy}.}
\label{fig-galaxyreso}
\end{figure}

From Table~\ref{galaxycase}, it is evident that the dark matter halo has a negligible effect on the global spiral modes for the Galaxy, the difference can be seen only in the change of the specific values of pattern speed $\Omega_{\rm p}$. This result is at par with our expectation, since the dark matter halo is known to be not important in the inner regions of the Galaxy \citep [e.g., ][]{Sac97,dBM01}. These results are consistent with the results of Paper 1 where the Galactic disk was modeled as a fluid disk.

The key difference with results from  Paper 1 is that the specific values of pattern speed $\Omega_{\rm p}$ which give the global modes are different   
(for comparison, see Table 3 in Paper 1). \citet {Sie12} found a range of $18-24$ km s$^{-1}$ kpc$^{-1}$ for the pattern speed of spiral arms for the Galaxy. On comparison the pattern speed values obtained in this paper lie outside the observed range of pattern speed. 

The shape of the closed contours in this work (Figures~ \ref{fig:UGC7321},~\ref{diskalone-ugc7321},~\ref{fig-galaxy}) is different from that in Paper 1. A curve of constant $\Omega_{\rm p}$ had nearly identical $k$ values at the turning points of a closed curve of {\it {Type B}} in Paper 1, whereas the contours look slanted in the present paper, meaning that the $k$ values at the turning points are substantially different. This is due to the difference in the dispersion relation for the
two cases. A more detailed interpretation is not possible in a simple analytical form, since the dispersion relation for the collisionless case (equation~(\ref{disp-equation})) has a transcendental form.

In this paper for simplicity we have not included the low velocity dispersion component, namely, gas in the system which could further modify the obtained range of pattern speeds. Note that any late-type spiral galaxy contains non-negligible amount of gas, and it has been shown that the gas plays a  significant role in various dynamical issues \citep[e.g.,][]{JS84a,JS84b,Jog92,Raf01,GJ15,GJ16}. The amount of gas present in UGC~7321 is quite small in comparison to that in HSB galaxies \citep{UM03}. Therefore, the results for UGC~7321 obtained in this paper, in particular the suppression of global modes by the dominant dark matter halo is unlikely to change significantly if gas is included in the calculation. The effect of gas on the global spiral modes in a gravitationally coupled two-component (stars and gas) system, as applicable to the gas-rich HSB galaxies, will be taken up in a future paper.

\section{Discussion}
\label{dis7}
The results presented here are based on the assumption that the grand-design spiral structure seen in disk galaxies is due to the density waves, as proposed by \citet{LS64,LS66}. However this hypothesis has not yet been fully confirmed observationally. For example, the angular off-set in age of the stellar population, as predicted by the classical density wave theory has not been found in studies by \citet{Foy11,Fer12} in their sample of late-type disk galaxies. Also the study of $CO$ observation of NGC~1068 using generalized Tremaine--Weinberg method by \citet{MRM06} has revealed that the pattern speed of the spiral structure of this galaxy varies rapidly with radius, indicative of short-lived spiral features. A study of the gas content and the $H\alpha$ line of sight velocity distribution in NGC~6754 by \citet{San16} revealed a different sense of streaming motion in the trailing and leading edge of this galaxy, indicating that the spiral arms of this galaxy is likely to be a transient. 

On the other hand, it is also worth noting that \citet{Im16} have recently presented a study of measuring pitch angle from the images taken in different wavelengths for a large sample of disk galaxies and found that the pitch angle varies from one wavelength to another as predicted by the classical density wave theory, and thus furnishes strong observational evidence for the validity of the density-wave theory of spiral structure in disk galaxies.

Generally the various $N$-body studies so far do not show the evidence of long-lived spiral structure. Several $N$-body simulations have shown that the spiral arms are transient and get wound up quickly \citep{Sel11,Fuji11,Gra12,Don13}, which goes against the classical density wave picture. Interestingly an opposite trend is shown in a recent work by \citet{SaEl16} who reported long-lived spiral structure in a $N$-body study of models for galaxies with intermediate bulges.

\section {Conclusion}
\label{con7}

We have investigated the effect of a dominant dark matter halo on the possible existence of global spiral arms while modeling the galactic disk as a \emph {collisionless} system. The WKB dispersion relation and  the Bohr-sommerfeld quantization condition were used to obtain discrete global spiral modes present in any model. We have analysed a superthin LSB galaxy UGC~7321, and the Galaxy, for this work. We found that for UGC~7321 both the disk-alone and the disk plus dark matter halo cases did not yield any discrete global spiral modes. Even an increase in the stellar central surface density by a factor of few failed to produce any global spiral modes. Thus our findings provide a natural explanation for the observed dearth of  strong large-scale spiral structure in the LSBs. Our results differ from those obtained in Paper 1 where the galactic disk was treated as a fluid, where it was found that the dominant dark matter halo had a  negligible 
constraining effect on the existence of global spiral modes in these galaxies. This difference is due to the different dispersion relation for fluid and collisionless systems, which played a pivotal role in determining the discrete global mode(s) present in a model. 
Thus the correct collisionless treatment for stars as done here has led us to a result for the non-existence of global spiral modes which
is in agreement with the observations.  
As a check, for the Galaxy we carried out  a similar modal analysis and found that the dark matter halo has a negligible effect on the global spiral modes, as expected since the Galaxy is not dark matter dominated in the inner regions, in contrast to UGC~7321. 

Thus, the dark matter halo that dominates from the innermost regions is shown to suppress the growth of local, swing-amplified non-axisymmetric features \citep{GJ14} as well as the global spiral modes as shown here.

\thispagestyle{empty}
\chapter[Conclusions \& Future Work]{Conclusions \& Future Work}
\chaptermark{\it Conclusions \& Future Work}
\vspace {2.5cm}

\section{Concluding Remarks}

This thesis deals with a study of the dynamical effect of dark matter halo and interstellar gas on the spiral structure of different scales as seen in the disks of the late-type spiral galaxies.

\medskip

In the first part of the thesis, we studied the dynamical effect of the dark matter halo on the small-scale spiral structure in some galaxies where the dark matter halo is known to dominate the dynamics over most the disk regions. We chose UGC~7321, a typical superthin Low surface brightness (LSB) galaxy and a set of five late-type dwarf irregular galaxies with extended $HI$ disks to study the role of the dark matter on small-scale spiral arms. Since all of these above-mentioned galaxies are dark matter dominated right from the very inner regions, therefore they naturally provide an ideal `laboratory' to test the effect of dark matter halo on various dynamical aspects.

We carried out the local, non-axisymmetric perturbation analysis (swing amplification) in the disks of such galaxies. Using the 
observed input parameters for UGC 7321, a typical superthin LSB galaxy, we showed that when only 
stellar disk is taken into account, the system exhibits finite growth of non-axisymmetric perturbations, thus indicating that the system can show small-scale spiral features. But, when the dark matter halo is added
 to the system, it damps the growth of the perturbations, almost completely. Even when the low velocity dispersion
 component, namely, the interstellar gas is added in the system, this finding remains unchanged ({\bf Chapter~2}). Similarly we studied the dwarf irregulars with extended $HI$ disk where interstellar gas dominates the main baryonic contribution, and not the stars.  Here also, we found that when only the gas disk is taken into account, it allows finite amplification, indicative of having small-scale spiral arms, but the inclusion of dark matter prevents small-scale spiral features almost completely ({\bf Chapter~3}). Thus we showed that the dark matter
 halo plays a pivotal role in the suppression of small-scale spiral features in parts of disks where it dominates, in fair agreement with the observations.

\medskip

In the second part of the thesis, we studied how the interstellar gas could alter the longevity of the `grand-design' (large-scale) spiral structure by treating the large-scale spiral arms as density waves in a galactic disk. In the past, it was shown for a pure stellar (collisionless) disk that such a wavepacket of density wave would travel with a group velocity that is sufficient to destroy the wavepacket and hence the large-scale spiral structure, on a time-scale of the order of $10^9$ years. Since in the past literatures, it was also shown that the interstellar gas plays important role in several dynamical contexts, therefore we investigated how the interstellar gas would affect the longevity of large-scale spiral structure in a gravitationally-coupled two-component (stars plus gas) system.

We showed that the group velocity
 of a wavepacket in Milky Way-like disk galaxies decreases steadily (by a factor of few) with the inclusion of gas, implying 
that  the spiral pattern will survive for a longer time-scale ({\bf Chapter~4}).
 Also using the observed rotation curve and measured pattern speed of spiral arm for three late-type, gas-rich galaxies, namely, NGC~6946, NGC~2997 and M~51, we showed that
 the interstellar gas is necessary in order to have a stable density wave corresponding to the observed pattern speed values. The same trend is also shown to be true for our Galaxy ({\bf Chapter~5}).

\medskip

In the last part of the thesis, we studied the dynamical effect of a dominant dark matter halo on the global spiral modes (used as a `proxy' for large-scale spiral arms) in the disks of LSB galaxies. The result we found earlier that in the disks of LSB galaxies, the small-scale spiral arms are prevented almost completely by the dominant dark matter halo naturally provided the motivation to study the dynamical effect of dark matter halo on the large-scale spiral structure also. Observationally it is also known that typical LSB galaxies (like UGC~7321) do not display prominent,well-defined large-scale spiral arms like those seen in case of normal, high surface brightness (HSB) galaxies such as our Milky Way. This very observational fact also provided the impetus to study the effect of dark matter halo on the large-scale spiral structure in the disks of LSB galaxies.

We investigated the influence of dominant dark matter halo on global spiral modes  in the disks of LSB galaxies by treating the galactic disk first as a fluid ({\bf Chapter~6}) and then as a collisionless system ({\bf Chapter~7}). The global modes in galaxies identified via a novel quantization rule. We used different input parameters of UGC~7321 which are obtained either observationally or by modeling.

 When the stars are treated as fluid system we found that the dark matter fails to prevent the global spiral modes. However, even if the global modes are present in our theoretical models, we argued that the lack of tidal interaction will not give rise to these modes in a realistic case ({\bf Chapter~6}). 

However, when the disk is modeled as more realistic collisionless system, we found that the large-scale spiral structure (global spiral modes) in LSBs is damped by the dominant DM halo, in fair agreement with the observations ({\bf Chapter~7}).

\section{Future works}

We have so far summarized the brief motivations and important results of the thesis. There are some more aspects related to the origin and persistence of spiral structure seen in different types of disk galaxies that can be studied further. Below I mention some of these issues.

\begin{itemize}
\item{{\bf Effect of dark matter halo on spiral structure in Giant LSBs :}  For typical LSBs, the dominant dark matter halo prevents both the small-scale and the large-scale spiral structure almost completely. There is also a class of galaxies called Giant LSB (GLSB) galaxies (e.g. Malin~1) and some of which show dynamical properties similar to normal, HSB galaxies in the inner parts while in the outer parts they exhibit properties similar to LSBs. Past observations revealed that some these GLSBs display non-axisymmetric features such as bars, and spiral structure in their disks (like the cases for Malin~1, UGC~6614). Therefore, it is worth investigating what give rise to the non-axisymmetric features in the disks of GLSBs and also study quantitatively the possible role of dark matter halo on spiral structure of different scales.
}
\end{itemize}

\begin{itemize}
\item{{\bf Dynamical effect of interstellar gas on pitch angle of the spiral arms : } The mass fraction contained in the interstellar gas increases along the Hubble sequence with more late-type galaxies (Sc-Scd type) having larger mass fraction contained in the interstellar gas -- it is well-known trend. Also it is observationally found that the pitch angle of spiral arm increases (at least the mean value for a sub-class) and the arms are more open as one moves along the Hubble sequence from Sa-type to Scd-type.

 In the past, several studies have shown the importance of gas in different contexts of galactic dynamics. However, existence of any possible correlation between these two above-mentioned trends has not been explored. Therefore, it is worth investigating whether increasing gas-fraction in disk galaxies could influence the observed trend of pitch angle of spiral arms with Hubble type in the disk galaxies.
}
\end{itemize}

\begin{itemize}
\item{{\bf Longevity of spiral arms -- interplay of bulge and interstellar gas : } The longevity of the spiral arms in disk galaxies is long-debated issue. Several numerical simulations have shown that the spiral arms which appear in $N$-body models of disk galaxies only persist for few dynamical time-scales. However, a recent study by Saha \& Elmegreen (2016) reported long-lived spiral arms in $N$-body study of models for disk galaxies where it was shown that the bulge plays a crucial role to maintain a long-lasting spiral pattern in the disk of galaxies. On the other hand, in this thesis we showed that the inclusion of interstellar gas helps the spiral patterns to persist for longer time-scale (few billion years). Therefore, it would be interesting to investigate by means of both semi-analytic models and $N$-body simulations the dynamical effect of both the bulge and the interstellar gas in context of the longevity of spiral arms in the disk galaxies.
}
\end{itemize}

\cleardoublepage






\end{document}